# Electrogravity: On a scalar field of time and electrogmagnetism

Eytan H. Suchard, Metivity Ltd, email: eytansuchard@gmail.com

## Abstract

It is possible to describe a universal scalar field of time but not a universal coordinate of time and to attribute its non-geodesic alignment to the electromagnetic phenomena. A very surprising outcome is that not only mass generates gravity, but also electric charge does. Charge is, however, coupled to a non-geodesic vector field and thus is not totally equivalent to inertial mass. Only the entire "Energy-Momentum4" tensor has a vanishing divergence. The model can be seen as misalignment of physically accessible events in an observer spacetime and of gravity as a controlling response by volumetric contraction of the observer spacetime in the direction where events bend or accelerate to. This non geodesic acceleration is described by a generalization of the Reeb class vector. Misalignment of events can be described by 1, 2, and 3 such vectors. The paper presents a term with 4 vectors but does not discuss its physical meaning. The paper also discusses particle mass ratios and the Fine Structure Constant where added or subtracted area in relation to a disk does not involve a ratio $\frac{1}{24}$ but $\frac{1}{96}$ due to the physical meaning of the orientation of a space foliation which is perpendicular to a time-like vector $\alpha$ and due to the orientation of a plane which is perpendicular to a time-like vector $\alpha$ and its Reeb class vector $\eta$ where $\alpha$ is mapped to a 1-Form, $d\alpha = \pm\eta^\wedge\alpha$. This forgotten definition of the Reeb class vector $\eta$ is not limited to contact manifolds. These two orientations mean that only one side of a 3-dimensional foliation has a physical meaning and only one side of a sub-plane of that foliation has a physical meaning then $\frac{1}{2}\frac{1}{2}\frac{1}{24} = \frac{1}{96}$. Another interpretation of the factor $\frac{1}{4}$ is the Bekenstein - Hawking entropy to area constant. An additional coefficient $\frac{4}{\pi}$ describes an acceleration field strength and has a compelling source in mainstream physics. Other two field strength coefficients have compelling explanations, these are $\frac{95}{96}$ and a critical value due to an imbalance equation between gravity and anti-gravity ~1.556198537190348396563877031439915299.

## 1. Introduction – measurement of non - geodesic deviation

### Notations

In this paper the Einstein summation convention is used. Derivative is denoted by comma and covariant derivative by semicolon. E.g, $V_{\mu},_{\nu} \equiv \frac{dV_{\mu}}{dx^{\nu}}$ with the coordinates $x^{\nu}$. A covariant derivative example is $V_{\mu};_{\nu} \equiv \frac{dV_{\mu}}{dx^{\nu}} - \Gamma^{\lambda}_{\mu\nu} V_{\lambda}$ where $\Gamma^{\lambda}_{\mu\nu}$ are the Christoffel symbols.

The Lie bracket is used either to define the usual Lie derivative or to define a commutator of indices, $V_{[\mu} U_{\nu]} - V_{[\nu} U_{\mu]}$. Where the term Reeb Class vector is used, it does not mean the usual Reeb vector which is known in Contact forms. This vector will be explained as an acceleration vector in 4 dimensions.

### The Main idea

By making the coordinates also fields, through allowing non-geodesic accelerations to be part of the Lagrangian of the theory, the presented theory, which is named by the author and by his colleague, Dr. Sam Vaknin, "Geometric Chronon Field Theory" or "Chronon Field Theory" becomes self-contained just as causal-set based theories are, in which the variational principle is defined on the causal set itself.

It means the entire structure of the spacetime manifold, including its dimension and the arrow of time, is the manifold itself and nothing else. The dimensions and the arrow of time through theorems 0 and 7 that are presented in this paper are a consequence of the minimal dimension which is required for the arrow of time through non-degeneracy of the acceleration fields that will be presented. The fields and the geometry are all one self-contained geometric object just as a causal set is.

This paper tries to implement the following thought methodology as the author first encountered in high school when taught physics by an Austrian physics teacher Victor Sela (Stein or Steinhart):

1) Think big.
2) Keep it simple.
3) Search for the patterns.

**Think big:** In 2004, thinking big was several paths that could be taken. First was Causal Sets and Relational Observables, which mean points of space-time and their global and local interrelations can derive spacetime. This first approach looked promising to the author, however, there were problems with this approach. A) Time may not be fully relational because in FRWL geometry, a universal time can be defined although it is not traceable, i.e. the same proper time from an initial Cauchy surface or the Big Bang as a limit, can be measured along different curves, to the same event. Universal time is, therefore, a scalar function, not a global coordinate of time but still not a fully relational observable. B) Time seems to be fundamental because even formal symbolic language, including “physical laws”, requires a sequential order. So, the sequential order must be fundamental and not injected into the structure of points of a causal set.

These two points (A) and (B) meant that the action principle must define spacetime in its entirety including matter and forcefields and best, without resorting to causal sets and external structures from which spacetime is derived. Fields within spacetime, therefore, need to be also local coordinates, which means they cannot be always geodesic, otherwise, the simplest solution would be local geodesic coordinates and spacetime without matter and curvature. This idea means that non-geodesic accelerations must describe matter and forces and therefore must entirely replace gauge fields unless the gauge field is a symmetry of the coordinates themselves.

**Keep it simple:** When keeping the theory simple, it must not be too simple. It must be sufficiently rich to describe all forces of Nature. Although a lot of work still needs to be done, the idea of an action which is based on non-geodesic accelerations is well understood.

**Search for the patterns:** Perhaps this is the most difficult principle to implement because it is not clear whether geometry is fundamental or algebra is fundamental. If measure such as length, area, volume and hyper-volume in higher dimensions, is fundamental, then algebraic relations alone between physical objects are insufficient or not fundamental. For example, symmetry groups are the result of geometry and not the other way around. For this reason, the author chose a starting point which is not the Dirac equation and spinors but rather, the idea of acceleration. There is more to choosing acceleration. It is mainly the following crucial property of 4 dimensions. The following is the most important pattern as a guiding principle.

**Crucial:** Only in 4 dimensions there is an infinite number of inequivalent smoothness structures [1]. A smoothness structure is a collection of smoothly-equivalent smooth atlases. In 1, 2, 3, and 5 dimensions there is only one smoothness structure. In all other dimensions, which are greater than 4, there is a finite number of inequivalent smoothness structures. An atlas on a manifold offers open sets and coordinate transitions between the open sets such that these open sets cover the entire manifold. There are local maps $\varphi(i)$ between open sets $W(i)$ that cover a manifold M and $\mathbb{R}^4$, $\varphi(i)W(i) \to \mathbb{R}^4$ such that if $W(i) \cap W(k) \neq \emptyset$, $\varphi^{-1}(i)\varphi(k)(W(i) \cap W(k))$ is smooth and the inverse of $\varphi^{-1}(i)\varphi(k)$ is also smooth. The failure of 4 dimensional manifolds to have a finite set of smoothness structures is a very strong motivation to represent the coordinates themselves as fields in the theory’s Lagrangian and to include them as fields, of course, without violating the principle of general relativity. Such Lagrangians can be achieved without resorting to Hilbert spaces and to self-adjoint operators if they use non-geodesic accelerations.

**Back to the main idea:** The main idea is to use actions which are based on accelerations of a unit vector. Suppose we have a scalar field $p$ then its derivative by the coordinates $x^{\nu}$ is denoted by $p_{\mu}$ and the unit vector if it exists is $\frac{p_{\mu}}{\sqrt{|p^{\lambda}p_{\lambda}|}}$ or in the complex formalism $\frac{p_{\mu}}{\sqrt{\left|\frac{1}{2}(p^{*\lambda}p_{\lambda}+p^{\lambda}p^{*}{}_{\lambda})\right|}}$. In

most of the paper $Z \equiv p^{\lambda}p_{\lambda}$ and $Z \equiv \frac{1}{2}(p^{*\lambda}p_{\lambda}+p^{\lambda}p^{*}{}_{\lambda})$ is typically used and therefore the unit vectors are written as $\frac{p_{\mu}}{\sqrt{|Z|}}$ or as $\frac{P_{\mu}}{\sqrt{|Z|}}$ with a capital letter. Another goal is to be able to reconstruct spacetime from points such as in causal set theories but without violating the principle of locality which is part of the covariance principle. This goal will be presented in "Cartan formalism of the Ricci action with (p_μ (a))/√(Z(a)) Fields", "Appendix C: Generalization to more than one generalized Reeb class vector", "Why spacetime with an arrow of time must have 4 dimensions?" and "Uniform gravity and forces formalism, challenges, and an open problem". The points will be locally, and in the case where $P(0)P^{*}(0)$ is not global, only locally, of 5 complex values $P(0)P^{*}(0),\ P(4)P^{*}(4),\ P(5)P^{*}(5), P(6)P^{*}(6)$ and $P(1)P^{*}(1)$ which with the previous 4 values derive the arrow of time. There are additional optional two functions $P(2), P(3)$ which if added to the 5-tuples, define a system of points which is not parsimonious.

If the absolute value is omitted, it is assumed that in a correct metric convention (+,-,-,-) or (-,+,+,+), $\sqrt{Z}$ is well defined. The paper assumed $p^{2}$ or in the complex formalism $pp^{*}$ to have a meaning of time, i.e. $p_{\mu}$ is time-like as will be explained. Acceleration vectors will always be perpendicular to the vector $p_{\mu}$. The theory will be developed for 5 such acceleration fields but will focus on one such acceleration. $\frac{p_{\mu}}{\sqrt{|Z|}}$ and its acceleration which will be denoted $\frac{U_{\mu}}{2}$ or $\frac{u_{\mu}}{2}$ can at most span one plane in spacetime. Extending theory to more than one such acceleration can be either for the plane perpendicular to $\frac{p_{\mu}}{\sqrt{|Z|}}$ and $\frac{u_{\mu}}{2}$ or for the three-dimensional foliation which is perpendicular only to $\frac{p_{\mu}}{\sqrt{|Z|}}$. The generalization to more than one acceleration vector will be covered in "Appendix C: Generalization to more than one generalized Reeb class vector".

**Adhering to the methodology of Physics**

Since the vector $\frac{U_{\mu}}{2}$ that will be discussed is a space-like vector, in the non-covariant classical limit it is an excellent candidate to describe the electric field. In the complex case, this will be ≈ $E \approx \frac{Constant}{2}(\frac{U_{\mu}}{2}+\frac{U^{*}{}_{\mu}}{2})$ where $\mu = 1,2,3$ and $E$ is the classical non-covariant electric field. Therefore, charge must be proportional to the divergence $-U_{\mu};^{\mu}$ and $-U_{\mu}U^{\mu}$ must be proportional to the energy density of the electric field in (+,-,-,-) metric convention or $+U_{\mu};^{\mu}$ and $+U_{\mu}U^{\mu}$ in the (-,+,+,+) metric convention. When speaking on force fields, it is essential to define what measurement is. Measurement in this theory is an accessible event. Suppose the theory is based on 6 complex scalar fields, one such scalar field is $P$ and we want $PP^{*}$ to describe proper time if $PP^{*}$ is measured as proper time along some curves, not necessarily unique, so $PP^{*}$ is not necessarily a coordinate of time as it lacks a specific direction and can be measured along different curves, from a sub-manifold of spacetime to the same event where $PP^{*}$ is measured. We also want to use other scalar functions $P(4), P(5)$ and $P(6)$. Then a nice goal of this theory should be that either $ff^{*} = PP^{*}P(4)P^{*}(4)P(5)P^{*}(5)P(6)P^{*}(6)$ will be a probability density or $ff^* = P(4)P^*(4)P(5)P^*(5)P(6)P^*(6)$ will be a probability density and $PP^*$ will describe time. In both cases, $ff^*$ describes the probability of a reachable event in spacetime. The latter option where $PP^*$is time, can be realized in a causal spacetime as a Geroch function [2] and it means that time $PP^*$ is an increasing function and not only a result of measurement while the other functions describe which events are reachable. Notice the omission of $P(1)$. This omission is because, as we shall see $P(1)$ is dependent on $P$. $P(2), P(3)$ are mentioned in this paper in (64.a) but are not necessary due to Occam's razor. (64.b) is preferrable.

**Introduction**

The Result of the Geroch Splitting Theorem [2] is that a field of time can be defined. In simple geometries such as FRWL, which are Big Bang geometries, such time also has an intuitive meaning; it is a scalar field and not a coordinate of time. It is the maximal time between each event of space-time and the Big Bang as a limit, measured by a physical clock that may experience forces. Such proper time can be measured along different curves and is therefore not traceable, not geodesic under forces and cannot be a coordinate that also requires a 4-direction. The existence of a non – traceable time is not a new idea and was postulated by the philosopher R. Joseph Albo [3] in the 14th century. The approach that will be presented to make peace between General Relativity and Quantum Mechanics is not to describe Space-Time as emergent out of huge matrices and to preserve the particles approach [4], but to replace particles with events. In non-hyperbolic spacetime, a scalar field can still be defined as universal clock but will no longer be an upper limit of measurable time to an event from a Cauchy surface as an interpretation to [2].

What information can a scalar field encode, that is not already predicted by the metric tensor of space time $g_{\mu\nu}$ ? The answer is non - geodesic motion. The motion equations of the theory of General Relativity predict only geodesic motion. This theory is based on two assumptions,

1) The basic assumption is that matter can be described via acceleration in the gradients of scalar fields, more specifically, the electromagnetic phenomena can be described by a non-zero weak acceleration of the gradient of a Geroch function [2], $P^2$ in hyperbolic space-time or PP* if P is complex. The motivation for this is the vanishing of the rotor of such an acceleration when reduced to 3 dimensional foliations like the electric field. This will be shown in theorem 3. This acceleration is known as a Reeb class vector field [5] in odd dimensions but can also be defined in 4 dimensions via a 1-Form $\alpha$, $d\alpha = \eta^\wedge\alpha$ where $\eta$ is the Reeb class vector.
   Important: In odd dimensions, the Reeb field can be defined in a way that it is not the acceleration of a unit vector field [6]. In two dimensions, the generalized Reeb class vector is not geodesic. That is an important difference that has been missed all these years. $\mp\eta$ from $d\alpha = \eta^\wedge\alpha$ is the forgotten definition of a Reeb class vector which is used in the definition of the Reeb class [7] and which is not limited to contact manifolds but is also defined on Symplectic manifolds. Reeb class vectors and Reeb vectors are not the same objects.

**Important:** Another problem with most papers on Reeb class vectors is that they ignore divergence points.
Actions are defined for 1 Reeb class field, "electromagnetic", 2 Reeb class fields "electro-weak", 3 Reeb class fields, "Strong" and 4 Reeb class fields as a “Fifth Force” or massive gravity. A definition can be made also for 4 Reeb class fields but its physical meaning is not discussed in this paper. See appendix C, (65). The motivation to use Reeb class vector fields, including a complex formalism, can be seen in the paper by Yaakov Friedman [8]. To complete assumption 1, energy density is $\frac{a_\mu a^\mu}{8\pi K}$ where $K$ is Newton’s constant of gravity and $a^\mu$ describes an acceleration of a normalized vector $X = c\frac{p_\mu}{\sqrt{P_\lambda P^\lambda}}$ where $p_\mu = \frac{dp}{dx^\mu}$ where $p$ is a scalar field, $x^\mu$ are the coordinates of the spacetime manifold and c is the speed of light. In simple words, what is claimed in this paper as one possible interpretation of the theory is that starting from the field X, which is derived from a Geroch function, there exists a physical test clock type which moves along X and which will continue to move along X also when X is not geodesic. Another more mathematical interpretation is that an acceleration of the field X does not mean a physical acceleration of a material clock but that such an acceleration is merely a mathematical object, more precisely, a field. The field interpretation is consistent with the complex formalism of this theory where $p$ can be a complex scalar field. That is to say that $a_\mu$ is a field which prohibits X from being geodesic. The paper will show a way to define such a field regardless of the direction of motion of the test clock in the field. $X$ and $a_\mu$ span only one two-dimensional hyperplane of spacetime. The field must be defined in 4 dimensions. If such a field is the reason for the energy of the electric field, then the components of $a_\mu$ must be very small, otherwise the simplest interpretation of such acceleration as of neutral particles in a strong electromagnetic field would be easily noticeable, however, the simplest interpretation may not be correct simply because of a complex formalism of the theory.
The strongest claim against the non-geodesic acceleration being more than a field will come later when it will be argued that such an acceleration would be towards negative charge and outwards from a positive charge with a new type of charge-based gravity pseudo acceleration half that size and in the opposite direction. It would mean by the conventional theory of virtual photons, that negative charge emits virtual photons with negative momentum. If that was the case, short lived photons with negative mass would also be possible and the existence of such photons is not supported by empirical evidence.

2) The scalar fields quantization is $P = \sum_{k=1}^{\infty} P(k)$ such that $\int_\Omega \frac{P(k)P^*(j)+P(j)P^*(k)}{2}\sqrt{-g}d\Omega = 0$ if $k \neq j$ and $\int_\Omega \frac{1}{2}(P(k)P^*(j) + P(j)P^*(k))\sqrt{-g}d\Omega = 1$ if $k = j$ where $\sqrt{-g}$ is the volume element of space-time, where $g$ is the determinant of the metric tensor. In other words, instead of a Geroch function, $PP^*$ can be replaced by a scalar $PP^*$ that integrates to 1 on reference spacetime manifold and the Lagrangians of the theory will be defined almost-everywhere in terms of measure theory.

**Note:** The mathematical foundation of this paper is the Geroch function [2], [3], Reeb class vector fields [5], not the usual Reeb vector, for encoding trajectory curvature, symplectic

geometry directly on spacetime and not on any phase space due to [8], and the idea of physically accessible events in an embedding spacetime, an idea very similar to Hartland Snyder's quantized spacetime [9] but without any assumed non-commutative relation. The Lagrangians of this paper are based only on acceleration vectors of normalized gradients of scalar fields.

**Challenges to the reader:** The challenges to the reader are to understand Reeb class vectors in their original formalism with the meaning of non-geodesic acceleration, which unlike the usual Reeb vector, is not limited to contact manifolds but describes how much a gradient of a scalar field is not geodesic, to understand how two scalar fields and two Reeb class vectors describe a Scarr – Friedman acceleration matrix [11] as a field and not as a uniform acceleration as originally proposed in their paper, with the differences from the Scarr-Friedman [11] matrix which are +1,-1 handedness of a second acceleration plane, the ability to describe spin through y, z rotation when the acceleration is in the t, x axes plane and the ability to describe zero charge when one divergence of the acceleration in the complex plane is positive and the other is negative so that adding conjugates of divergences nullifies. These are not trivial properties and they do not exist in the Scarr Friedman matrix [11]. Another challenge is to understand how such an acceleration matrix can serve as a Symplectic form that acts directly on spacetime and not on any phase space as is the usual case in mainstream physics. Another challenge, which is somewhat a quantum leap, is to understand the use of non-geodesic geometry of foliations of spacetime and its meaning as matter. The Scarr-Friedman formalism [11] will be discussed shortly in this paper and is essential to the understanding of this paper. This paper does not, however, take the path of Tzvi-Scarr and Yaakov Friedman [11] because the acceleration matrix which is used in this paper has different properties and a different goal. The paper purports to reach a description of a field that rotates the gradients of scalar fields in order to be able to describe spin for example. Most theories in mainstream physics deal with geodesic curves and not with accelerated curves, unlike this paper which speaks of both. Another challenge is to accept that lack of collaboration in solving the field equations of this paper (4), (64.b) vs (64.15) which is not considered as physical but mathematically interesting, (64.01) require educated guess of field strength coefficients for Leptons. It is responsible to say that $\frac{95}{96}$ for the electron is better understood than before and that the Tau field strength coefficient is better understood too though more research and collaboration would greatly benefit the paper. The muon field strength coefficient $\frac{4}{\pi}$ is, however, from a critical field value of Quantum Mechanics and not directly from the presented theory.

We can describe non geodesic integral curves along a field $P_\mu \equiv \frac{dP}{dx^\mu}$ for the coordinates $x^\mu$, also, $P_\mu$ need not be time-like in all events of space-time. We now define the square norm for real numbers as $Z \equiv |P_\lambda P^\lambda|$ and its gradient $Z_\mu \equiv \frac{dZ}{dx^\mu}$. We define a geometric object $\frac{U_\mu}{2}$ that will measure how much the field $P_\mu$ is not geodesic.

When $c\tau$ describes the evolution of the vector $X = c\frac{p_\mu}{\sqrt{P_\lambda P^\lambda}}$ along the integral curves which are formed by the field $X$, $\frac{dX}{d\tau}$ must be perpendicular to $X$ because $X_\mu X^\mu = c^2$ and then $\frac{d(X_\mu X^\mu)}{cd\tau} = \frac{1}{c}(\dot{X}_\mu X^\mu + X_\mu \dot{X}^\mu) = 0$ which implies $X^\mu \dot{X}_\mu = 0$ since $d\tau$ is a scalar. Now writing $Z = P_\lambda P^\lambda$ we have,

$$\frac{d}{d\tau}\frac{p_\mu}{\sqrt{P_\lambda P^\lambda}} = \frac{d}{d\tau}\frac{p_\mu}{\sqrt{Z}} = \frac{\dot{p}_\mu}{\sqrt{Z}} - \frac{p_\mu \dot{Z}}{2z^{\frac{3}{2}}} = \frac{P_{\mu;\nu}}{\sqrt{Z}}\frac{dx^\nu}{d\tau} - \frac{p_\mu Z_{;\nu}}{2z^{\frac{3}{2}}}\frac{dx^\nu}{d\tau} = \frac{P_{\mu;\nu}}{\sqrt{Z}}\frac{p^\nu}{\sqrt{Z}} - \frac{p_\mu Z_{;\nu}}{2z^{\frac{3}{2}}}\frac{p^\nu}{\sqrt{Z}} = \tag{1}$$

$$= \frac{P_{\mu;\nu}\, p^\nu}{Z} - \frac{p_\mu Z_{;\nu}\, p^\nu}{2z^2} = \frac{P_{\nu;\mu}\, p^\nu}{Z} - \frac{p_\mu Z_{;\nu}\, p^\nu}{2z^2} = \frac{Z_\mu}{2Z} - \frac{Z_\nu p^\nu p_\mu}{2z^2}$$

**Important:** (1) is a direct mathematical interpretation of the acceleration in assumption 1 of this theory. Writing $\frac{U_\mu}{2} \equiv \frac{Z_\mu}{2Z} - \frac{Z_\nu p^\nu p_\mu}{2z^2}$ when $p_\mu$ is restricted to be over the real field, means that the action of the field is dictated to be $-\frac{U_\mu U^\mu}{4}\sqrt{-g}$ in (+,-,-,-) metric convention, g is the determinant of the metric tensor. The minus sign is because $U_\mu$ is a space-like vector which is easily shown to satisfy $U_\mu P^\mu = 0$. It is also easy to see that $A_{\mu\nu} = \frac{U_\mu}{2}\frac{P_\nu}{\sqrt{Z}} - \frac{U_\nu}{2}\frac{P_\mu}{\sqrt{Z}}$ satisfies $A_{\mu\nu}\frac{P^\nu}{\sqrt{Z}} = \frac{U_\mu}{2}$ and that $A_{\mu\nu}$ can be represented as a matrix,

$$A = \begin{pmatrix} 0 & -a \\ a & 0 \end{pmatrix} \Rightarrow Det(A) = -a^2 = \frac{U^\mu U_\mu}{4} \tag{1.1.1}$$

in the real case with two basis vectors $\frac{P^\nu}{\sqrt{Z}}$ and $\frac{U^\nu}{\sqrt{\|U^\lambda U_\lambda\|}}$, where $\|U^\lambda U_\lambda\| \neq 0$.

**Important:** Understanding (1) is all which is needed to understand this paper. There could be one or more such fields as $\frac{U_\mu}{2} = \frac{Z_\mu}{2Z} - \frac{Z_\nu p^\nu p_\mu}{2z^2}$ and there is also a complex numbers formalism of $\frac{U_\mu}{2}$ however, all the Lagrangians in this paper are based on one or more such fields. (1) is consistent with the Reeb class vector, not with the ordinary Reeb vector and it means acceleration of a unit vector in Minkowski spacetime, while $P^2$ or $PP^*$ in the complex case is a Geroch function [2]. When describing space as a foliation of spacetime, except for a Geroch function, there are also 3 gauge fields that describe the foliation and one additional gauge field due to the fact that acceleration can be described in two perpendicular planes. These are not gauge fields in the sense that they impose interaction between particles that otherwise would not interact as in the Standard Model. They "gauge" fields in terms of symmetry. The field interacts with itself.

Defining: $U_\mu \equiv \frac{Z_\mu}{Z} - \frac{Z_k P^k}{Z^2}P_\mu$ consider,

$$\frac{d}{dx^\nu}\frac{P_\mu}{\sqrt{Z}} - \frac{d}{dx^\mu}\frac{P_\nu}{\sqrt{Z}} = \tag{1.1.2}$$

$$\frac{P_{\mu,\nu}}{\sqrt{Z}} - \frac{P_\mu Z_\nu}{2Z^{\frac{3}{2}}} - \frac{P_{\nu,\mu}}{\sqrt{Z}} + \frac{P_\nu Z_\mu}{2Z^{\frac{3}{2}}} =$$

$$\frac{P_\nu Z_\mu}{2Z^{\frac{3}{2}}} - \frac{P_\mu Z_\nu}{2Z^{\frac{3}{2}}} =$$

$$\frac{1}{2}\left(\frac{Z_\mu}{Z}\frac{P_\nu}{\sqrt{Z}} - \frac{Z_k P^k}{Z^2}P_\mu\frac{P_\nu}{\sqrt{Z}}\right) - \frac{1}{2}\left(\frac{Z_\nu}{Z}\frac{P_\mu}{\sqrt{Z}} - \frac{Z_k P^k}{Z^2}P_\nu\frac{P_\mu}{\sqrt{Z}}\right) = \frac{U_\mu}{2}\frac{P_\nu}{\sqrt{Z}} - \frac{U_\nu}{2}\frac{P_\mu}{\sqrt{Z}}$$

But why to use, $\frac{1}{2}U_\mu = \frac{1}{2}\left(\frac{Z_\mu}{Z} - \frac{Z_k P^k P_\mu}{Z^2}\right)$ and not simply, $\frac{Z_\mu}{Z}$ ? The reason is that $\frac{U_\mu}{2}\frac{P^\mu}{\sqrt{Z}} = 0$.

It is easy to show that $\frac{U_\mu}{2}$ behaves as the acceleration of the unit vector $\frac{P_\mu}{\sqrt{Z}}$. See Appendix D for another way to derive the Reeb class vector. In terms of a 4-acceleration $a_\mu$, it is easy to see:

$$\frac{U_\mu}{2} = \frac{dc^{-1}X^\mu}{cd\tau} = \frac{a_\mu}{c^2} \tag{2}$$

Where $c$ is the speed of light, $\tau$ is proper time. $\frac{U_\mu}{2}$ is the generalization of a Reeb class vector [5] to 4 dimensions. Can this $a_\mu$ have a simple physical meaning of accelerating any neutral mass? There is an experimental way to find out, once we analyze the electric field in the coming sections. More rational is to see the acceleration as a mathematical object, as a field.

Defining $A_{\mu\nu} \equiv \frac{U_\mu}{2}\frac{P_\nu}{\sqrt{Z}} - \frac{U_\nu}{2}\frac{P_\mu}{\sqrt{Z}}$ we get $A_{\mu\nu}\frac{p^\nu}{\sqrt{Z}} = \frac{U_\mu}{2}$ which means that $A_{\mu\nu}$ is a rotation and scaling matrix, however, as a linear operator it acts only on one of two hyper-planes of spacetime.

**The non-covariant classical limit**

From the first assumption 1, and (2) the implication regarding the classical non-covariant limit of the energy of the electric field is as follows:

$Action\ density = -\frac{U_\mu U^\mu}{4}\sqrt{-g}$ in (+,-,-,-) convention and by (2) and the matter action of General Relativity ,

$$Energy\ density = -\frac{c^4}{8\pi K}\frac{U_\mu U^\mu}{4} = -\frac{c^4}{8\pi K}\frac{a_\mu a^\mu}{c^4} = -\frac{a_\mu a^\mu}{8\pi K} \tag{2.0.1}$$

Where $K$ is the gravitational constant. The explanation for $8\pi K$ will follow.

Comparing (2.0.1) to the classical limit of the energy density of the non-covariant electric field, $\frac{1}{2}\varepsilon_0 E^2$ where $E^2$ is the classical non-covariant square norm of the electric field $E$ and $\varepsilon_0$ is the permittivity of vacuum, we have for charge density $\rho$,

$$-\frac{a_\mu a^\mu}{8\pi K} \approx \frac{1}{2}\varepsilon_0 E^2 \Longrightarrow \|a\| = \sqrt{-a_\mu a^\mu} \approx \|E\|\sqrt{4\pi K\varepsilon_0} \tag{2.0.2}$$

$$\frac{1}{2}U^k{}_{;k} = \frac{a^k{}_{;k}}{c^2} \approx \sqrt{4\pi K\varepsilon_0}\, E^s{}_{,s} = \sqrt{4\pi K\varepsilon_0}\,\frac{\rho}{\varepsilon_0 c^2} = \sqrt{\frac{4\pi K}{\varepsilon_0}}\frac{\rho}{c^2} \tag{2.0.3}$$

**Crucial:** When the filed is dynamic then $(0, E) \approx \frac{c^2}{\sqrt{4\pi K\varepsilon_0}}\frac{1}{4}(U_\mu + U_\mu^*)$ and $E^2 \neq -\frac{c^4}{4\pi K\varepsilon_0}\frac{1}{8}(U_\mu U^{*\mu} + U_\mu^* U^\mu)$ because the imaginary part cannot be neglected!.

The calculation (2.0.3) will be repeated in the description of surprising charge-based gravity and anti-gravity.

**Caveat:** Notice that the divergence of the classical non-covariant electric field $E^s{}_{;s}$ sums only on indices 1,2,3 and does not include the time index 0.

**Caveat:** The action $-\frac{U_\mu U^\mu}{4}\sqrt{-g}$ may need an additional constant to be related to the action we know from general relativity although there are some considerations that suggest that (2.0.2), (2.0.3) should not be modified.

In the classical non-covariant limit, being at rest in a gravitational field means to be accelerated against the gravitational field. By (2.0.2) we can write in non-covariant formalism $\frac{-a_\mu a^\mu}{8\pi K} = \frac{a^2}{8\pi K}$ where in this case $a = \sqrt{-a_\mu a^\mu}$. It is logical that the energy of the acceleration field that resists gravity will approximate the energy of the non-covariant gravitational field. Starting with radius r this energy is

$$-\frac{KM^2}{2r} = \frac{1}{8\pi K}\int \frac{K^2 M^2}{r^4} 4\pi r^2 dr = \int \frac{g^2}{8\pi K} dVolume \approx \int \frac{-a^2}{8\pi K} dVolume \tag{2.0.4}$$

Where here $g$ is the classical non-covariant gravitational acceleration. So, the induced acceleration of keeping all test masses from falling from radius r to infinity becomes equal to minus the classical non-covariant density of the gravitational field. This is a qualitative explanation to the constant $8\pi K$ where gravity and a field of not geodesic acceleration opposite to gravity are in balance in the classical non-covariant limit.

**Caveat:** The acceleration of a unit vector field need not be realizable as a physical acceleration of every type of a material clock or of material clocks in general. It is a mathematical object.

**Hodge star extension:** To extend $A_{\mu\nu}$ from a rotation and scaling matrix (1.1.1) to the entire tangent bundle $T(M)$ of the spacetime manifold $M$ there is a need to use a contraction of $A_{\mu\nu}$ with an antisymmetric tensor. The reason for the difference between the Scarr-Friedman acceleration matrix [11] and a field of acceleration can be viewed in the light of rotations in spacetime when both indices of the acceleration matrix are either lower or upper.

The tangent space at the identity of a Lie group is a Lie Algebra and it follows from a differentiation of the Lie Group left action at the identity. Consider that $A_{\mu\nu}$ is extended to a second plane in order for $A_{\mu\nu}$ to become a regular matrix so now $A_{\mu\nu} = A_{\mu\nu}(1) + A_{\mu\nu}(2)$ and in local base $A(1) = \begin{pmatrix} 0 & -a & 0 & 0 \\ a & 0 & 0 & 0 \\ 0 & 0 & 0 & 0 \\ 0 & 0 & 0 & 0 \end{pmatrix}$ and $A(2) = \begin{pmatrix} 0 & 0 & 0 & 0 \\ 0 & 0 & 0 & 0 \\ 0 & 0 & 0 & -a \\ 0 & 0 & a & 0 \end{pmatrix}$ for some field $a$.

For regular orthogonal matrices we have $A(\tau)A^{trasposed}(\tau) = I$ where $I$ is the identity matrix and with orbits crossing $\tau = 0$, differentiating at Tau = 0 and remembering that an exponent of a transposed matrix is the transposed of the exponent, we get from $A(\tau) = e^{\tau A'}$, and $A^{transposed}(\tau) = e^{\tau A'^{trasposed}}$.

$$(A(\tau)A^{trasposed}(\tau))' = (I)' = 0 \tag{2.1}$$

$$\left(A(\tau)A^{trasposed}(\tau)\right)' =$$

$$A'e^{\tau A'}A^{trasposed}(\tau) + A(\tau)A'^{transposed}e^{\tau A'^{transposed}}$$

Where $A'$ is a Lie Algebra matrix. Setting

$$A^{trasposed}(\tau = 0) = e^{0A'^{transposed}} = A(\tau = 0) = e^{0A'} = I \tag{2.2}$$

$$A'II + IA'^{transposed}I = A' + A'^{transposed} = 0$$

Which means that the Lie Algebra of orthogonal matrices is antisymmetric matrices.

**Cartan subalgebras**

There are 6 ways to split the tangent space of spacetime into 2 rotation and acceleration planes.

Without loss of generality, consider for some real numbers a, b:

$$\begin{pmatrix} 0 & -a & 0 & 0 \\ a & 0 & 0 & 0 \\ 0 & 0 & 0 & -b \\ 0 & 0 & b & 0 \end{pmatrix} = \begin{pmatrix} 0 & -a & 0 & 0 \\ a & 0 & 0 & 0 \\ 0 & 0 & 0 & 0 \\ 0 & 0 & 0 & 0 \end{pmatrix} + \begin{pmatrix} 0 & 0 & 0 & 0 \\ 0 & 0 & 0 & 0 \\ 0 & 0 & 0 & -b \\ 0 & 0 & b & 0 \end{pmatrix} = \tag{2.3}$$

$$\begin{pmatrix} 1 & 1 & 0 & 0 \\ i & -i & 0 & 0 \\ 0 & 0 & 1 & 1 \\ 0 & 0 & i & -i \end{pmatrix} \begin{pmatrix} -ai & 0 & 0 & 0 \\ 0 & ai & 0 & 0 \\ 0 & 0 & -bi & 0 \\ 0 & 0 & 0 & bi \end{pmatrix} \begin{pmatrix} \frac{1}{2} & -\frac{1}{2}i & 0 & 0 \\ \frac{1}{2} & \frac{1}{2}i & 0 & 0 \\ 0 & 0 & \frac{1}{2} & -\frac{1}{2}i \\ 0 & 0 & \frac{1}{2} & \frac{1}{2}i \end{pmatrix}$$

The eigenvectors are the columns of

$$\begin{pmatrix} 1 & 1 & 0 & 0 \\ i & -i & 0 & 0 \\ 0 & 0 & 1 & 1 \\ 0 & 0 & i & -i \end{pmatrix} \tag{2.4}$$

and

$$\begin{pmatrix} 1 & 1 & 0 & 0 \\ i & -i & 0 & 0 \\ 0 & 0 & 1 & 1 \\ 0 & 0 & i & -i \end{pmatrix}^{-1} = \begin{pmatrix} \frac{1}{2} & -\frac{1}{2}i & 0 & 0 \\ \frac{1}{2} & \frac{1}{2}i & 0 & 0 \\ 0 & 0 & \frac{1}{2} & -\frac{1}{2}i \\ 0 & 0 & \frac{1}{2} & \frac{1}{2}i \end{pmatrix}$$

Each one of the six Cartan subalgebras is a maximal Abelian set of matrices which are diagonalizable to a purely imaginary trace zero matrix, quite like skew-Hermitian matrices:

$$D = \begin{pmatrix} -ai & 0 & 0 & 0 \\ 0 & ai & 0 & 0 \\ 0 & 0 & -bi & 0 \\ 0 & 0 & 0 & bi \end{pmatrix} \Longrightarrow Det(D)^{\frac{1}{2}} = ab \tag{2.5}$$

Where $i = \sqrt{-1}$. (2.5) chooses rows and columns 0,1 for the first matrix and the rest for the second matrix. There are 6 such choices which make 6 Cartan subalgebras [10]. Due to (1.1.1) The action of (2.5) can be defined as

$$Det(D)^{\frac{1}{2}}\sqrt{-g} \text{ and } a = b \Longrightarrow Det(D)^{\frac{1}{2}}\sqrt{-g} = a^2\sqrt{-g} = \left|\frac{U^{\lambda}U_{\lambda}}{4}\right|\sqrt{-g} \tag{2.5.1}$$

in the real case, where $g$ is the determinant of the metric tensor.

**Fig. 1.** – The generalized Reeb class vector, not the usual Reeb vector, as an acceleration vector.

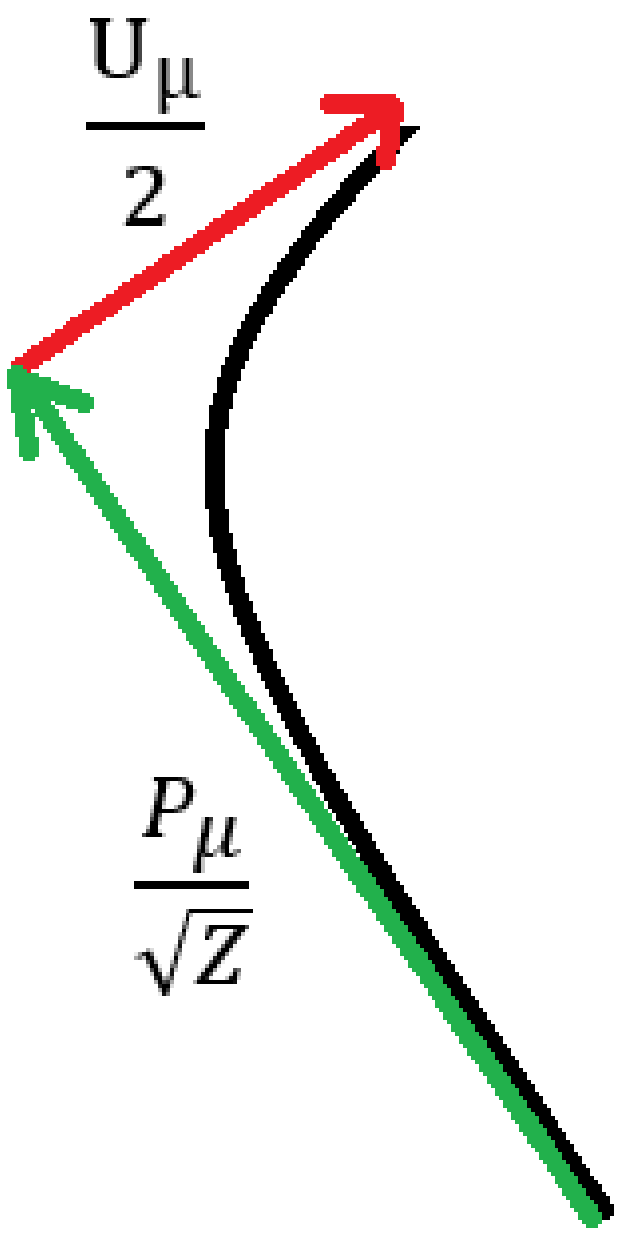

To describe a field that accelerates any unit vector, we need an anti-symmetric matrix of acceleration similar to the Tzvi Scarr & Yaakov Friedman's acceleration matrix [11] but with the mentioned important differences.

Considering A(1) in the $x^0, x^1$ plane, say $ct, x$ in Special Relativity and A(2) in the $x^2, x^3$ plane, say $y, z$ in Special Relativity, the second rotation and scaling matrix means spin while the x direction is a boost. So, looking at a radial source of such a field, the field perpendicular to the radius appears rotating from every angle of view and can have two real valued orientations A(2) and -A(2). Both A(1) and A(2) can be complex, however, there is a problem using skew-Hermitian matrices because skew-Hermitian matrices allow non-zero imaginary diagonal values in the complex plane. Diagonal elements should be zero if A(1), A(2) describe an acceleration field, unless only the real value of $VAV^*$is considered, where A is skew-Hermitian, and $V$ is a complex vector,

$$Real(VAV^*) = \frac{1}{2}(VAV^* + (VAV^*)^*) = \frac{1}{2}(VAV^* + VA^*V^*) = \frac{1}{2}(VAV^* - VAV^*) = 0 \quad (2.6)$$

And therefore, in the skew-Hermitian case $VAV^*$ is purely imaginary.

In that case The matrix $A_{\mu\nu} = \frac{U_\mu}{2}\frac{P_\nu}{\sqrt{Z}} - \frac{U_\nu}{2}\frac{P_\mu}{\sqrt{Z}}$ is insufficient for that purpose; however, it can be extended quite easily, by using the Levi-Civita alternating tensor [12], not the alternating Levi-Civita symbol.

**The problem of chirality relative to the direction of the acceleration field** – can be skipped except for (2.7), (2.8) and up to Theorem 0

Next, we would like to see if there is a mathematical reason to prefer right handedness, as an extension of (1.1.1),

$$A1 = \begin{pmatrix} 0 & -a & 0 & 0 \\ a & 0 & 0 & 0 \\ 0 & 0 & 0 & -a \\ 0 & 0 & a & 0 \end{pmatrix} \tag{2.7}$$

or left handedness of the acceleration fields $A1\frac{V}{c} = \frac{a}{c^2}$ or $A2\frac{V}{c} = \frac{a}{c^2}$ where $\frac{V}{c}$ is a unit 4-vector in spacetime, $\frac{a}{c^2}$ is its 4-acceleration and $c$ is a speed of light, where $\frac{V}{c}$ is a unit 4-vector in spacetime, $\frac{a}{c^2}$ is its 4-acceleration and $c$ is a speed of light. The action of $A1$, $A2$ takes the form of the Scarr-Friedmann uniform acceleration [11] although it has a very different meaning in this paper.

$$A2 = \begin{pmatrix} 0 & -a & 0 & 0 \\ a & 0 & 0 & 0 \\ 0 & 0 & 0 & a \\ 0 & 0 & -a & 0 \end{pmatrix} \tag{2.8}$$

Both matrices A1 and A2 implement a possible way to extend the matrix $A_{\mu\nu} \equiv \frac{U_\mu}{2}\frac{P_\nu}{\sqrt{Z}} - \frac{U_\nu}{2}\frac{P_\mu}{\sqrt{Z}}$ from$\begin{pmatrix} 0 & -a \\ -a & 0 \end{pmatrix}$ to a 4-dimensional matrix. The preference of $A1$ is discussed.

Obviously when

$$A = \begin{pmatrix} 0 & -a & 0 & 0 \\ a & 0 & 0 & 0 \\ 0 & 0 & 0 & -b \\ 0 & 0 & b & 0 \end{pmatrix}, \; C = \begin{pmatrix} 0 & -c & 0 & 0 \\ c & 0 & 0 & 0 \\ 0 & 0 & 0 & -d \\ 0 & 0 & d & 0 \end{pmatrix} \tag{2.9}$$

AC-CA = 0, therefore A(B+C)-(B+C)A = AB -BA such that

$$B = \begin{pmatrix} 0 & 0 & -x & -z \\ 0 & 0 & -y & -w \\ x & y & 0 & 0 \\ z & w & 0 & 0 \end{pmatrix} \tag{2.10}$$

To find the root decomposition of the Lie algebra of the skew-symmetric matrices by the Cartan subalgebra which is described by matrix A we need to solve for some eigenvalues $\lambda$ and eigenvectors of the ad() operator $ad(A)B = [A, B] = AB - BA$.

$$ad(A)B = \lambda B \tag{2.11}$$

$$\begin{pmatrix} 0 & -a & 0 & 0 \\ a & 0 & 0 & 0 \\ 0 & 0 & 0 & -b \\ 0 & 0 & b & 0 \end{pmatrix}\begin{pmatrix} 0 & 0 & -x & -z \\ 0 & 0 & -y & -w \\ x & y & 0 & 0 \\ z & w & 0 & 0 \end{pmatrix} -$$
$$\begin{pmatrix} 0 & 0 & -x & -z \\ 0 & 0 & -y & -w \\ x & y & 0 & 0 \\ z & w & 0 & 0 \end{pmatrix}\begin{pmatrix} 0 & -a & 0 & 0 \\ a & 0 & 0 & 0 \\ 0 & 0 & 0 & -b \\ 0 & 0 & b & 0 \end{pmatrix} = \lambda \begin{pmatrix} 0 & 0 & -x & -z \\ 0 & 0 & -y & -w \\ x & y & 0 & 0 \\ z & w & 0 & 0 \end{pmatrix} \tag{2.12}$$

$$\begin{pmatrix} 0 & 0 & ay & aw \\ 0 & 0 & -ax & -az \\ -bz & -bw & 0 & 0 \\ bx & by & 0 & 0 \end{pmatrix} -$$
$$\begin{pmatrix} 0 & 0 & -bz & bx \\ 0 & 0 & -bw & by \\ ay & -ax & 0 & 0 \\ aw & -az & 0 & 0 \end{pmatrix} = \begin{pmatrix} 0 & 0 & ay + bz & aw - bx \\ 0 & 0 & -ax + bw & -az - by \\ -ay - bz & ax - bw & 0 & 0 \\ -aw + bx & az + by & 0 & 0 \end{pmatrix} \tag{2.13}$$

So, we have the following equations:

$$\begin{pmatrix} -ay - bz \\ ax - bw \\ bx - aw \\ az + by \end{pmatrix} = \lambda \begin{pmatrix} x \\ y \\ z \\ w \end{pmatrix} \tag{2.14}$$

Therefor we need to find solutions to the eigenvectors and values equation:

$$\begin{pmatrix} 0 & -a & -b & 0 \\ a & 0 & 0 & -b \\ b & 0 & 0 & -a \\ 0 & b & a & 0 \end{pmatrix}\begin{pmatrix} x \\ y \\ z \\ w \end{pmatrix} = \lambda \begin{pmatrix} x \\ y \\ z \\ w \end{pmatrix} \tag{2.15}$$

Bearing in mind that the skew symmetric matrices are a Lie algebra also over the complex numbers, consider the root system of the Cartan subalgebra of skew-symmetric matrices (2.3):

$$S = \begin{pmatrix} -1 & 1 & 1 & -1 \\ -i & -i & i & i \\ -i & i & -i & i \\ 1 & 1 & 1 & 1 \end{pmatrix} \tag{2.16}$$

$$S^{-1} = \begin{pmatrix} -\frac{1}{4} & \frac{i}{4} & \frac{i}{4} & \frac{1}{4} \\ \frac{1}{4} & \frac{i}{4} & -\frac{i}{4} & \frac{1}{4} \\ \frac{1}{4} & -\frac{i}{4} & \frac{i}{4} & \frac{1}{4} \\ -\frac{1}{4} & -\frac{i}{4} & -\frac{i}{4} & \frac{1}{4} \end{pmatrix}$$

$$D = \begin{pmatrix} -i(a+b) & 0 & 0 & 0 \\ 0 & i(a-b) & 0 & 0 \\ 0 & 0 & i(b-a) & 0 \\ 0 & 0 & 0 & i(a+b) \end{pmatrix}$$

$$\begin{pmatrix} 0 & -a & -b & 0 \\ a & 0 & 0 & -b \\ b & 0 & 0 & -a \\ 0 & b & a & 0 \end{pmatrix} = SDS^{-1}$$

In view of the action of $A$ in (2.5.1) and due to (1.1.1), (2.9), the action of $ad(A)$ is:

$$Det(D)^{\frac{1}{2}}\sqrt{-g} = |(a+b)(a-b)|\sqrt{-g} \tag{2.16.1}$$

**Important:** $Det(D)^{\frac{1}{2}}\sqrt{-g}$ is zero where $g$ is the determinant of the metric tensor and where $a = b$ or $-a = b$. This is a good motivation for considering only $a = b$ or $-a = b$. The reason is that the sum of the action of the roots of the Cartan subalgebra and the action of the acceleration matrix (2.5.1) when $a = b$ or $-a = b$ depends only on (1.1.1) and is well defined in (2.5.1).

We also touched on another possible point which is the joint action of

$$Action(A) + Action(ad(A)) = (|ab| + |(a+b)(a-b)|)\sqrt{-g} \tag{2.16.2}$$

$$a = b \ \vee -a = b \Longrightarrow Action(A) + Action\big(ad(A)\big) = a^2\sqrt{-g} = -\frac{U^{\lambda}U_{\lambda}}{4}$$

in the metric convention (+, -, -, -) or $\frac{U^{\lambda}U_{\lambda}}{4}$ in the metric convention (-, +, +, +).

And consider the root system over the Cartan sub-algebra of the skew-symmetric matrices,

$$A = \begin{pmatrix} 0 & -a & 0 & 0 \\ a & 0 & 0 & 0 \\ 0 & 0 & 0 & -b \\ 0 & 0 & b & 0 \end{pmatrix} \tag{2.17}$$

$$Root = \begin{pmatrix} 0 & 0 & -x & -z \\ 0 & 0 & -y & -w \\ x & y & 0 & 0 \\ z & w & 0 & 0 \end{pmatrix} \in \tag{2.18}$$

$$\left\{ \begin{pmatrix} 0 & 0 & 1 & i \\ 0 & 0 & i & -1 \\ -1 & -i & 0 & 0 \\ -i & 1 & 0 & 0 \end{pmatrix}, \begin{pmatrix} 0 & 0 & -1 & -i \\ 0 & 0 & i & -1 \\ 1 & -i & 0 & 0 \\ i & 1 & 0 & 0 \end{pmatrix}, \begin{pmatrix} 0 & 0 & -1 & i \\ 0 & 0 & -i & -1 \\ 1 & i & 0 & 0 \\ -i & 1 & 0 & 0 \end{pmatrix}, \begin{pmatrix} 0 & 0 & 1 & -i \\ 0 & 0 & -i & -1 \\ -1 & i & 0 & 0 \\ i & 1 & 0 & 0 \end{pmatrix} \right\}$$

A physical meaning in the classical sense - because roots are linear operators and are therefore acceleration matrices - is that the real part of the resulting matrix should consist of the following representations $\begin{pmatrix} 0 & \dots & -a \\ \dots & \dots & \dots \\ a & \dots & 0 \end{pmatrix}$. The Cartan algebra has no real eigenvalues except for zero, as expected from an acceleration matrix $\frac{1}{2}(Root + Root^*)$ where the * operator does not represent transposed conjugation but conjugation over the complex numbers

$$\frac{1}{2}(Root + Root^*)\frac{V}{c} = \frac{a}{c^2} \tag{2.19}$$

Where $\frac{V}{c}$ is a unit 4-vector in spacetime, $\frac{a}{c^2}$ is its 4-acceleration and $c$ is a speed of light.

The meaning of (2.19) which leads to (2.7) or (2.8) is that a subspace of the skew-symmetric matrices can be used to represent a physical acceleration in addition to (2.3). This subspace is represented as a direct sum of the Cartan algebra [10] and two roots out of 4 when a=b or -a=b. The linear combination of only two roots obviously does not cover all the skew-symmetric matrices over the complex field.

In conventional particle - based physics, there is no meaning to the chirality of an electric field of an electron although existing models do say that an electric charge emits virtual photons and obviously such virtual photons should have a chirality. However, there should be a fundamental difference between the chirality of a spin and the chirality of the acceleration field, which is discussed in this section.

The result of (2.19), (2.7), (2.8) is an equivalence between the orientability in space and the asymmetry of time if either (2.7) or (2.8) is maintained in the same form along world lines.

An orientation on a manifold is the sign of the determinant of an atlas of coordinate systems. When Dr. Sam Vaknin was shown the result in (2.19) he made an important remark that (2.19), (2.7), (2.8) leads to time asymmetry. Obviously due to “Appendix H – Causality conservation theorem”, the Geroch time function $PP^*$ can be either monotonically increasing or monotonically decreasing except for a set of measure zero. If we assume that the cosmos is a “Big Bang” cosmos then the Geroch function must be increasing, however, it is preferable not to make such an assumption.

**Theorem 0: Time asymmetry special theorem (Suchard - Vaknin)**: The local time coordinate $\frac{1}{2}(P_\mu + P_\mu^*)$ must have only one possible direction, when $\frac{u_\mu}{2}$ is not zero, or in the real case, $P_\mu$ must have only one possible direction if and only if space is orientable. This theorem has a meaning and is not trivial due to "Appendix H – Causality conservation theorem". For the sake of simplicity, instead of dealing with $\frac{1}{2}(P_\mu + P_\mu^*)$ and $\frac{1}{4}(U_\mu + U_\mu^*)$ this theorem is proven for real numbered vectors.

**Note:** Implicitly the theorem assumes that $\frac{u_\mu}{2}$ is smooth and that A1 in (2.7) or A2 in (2.8) describes a non-degenerate smooth matrix along the integral curves of $P_\mu$ and in small neighborhoods around this curve in each foliation perpendicular to $P_\mu$ in the real case or $\frac{1}{2}(P_\mu + P_\mu^*)$ in the complex case. Without loss of generality, we assume A1 as in (2.7) is maintained along continuous open time-like curves.

**Proof**: Notice that if we choose $x^0 = \frac{p^\nu}{\sqrt{Z}}$, $x^1 = \frac{u^\nu}{\sqrt{u_\lambda \mathrm{u}^\lambda}}$ or $x^0 = \frac{p^\nu}{\sqrt{Z}}$, $x^1 = -\frac{u^\nu}{\sqrt{u_\lambda \mathrm{u}^\lambda}}$ the acceleration matrix restricted to the plane spanned by $x^0, x^1$ will be $\begin{pmatrix} 0 & -a \\ a & 0 \end{pmatrix}$ and $\begin{pmatrix} 0 & a \\ -a & 0 \end{pmatrix}$ respectively and (2.7) is either $A1$ or $-A1$. It is easier to check how the sign of alternating forms changes instead of using determinants. For example, consider $x^0 \wedge x^1 \wedge x^2 \wedge x^3$ with locally perpendicular coordinates $x^0, x^1, x^2, x^3$. The theorem proof is immediate from the orientability of the space foliation which is perpendicular to $\frac{1}{2}(P_\mu + P_\mu^*)$ or to $P_\mu$ in the real case. For simplicity, proceed with the real case; one direction of the proof is easy, if $P_\mu$ is asymmetrical and therefore $-P_\mu$ is not a valid direction of time, then (2.7) dictates the orientation of space when $\frac{u_\mu}{2}$ is not zero, to see why, consider that the alternating form $(\frac{p_\nu}{\sqrt{Z}}\frac{u_\mu}{2} - \frac{p_\mu}{\sqrt{Z}}\frac{u_\nu}{2})dx^\mu \wedge dx^\nu$ defines the orientation also of the perpendicular plane by using the 2-form $dx^2 \wedge dx^3$ and by (2.7). We will later write such an extension to the perpendicular plane by using the form $(\frac{p_\nu}{\sqrt{Z}}\frac{u_\mu}{2} - \frac{p_\mu}{\sqrt{Z}}\frac{u_\nu}{2})dx^\mu \wedge dx^\nu$ in a tensorial way. For simplicity, by choosing $x^0 = \frac{p^\nu}{\sqrt{Z}}$ and $x^1 = \frac{u^\nu}{\sqrt{u_\lambda \mathrm{u}^\lambda}}$, two possible orientations of a perpendicular form $x^2 \wedge x^3$ are dictated by (2.7) which depend only on the sign of $u_\nu$, $(\pm)u_\mu$, but that means that the sign of $x^1 \wedge x^2 \wedge x^3$ can be only one. The 3 cases we need to consider that are dictated by (2.7) are,

$$x^1 -> -x^1 \wedge (x^2 -> -x^2 \ \vee \ x^3 -> -x^3) \Longrightarrow A1 = \begin{pmatrix} 0 & a & 0 & 0 \\ -a & 0 & 0 & 0 \\ 0 & 0 & 0 & a \\ 0 & 0 & -a & 0 \end{pmatrix} \qquad (2.20)$$

And

$$x^2 -> -x^2 \wedge x^3 -> -x^3 \Longrightarrow A1 = \begin{pmatrix} 0 & -a & 0 & 0 \\ a & 0 & 0 & 0 \\ 0 & 0 & 0 & -a \\ 0 & 0 & a & 0 \end{pmatrix} \quad (2.21)$$

In both cases the orientation of the space spanned by $x^1, x^2, x^3$ is maintained. The converse starts with space foliations which are orientable so the sign of $x^1 \wedge x^2 \wedge x^3$ is determined but then if we change the sign of $x^1 = \frac{u_\nu}{\sqrt{u_\lambda u^\lambda}}$ we also must change the sign of either $x^2$ or $x^3$ but not both or to swap their order and then by (2.7), the sign of $x^0 = \frac{p_\nu}{\sqrt{Z}}$ cannot change which means time asymmetry. The orientation $x^1, x^2, x^3$ needs clarification. Locally we can write the following vectors $\frac{p_\mu(0)}{\sqrt{Z(0)}} = \frac{p_\mu}{\sqrt{p^\lambda p_\lambda}}, \frac{u_\mu(0)}{2}, \frac{p_\mu(1)}{\sqrt{Z(1)}}, \frac{u_\mu(1)}{2}$, to represent two perpendicular planes of rotation and scaling which are two perpendicular planes where unit vectors are accelerated. By using the Levi-Civita tensor [12], two different orientations of $\frac{p_\mu(1)}{\sqrt{Z(1)}}, \frac{u_\mu(1)}{2}$ can be defined as two possible signs

$$sign(E^{\mu\nu\lambda\xi} \frac{p_\mu(0)}{\sqrt{Z(0)}} \frac{u_\nu(0)}{2} \frac{p_\lambda(1)}{\sqrt{Z(1)}} \frac{u_\xi(1)}{2}) \quad (2.22)$$

which is either +1 or -1 when $E^{\mu\nu\lambda\xi} \frac{p_\mu(0)}{\sqrt{Z(0)}} \frac{u_\nu(0)}{2} \frac{p_\lambda(1)}{\sqrt{Z(1)}} \frac{u_\xi(1)}{2}$ is non-degenerate. Now, having other three functions $\frac{p_\mu(2)}{\sqrt{Z(2)}}, \frac{p_\mu(3)}{\sqrt{Z(3)}}, \frac{p_\mu(4)}{\sqrt{Z(4)}}$, which span the same tangent space as the one spanned by $\frac{u_\mu(0)}{2}, \frac{p_\mu(1)}{\sqrt{Z(1)}}, \frac{u_\mu(1)}{2}$ another orientation can be defined

$$\text{Sign}(E^{\mu\nu\lambda\xi} \frac{u_\nu(0)}{2} \frac{p_\lambda(1)}{\sqrt{Z(1)}} \frac{u_\xi(1)}{2} E_{\mu\alpha\beta\zeta} \frac{p^\alpha(2)}{\sqrt{Z(2)}} \frac{p^\beta(3)}{\sqrt{Z(3)}} \frac{p^\zeta(4)}{\sqrt{Z(4)}}) \quad (2.23)$$

So maintaining both the sign (2.22) and the sign (2.23) consistently along the worldlines by $\frac{p_\mu(0)}{\sqrt{Z(0)}}$ forces $\frac{p_\mu(0)}{\sqrt{Z(0)}}$ to have one possible direction along these worldlines. The vectors $\frac{p^\alpha(2)}{\sqrt{Z(2)}}, \frac{p^\beta(3)}{\sqrt{Z(3)}}, \frac{p^\zeta(4)}{\sqrt{Z(4)}}$ will be discussed in (64.b) and in Theorem 7. They are the basis for a chromodynamic theory which means it has a Lagrangian which depends on 3 acceleration vectors.

Q.E.D.

**Note:** Not to be ungrateful it is important to mention that (2.16) - (2.18) was checked by using the online Wolfram Equations internet site.

**A 3-Chirality tensor from $A_{\mu\nu} = -A_{\nu\mu}$ where $A_{\mu\nu} \neq D\zeta_\mu$, the exterior derivative for some $\zeta_\mu$** – can be skipped

$$\bar{A}_{ab,c} = \frac{\partial}{\partial y^c}\left(A_{\mu\nu}\frac{\partial x^\mu}{\partial y^a}\frac{\partial x^\nu}{\partial y^b}\right) = A_{\mu\nu,\lambda}\frac{\partial x^\lambda}{\partial y^c}\frac{\partial x^\mu}{\partial y^a}\frac{\partial x^\nu}{\partial y^b} + A_{\mu\nu}\frac{\partial^2 x^\mu}{\partial y^a \partial y^c}\frac{\partial x^\nu}{\partial y^b} + A_{\mu\nu}\frac{\partial x^\mu}{\partial y^a}\frac{\partial^2 x^\nu}{\partial y^b \partial y^c} \tag{2.24}$$

Now

$$A_{\mu\nu}\frac{\partial x^\mu}{\partial y^a}\frac{\partial^2 x^\nu}{\partial y^b \partial y^c} = -A_{\mu\nu}\frac{\partial^2 x^\mu}{\partial y^b \partial y^c}\frac{\partial x^\nu}{\partial y^a} \tag{2.25}$$

$$\bar{A}_{ab,c} = \frac{\partial}{\partial y^c}\left(A_{\mu\nu}\frac{\partial x^\mu}{\partial y^a}\frac{\partial x^\nu}{\partial y^b}\right) = A_{\mu\nu,\lambda}\frac{\partial x^\lambda}{\partial y^c}\frac{\partial x^\mu}{\partial y^a}\frac{\partial x^\nu}{\partial y^b} + A_{\mu\nu}\frac{\partial^2 x^\mu}{\partial y^a \partial y^c}\frac{\partial x^\nu}{\partial y^b} - A_{\mu\nu}\frac{\partial^2 x^\mu}{\partial y^b \partial y^c}\frac{\partial x^\nu}{\partial y^a} \tag{2.26}$$

Writing $\gamma_{acb} = A_{\mu\nu}\frac{\partial^2 x^\mu}{\partial y^a \partial y^c}\frac{\partial x^\nu}{\partial y^b}$

$$\frac{\partial}{\partial y^c}\left(A_{\mu\nu}\frac{\partial x^\mu}{\partial y^a}\frac{\partial x^\nu}{\partial y^b}\right) = A_{\mu\nu,\lambda}\frac{\partial x^\lambda}{\partial y^c}\frac{\partial x^\mu}{\partial y^a}\frac{\partial x^\nu}{\partial y^b} + \gamma_{acb} - \gamma_{bca} \tag{2.27}$$

And adding cyclically we have

$$\gamma_{acb} - \gamma_{bca} + \gamma_{cba} - \gamma_{abc} + \gamma_{bac} - \gamma_{cab} = 0 \tag{2.28}$$

Because $\gamma_{acb} = \gamma_{cab}$ as $\frac{\partial^2 x^\mu}{\partial y^a \partial y^c} = \frac{\partial^2 x^\mu}{\partial y^c \partial y^a}$ so $\gamma_{acb} - \gamma_{cab} = -\gamma_{bca} + \gamma_{bac} = \gamma_{cba} - \gamma_{bca} = 0$.

We therefore have

$$\frac{\partial}{\partial y^c}\left(A_{\mu\nu}\frac{\partial x^\mu}{\partial y^a}\frac{\partial x^\nu}{\partial y^b}\right) + \frac{\partial}{\partial y^b}\left(A_{\mu\nu}\frac{\partial x^\mu}{\partial y^c}\frac{\partial x^\nu}{\partial y^a}\right) + \frac{\partial}{\partial y^a}\left(A_{\mu\nu}\frac{\partial x^\mu}{\partial y^b}\frac{\partial x^\nu}{\partial y^c}\right) = \tag{2.29}$$

$$\left(A_{\mu\nu,\lambda} + A_{\lambda\mu,\nu} + A_{\nu\lambda,\mu}\right)\frac{\partial x^\lambda}{\partial y^c}\frac{\partial x^\mu}{\partial y^a}\frac{\partial x^\nu}{\partial y^b} = \bar{A}_{ab,c} + \bar{A}_{ca,b} + \bar{A}_{bc,a} \tag{2.30}$$

Which proves that $A_{\mu\nu,\lambda} + A_{\lambda\mu,\nu} + A_{\nu\lambda,\mu}$ is a tensor when $A_{\mu\nu} = -A_{\nu\mu}$.

**The 3-Chirality tensor vanishes where $A_{\mu\nu}$ is exact** – can be skipped

When $A_{\mu\nu}$ is an exact form then $\mathrm{D}A_{\mu\nu}dx^\mu \wedge dx^\nu = 0 \Longrightarrow A_{\mu\nu,\lambda} + A_{\lambda\mu,\nu} + A_{\nu\lambda,\mu} = 0$. The proof is not difficult.

$$A_{\mu\nu,\lambda} - A_{\mu\lambda,\nu} = \omega_{\mu,\nu,\lambda} - \omega_{\mu,\lambda,\nu} = 0 \Longrightarrow A_{\mu\nu,\lambda}\, dx^\lambda \wedge dx^\mu \wedge dx^\nu = \mathrm{D}A_{\mu\nu}dx^\mu \wedge dx^\nu = 0 \tag{2.31}$$

$$A_{\mu\nu,\lambda} + A_{\lambda\mu,\nu} + A_{\nu\lambda,\mu} = \omega_{\mu,\nu,\lambda} - \omega_{\nu,\mu,\lambda} + \omega_{\lambda,\mu,\nu} - \omega_{\mu,\lambda,\nu} + \omega_{\nu,\lambda,\mu} - \omega_{\lambda,\nu,\mu} =$$

$$\omega_{\mu,\nu,\lambda} - \omega_{\mu,\lambda,\nu} + \omega_{\lambda,\mu,\nu} - \omega_{\lambda,\nu,\mu} + \omega_{\nu,\lambda,\mu} - \omega_{\nu,\mu,\lambda} = 0$$

**The 3-Chirality tensor that does not vanish** – can be skipped

$$H_{\mu\nu} \equiv \frac{P_\mu(0)}{\sqrt{|P^s(0)P_s(0)|}}\frac{P_\nu(1)}{\sqrt{|P^s(1)P_s(1)|}} - \frac{P_\mu(1)}{\sqrt{|P^s(1)P_s(1)|}}\frac{P_\nu(0)}{\sqrt{|P^s(0)P_s(0)|}} \tag{2.32}$$

$$H_{\mu\nu,\lambda} + H_{\lambda\mu,\nu} + H_{\nu\lambda,\mu} \neq 0 \tag{2.33}$$

If $P_\mu(0)$ and $P_\mu(1)$ are independent vectors.

**Hodge star spin-like field extension and zero charge:** Regarding (2.7) or (2.8) We have $B^{\mu\nu} = \frac{1}{2}E^{\mu\nu\alpha\beta}A_{\alpha\beta}$ which define an acceleration matrix in a perpendicular plane to the plane spanned by $\frac{P_\mu}{\sqrt{Z}}$ and $\frac{U_\mu}{2}$. In the complex case we define the acceleration matrix: $F_{\mu\nu} = A_{\mu\nu} + \gamma B_{\mu\nu}$ where $\gamma \epsilon U(1)$. With a vector $w^\nu$, $w^\nu w_\nu = c^2$, we derive its acceleration,

$$F_{\mu\nu}\frac{w^\nu}{c} = \frac{a_{\mu(w)}}{c^2}, B^{\mu\nu} = \mp\frac{1}{2}E^{\mu\nu\alpha\beta}A_{\alpha\beta} \tag{3}$$

$$\frac{1}{4}F_{\mu\nu}F^{\mu\nu} = \frac{U_\mu U^\mu}{4}$$

**Note:** If spacetime could have only one orientation as hinted in (2.19), then either $+\frac{1}{2}E^{\mu\nu\alpha\beta}A_{\alpha\beta}$ would result in $\frac{U_\mu U^\mu}{4} = 0$ or $-\frac{1}{2}E^{\mu\nu\alpha\beta}A_{\alpha\beta}$ would result in $\frac{U_\mu U^\mu}{4} = 0$. In this case, the action could be described differently as $\frac{1}{4}\left(A_{\mu\nu}A^{\mu\nu} \mp B_{\mu\nu}B^{\mu\nu}\right)$. Such a possibility would lead to only one type of neutrino orientation but there should be another orientation too.

**Zero charge which can work for both complex and real formalisms**

This idea of spacetime orientation leads to an expression of zero charge even without using any complex formalism.

$$A^{\mu\nu} = \frac{1}{2}(\bar{A}^{\mu\nu} + \frac{1}{2}E^{\mu\nu\alpha\beta}\bar{B}_{\alpha\beta}),\ B^{\mu\nu} = \frac{1}{2}(\bar{B}^{\mu\nu} - \frac{1}{2}E^{\mu\nu\alpha\beta}\bar{A}_{\alpha\beta}) \tag{3.1}$$

With field generators $\bar{A}^{\mu\nu} = \frac{U_\mu(0)}{2}\frac{P_\nu(0)}{\sqrt{Z(0)}} - \frac{U_\nu(0)}{2}\frac{P_\mu(0)}{\sqrt{Z(0)}}$ and $\bar{B}^{\mu\nu} = \frac{U_\mu(1)}{2}\frac{P_\nu(1)}{\sqrt{Z(1)}} - \frac{U_\nu(1)}{2}\frac{P_\mu(1)}{\sqrt{Z(1)}}$. Notice that it is possible to have

$$A^{\mu\nu} = \frac{1}{2}(\bar{A}^{\mu\nu} + \frac{1}{2}E^{\mu\nu\alpha\beta}\bar{B}_{\alpha\beta}) = 0 \ \vee \ A^{\mu\nu} = \frac{1}{2}(\bar{A}^{\mu\nu} + \frac{1}{2}E^{\mu\nu\alpha\beta}\bar{B}_{\alpha\beta}) \neq 0 \tag{3.1.1}$$

Moreover (3.1.1) should work also if $U_\mu(0)U^\mu(0) = 0$, but $U_\mu(0) \neq 0$.

Exercise to the reader: show that the Reeb class vector $\frac{U_\mu}{2}$ of $\frac{P_\mu}{\sqrt{|Z|}}$ is the same as for $\frac{P_\mu}{\sqrt{|Z|}}e^{i\theta}$ for $i = \sqrt{-1}$ and a smooth scalar $\theta$. See that you understand the idea of a field of acceleration that maps 4-velocity to 4-acceleration by multiplication with an anti-symmetric matrix [11], $F_{\mu\nu}\frac{w^\nu}{c} = \frac{a_{\mu(w)}}{c^2}$.

In Special Relativity, 4-velocity is perpendicular to 4-accleration and $w^{\nu}w_{\nu} = c^2$. $F_{\mu\nu}$ is then an Acceleration Field and it can be deconstructed into the sum of two matrices which act on two perpendicular two-dimensional hyperplanes in spacetime. $B_{\mu\nu} = \frac{1}{2}E^{\mu\nu\alpha\beta}A_{\alpha\beta}$ and $B_{\mu\nu} = -\frac{1}{2}E^{\mu\nu\alpha\beta}A_{\alpha\beta}$ yield the same result in (3). When reduced to the three-dimensional foliation which is perpendicular to $P_{\mu}$, If $\frac{U_{\mu}}{2}$ has a divergence point, say Q, then the choice $B_{\mu\nu} = -\frac{1}{2}E^{\mu\nu\alpha\beta}A_{\alpha\beta}$ mean that $B_{\mu\nu} = \frac{P(2)_{\nu}}{\sqrt{|Z|}}\frac{U(2)_{\mu}}{2} - \frac{P(2)_{\mu}}{\sqrt{|Z|}}\frac{U(2)_{\nu}}{2}$, where $P(2)_{\nu}P^{\nu} = 0$, $U(2)_{\nu}P^{\nu} = 0$, $U(2)_{\nu}P(2)^{\nu} = 0$, and finally the complex numbers case can also yield the following,

$$U_{\mu};^{\mu} + U^{*}{}_{\mu};^{\mu} = 0 \tag{3.2}$$

With two independent Geroch functions [2] and the root of the determinant of a Gram matrix, it is easy to develop a theory of zero charge also when (3.2) does not hold. Such an offer will be presented in (64.b), (64.01).

**Lagrangian generalization offer and further research offer** – can be skipped up to "What is this paper's goal?".

It is possible to define a Lagrangian for two independent acceleration vectors that are related to each other by multiplication, here it is presented in a complex formalism, with a volume element $\sqrt{-g}$, and with Reeb class vectors, not Reeb vectors, which are perpendicular to $\frac{P_{\mu}}{\sqrt{Z}}$, and see the definition of G(3) in (64.a), (64.b) against (64.2)-(64.12).

$$\left|\frac{\zeta^{*\lambda}\zeta_{\lambda}+\zeta^{\lambda}\zeta^{*}{}_{\lambda}}{8}\right|\sqrt{-g} = \begin{vmatrix} 1 & \frac{\mathrm{P_k}P(1)^{*k}+\mathrm{P^*}_{\mathrm{k}}P(1)^{k}}{2\sqrt{\mathrm{ZZ(1)}}} & 0 & \frac{\mathrm{P_k}U(1)^{*k}+\mathrm{P^*}_{\mathrm{k}}U(1)^{k}}{2\sqrt{2\mathrm{Z}}} \\ \frac{\mathrm{P_k}P(1)^{*k}+\mathrm{P^*}_{\mathrm{k}}P(1)^{k}}{2\sqrt{\mathrm{ZZ(1)}}} & 1 & \frac{\mathrm{P(1)_k}U^{*k}+\mathrm{P(1)^*}_{\mathrm{k}}U^{k}}{2\sqrt{2\mathrm{Z}}} & 0 \\ 0 & \frac{\mathrm{P(1)_k}U^{*k}+\mathrm{P(1)^*}_{\mathrm{k}}U^{k}}{2\sqrt{2\mathrm{Z}}} & \frac{U^{k}U_{k}^{*}+U^{*k}U_{k}}{8} & \frac{U(1)^{k}U_{k}^{*}+U(1)^{*k}U_{k}}{8} \\ \frac{\mathrm{P_k}U(1)^{*k}+\mathrm{P^*}_{\mathrm{k}}U(1)^{k}}{2\sqrt{2\mathrm{Z}}} & 0 & \frac{U(1)^{k}U_{k}^{*}+U(1)^{*k}U_{k}}{8} & \frac{U(1)^{k}U(1)_{k}^{*}+U(1)^{*k}U(1)_{k}}{8} \end{vmatrix}^{\frac{1}{2}} \mathrm{G(3)}^{-\frac{1}{2}}\sqrt{-g} \tag{3.2.1}$$

The meaning of (3.2.1) is of a squared acceleration which is the Minkowski squared norm of a spacelike vector. In (+,-,-,-) metric convention, a negative sign has to be added, $-\frac{\zeta^{*\lambda}\zeta_{\lambda}+\zeta^{\lambda}\zeta^{*}{}_{\lambda}}{8}$.

The following norm calculates a physical non-geodesic acceleration, $\sqrt{\frac{\zeta^{*\lambda}\zeta_{\lambda}+\zeta^{\lambda}\zeta^{*}{}_{\lambda}}{8}}$. Since the 3 forces in Nature seem to be aligned with the electric field, it is reasonable to assume that $\zeta_{\lambda}$ must be either aligned or anti-aligned with $U_{\lambda}$, or in other words,

$$\zeta_\lambda = f(x^\mu) U_\lambda \tag{3.2.2}$$

$$\frac{\zeta_\lambda + \zeta^*{}_\lambda}{4} = f(x^\mu) \frac{U_\lambda + U^*{}_\lambda}{4}$$

for some scalar function of the coordinates $f(x^\mu)$. A real valued vector is then $\frac{U_\mu + U^*{}_\mu}{4}$ but to assume this expression is the direction of an acceleration vector, by Occam's razor must be inferred from a variation of the Lagrangian L=$\frac{U_\mu U^{*\mu} + U^*{}_\mu U^\mu}{8}\sqrt{-g}$. Such a variation indeed involves the divergence, $\left(\frac{U^\mu + U^{*\mu}}{4}\right)_{;\mu}$ which implies that $\frac{U_\mu + U^*{}_\mu}{4}$ has indeed a meaning of an acceleration of a unit vector. The zeros in (3.2) mean that the acceleration vector $\frac{U_k}{2}$ is perpendicular to the unit vector $\frac{P^*{}_k}{\sqrt{Z}}$, $\left(\frac{Z_\mu}{2Z} - \frac{Z_\lambda P^{*\lambda} P_\mu}{2Z^2}\right)\frac{P^{*\mu}}{\sqrt{Z}} = \left(\frac{Z_\mu P^{*\mu}}{2Z} - \frac{Z_\lambda P^{*\lambda}}{2Z^2} Z\right)\frac{1}{\sqrt{Z}} = 0$ and then the term $\frac{P_k U^{*k} + P^*{}_k U^{*k}}{2\sqrt{2Z}} = 0$. If $P_k$ and $U_k$ are perpendicular to $P(2)_k$ and $U(2)_k$ then of course $\frac{P_k U(2)^{*k} + P^*{}_k U(2)^{*k}}{2\sqrt{2Z}} = 0$, however, this Lagrangian can define an action operator even without such an orthogonality as a prerequisite and is therefore more general. The Lagrangian above has symmetry SU(2) and is therefore offered as a generalization of this paper with properties of the "electroweak" field. To summarize the motivation of this section, saying that an energy density can be described as the negative squared norm of an acceleration of unit vectors in (+,-,-,-) metric does not mean such acceleration field can't be a result of other Reeb class fields. The description of the electric field as the simplest example is discussed later.

**Quantum Gravity - Could non-geodesic acceleration vectors also explain the gravitational field?** – can be skipped up to "What is this paper's goal?".

The pseudo-acceleration of a test particle with velocity $\frac{dx^\mu}{d\tau}$ in weak gravity satisfies,

$$\frac{d^2x^\mu}{d\tau^2} + \Gamma^\mu_{\alpha\beta}\frac{dx^\alpha}{d\tau}\frac{dx^\beta}{d\tau} = 0 \tag{3.3}$$

$$\frac{d^2x^\mu}{dt^2} + \Gamma^\mu_{\alpha\beta}\frac{dx^\alpha}{dt}\frac{dx^\beta}{dt} = 0 \tag{3.4}$$

With weak gravity:

$$g_{\mu\nu} = \eta_{\mu\nu} + \epsilon h_{\mu\nu} \tag{3.5}$$

$$O\left(\frac{dx^\mu}{dt}\right) = O(\epsilon) \tag{3.6}$$

And then space terms are neglected, which reduces the equation of motion to

$$\frac{d^2x^\mu}{dt^2} + \Gamma^\mu_{\alpha\beta}\frac{dx^0}{dt}\frac{dx^0}{dt} \approx 0 \tag{3.7}$$

While $\frac{dx^0}{dt} = c$, the speed of light, we get,

$$\frac{d^2x^\mu}{dt^2} \approx -c^2\Gamma^\mu_{00} = -\frac{1}{2}c^2\epsilon({h^\mu}_{0,0} + {h^\mu}_{0,0} - h_{00},^\mu) \tag{3.8}$$

And in a static field

$$\Gamma^\mu_{00} = -\frac{1}{2}h_{00},^\mu \tag{3.9}$$

$$\frac{d^2x^\mu}{dt^2} \approx \frac{1}{2}c^2\epsilon h_{00},^\mu \tag{3.10}$$

Consider the following representation of the metric tensor,

$$g_{\mu\nu} = \frac{P(0)_\mu P(0)_\nu}{Z(0)} - \frac{P(1)_\mu P(1)_\nu}{Z(1)} - \frac{P(2)_\mu P(2)_\nu}{Z(2)} - \frac{P(3)_\mu P(3)_\nu}{Z(3)} \tag{3.11}$$

At this point no full tetradic representation is considered and $P(i)_\mu P(j)^\mu \neq 0$ for $i \neq j$ or $P(i)_\mu P(j)^\mu = 0$ for $i \neq j$.

Consider the weak field equation of motion while focusing on the contribution of $\frac{P(0)_\mu P(0)_\nu}{Z(0)}$ so in that case it is necessary to say that we account for only $\frac{1}{4}$ of the gravity if the contribution from all fields, $P(i)$ is equal, in that case, $P(i)^2$ represents time,

$$-\frac{1}{2}\frac{(p(0)_0)^2}{Z(0)},^\mu = -\frac{1}{2}\frac{2p(0)_0}{\sqrt{Z(0)}}\left(\frac{p(0)_0,^\mu}{\sqrt{Z(0)}} - \frac{p(0)_0 Z(0),^\mu}{2Z^{\frac{3}{2}}}\right) \approx \frac{1}{4}\Gamma^\mu_{00} \tag{3.12}$$

$$-\frac{1}{2}\frac{2p(0)_0}{\sqrt{Z(0)}}\left(\frac{p(0)_0,^\mu}{\sqrt{Z(0)}} - \frac{p(0)_0 Z(0),^\mu}{2Z^{\frac{3}{2}}}\right) = \frac{p(0)_0}{\sqrt{Z(0)}}\left(\frac{p(0)_0 Z(0),^\mu}{2Z^{\frac{3}{2}}} - \frac{p(0)_0,^\mu}{\sqrt{Z(0)}}\right) \approx \frac{1}{4}\Gamma^\mu_{00}$$

$$p(0)_0,^\mu \approx {p(0)^\mu},_0 \Longrightarrow \frac{p(0)_0}{\sqrt{Z(0)}}\left(\frac{p(0)_0 Z(0),^\mu}{2Z^{\frac{3}{2}}} - \frac{{p(0)^\mu},_0}{\sqrt{Z(0)}}\right) \approx \frac{Z(0),^\mu}{2Z} \approx \frac{U(0)^\mu}{2} \approx \frac{1}{4}\Gamma^\mu_{00}$$

The latter result is due to $p(0)_1, p(0)_2, p(0)_3$ being neglected but not their derivatives and due to (1). The conclusion of (3.12) is that (3.11) can describe weak gravity, however the scalar fields $P(0), P(1), P(2), P(3)$ in this case, do not represent force fields but Gauge fields.

**Caveat:** An important caveat is that even if the complex formalism of (3.12) is used, 8 complex scalars may not be able to describe gravity. Contribution from additional fields may be needed.

**Caveat:** Do not confuse pseudo-acceleration and non-geodesic acceleration, which is a generalized Reeb class vector, here used to describe the energy of force fields. In general, geodesic curves are not geodesic when mapped to a flat spacetime. The meaning of (3.12) was simply to show the possibility of using non-geodesic curves as the underlying field that drives gravity too, and not only other force fields. This can be achieved by mapping geodesic curves to non-geodesic curves in flat spacetime. Even the complex formalism may not be sufficient as mentioned in the previous caveat note:

$$g_{\mu\nu} = \frac{P(0)_\mu P^*(0)_\nu + P^*(0)_\mu P(0)_\nu}{2} - \sum_{i=1}^{3} \frac{P(i)_\mu P^*(i)_\nu + P^*(i)_\mu P(i)_\nu}{2} \tag{3.13}$$

**Important:** $P(0)_\mu$ is interesting when it is not geodesic also in curved geometry, when it is not only pseudo non-geodesic simply by omission of the Christoffel symbols.

**What is this paper's goal?**

This theory represents energy density as a Lagrangian of accelerations of normalized gradients of scalar fields. The most interesting case is when these scalar fields are over the complex field. The first scalar field $P$ is such that locally $PP^*$, or $P^2$ in the real case, can be considered as a time coordinate through a time-like vector $(PP^*)_{,\mu} = \frac{d(PP^*)}{dx^\mu}$. This paper will also offer a Lagrangian for 3 accelerations which result from three scalar functions, $P1, P2, P3$ such that $P1P1^*, P2P2^*, P3P3^*$ are "local coordinates" of the foliation perpendicular to $(PP^*)_{,\mu}$. The offered Lagrangian will be offered such that $P_{,i}$ , $P1_{,i}$ , $P2_{,i}$ , $P3_{,i}$ need not be perpendicular, by getting rid of the non-perpendicular components in the Lagrangian calculation. Such a definition does not require cumbersome spin-connections and simplifies the theory. The case for two accelerations can be defined in two different ways. One is to define $P_{,\mu}$ and its acceleration $\frac{U_\nu}{2}$ and then to use the Levi-Civita tensor to calculate the plane perpendicular to $P_{,\mu}$ and its acceleration $\frac{U_\nu}{2}$. The second is to use a Lagrangian formalism for two perpendicular planes. The first formalism extends the acceleration from one plane to the other. These planes are known as Lagrangian Planes in the theory of Symplectic Geometry. The second case represents two independent accelerations fields. The theory is also well defined if $PP^*, P1P1^*, P2P2^*, P3P3^*$ integrate to 1 on a spacetime manifold as long as the gradients $P_{,i}$ , $P1_{,i}$ , $P2_{,i}$ , $P3_{,i}$ vanish on a set whose measure is zero. In such a case, it is said that $P_{,i}$ , $P1_{,i}$ , $P2_{,i}$ , $P3_{,i}$ are geometric chronon scalar fields. As a last caveat it is important to distinguish between an acceleration of a normalized velocity as in Special Relativity and the acceleration of a normalized gradient of a scalar field as it is described in this paper. In the following expression for normalized acceleration of a moving frame, the dot above the coordinates means derivative in relation to time,

$$\frac{d\frac{\dot{x}^\mu}{c\sqrt{1-\frac{V^2}{c^2}}}}{cd\tau} = \frac{d\frac{\left(1,\frac{\dot{x}^1}{c},\frac{\dot{x}^2}{c},\frac{\dot{x}^3}{c}\right)}{\sqrt{1-\frac{\dot{x}^{1^2}+\dot{x}^{2^2}+\dot{x}^{3^2}}{c^2}}}}{\sqrt{1-\frac{\dot{x}^{1^2}+\dot{x}^{2^2}+\dot{x}^{3^2}}{c^2}}cdt} = \frac{\frac{1}{c^2}(0,\ddot{x}^1,\ddot{x}^2,\ddot{x}^3)}{1-\frac{\dot{x}^{1^2}+\dot{x}^{2^2}+\dot{x}^{3^2}}{c^2}} + \frac{\frac{1}{c^2}(c,\dot{x}^1,\dot{x}^2,\dot{x}^3)\left(\frac{\dot{x}^1\ddot{x}^1+\dot{x}^2\ddot{x}^2+\dot{x}^3\ddot{x}^3}{c^2}\right)}{\left(1-\frac{\dot{x}^{1^2}+\dot{x}^{2^2}+\dot{x}^{3^2}}{c^2}\right)^2} \tag{3.14}$$

Which is not solely dependent on the normalized velocity as is, $\frac{\dot{x}^{\mu}}{c\sqrt{1-\frac{V^2}{c^2}}}$.

From (3) a generalization for multiple event fields $\int_{\Omega} P(i)P^*(i)d\Omega = 1$ where $P(i)P^*(i)$ is no longer a Geroch function is

$$A_{\mu\nu}(i) = \left(\frac{P_{\mu}(i)}{\sqrt{Z(i)}}\right)_{,\nu} - \left(\frac{P_{\nu}(i)}{\sqrt{Z(i)}}\right)_{,\mu} \text{ and } i \in \mathbb{N} \tag{3.15}$$

And the action is

$$L = \frac{1}{8}(\textstyle\sum_{i,j}(A_{\mu\nu}(i)A^{*\mu\nu}(j) + A^{*}_{\mu\nu}(i)A^{\mu\nu}(j))(P(i)P^*(j) + P^*(i)P(j))\sqrt{-g} \tag{3.16}$$

It will be clearer, as this paper develops, that the mixed terms in (3.16) are due to the non-covariant classical limit of the electrostatic field, specifically $\|E(1) + E(2)\|^2 = \|E(1)\|^2 + \|E(2)\|^2 + 2\,E(1)\cdot E(2)$ for two non-covariant electric fields $E(1)$ and $E(2)$.

With Einstein Hilbert Langrangian, which should not change also when considering other fields such as will be seen in (64.b), (64.01). This is because other fields are emergent from time.

$$L = \frac{1}{2}\textstyle\sum_{i,j} R(P(i)P^*(j) + P^*(i)P(j))\sqrt{-g} \tag{3.16.1}$$

Where g is the determinant of the metric tensor.

**Important:** By (3.16) and (3.16.1) even if the embedding spacetime of events $P(i)$ is deterministic, gravity becomes quantum, however, there is another way to describe quantum gravity as in (3.11) and as will be seen in (65), where (65) is possibly the action of massive gravity.

In the real case and from (3),

$$L = \frac{1}{2}\textstyle\sum_{i,j} A_{\mu\nu}(i)A^{\mu\nu}(j)P(i)P(j)\sqrt{-g} = \frac{1}{4}\sum_{i,j} F_{\mu\nu}(i)F^{\mu\nu}(j)P(i)P(j)\sqrt{-g} \tag{3.17}$$

Or as energy density

$$\frac{c^4}{8\pi K}\frac{1}{4}\textstyle\sum_{i,j} F_{\mu\nu}(i)F^{\mu\nu}(j)P(i)P(j)\sqrt{-g} \tag{3.18}$$

**Mass mechanism for null acceleration fields**

Suppose $\frac{U_{\mu}(2)}{2}$ is a null vector, the condition $\frac{U_{\mu}}{2}\frac{P^{\mu}}{\sqrt{Z}} = 0$ which is mentioned before (2) or in the complex form $\frac{U_{\mu}}{2}\frac{P^{*\mu}}{\sqrt{Z}} = 0$, dictates in this case that $P_{\lambda}$ cannot be the gradient of a Geroch function [2] because it is easy to see that in this specific case $P_\lambda$ is space-like and not time-like. However (3.16) allows an energy density to be non-zero through the following mass mechanism,

$$\frac{c^4}{8\pi K}\frac{1}{2}\left(\frac{U_\mu(2)}{2}\frac{U^{*\mu}(1)}{2}+\frac{U_\mu^*(2)}{2}\frac{U^\mu(1)}{2}\right)\sqrt{-g} \tag{3.19}$$

Where $\frac{U_\mu(1)}{2}$ is the Reeb class vector, i.e. acceleration of the unit vector $\frac{P^\mu(1)}{\sqrt{Z(1)}}$, not the usual Reeb vector where $PP^*$ is a Geroch function [2] or $P^2$ is a Geroch function. It will be evident in (4) that the field $\frac{U_\mu(1)}{2}$ must have a zero-sum divergence. In the real case the integration of $U;_k^k(1)$ must be zero, which means the field will have a zero-sum electric charge. This condition limits $\frac{U_\mu(1)}{2}$ in such a case to appear as a short-lived dipole but to give mass to a null field $\frac{U_\mu(2)}{2}$.

**Electro-gravity**

The action of gravity is defined as: $Action = Min \int_\Omega \left(R - \frac{1}{4ב}U^k U_k\right)\sqrt{-g}\, d\Omega$

The Euler Lagrange equations by the metric $g_{\mu\nu}$, by the scalar field of time P yield, Appendix A or [13]:

$$\frac{1}{4ב}\left(U_\mu U_\nu - \frac{1}{2}g_{\mu\nu}U_\lambda U^\lambda - 2U^k;_k\frac{P_\mu P_\nu}{Z}\right) = R_{\mu\nu} - \frac{1}{2}Rg_{\mu\nu} \tag{4}$$

$$W^\mu;_\mu = \left(-4U^k;_k\frac{P^\mu}{Z} - 2\frac{Z_\nu P^\nu}{Z^2}U^\mu\right);_\mu = 0$$

**Caveat:** (3.11), (3.16), (3.16.1) and possibly also (65) entail quantum gravity. (4) is therefore an approximation. The embedding spacetime is not the physically accessible events.

It is easy to prove without the right hand side that $\frac{1}{4ב}\left(U_\mu U_\nu - \frac{1}{2}g_{\mu\nu}U_\lambda U^\lambda - 2U^k;_k\frac{P_\mu P_\nu}{Z}\right);^\nu = 0$ see Appendix B or [13]. (4) assumes $ב = 1$.

Consider $\rho = \frac{1}{2}U^k;_k$ to be stationary along $p^\mu$, with local coordinates such that only $p_0$ is numerically significant. We will neglect all small terms that are multiplied by $U^\mu$ and its derivatives. with $2\left(\frac{p^\mu}{Z}\right);_\mu \approx \frac{Z_\nu p^\nu}{Z^2}$ and $\left(\frac{Z_\nu p^\nu}{Z^2}\right);_\mu U^\mu \ll 1$, the first approximation is the result of $2\left(\frac{p^\mu}{Z}\right);_\mu \approx 2\left(\frac{p^0}{p^0 p_0}\right),_0$ and $\frac{Z_\nu p^\nu}{Z^2} \approx \frac{p^0(p^0 p_0),_0}{(p^0 p_0)^2} \approx 2\left(\frac{p^0}{p^0 p_0}\right),_0$ the last approximation $\left(\frac{Z_\nu p^\nu}{Z^2}\right);_\mu U^\mu \ll 1$ is due to $\frac{Z_\nu p^\nu}{Z^2} \approx \frac{p^0(p^0 p_0),_0}{(p^0 p_0)^2}$ and the fact that $U_\mu$ is spacelike. Then restricting $k \in \{1,2,3\}$, when the metric tensor is nearly diagonal $\eta_{\mu\nu}$, we have,

$$k \in \{1,2,3\},\ g_{\mu\nu} \approx \eta_{\mu\nu} \Longrightarrow \left(-4U^k;_k\frac{P^\mu}{Z} - 2\frac{Z_\nu P^\nu}{Z^2}U^\mu\right);_\mu = 0 \Longrightarrow 2\rho \approx U^k,_k \tag{4.1}$$

Dynamics: (4.1) implies the dynamics of the electric field of points of divergence $U^k{}_{;k} \neq 0$.

**Theorem 1**: If non-geodesic curves are prescribed to motion in material fields, then zero Einstein tensor implies $\frac{1}{2}U_\mu = 0$, i.e. $R_{\mu\nu} - \frac{1}{2}Rg_{\mu\nu} = 0 \Longrightarrow \frac{1}{2}U_\mu = 0$ i.e. geodesic motion.

**Proof**: We contract both sides of (4) with $U^\mu U^\nu$ so $\left(U_\mu U_\nu - \frac{1}{2}g_{\mu\nu}U_\lambda U^\lambda - 2U^k{}_{;k}\frac{P_\mu P_\nu}{Z}\right)U^\mu U^\nu = 0 \Longrightarrow U_\lambda U^\lambda = 0$ because $U^\mu P_\mu = 0$ and now we contract both sides of (4) with $\frac{P^\mu P^\nu}{Z}$ so we have $\frac{P^\mu P^\nu}{Z}\left(U_\mu U_\nu - \frac{1}{2}g_{\mu\nu}U_\lambda U^\lambda - 2U^k{}_{;k}\frac{P_\mu P_\nu}{Z}\right) = -\frac{1}{2}U_\lambda U^\lambda - 2U^k{}_{;k} = 2U^k{}_{;k} = 0$ because $U_\lambda U^\lambda = 0$ and $\frac{P^\lambda P_\lambda}{Z} = 1$ so we get $U_\mu U_\nu - \frac{1}{2}g_{\mu\nu}U_\lambda U^\lambda - 2U^k{}_{;k}\frac{P_\mu P_\nu}{Z} = U_\mu U_\nu = 0 \Longrightarrow U_\mu = 0$. In other words, motion must be geodesic and we are done.

Remember $\frac{U_\mu}{2} = \frac{a_\mu}{c^2}$ as acceleration and the equation of gravity by Einstein, using the dust energy momentum tensor from General Relativity,

$$\frac{8\pi K}{c^4}T_{\mu\nu} = R_{\mu\nu} - \frac{1}{2}Rg_{\mu\nu} \tag{5}$$

in (-,+,+,+) convention, we will use (5) further on, to show unique gravity by electric charge.

**A surprising result (6)-(13) and explanation of why charge generates gravity and anti-gravity**

**Critical:** This section requires a high concentration from the reader. It uses assumption 1 and its interpretation (1). We will start by noticing that $\frac{U_\lambda}{2}$ is an acceleration of a unit vector and as such, it can be written as an acceleration of some scaled velocity $\frac{V_\lambda}{c}$ where $c$ is the speed of light.

$$\frac{1}{4}U^\lambda U_\lambda = \frac{a^\lambda a_\lambda}{c^4} \tag{6}$$

(6) compared to Einstein's tensor means that the energy density in old physics terms can be seen when the metric convention is (-, +, +, +) as:

$$\frac{a^\lambda a_\lambda}{8\pi K \beth} = EnergyDensity \Longrightarrow \frac{8\pi K}{c^4}EnergyDensity = \frac{a^\lambda a_\lambda}{\beth c^4} = \frac{1}{4\beth}U^\lambda U_\lambda \tag{7}$$

Where $\beth = 1$ relates non geodesic acceleration to geometry, direct outcomes of (7) will be shown in (13) and (43). (7) means that the energy of the classical non-covariant electric field $E$ must be hidden in a very weak acceleration field and restricting $k \in \{1,2,3\}$ yields,

$$\frac{a^k a_k}{8\pi K \beth} \approx \frac{1}{2}\varepsilon_0 E^2 \tag{8}$$

$\varepsilon_0$ is the permittivity of vacuum, K is Newton's constant of gravity, which means

$$|a|^2 = 4\pi K\varepsilon_0 ב E^2 \tag{9}$$

and

$$\|a^\mu\| = \sqrt{4\pi K\varepsilon_0 ב}\|E\| \tag{10}$$

Indeed, a very weak acceleration if $ב = 1$. However, there is a surprise:

$$\frac{1}{4ב}(U_\mu U_\nu - \frac{1}{2}g_{\mu\nu}U_\lambda U^\lambda - 2U^k{}_{;k}\frac{P_\mu P_\nu}{Z}) = R_{\mu\nu} - \frac{1}{2}Rg_{\mu\nu} \tag{11}$$

Means that $\frac{1}{2ב}U^k{}_{;k} = \frac{a^k{}_{;k}}{c^2} = \sqrt{\frac{4\pi K\varepsilon_0 ב}{ב^2}}\frac{\rho}{\varepsilon_0 c^2} = \sqrt{\frac{4\pi K}{ב\varepsilon_0}}\frac{\rho}{c^2}$ where $\rho$ is charge density.

Now remember the term $\frac{1}{4ב}(-2U^k{}_{;k}\frac{P_\mu P_\nu}{Z})$ and the relation $\frac{P^\mu P^\nu}{Z} \approx \frac{V^\mu V^\nu}{c^2}$ where $\frac{P^\mu}{\sqrt{Z}}$ is equivalent to a normalized velocity vector $\frac{V^\mu}{c}$, in Special Relativity $V^\mu = \frac{(c,v_x,v_y,v_z)}{\sqrt{1-v^2/c^2}}$, so we get

$$\frac{1}{8\pi K}\frac{U^\mu{}_{;\mu}}{2ב}\frac{P^\mu P^\nu}{Z^2} \approx \frac{1}{8\pi K}\sqrt{\frac{4\pi K ב}{ב^2\varepsilon_0}} \cdot \frac{\rho_{charge}V^\mu V^\nu}{c^4} = \frac{1}{8\pi K c^4}\sqrt{\frac{4\pi K}{ב\varepsilon_0}}\rho_{charge}V^\mu V^\nu \tag{12}$$

But that can only mean that charge density behaves like mass density except for the fact that $\frac{P^\mu}{\sqrt{Z}}$ is not geodesic and is not the velocity of the charge and therefore from (12) we have:

$$M = \frac{Q}{\sqrt{16\pi K\varepsilon_0 ב}}, ב = 1 \Rightarrow \pm 1\ Coulombs \Leftrightarrow \sim \pm \mathbf{5.802135215 * 10^9\ Kg} \tag{13}$$

Assuming $ב = 1$ where $\varepsilon_0$ is the permittivity of vacuum and K is Newton's constant of gravity, M is a gravitational mass. The following is a surprising result due to the distribution of electric charge in the real world, which appears to be at probabilistically non-local centers of charged particles.

**Why does electric charge have zero inertial mass?**

**Crucial:** $\frac{1}{4}2U^\lambda{}_{;\lambda}\frac{P_\mu P_\nu}{Z}$ in which $\frac{1}{2}U^\lambda{}_{;\lambda}$ represents electric charge, does not behave as inertial mass not only because the unit vector $\frac{P_\mu}{\sqrt{|Z|}}$ is not geodesic but because $P_\mu$ is derived from a Geroch function [2]. In theorem 3, it will be seen that when restricted to a 3D foliation perpendicular to $P_\mu$ and where $U^\lambda{}_{;\lambda} \neq 0$, $U_\mu$ has points of drains and sources. Taking the quantum approach around such points, then due to symmetry, $P_\mu$ is not accelerated at such points, in other words the average of $U^\lambda{}_{;\lambda}\frac{P_\mu P_\nu}{Z}$ in a small sphere in the foliation perpendicular to $P_\mu$ is not affected by the velocity of the center of an electric charge and therefore charge must have zero inertial mass unlike its mass density $-\frac{1}{4}U^{\lambda}U_{\lambda}$ in (+,-,-,-) metric convention or $\frac{1}{4}U^{\lambda}U_{\lambda}$ in (-,+,+,+) convention. Gravitational mass is not identical to inertial mass. Equation (4) should be written as an average equation when it purports to describe quantum properties of small volumes as $r \ll 1$.

$$\frac{1}{4}\left(U_{\mu}U_{\nu} - \frac{1}{2}g_{\mu\nu}U_{\lambda}U^{\lambda} - 2U^{k};_{k}\frac{\overline{P_{\mu}P_{\nu}}}{Z}\right) = R_{\mu\nu} - \frac{1}{2}Rg_{\mu\nu} \qquad (13.01)$$

Fig. 2. Describes a particle-like charge center in motion in relation to the Geroch [2] time curves.

**Note:** The term $\frac{\overline{P_{\mu}P_{\nu}}}{Z}$ means we treat the electric charge is if it is concentrated in one point in the 3D foliation perpendicular to $P_{\mu}$ so we treat charge density as a Dirac delta function. This idea stems from the fact that charge is quantized which is strong evidence of the correctness of this assumption. The outcome of this assumption led us to the inevitable conclusion that electric charge, excluding the energy of its field, cannot have an inertial mass. There is still a legitimate question whether the unit vector $\frac{P_{\mu}}{\sqrt{Z}}$ can be affected by other points of non-zero divergence $U_{\lambda};^{\lambda}$. If $\|U_{\lambda}U^{\lambda}\| \gg 1$ near the center of the electric charge, then it is reasonable say that such an influence on $\frac{P_{\mu}}{\sqrt{Z}}$ at the center of a charge see Fig. 2. by another charge should be negligible and tend to zero. Near such points $\|U_{\lambda}U^{\lambda}\|$ is very large.

**Gravity is emergent from the impedance of charge at the Planck scale**

This section, like the previous one, also refers to Fig. 2. For the sake of simplicity, this discussion is limited to real numbers. The acceleration of $\frac{P_{\mu}}{\sqrt{Z}}$ is $\frac{U_{\mu}}{2} = \left(\frac{P_{\mu}}{\sqrt{Z}}\right);^{\nu}\frac{d\tau}{dx^{\nu}} = \left(\frac{P_{\mu}}{\sqrt{Z}}\right);^{\nu}\frac{P_{\nu}}{\sqrt{Z}}$ but at the center of charge $\frac{P_{\mu}}{\sqrt{Z}}$ must be geodesic and therefore $\frac{U_{\mu}}{2} = 0$. Due to theorem 3, which will be presented, points of $\frac{U_{\mu}}{2}$ divergence must exist when reduced to 3D foliations perpendicular to $P_{\mu}$, and we can describe $\frac{U_{\mu}}{2}$ in local spherical coordinates, $\frac{U_{r}}{2}$ where $r$ is radius and in this case only $\frac{P_{0}}{\sqrt{Z}} \neq 0$. So, either $U_{r}$ points towards the point of a drain divergence or away from the point of a source divergence. The divergence when $r \ll 1$, can be written as $Divergence(\frac{U_{r}}{2}) \approx \left(\frac{U_{r}}{2} - 0\right)\frac{1}{r}$ which is proportional to $r$ and we also assume that $r$ functions as an atom of length. Now contract (4) by $\frac{P^{\mu}P^{\nu}}{Z}$ and we have $(\pm)\frac{1}{2}U^{\lambda};_{\lambda} - \frac{1}{8}U^{\zeta}U_{\zeta} = (R_{\mu\nu} - \frac{1}{2}g_{\mu\nu}R)\frac{P^{\mu}P^{\nu}}{\sqrt{Z}}$ in (-,+,+,+) metric convention. We can already see that $(\pm)\frac{1}{2}U^{\lambda};_{\lambda} \approx \left(\frac{U_{r}}{2} - 0\right)\frac{1}{r}$ which means that up to scaling by a factor $\frac{1}{r}$ gravity due to divergence of $\frac{U_{r}}{2}$ becomes proportional to $\frac{U_{r}}{2}$. In plain English, a volumetric curvature perpendicular to the time axis $\frac{P_{\mu}}{\sqrt{Z}}$ due to $Divergence(\frac{U_{r}}{2})$ is opposite in direction to $\frac{U_{r}}{2}$. We also have the contribution to gravity from $-\frac{1}{8}U^{\zeta}U_{\zeta}$ also in (-,+,+,+) metric convention. Instead of seeing $(\pm)\frac{1}{2}U^{\lambda};_{\lambda}$ as a residual gravity because the action by $U_\zeta$ is $\frac{1}{4}U^\zeta U_\zeta\sqrt{-g}$ we can see $-\frac{1}{8}U^\zeta U_\zeta$ as a residual contribution to gravity while the anti-gravity and gravity due to $(\pm)\frac{1}{2}U^{\lambda};_{\lambda} \approx \left(\frac{U_r}{2}-0\right)\frac{1}{r}$ is an impedance by volumetric curvature. The claim that $\frac{1}{r}$ is in the Planck scale has evidence that will be shown in (43) in "7. The mass hierarchy – a possible exponential relationship". Putting it all together we have the next section.

## Einstein's field equations from the Geometric Chronon Field Theory

In this section it will be shown that an impedance response of the metric to charge as a source or a drain of non-geodesic acceleration derives the Einstein – Hilbert – Grossmann action.

For this section it is recommended to read John Baez's paper [14].

Unlike John Baez's metric convention, here we use the (+1,-1,-1,-1) metric convention. We get the time component for the Energy Momentum tensor $T_{\mu\nu}$:

$$R_{tt} = T_{tt} - \frac{1}{2}T^{a}_{\alpha}g_{tt} \tag{13.01.1}$$

From (4),

$$\begin{aligned}
R_{\mu\nu} &= \frac{1}{4}\left(U_\mu U_\nu - \frac{1}{2}g_{\mu\nu}U_\lambda U^\lambda - 2U_\lambda;^\lambda\frac{P_\mu P_\nu}{Z}\right) - \frac{1}{8}\left(U_\lambda U^\lambda - 2U_\lambda U^\lambda - 2U_\lambda;^\lambda\right)g_{\mu\nu} \\
&= \frac{1}{4}\left(U_\mu U_\nu - \frac{1}{2}g_{\mu\nu}U_\lambda U^\lambda - 2U_\lambda;^\lambda\frac{P_\mu P_\nu}{Z}\right) - \frac{1}{8}\left(-U_\lambda U^\lambda - 2U_\lambda;^\lambda\right)g_{\mu\nu} \\
&= \frac{1}{4}\left(U_\mu U_\nu - \frac{1}{2}g_{\mu\nu}U_\lambda U^\lambda - 2U_\lambda;^\lambda\frac{P_\mu P_\nu}{Z}\right) + \frac{1}{4}\left(\frac{1}{2}U_\lambda U^\lambda + U_\lambda;^\lambda\right)g_{\mu\nu} \\
&= \frac{1}{4}\left(U_\mu U_\nu - 2U_\lambda;^\lambda\frac{P_\mu P_\nu}{Z}\right) + \frac{1}{4}U_\lambda;^\lambda g_{\mu\nu} = \frac{1}{4}\left(U_\mu U_\nu - 2U_\lambda;^\lambda\frac{P_\mu P_\nu}{Z} + U_\lambda;^\lambda g_{\mu\nu}\right)
\end{aligned} \tag{13.01.2}$$

$\frac{U_\mu}{2}$ is the acceleration of $\frac{P_\mu}{\sqrt{\|Z\|}}$ in the real case.

Notice the term $\frac{1}{4}\left(-2U_\lambda;^\lambda\frac{P_\mu P_\nu}{Z} + U_\lambda;^\lambda g_{\mu\nu}\right) \approx -\frac{1}{4}U_\lambda;^\lambda\frac{P_\mu P_\nu}{Z}$ where $-c^2\frac{U_\lambda;^\lambda}{2} = -\frac{a_\lambda;^\lambda}{2}$ such that $a_\lambda$ is a non-geodesic acceleration.

Returning back to John Baez's paper,

$$R_{tt} = \frac{\ddot{V}}{V} = \frac{4\pi r^2\ddot{r}}{\frac{4}{3}\pi r^3} = 3\frac{\ddot{r}}{r} \tag{13.01.3}$$

Notice that $\frac{\ddot{r}}{r}$ is the divergence of a pseudo-acceleration by converging geodesic curves along the time direction, where $V$ is a volume of ball with test particles at small radius $r$ starting from $\dot{r}(r) = 0$.

Recall that by (13.01), the center of charge carrier has a geodesic $\frac{P_\mu}{\sqrt{\|Z\|}}$ and that $\frac{U_\mu}{2}$ is the acceleration of $\frac{P_\mu}{\sqrt{\|Z\|}}$, which satisfies the conditions for theorem 3 in Chronon Field Theory of a vanishing rotor when restricted to 3D foliations perpendicular to $\frac{P_\mu}{\sqrt{\|Z\|}}$. Then when the radius $r \ll 1$, checking $-2U_\lambda;^\lambda \frac{P_t P_t}{Z} + U_\lambda;^\lambda g_{tt}$, $-2U_\lambda;^\lambda \frac{P_t P_t}{Z} + U_\lambda;^\lambda g_{tt} \cong -U_\lambda;^\lambda g_{tt}$ and without loss of generality because $-\frac{1}{2}U_\lambda;^\lambda$ can be positive or negative, in the general case where $-\frac{(U_\lambda;^\lambda + U^*{}_\lambda;^\lambda)}{4}$ replaces $-\frac{1}{2}U_\lambda;^\lambda$ , we can write:

$$\frac{1}{2}c^2 \left\| -\frac{(U_\lambda;^\lambda + U^*{}_\lambda;^\lambda)}{4} \right\| + other\ terms = \frac{1}{2}\left\| -\frac{(a_\lambda;^\lambda + a^*{}_\lambda;^\lambda)}{2} \right\| + other\ terms = \frac{1}{2}\frac{\|a\|}{r} + other\ terms = -Constant * 3\frac{\ddot{r}}{r} = -Constant *\ R_{tt} \qquad (13.01.4)$$

Where the divergence of the non-geodesic acceleration field is $\frac{\|a\| - 0}{r}$ because $r = 0 \Longrightarrow \|a\| = 0$. Then we have established, where c is the speed of light,

$$\frac{1}{2}c^2 \left\| -\frac{(U_\lambda;^\lambda + U^*{}_\lambda;^\lambda)}{4} \right\| + other\ terms = -Constant * 3\frac{\ddot{r}}{r} = -Constant *\ R_{tt} \qquad (13.01.5)$$

Or more precisely

$$-c^2 \frac{(U_\lambda;^\lambda + U^*{}_\lambda;^\lambda)}{4} + 2 * other\ terms = -Constant * 6 * \frac{\ddot{r}}{r} = -2 * Constant *\ R_{tt} \qquad (13.01.6)$$

The last equation means that indeed gravity in the electromagnetic case, is due to an impedance response to the divergence of the electric field + other terms. Now consider,

$$\frac{\ddot{r}}{r} = \frac{\ddot{r}(r) - \ddot{r}(0)}{r} \qquad (13.01.7)$$

This term is the divergence of the pseudo acceleration which proves that indeed an impedance response of the metric to charge $\frac{(U_\lambda;^\lambda + U^*{}_\lambda;^\lambda)}{4}$ must be Ricci curvature. Due to the principle of parsimony, the simplest action that yields such a term in the Euler-Lagrange equations is the action $-\frac{U^\lambda U_\lambda}{4}$ or in the complex case $-\frac{U^{*\lambda}U_\lambda + U^\lambda U^*{}_\lambda}{8}$. The extremal variation of $-\frac{U^\lambda U_\lambda}{4}$ yields in the real case $-\frac{1}{2}U_\lambda;^\lambda \frac{P_\mu P_\nu}{Z} + other\ terms$. The other terms therefore generate residual gravity and the divergence $-\frac{1}{2}U_{\lambda};^{\lambda}$ is either positive or negative, which means both gravity and antigravity are inevitable, while macroscopic gravity is weak because it is a residual gravity.

**Fig. 2.** – An over simplified charge in x-spatial horizontal and local Geroch-time axes $\frac{P_{\mu}}{\sqrt{|Z|}}$

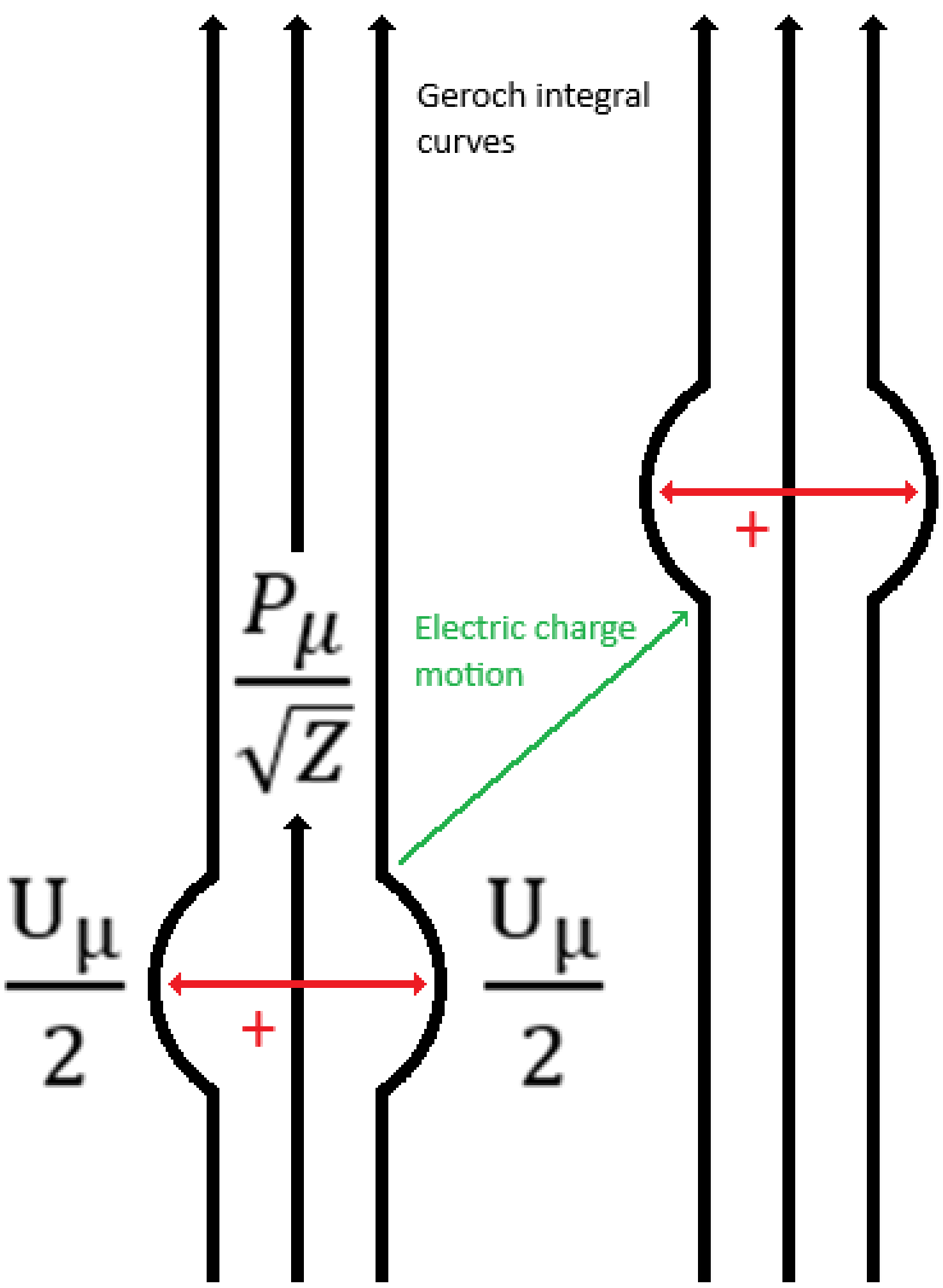

**The gravitational field due to positive electric charge near a charged sphere**

Consider charging a sphere of radius 0.1 Meters, under V=1,000,000 volts then by (13),

$4\pi\varepsilon_0 * 1{,}000{,}000 * 0.1 = Voltage * Capacitance = Charge$ ≅1.11265E-05 Coulomb.

Dividing by $M = \frac{Q}{\sqrt{16\pi K\varepsilon_0}}$ yields ~64557.46071 Kg of gravity and 0.000430876 M/Sec$^2$ acceleration. Such a low acceleration is orders of magnitude less than any acceleration that can be measured due to the electric interaction and is therefore hard to measure.

**Challenge:** The biggest problem is that the detector is not allowed to polarize. Any such polarization is a gravitational dipole that will cancel out the effect of the external charge-based gravitational field.

**Questions and answers to test the understanding of the reader**

The following are questions and answers to test the understanding of the reader.

1) Is this quantity $U_{\lambda};^{\lambda}$ conserved?

2) Why is this charge measured in Coulombs?

3) When an electrical potential difference of 1kV is applied on a 1 millifarad capacitor (+/-) 1 C electrical appears on each of the two electrodes but the total remains zero. Along with the above conventional charge, how much of this new charge is also induced in the capacitor?

4) Does this new charge in motion experience the conventional Lorentz force?

5) How are the old results of Milliken's charged oil drop experiment and the thousand other confirmations using standard electron mass tally with the proposed new gravitational effects?

Answers:

1) Yes. The divergence of the acceleration field appears in Euler Lagrange equations not in the action itself which means the sum of charge must be always zero.

2) Because $\frac{a^2}{8\pi K} = -\frac{U^{\lambda}U_{\lambda}}{4}\frac{c^4}{8\pi K}$ is energy density and the non-covariant classical energy density of the electric field E is $\frac{1}{2}\varepsilon_0\|E\|^2$ so $\frac{a^2}{8\pi K} \approx \frac{1}{2}\varepsilon_0\|E\|^2 \Longrightarrow a \approx \sqrt{4\pi K\varepsilon_0}$. K is Newton's gravity constant, $\varepsilon_0$ is the permittivity of vacuum. In complex formalism when the imaginary part cannot be neglected as in the static case, $(0, E) \approx \frac{c^2}{\sqrt{4\pi K\varepsilon_0}}\frac{1}{4}\left(U_\mu + U_\mu^*\right)$ and $\|E\|^2 \approx \frac{c^4}{4\pi K\varepsilon_0}\frac{1}{4}(Re(U_\mu)^2)$.

   The basic assumption of the theory is that electromagnetic energy is stored in an acceleration field of a unit vector. This assumption is due to the idea that matter must be derived from the physically accessible space time and cannot be a separate field from matter, which is coupled to matter. The motivation for this idea is the principle of parsimony. The other motivation is Einstein's principle that even matter must emerge out of geometry and therefore not only Ricci curvature is geometrical but also the energy momentum tensor must emerge from geometry. The simplest form of geometrical information which is not already included in General Relativity is non-geodesic acceleration of curves. By the principle that even energy must come out of the physically accessible spacetime and by Einstein's geometrical principle, the energy of the electromagnetic field must be derived from accelerations of curves of time, however, a preferred time coordinate violates the principle of relativity which leaves only one option that such time must be a scalar and not a coordinate. A scalar of time can be easily defined on a causal big bang manifold as the maximal time between each event and back to the big bang as a limit. Such a requirement is, however, redundant if the scalar function can come out of a minimal action principle. Notice that in a big bang FRWL geometry an event can be connected to the big bang limit or to any other 3D foliation of spacetime by more than one maximal proper time curve. This means that even by using only geodesic curves, it is not guaranteed that the gradient of the maximal proper time, as a scalar function, will be geodesic. The purpose of an action based on acceleration of a gradient, however, is to allow non-zero acceleration with or without defining the underlying scalar function as the original Geroch function. In general, a Geroch function need not be geodesic and need not be even unique. It is merely a Morse function in spacetime. Its value must come out of boundary conditions and a minimum action just as in any other physics PDE.

3) The same amount.

4) Yes. The reason is that the acceleration field can be viewed via local summation of fields. The Geroch function is universal, but its acceleration can be decomposed into a sum of acceleration vectors. An easy mistake is to add Geroch functions when summing two fields and forgetting that at each event in spacetime the Geroch function is one. Decomposition of the Geroch function is also possible but not through thinking of different charge centers but of event wave functions / chronons, the building blocks of the time function.

5) As you can see after (13) in the paper, the mass of the charge itself, not of its energy, must be zero. For example, there must be no difference between the mass of the electron and the mass of the positron. As for gravitational effects there are two reasons for not detecting charge-based gravity and anti-gravity.

   5.1) It is easy to show that forces due to charge-based gravity are many orders of magnitude less than the electrostatic force. For example, in a parallel plates capacitor with 1,000,000 volts with 1mm gap, the gravitational acceleration is 4.305 cm / Sec^2. The ratio between this acceleration to 1g is 4.305 * 10^-2 / 9.807 ~= 0.00438972 g and 1,000,000 volts over 1mm gap is an immense field.
   5.2) Dielectric materials even with a low dielectric constant have sufficient local charge to generate oppositely aligned dipoles to cancel out any Bondi dipole external field.

**Usable gravitational energy**

**Note:** It is easy to see that in a weak electric field, if all the energy is of an electric field, then when taking (13) into account, the component of pseudo-gravitational acceleration $g$ by the divergence of the classical non-covariant electric field $E$ is

$$\|a_\mu\| \approx \sqrt{4\pi K\varepsilon_0}\|E\| \Longrightarrow \|g\| \approx \frac{1}{2}\|a_\mu\|, \sqrt{\pi K\varepsilon_0}\|E\| \approx -g \approx \frac{1}{2}a_\mu \qquad (13.02)$$

In other words, at rest in relation to an electric charge, the acceleration of $\|a^\mu\|$ is partly due to the metric change, because of charge-based gravity (13). Consider that the usable gravitational energy depends on $g$ as it is not part of a preserved gravitational field because charge can be annihilated,

$$\frac{1}{4}\frac{1}{8\pi K}\int\|a^\mu\|^2 dVolume = Usable\ gravitational\ energy \qquad (13.03)$$

(13.03) may look vague right now, however, when describing decay processes of charged particles, it is inevitable that the same portion $\frac{1}{4}$ of the added and subtracted area around negative and positive charge, should account for usable gravitational energy. The term "usable energy" is a concept from thermodynamics. Also note that the sign of $g$ is opposite to the sign of the weak electric acceleration.

**Crucial**: As a remark on (13.03), the factor $\frac{1}{4}$ is very important. It will be assumed in this paper that when a particle like the Muon decays. A quarter of the gravitational energy of the Muon can be used to create the neutrinos and the electrons which the products of Muons decay. Notice that in this case, the factor $\frac{1}{4}$ is generalized to the entire gravitational field. Generalizing the factor $\frac{1}{4}$ also to $a_\mu;^\mu$ instead of the factor $\frac{1}{2}$ is not trivial but does make sense if the field decays in particle annihilations. In such a case, $\frac{1}{4}$ can be seen as Impedance [15] Matching. As gravity drops to a quarter of its value, a portion of this energy can be transformed into other forms of energy.

**Caveat:** $\frac{P^\mu}{\sqrt{Z}}$ is not geodesic unless $\frac{1}{2}U_\mu = 0$. So $\rho_{charge}\frac{P_\mu P_\nu}{Z}$ does not behave as inertial mass.

**Separation of charge in the Bullet cluster – Possible misinterpretation as Dark Matter**

The center of the collision in the Bullet Cluster has a strong magnetic field between 0.2 and 2.8 micro-Gauss. The collision is at relative velocity of about 3000 Km/Sec. See Bullet Cluster [16].

In the special relativistic case, the Larmor radius is,

$$rL = \frac{momentum\perp}{|q|B} = \frac{\gamma m V\perp}{|q|B} \qquad (13.04)$$

where $q$ is charge, B is the magnetic field, $V\perp$ is the velocity of the charge, which is perpendicular to B, $\gamma^{-1} = \sqrt{1 - V^2/c^2}$ where V is the 3D norm of the velocity. $1 - \gamma \ll 1$

For the electron:

$$rL \cong \frac{mV\perp}{|q|B_{gauss}10^{-4}} \leq \frac{9.1093837 \times 10^{-31}*3*10^{6}}{1.60217663 \times 10^{-19}*2*10^{-7}*10^{-4}} \cong 852{,}844.5175\ Meters = 852.84451752\ Km. \tag{13.05}$$

In astronomical perspective, even if this calculation is mistaken by 6 orders of magnitude due to electron acceleration by EM fields, it is still a small scale. This result means that only heavy element atoms, rocks, meteorites and galaxies that could escape the collision would have a net positive charge and indeed, the Dark Matter effect in the Bullet Cluster does coincide with a small portion of known baryonic matter. Due to friction, this portion must be positively charged due to the magnetic field that trapped free electrons. With the prediction of (13) of 5.802135 * $10^9$ Kg * Coulomb$^{-1}$ at least part of the Dark Matter effect of the Bullet Cluster can be explained as generated by positive charge. This claim by no means rules out the existence of weakly interacting particles, however it does mean they do not generate all the Dark Matter effect.

Electric field to acceleration from far observer coordinates – the following is not the way to derive the relation between gravitational mass and charge, not only because charge is coupled to a non-geodesic bivector, however, it does serve as an indication that the results are correct.

$$\frac{e}{4\pi\varepsilon_0 r^2}(4\pi\varepsilon_0 K)^{\frac{1}{2}} = \frac{c^2}{r} \tag{13.1}$$

Where the right-hand side stands for acceleration or the norm of the Reeb class vector multiplied by the squared speed of light. 'e' is the charge of the electron, $\varepsilon_0$ the permittivity of vacuum and $K$ is the gravity constant of Newton. (13.1) is a result of (10).

$$\frac{e}{c^2}\left(\frac{K}{4\pi\varepsilon_0}\right)^{\frac{1}{2}} = r \tag{13.2}$$

We will equate the right-hand side to the Schwarzschild radius of some mass,

$$\frac{e}{c^2}\left(\frac{K}{4\pi\varepsilon_0}\right)^{\frac{1}{2}} = \frac{2Km}{c^2} \tag{13.3}$$

From which

$$\mathrm{e}\left(\frac{1}{16\pi \mathrm{K}\varepsilon_0}\right)^{\frac{1}{2}} = \mathrm{m} \tag{13.4}$$

This is a very surprising result although it is not derived from the Euler Lagrange equations but just agrees with them 100% for the choice $ב = 1$ in (13).

We are now set to derive the inverse Fine Structure Constant from (13) and from a spin term. We sloppily do this by mixing ideas from General Relativity and Quantum mechanics and (13). From Quantum Mechanics, the angular momentum of the electron is,

$$\sqrt{s(s+1)}\hbar \tag{13.5}$$

where the spin number is $s=\frac{1}{2}$ for the electron, where $\hbar$ is the reduced Planck constant and $\sqrt{s(s+1)}$ is specific to a particle's spin. Suppose that a positive charge with $\frac{1}{8}\left(U^{*\mu}U_{\mu}+U^{\mu}U^{*}{}_{\mu}\right)=0$ or in the real case $\frac{1}{4}U^{\mu}U_{\mu}=0$, is spinning near the speed of light at twice the Schwarzschild radius created by the charge in (13), where this radius is known as the radius of the Marginally Bound [unstable] Orbit, then by (13) this radius should be $2\frac{2eK}{\sqrt{16\pi\varepsilon_0 K}c^2}=2\frac{2Km}{c^2}$, where $m=\frac{e}{\sqrt{16\pi K\varepsilon_0}}$ and e is the charge of the positron. Of course, we need to remember that $\frac{P_\mu}{\sqrt{Z}}$ is not velocity and therefore the interpretation of m is not as the familiar inertial mass, moreover, $\frac{P_\mu}{\sqrt{Z}}$ is not a geodesic vector field if $\frac{U_\mu}{2}$ is not zero. The radial metric coefficient is 1 due to Schwarzschild metric, not Kerr metric, then by (13) and remembering that the angular momentum does not mean a classical rotation, the angular momentum should be,

$$J=\frac{ec}{\sqrt{16\pi\varepsilon_0 K}}\frac{2*2eK}{\sqrt{16\pi\varepsilon_0 K}c^2}=\frac{e^2}{4\pi\varepsilon_0 c} \tag{13.6}$$

and we ignore any Kerr metric because the spin effect on spacetime is not identical to the classical rotation of a black hole, otherwise positrons would dissipate their spin energy. We also assume that our field is a fundamental field to all charged particles and therefore omit the $\sqrt{s(s+1)}$ which is specific to the spin number s.

Now consider the ratio between $J$ and the spin independent coefficient $\hbar$, we get,

$$\frac{J}{\hbar}=\frac{e^2}{4\pi\varepsilon_0 c\hbar} \tag{13.6.1}$$

Which is the known term for the Fine Structure Constant as an upper limit on a ratio between classical angular momentum and Quantum angular momentum.

**Theorem 2:** If the electromagnetic energy is not zero and the charge density $U^{k}{}_{;k}$ is zero in a domain D of space-time then $U_0$ is never 0 in all events of D.

**Proof:**

We write the Einstein - Grossmann equation (4) in its dual form, $R_{\mu\nu} = T_{\mu\nu} - \frac{1}{2} g_{\mu\nu} T_{\alpha}^{\alpha} =$ $\frac{1}{4\beth}\left(U_\mu U_\nu - \frac{1}{2} g_{\mu\nu} U_\lambda U^\lambda - 2U^k{}_{;k} \frac{P_\mu P_\nu}{Z} - \frac{1}{2} g_{\mu\nu} g^{ij}\left(U_i U_j - \frac{1}{2} g_{ij} U_\lambda U^\lambda - 2U^k{}_{;k} \frac{P_i P_j}{Z}\right)\right) =$ $\frac{1}{4\beth}\left(U_\mu U_\nu - \frac{1}{2} g_{\mu\nu} U_\lambda U^\lambda - 2U^k{}_{;k} \frac{P_\mu P_\nu}{Z} - \frac{1}{2} g_{\mu\nu} U^\lambda U_\lambda + g_{\mu\nu} U_\lambda U^\lambda + g_{\mu\nu} U^k{}_{;k}\right) = \frac{1}{4\beth}(U_\mu U_\nu +$ $U^k{}_{;k}\,(g_{\mu\nu} - 2\frac{P_\mu P_\nu}{Z}))$. If $U_0 = 0$ in D then there exist local coordinates such that only the $P_0$ component of $P_\mu$ is not zero. We assumed $U^k{}_{;k} = 0$. Since $U_0 = 0$, $R_{00} = 0$ so the electromagnetic energy is zero. On the other hand, since $U_\mu$ is not zero, $P_\mu$ cannot be geodesic and therefore $P_0$ cannot be the only component of $P_\mu$ which is not zero along geodesic coordinates. Note: If there is a time-like curve $\gamma$ around which $U_\mu$ is in relative motion in different events of every small D that contains $\gamma$, then $R_{00}$ is not zero in D.

**Note:** There is one obvious peculiarity about charge generated gravity, $\frac{P^\mu}{\sqrt{Z}}$ is not the velocity of the charge. It is dictated by a scalar field of space-time!

**Note – physical interpretation**: From (10) and (13), if $a^\mu$ has a simple physical interpretation as a field that accelerates any neutral mass then we have to take (13) into account as an opposite effect. The result is that a field of 1,000,000 volts over 1 mm distance will accelerate any neutral particle at 8.61 cm * $sec^{-2}$ and with taking into account (13), (13.02) it will be less, due to an opposite gravitational effect, see (14), will be reduced to 4.305 cm * $sec^{-2}$. If the acceleration is merely a mathematical object, a field, which is consistent with a complex formalism then only the gravitational acceleration has a physical meaning as pseudo acceleration of 4.305 cm * $sec^{-2}$.

**Caveat:** In this theory, the most logical explanation is that $a^\mu$ is a field, not necessarily a mechanical acceleration.

The quantization of P is into a sum of event wave functions and has the physical meaning of Sam Vaknin's realization chronons [17]. The theory is easily expanded to 2 and to 3 Reeb class vectors where the Lagrangian has U(1) SU(2) SU(3) symmetry if orientation is preserved, otherwise the symmetry group contains also reflections, see also an SU(4) Lagrangian, Appendix C. It is important to say that Vaknin's approach [17] is diametrically opposed to that of Jungjai Lee and Hyun Seok Yang [4].

**A conformal map vs. gauge transformation** – can be skipped up to "Ceramic capacitors and a Hermann Bondi gravitational dipole".

Consider a local gauge transformation G, and the acceleration matrix (1.1.1), (1.1.2),

$$A_{\mu\nu} = \left(\frac{P_\mu}{\sqrt{Z}}\right)_{,\nu} - \left(\frac{P_\nu}{\sqrt{Z}}\right)_{,\mu} \tag{13.7}$$

And the gauge transformation

$\frac{P_\lambda}{\sqrt{Z}} \longrightarrow G^\lambda{}_\mu \frac{P_\lambda}{\sqrt{Z}}$

Then

$$\left(G^\lambda{}_\mu \frac{P_\lambda}{\sqrt{Z}}\right);_\nu \neq G^\lambda{}_\mu \left(\frac{P_\lambda}{\sqrt{Z}}\right);_\upsilon \tag{13.8}$$

And therefore, the Gauge-Covariant derivative operator
$\begin{bmatrix} g \\ ; \end{bmatrix}_\nu$ would be used with the ordinary gauge field $G^\lambda{}_\mu$

$$\left(G^\lambda{}_\mu \frac{P_\lambda}{\sqrt{Z}}\right)\left[\begin{matrix} g \\ ; \end{matrix}\right]_\nu \equiv \left(G^\lambda{}_\mu \Delta_\nu - igS^\lambda{}_{\mu\nu}\right)\left(\frac{P_\lambda}{\sqrt{Z}}\right) \tag{13.9}$$

Where Δ is the covariant derivative and where g is the coupling coefficient and $S^\lambda{}_{\mu\nu}$ is expressed by vectors of the Lie Algebra of $G$.

When choosing $G_{kj} dx^k \wedge dx^j$ to be an exact form, we do not have a Lie group because multiplication of anti-symmetric matrices is not closed, and the transformation is usually not a gauge transformation but a conformal map when non-degenerate, however, similar to $\left(\left(\frac{P_k}{\sqrt{Z}}\right),_j - \left(\frac{P_j}{\sqrt{Z}}\right),_k\right) dx^k \wedge dx^j$, with $A_{kj} = \left(\frac{P_k}{\sqrt{Z}}\right),_j - \left(\frac{P_j}{\sqrt{Z}}\right),_k = \frac{U_k}{2}\frac{P_j}{\sqrt{Z}} - \frac{U_j}{2}\frac{P_k}{\sqrt{Z}}$

Demanding non-dependence of $\left(\frac{G_{\mu\lambda}P^\lambda}{\sqrt{Z}}\right);^s$ on derivatives of $G_{\mu\lambda} = \omega_\mu;_\lambda - \omega_\lambda;_\mu$ leads to

$$\frac{P^\lambda}{\sqrt{Z}}\left(G_{\mu\lambda};_\nu - G_{\nu\lambda};_\mu\right) = \frac{P^\lambda}{\sqrt{Z}}\left(-G_{\lambda\mu};_\nu + G_{\lambda\nu};_\mu\right) = 0 \tag{13.10}$$

$$G_{\lambda\nu};_\mu - G_{\lambda\mu};_\nu = \omega_\lambda;_\nu;_\mu - \omega_\nu;_\lambda;_\mu - \omega_\lambda;_\mu;_\nu + \omega_\mu;_\lambda;_\nu = \omega_\beta R^\beta_{\lambda\nu\mu} - \omega_\nu;_\lambda;_\mu + \omega_\mu;_\lambda;_\nu$$

$$-\omega_\nu;_\lambda;_\mu = -\omega_\nu;_\lambda;_\mu + \omega_\nu;_\mu;_\lambda - \omega_\nu;_\mu;_\lambda = -\omega_\beta R^\beta_{\nu\lambda\mu} - \omega_\nu;_\mu;_\lambda$$

$$\omega_\mu;_\lambda;_\nu - \omega_\mu;_\nu;_\lambda + \omega_\mu;_\nu;_\lambda = \omega_\beta R^\beta_{\mu\lambda\nu} + \omega_\mu;_\nu;_\lambda$$

$$\omega_\beta\left(R^\beta_{\lambda\nu\mu} - R^\beta_{\nu\lambda\mu} + R^\beta_{\mu\lambda\nu}\right) + \omega_\mu;_\nu;_\lambda - \omega_\nu;_\mu;_\lambda = \omega_\beta\left(R^\beta_{\lambda\nu\mu} + R^\beta_{\nu\mu\lambda} + R^\beta_{\mu\lambda\nu}\right) + \omega_\mu;_\nu;_\lambda - \omega_\nu;_\mu;_\lambda$$

And by the first Bianchi identity $R^\beta_{\lambda\nu\mu} + R^\beta_{\nu\mu\lambda} + R^\beta_{\mu\lambda\nu} = 0$

$$\frac{P^\lambda}{\sqrt{Z}}\left(G_{\lambda\nu};_\mu - G_{\lambda\mu};_\nu\right) = \frac{P^\lambda}{\sqrt{Z}}\left(\omega_\mu;_\nu;_\lambda - \omega_\nu;_\mu;_\lambda\right) = G_{\mu\nu};_\lambda \frac{P^\lambda}{\sqrt{Z}} = 0 \tag{13.11}$$

$$\left(\frac{G_{\mu\lambda}P^\lambda}{\sqrt{Z}}\right);_\nu - \left(\frac{G_{\nu\lambda}P^\lambda}{\sqrt{Z}}\right);_\mu = G_{\mu\lambda}\left(\frac{P^\lambda}{\sqrt{Z}}\right);_\nu - G_{\nu\lambda}\left(\frac{P^\lambda}{\sqrt{Z}}\right);_\mu + G_{\mu\nu};_\lambda \frac{P^\lambda}{\sqrt{Z}}$$

$G_{\nu s}$ acts as rotation and scaling on both indices $\lambda$ and $s$ which means that,

$$G_{\nu\lambda}G_{\mu s}\frac{P^{\lambda}Z^{s}}{2\sqrt{Z}} - G_{\mu\lambda}G_{\nu s}\frac{P^{\lambda}Z^{s}}{2\sqrt{Z}} = G_{\nu\lambda}G_{\mu s}\left(\frac{P^{\lambda}Z^{s}}{2\sqrt{Z}} - \frac{P^{s}Z^{\lambda}}{2\sqrt{Z}}\right) = w^{2}\tilde{G}_{\nu\lambda}\tilde{G}_{\mu s}\left(\frac{P^{\lambda}Z^{s}}{2\sqrt{Z}} - \frac{P^{s}Z^{\lambda}}{2\sqrt{Z}}\right) \qquad (13.12)$$

where $\tilde{G}_{kj}$ are rotation matrices and $w$ is a scalar function.

$$\frac{1}{2}w^{2}\tilde{G}_{\nu k}\tilde{G}_{\mu j}A^{kj}w^{2}A_{sr}\tilde{G}^{s\nu}\tilde{G}^{r\mu} = w^{4}\frac{U_{\mu}U^{\mu}}{4} \qquad (13.13)$$

which means that $G_{\mu\nu}$ acts as a scaling on the action by the Reeb class vector as expected from a Scarr-Friedman type of matrix [11] acting twice on a vector. The addition $G_{\mu\nu};_{\lambda}\frac{P^{\lambda}}{\sqrt{Z}}$ was first missed by the author and was later corrected. (3.18) offers coupling between $G_{\mu s}$ tensors that is reducible to a classical non-covariant energy of the electric field.

Now, replacing $P_{\mu}$ in which case $\tau = PP^{*}$ is a Geroch function, with $\psi$ such that $\psi\psi^{*}$ integrates to 1. i.e. it is an event wave function, we get a theory like Sam Vaknin's chronon theory [17], providing $\psi_{\mu} = \frac{\partial\psi}{\partial x^{\mu}}$ is Almost Everywhere smooth and non-degenerate. So, this theory lacks bosons as carriers of interactions because the chronons themselves are the reason for interactions.

**Full gauge field as the normalized acceleration field itself**

When the acceleration matrix A is in the form $A = \begin{pmatrix} 0 & -a \\ a & 0 \end{pmatrix}$ every matrix M with determinant 1 is a Symplectic matric, i.e. the following is invariant: $M^{t}AM = A$. In block diagram, $M = \begin{pmatrix} B & C \\ D & E \end{pmatrix}$ and $B^{t}C = C^{t}B$, $D^{t}E = E^{t}D$ and $B^{t}E - D^{t}C = I$, where $I$ is the identity matrix.

In the complex case, the conjugate transposed matrix $M^{*t}$ replaces the transposed matrix $M^{t}$.

We can immediately see that the acceleration matrices are in this case the gauge fields of themselves because:

$$\begin{pmatrix} 0 & \theta^{*} \\ -\theta^{*} & 0 \end{pmatrix}\begin{pmatrix} 0 & -a \\ a & 0 \end{pmatrix}\begin{pmatrix} 0 & -\theta \\ \theta & 0 \end{pmatrix} = \begin{pmatrix} 0 & -a \\ a & 0 \end{pmatrix} \qquad (13.13.1)$$

In this theory therefore, gauge fields are redundant because their action is already included in the Lagrangian equations (4). If $\frac{U_{\mu}}{2}$ is not a null vector and $\frac{U_{\mu}}{2} = \theta\frac{w_{\mu}}{2}$ where $w_{\mu}$ is real then there are coordinates in which

$$\begin{pmatrix} 0 & -\theta \\ \theta & 0 \end{pmatrix} = \left(\frac{\|w_{\lambda}w^{\lambda}\|}{4}\right)^{-\frac{1}{2}}\left(\frac{p_{\mu}}{\sqrt{\|Z\|}}\frac{U_{\nu}}{2} - \frac{p_{\nu}}{\sqrt{\|Z\|}}\frac{U_{\mu}}{2}\right) \qquad (13.13.2)$$

Also see action before (4) as $\frac{1}{4}U^{k}U_{k}\sqrt{-g}$ in the real case.

From (13.13.2) we can obtain three fields when $\theta_1 = \theta_2$ and reach a more generalized gauge field,

$$\theta^*{}_2\theta_2 = 1, \qquad \theta^*{}_1\theta_1 = 1$$

$$\begin{pmatrix} 0 & 0 & 0 & \theta^*{}_1 \\ 0 & 0 & -\theta_1{}^* & 0 \\ 0 & \theta^*{}_2 & 0 & 0 \\ -\theta^*{}_2 & 0 & 0 & 0 \end{pmatrix}\begin{pmatrix} 0 & -a & 0 & 0 \\ a & 0 & 0 & 0 \\ 0 & 0 & 0 & -a \\ 0 & 0 & a & 0 \end{pmatrix}\begin{pmatrix} 0 & 0 & 0 & -\theta_1 \\ 0 & 0 & \theta_1 & 0 \\ 0 & -\theta_2 & 0 & 0 \\ \theta_2 & 0 & 0 & 0 \end{pmatrix} = \begin{pmatrix} 0 & -a & 0 & 0 \\ a & 0 & 0 & 0 \\ 0 & 0 & 0 & -a \\ 0 & 0 & a & 0 \end{pmatrix} \tag{13.13.3}$$

$$\begin{pmatrix} 0 & \theta^*{}_2 & 0 & 0 \\ -\theta^*{}_2 & 0 & 0 & 0 \\ 0 & 0 & 0 & \theta^*{}_1 \\ 0 & 0 & -\theta_1{}^* & 0 \end{pmatrix}\begin{pmatrix} 0 & -a & 0 & 0 \\ a & 0 & 0 & 0 \\ 0 & 0 & 0 & -a \\ 0 & 0 & a & 0 \end{pmatrix}\begin{pmatrix} 0 & -\theta_2 & 0 & 0 \\ \theta_2 & 0 & 0 & 0 \\ 0 & 0 & 0 & -\theta_1 \\ 0 & 0 & \theta_1 & 0 \end{pmatrix} = \begin{pmatrix} 0 & -a & 0 & 0 \\ a & 0 & 0 & 0 \\ 0 & 0 & 0 & -a \\ 0 & 0 & a & 0 \end{pmatrix} \tag{13.13.4}$$

Notice that $\begin{pmatrix} 0 & 0 & 0 & -\theta_1 \\ 0 & 0 & \theta_1 & 0 \\ 0 & -\theta_2 & 0 & 0 \\ \theta_2 & 0 & 0 & 0 \end{pmatrix}$ is not antisymmetric and does not look like an ordinary acceleration matrix. $\begin{pmatrix} 0 & -\theta_2 & 0 & 0 \\ \theta_2 & 0 & 0 & 0 \\ 0 & 0 & 0 & -\theta_1 \\ 0 & 0 & \theta_1 & 0 \end{pmatrix}$ does have the ordinary structure of an acceleration matrix.

Unlike in symplectic transformations, the sign $C = \begin{pmatrix} 0 & -a & 0 & 0 \\ a & 0 & 0 & 0 \\ 0 & 0 & 0 & -a \\ 0 & 0 & a & 0 \end{pmatrix}$ can change to $-C = \begin{pmatrix} 0 & a & 0 & 0 \\ -a & 0 & 0 & 0 \\ 0 & 0 & 0 & a \\ 0 & 0 & -a & 0 \end{pmatrix}$ because when $a$ is real, the Lagrangian $a^2 = \frac{U^\lambda U_\lambda}{4}$ stays invariant.

Changing the coordinates,

$$\begin{pmatrix} 0 & 0 & 0 & \theta_1{}^* \\ 0 & 0 & \theta_1{}^* & 0 \\ 0 & -\theta_2{}^* & 0 & 0 \\ -\theta_2{}^* & 0 & 0 & 0 \end{pmatrix}\begin{pmatrix} 0 & -a & 0 & 0 \\ a & 0 & 0 & 0 \\ 0 & 0 & 0 & -a \\ 0 & 0 & a & 0 \end{pmatrix}\begin{pmatrix} 0 & 0 & 0 & -\theta_1 \\ 0 & 0 & -\theta_1 & 0 \\ 0 & \theta_2 & 0 & 0 \\ \theta_2 & 0 & 0 & 0 \end{pmatrix} = \begin{pmatrix} 0 & a & 0 & 0 \\ -a & 0 & 0 & 0 \\ 0 & 0 & 0 & a \\ 0 & 0 & -a & 0 \end{pmatrix} \tag{13.13.5}$$

In this case the matrix M satisfies $M^{*T}CM = -C$.

(13.13.5) Is a Charge - Parity violation if we treat the columns 0 and 1 as describing the electric field and columns 2 and 3 as another field. Reversing time maps $\frac{p_\mu}{\sqrt{\|Z\|}}$ to $-\frac{p_\mu}{\sqrt{\|Z\|}}$ and therefore maps $\begin{pmatrix} 0 & -a & 0 & 0 \\ a & 0 & 0 & 0 \\ 0 & 0 & 0 & -a \\ 0 & 0 & a & 0 \end{pmatrix}$ to $\begin{pmatrix} 0 & a & 0 & 0 \\ -a & 0 & 0 & 0 \\ 0 & 0 & 0 & -a \\ 0 & 0 & a & 0 \end{pmatrix}$ which is a violation of the relative orientation between the plane 0, 1 and 2, 3. Please refer to theorem 0. Therefore, time reversal alone or Charge – Parity – Time violations are not allowed.

**Cartan formalism of the Ricci action with $\frac{p_\mu(a)}{\sqrt{Z(a)}}$ Fields** - can be skipped up to "Ceramic capacitors and a Hermann Bondi gravitational dipole".

The Geroch gradient norm is $p_\mu(0)p^\mu(0)$ or in the complex formalism $\frac{1}{2}(p^*_\mu(0)p^\mu(0) + p_\mu(0)p^{*\mu}(0))$. Please beware that unlike in Einstein's convention, summation of Roman indices occur also when two lower or two upper indices have the same letter.

$$e^a_\mu = \frac{p_\mu(a)}{\sqrt{Z(a)}} \tag{13.14}$$

$$g^{ab} = e^a_\mu e^b_\nu g^{\mu\nu} \tag{13.15}$$

We need upper tetrad indices because the Gradients we use are covariant.

Here the metric tensor is implicitly used because no orthogonality is imposed. With orthogonality, we can get rid of a variable metric tensor. This section will use the well-known Cartan's structure form $\omega^a_b$,

$$g^{ab} = \frac{p^\mu(a)}{\sqrt{Z(a)}}\frac{p_\mu(b)}{\sqrt{Z(b)}} \tag{13.16}$$

$$\partial_{[\nu}e^a_{\mu]} = \omega^a_{b[\mu}e^b_{\nu]}$$

Finally, here is the Ricci scalar which is our goal in order to reach the gravitational action:

$$R = 2g^{ab}e^\mu_a e^\nu_b(\partial_{[\nu}\omega^a_{b\mu]} + \omega^a_{c[\nu}\omega^c_{b\mu]}) \tag{13.17}$$

Which is the general case without using orthogonality. Orthogonality means

$$g^{ab} = \begin{pmatrix} 1 & 0 & 0 & 0 \\ 0 & -1 & 0 & 0 \\ 0 & 0 & -1 & 0 \\ 0 & 0 & 0 & -1 \end{pmatrix} \tag{13.18}$$

**Important:** As we shall see, with (64.b), (13.17) offers a theory of 4 spacetime vectors, however, the action in (64.b) allows a secondary acceleration plane. In that case we have 6 and not only 4 spacetime vectors or at least 5 and not 4. Therefore (64.b) and (13.17) are not compatible unless $\frac{P_\mu}{\sqrt{Z}}$ and $\frac{P_{(1)\mu}}{\sqrt{Z(1)}}$ are dependent as functions and orthogonal as vectors and $U(2)_\mu = U(3)_\mu = 0$, see (64.a) and preferrable (64.b).

## 2. Ceramic capacitors and a Hermann Bondi gravitational dipole

This section is based on the surprising result from (6)-(13) and the calculation after (13) and cannot be dismissed. Any theory which predicts the component of the electric charge in the energy momentum tensor to be $-\mathrm{U}^\lambda{}_{;\lambda} W_\mu W_\nu$ where $\mathrm{U}^\lambda$ is a spacelike vector, $W_\mu$ is a time-like unit vector and such that $\mathrm{U}^\lambda$ is derived from $W_\mu$, and such that the electric charge is point-like, means:

1) That the description of the location where $\mathrm{U}^\lambda{}_{;\lambda}$ is not zero is a Dirac delta.
2) Due to symmetry, $W_\mu$ must be geodesic at the center where the charge is. Therefore, electric charge itself, unlike its energy, does not have inertial mass because $(\mathrm{U}^\lambda{}_{;\lambda} W^\mu W^\nu)_{;\nu} = 0$, due to the symmetry of the charge distribution and due to symmetry of $W_\mu$ along the hyperplane perpendicular to $W_\mu$, $U_\mu$ should be zero at the center of the charge.
3) Charge generates gravity where $\mathrm{U}^\lambda{}_{;\lambda} > 0$ and anti-gravity where $\mathrm{U}^\lambda{}_{;\lambda} < 0$ in (+,-,-,-) metric convention.

The Geometric Chronon Field Theory predicts that not only does inertial mass generate gravity but also charge does. The gravitational mass by charge is predicted at about (-,+)5.802135 * 10^9 Kg / (-,+) Coulomb with negative charge generating weak anti-gravity and positive charge generating weak gravity but still many orders of magnitude more than predicted by conventional physics.

**Most critical no-go argument:** This argument will be repeated because it is fundamental. There is a no-go argument against static high voltage capacitors that will be discussed. Simply speaking in colloquial language, in the non-covariant classical limit, if the unexpected gravitational pseudo acceleration by charge-based gravity due to (13) and (13.01) is parallel to the electrostatic field when reduced to 3 dimensions, then the fact that charge carriers are averagely static and are also mass carriers means that their mass cannot be accelerated when summing up all gravitational pseudo accelerations. At that point, the reader can ask, then what is the point in this section? There is a point. First, by using ferroelectric materials and dynamic electric fields this obstacle can be removed. Second, in a very strong electric gradient, the nucleus of the atom can be electrically polarized which would end up in strong-force dipoles. Although a chromodynamic theory is not developed in this paper and although the strong force cannot be described by the Gaussian law, drains and sources of the 3 types of the strong charge could generate a Bondi dipole. Most of this section will try to ignore the no-go argument, assuming that there is some motion of charge, and discuss other obstacles. It is the author's view, however, that the no-go argument in the static case is a strong insurmountable argument.

**Caution:** The main drawback of this section is that it ignores the fact that in dielectric materials even with low relative permittivity, such as PTFE, the permanent dipoles have a local field stronger than the external field. If most of the molecular mass is within permanent dipoles, then the fact they oppositely align with an external electric field in high voltage capacitors, must cancel out any external gravitational dipole. The dipoles, before the external field was applied, cancel out and after the external field is applied are expected to reach equilibrium that cancels out the external dipole gravitational acceleration of the molecular mass within the molecular dipoles. A work around this problem is offered in this paper. In this section we will examine gravitational propulsion, not an Alcubierre's warp drive because the Alcubierre [18] extrinsic curvature condition $(K_i{}^i)^2 - K_{ij}K^{ij} < 0$ will not hold in the same geometry as in the Alcubierre warp drive bubble. However, a negative plate below and a positive plate above, will manifest weak acceleration upwards as the negative gravity will push the positive plate upwards and the negative plate will be pulled by the gravity of the positive plate above it. A claim of an apparent local violation of the conservation of energy and momentum ignores the fact that such a gravitational dipole interacts with the metric of spacetime and thus affects the trajectories of far bodies of mass. The main technological problem is that due to the dielectric material, the mass of the dielectric material will not be gravitationally repelled by the negative plate. Only a small portion of the mass of the capacitor will be affected in a highly dielectric material. Overcoming the anti-alignment, see Fig. 3.A., is a technological challenge which cannot be achieved without a dynamic electric field, see Fig. 3.B.

**Fig. 3.A.** – Only a small portion of the mass, in purple, is affected.

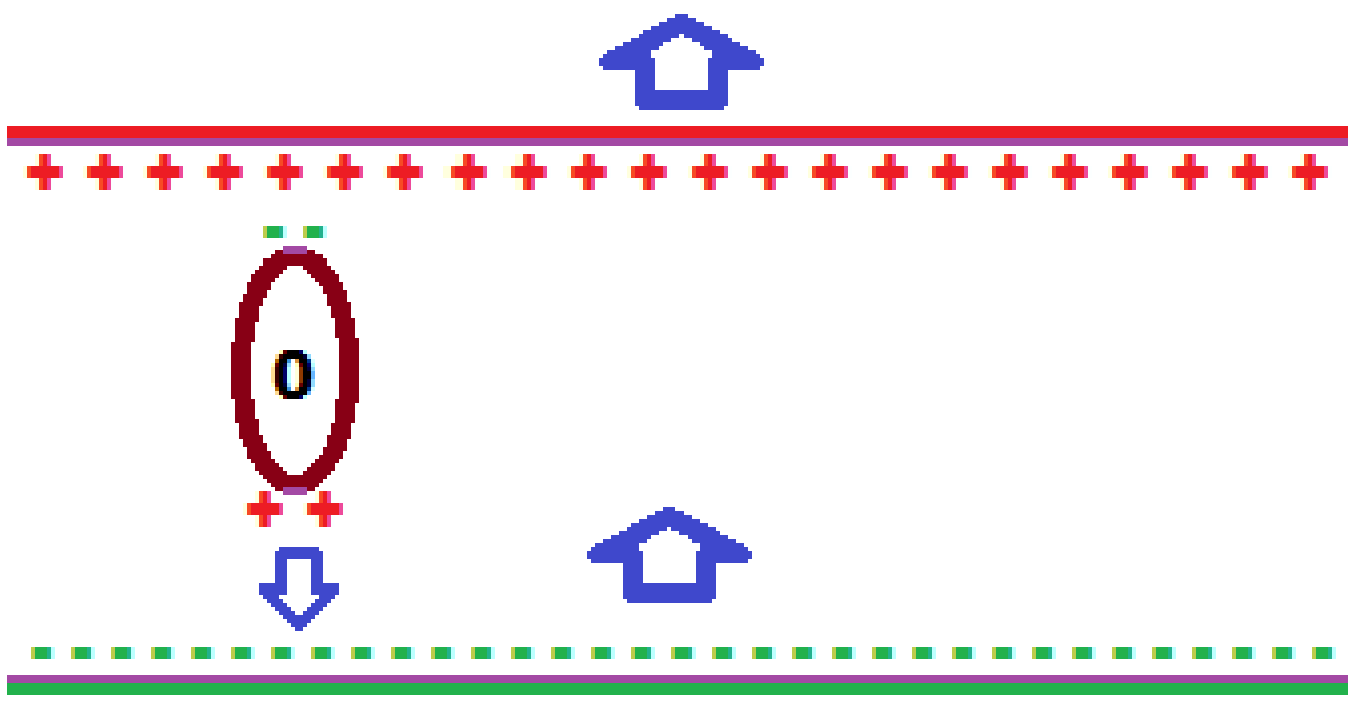

**Fig. 3.B.** – Electro-gravitational thrust engine with two capacitors and slow anti-alignment dielectric layer, which mitigates the anti-alignment by charging (right) and discharging (left) cycles. The capacitors rotate as depicted in the drawings. The ground direction is bottom. The arrows depict the direction in which the capacitors are rotated by an electric motor. Static field without dielectric anti-alignment requires about $2 * 10^{-4}$ Coulombs / $cm^2$ in order to accelerate the dielectric layer against the gravity of the Earth. This is why with the current technology, the offered thrust engine is insufficient for a commercial flight. Measurable thrust of up to 1 Newtons is expected with voltage above 2,000,000 volts, relative dielectric constant of above 1000, dielectric polarization time of a millisecond, heavy dielectric layers with mass density similar to $Ta_2O_5$ but with a higher dielectric constant, electric motor rotation of at least 3000 RPMs and capacitor areas of about 20 x 20 $cm^2$. Partners in this experiment are Jessica Lynne Suchard and Raviv Yatom. The next figure is **Fig. 3.B,**

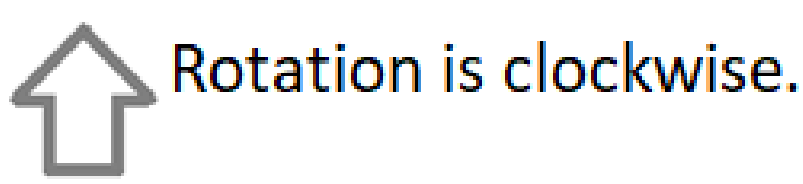

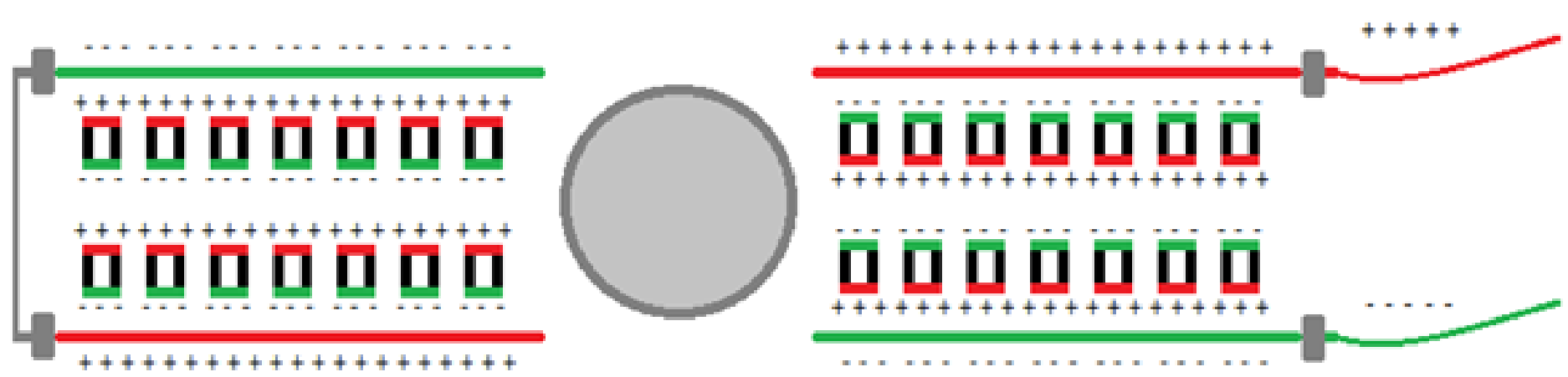

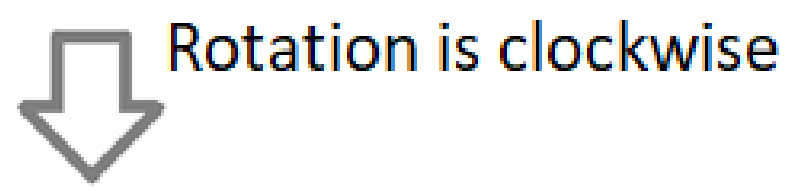

Discharge phase - for a short time of milliseconds, the dielectric layer provides weak bottom negative gravity and top weak positive gravity.

Charging phase:
The bottom plate provides weak anti-gravity and the top plate provides weak gravity. This gravitational dipole effect is weaker than the discharge effect by as much as 3 orders of magnitude.

Problem: commercial dielectric materials polarize and depolarize fast, not in milliseconds.

**Fig. 3.C. -** A less feasible proof of concept, the exposed dielectric layer generates a gravitational/inertial dipole which is not cancelled out

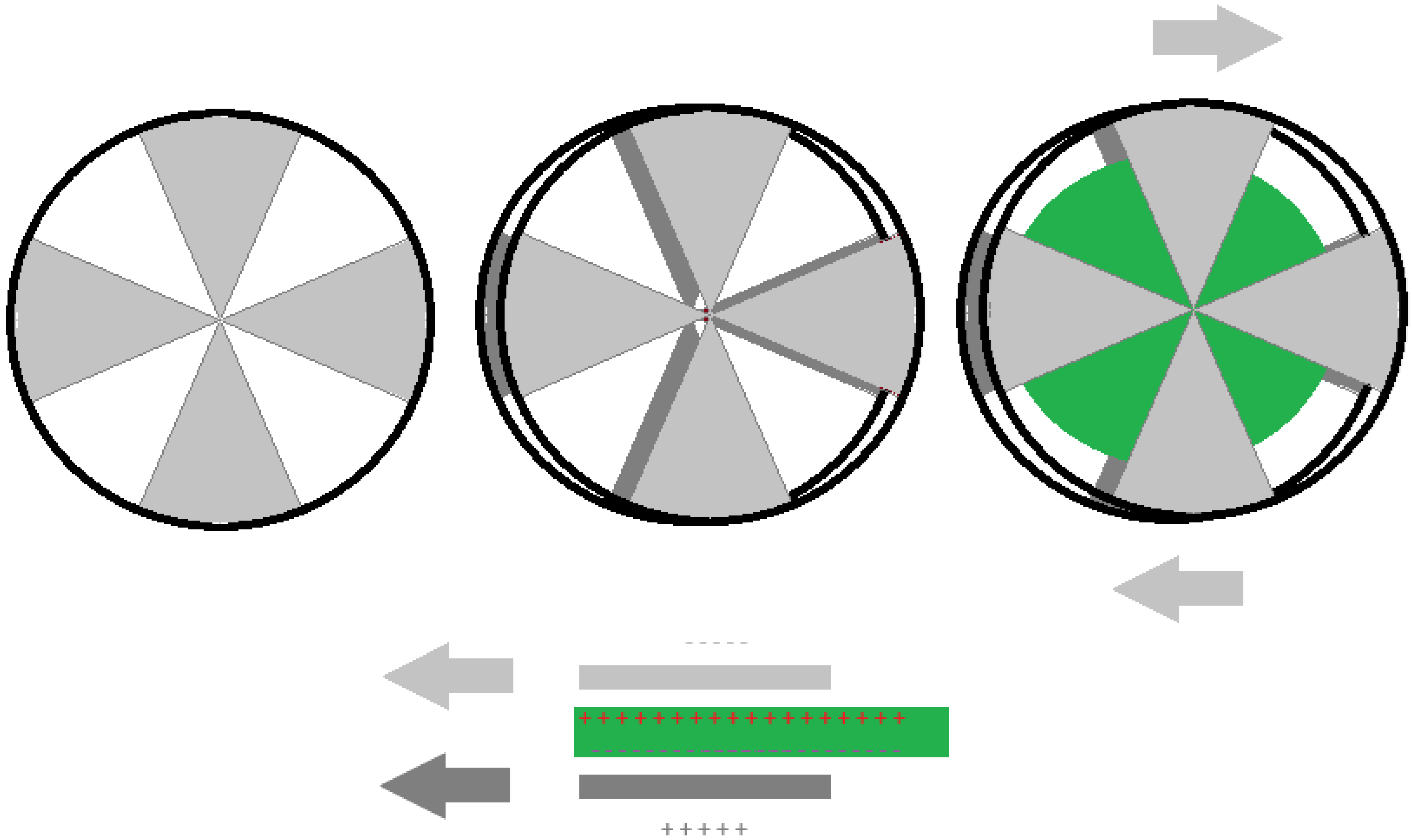

Fig. 3.C is of a top view of a capacitor with rotating plates in relation to a dielectric layer in green in the illustration. The bottom of the figure illustrates a side view of the dynamic capacitor. Both the top and the bottom plates have gaps in the rotating conductive metallic plates. The top plate is negatively charged, and the bottom plate is positively charged. When the plates rotate thousands of times per minute, the dielectric layer between the metallic sections of the plates align opposite to the field on the plates. When a dielectric portion of the dielectric layer is exposed to a gap in the external plates, the field vanishes but the opposite alignment of the dielectric material does not disappear at once. If the opposite dielectric alignment of induced and permanent dipoles appears much faster than it disappears when reaching the gaps, then a thrust upward can be measured. The problem with this design is the specific set of requirements from the dielectric material which are hard to achieve.

It is easy to see from (13) that in the classical limit near the plate, the gravitational field is mostly affected by charge density. By (13), and the classical non-covariant limit from (10) of the electrostatic field, $\|a^{\mu}\| = \sqrt{4\pi K\varepsilon_0 \beth}\|E\|$, $\left\|\frac{a^{\mu}}{c^2}\right\| = \left\|\frac{u^{\mu}}{2}\right\|$, the gravitational acceleration is

$$a \cong \frac{4\pi KQ}{A*\varepsilon*\sqrt{16\pi K\varepsilon_0 \beth}} = \frac{V}{d\sqrt{\beth}} * \sqrt{\pi K\varepsilon_0} \Longrightarrow \delta Weight \cong \frac{V}{d\sqrt{\beth}} * \frac{M_{dielectric}}{g}\sqrt{\pi K\varepsilon_0} = \frac{V\rho A}{g\sqrt{\beth}}\sqrt{\pi K\varepsilon_0} \quad (14)$$

where K is Newton's gravitational constant, Q is charge, A is area, $\rho$ is the dielectric layer's density and M is its mass and $\varepsilon_0$ is the permittivity of vacuum, $\varepsilon$ is the relative dielectric constant, assuming $\beth = 1$, g is the Earth surface acceleration. (14) is the result of $Q = V\varepsilon_0 \frac{A}{d} = \frac{V}{d}\varepsilon_0 A \Leftrightarrow \frac{Q}{\varepsilon_0 A} = 4\pi \frac{Q}{4\pi\varepsilon_0 A} E$ where E is the classical intensity of the electric field. We saw in (13): $M_{Gravitational} = \frac{Q}{\sqrt{16\pi K\varepsilon_0 \beth}}$ with $\beth = 1$. The gravitational acceleration by the charge is $a \cong \frac{4\pi KM}{A} = \frac{4\pi KQ}{A*\varepsilon*\sqrt{16\pi K\varepsilon_0\beth}}$, if we assume an attenuation by the dielectric layer's induced dipoles to be proportional to the attenuation of the electric field by the same induced dipoles. This assumption is problematic because the induced dipoles are the accelerated material by the gravitational dipole of the external plates, and they are in much closer proximity to local charge than to the charge on the external plates. $\delta Weight \cong \frac{V}{d\sqrt{\beth}} * \frac{M_{dielectric}}{g}\sqrt{\pi K\varepsilon_0} = \frac{V\rho A}{g\sqrt{\beth}}\sqrt{\pi K\varepsilon_0}$ is therefore a very optimistic model.

**Critical:** Obviously (14) ignores the local field of permanent dipoles which is usually much stronger than the field of the external capacitor plates. If most of the molecular mass is within such dipoles then (14) is unrealistic and the gravitational acceleration is totally cancelled out if the external field is static and nearly symmetrical.

The classical limit in (14) is suitable for a static charge density. For convenience, (3.3) - (3.6) are written again. In weak gravity and limited charge density,

$$g_{\mu\nu} = \eta_{\mu\nu} + \epsilon h_{\mu\nu} \tag{14.1}$$

Where the velocity of any test clock $\frac{dx^\lambda}{d\tau} \approx \frac{dx^\lambda}{dt} \ll c$ and $O\left(\frac{dx^\lambda}{d\tau}\right) = \epsilon$, c is the speed of light.

$$\Gamma^{\lambda}_{\mu\nu} = \frac{1}{2} g^{\lambda\alpha}\left(g_{\alpha\mu,\nu} + g_{\nu\alpha,\mu} - g_{\mu\nu,\alpha}\right) = \frac{1}{2} g^{\lambda\alpha}\left(g_{\alpha\mu,\nu} + g_{\alpha\nu,\mu} - g_{\mu\nu,\alpha}\right) \tag{14.2}$$

$$g^{\mu\alpha} g_{\nu\alpha} = \delta^{\mu}_{\nu} \Longrightarrow g^{\mu\nu} \approx \eta^{\mu\nu} - \epsilon h^{\mu\nu}$$

$$\Gamma^{\lambda}_{\mu\nu} \approx \frac{1}{2}\epsilon\eta^{\lambda\alpha}\left(h_{\alpha\mu,\nu} + h_{\nu\alpha,\mu} - h_{\mu\nu,\alpha}\right)$$

$$R_{00} \approx \Gamma^{\lambda}_{00,\lambda} - \Gamma^{\lambda}_{0\lambda,0} \approx$$

$$\frac{1}{2}\epsilon\eta^{\lambda\lambda}(h_{\lambda 0,0} + h_{0\lambda,0} - h_{00,\lambda}) - \frac{1}{2}\epsilon\eta^{\lambda\lambda}(h_{\lambda 0,\lambda} + h_{\lambda\lambda,0} - h_{0\lambda,\lambda})$$

$$= \frac{1}{2}\epsilon\eta^{\lambda\lambda}(2h_{\lambda 0,0} - h_{00,\lambda}) - \frac{1}{2}\epsilon\eta^{\lambda\lambda}(h_{\lambda\lambda,0})$$

Using the energy momentum tensor from (4),

$$T_{\mu\nu} = -\frac{1}{4}\left(U_\mu U_\nu - \frac{1}{2} g_{\mu\nu} U_\lambda U^\lambda - 2U^{k}{}_{;k}\frac{P_\mu P_\nu}{Z}\right) \tag{14.3}$$

$$-(T_{\mu\nu}-\frac{1}{2}g_{\mu\nu}T) = R_{\mu\nu}$$

$$\frac{1}{4}\left(U_\mu U_\nu-\frac{1}{2}g_{\mu\nu}U_\lambda U^\lambda-2U^k{}_{;k}\frac{P_\mu P_\nu}{Z}\right)+\frac{1}{4}\left(\frac{1}{2}U_\lambda U^\lambda+U^k{}_{;k}\right)g_{\mu\nu} = R_{\mu\nu}$$

$$\frac{1}{4}\left(U_\mu U_\nu-U^k{}_{;k}\,(2\frac{P_\mu P_\nu}{Z}-g_{\mu\nu})\right) = R_{\mu\nu}$$

$$\frac{1}{4}\left(U_0 U_0-U^k{}_{;k}\,(2\frac{P_0 P_0}{Z}-g_{00})\right) = R_{00}$$

$U_\mu$ is spacelike in relation to $P_\mu$, $U_0 \ll 1$, and assuming $\frac{P_\mu}{\sqrt{Z}} \ll 1, \mu \in \{1,2,3\}$, and with a minor abuse of Einstein's summation convention we have,

$$\tfrac{1}{4}U^k{}_{;k}\,(2\tfrac{P_0P_0}{Z}-g_{00}) \approx \tfrac{1}{4}U^k{}_{;k} \qquad (14.4)$$

$$-\frac{1}{4}U^k{}_{;k} \approx R_{00} \approx \frac{1}{2}\epsilon\eta^{\lambda\lambda}(2h_{\lambda 0,0}-h_{00,\lambda})-\frac{1}{2}\epsilon\eta^{\lambda\lambda}(h_{\lambda\lambda,0})$$

$$\left|U^k{}_{;k,0}\right| \ll 1 \Rightarrow -\frac{1}{4}U^k{}_{;k} \approx -R_{00} \approx \frac{1}{2}\epsilon\eta^{\lambda\lambda}h_{00,\lambda} = \frac{1}{2}\epsilon\nabla h_{00}$$

$$-\frac{1}{4}U^k{}_{;k} \approx \frac{1}{2}\epsilon\nabla h_{00}$$

And from (11), $\frac{1}{2}U^k{}_{;k} = \frac{a^k{}_{;k}}{c^2} = \sqrt{4\pi K\varepsilon_0}\,\frac{\rho}{\varepsilon_0 c^2} = \sqrt{\frac{4\pi K}{\varepsilon_0}}\frac{\rho}{c^2}$,

$$\frac{1}{4}U^k{}_{;k} = \frac{1}{2}\sqrt{\frac{4\pi K}{\varepsilon_0}}\frac{\rho}{c^2} \approx \frac{1}{2}\epsilon\nabla h_{00} \Rightarrow \sqrt{\frac{4\pi K}{\varepsilon_0}}\frac{\rho}{c^2} \approx \epsilon\nabla h_{00}$$

Where $\rho$ is the charge density.

**Hysteresis dielectric effect after removing the plates**

Under the following assumptions the gravitational effect on the dielectric layer can be simplified.

Instead of fighting the opposite alignment of permanent dielectric dipoles as they potentially totally cancel out the gravitational acceleration by the external plates, this opposite alignment can be harnessed to make new technology.

The assumption here is that a dielectric layer is placed between two parallel plates to which high voltage is applied and then the plates are removed or that the dielectric layer is fast to oppositely align with the external field and slow to lose this opposite alignment once the external field is collapsed. Units are $A$ = area, $Q$=charge, $C$=capacitance, $G$=Newton's constant, $\varepsilon_0$=permittivity of vacuum, $\varepsilon$ = relative dielectric constant, $E$ = classical non-covariant electric field, $g$=electro-gravitational acceleration by the external plates, $g_{dielectric}$= the electro-gravitational acceleration of the dielectric dipoles, $V$=voltage applied to the external plates, $d$ = the distance between the plates, $\pi$ = 3.14159265…

1) Most of the molecular mass of the dipoles is between the poles of each dipole.
2) After the dielectric material is oppositely aligned with an external electric field, other dipoles with other alignments have a negligible effect. That includes dipoles between molecular dipoles.
3) The mass of each molecule can be considered as if it is located at the center of each dipole.
4) After the external field collapses, the dielectric opposite alignment does not collapse too. This is known as ferroelectricity and is typical of materials such as Lead Zirconium Titanate – PZT.
   **Important:** The same cannot be said for materials such as electrets, e.g. glass, because in electrets the charge carriers appear near the surface while the deeper molecular dipoles totally cancel out the gravitational field by the dipole which is caused by the surface charge carriers.

In general, regarding the non-covariant classical electric field, the gravitational acceleration by unexpected charge-based gravity and anti-gravity is:

$$g \approx \|E\|\sqrt{\pi G \varepsilon_0} \tag{14.5}$$

In a parallel plates capacitor without any dielectric:

$$g \approx \frac{V}{d}\sqrt{\pi G \varepsilon_0} \tag{14.6}$$

And in terms of charge

$$\frac{Q}{Cd} = \frac{V}{d} \tag{14.7}$$

$$g \approx \frac{V}{d}\sqrt{\pi G \varepsilon_0} = \frac{Q}{\varepsilon_0 \frac{A}{d} d}\sqrt{\pi G \varepsilon_0} = \frac{Q}{A}\sqrt{\frac{\pi G}{\varepsilon_0}} \tag{14.8}$$

The last term depends only on the density of charge per area! It means that once placing a dielectric that negates this value, the gravitational effect by the dielectric becomes

$$g_{dielectric} \approx -\frac{Q}{A}\sqrt{\frac{\pi G}{\varepsilon_0}} \tag{14.9}$$

Consider that classically a dielectric material with a relative dielectric constat $\varepsilon$ reduces the electric field by a factor of $\frac{1}{\varepsilon}$

$$g + g_{dielectric} \approx \frac{1}{\varepsilon}\frac{Q}{A}\sqrt{\frac{\pi G}{\varepsilon_0}} \Longrightarrow g_{dielectric} \approx \frac{1}{\varepsilon}\frac{Q}{A}\sqrt{\frac{\pi G}{\varepsilon_0}} - g \approx (\frac{1}{\varepsilon} - 1)\,\frac{Q}{A}\sqrt{\frac{\pi G}{\varepsilon_0}} \qquad (14.10)$$

From which it is obvious that the higher the relative dielectric constant $\varepsilon$ is, the better, assuming claims 1-4 are correct.

$$g_{dielectric} \approx (\frac{1}{\varepsilon} - 1)\,\frac{Q}{A}\sqrt{\frac{\pi G}{\varepsilon_0}} \qquad (14.11)$$

The force can be calculated if the mass density $\rho$ of the dielectric layer is known,

$$Force \approx \rho A d\left(\frac{1}{\varepsilon} - 1\right)\frac{Q}{A}\sqrt{\frac{\pi G}{\varepsilon_0}} = \rho d Q\left(\frac{1}{\varepsilon} - 1\right)\sqrt{\frac{\pi G}{\varepsilon_0}} \qquad (14.12)$$

When assumptions 1 and 3 are not fulfilled, the measured force can be several orders of magnitude weaker than the one predicted in (14.12).

**Crucial:** Notice the amplification of the original voltage by $\approx \varepsilon$ for $\varepsilon \gg 1$. This is because if the dielectric layer is removed while the external plates have charge $Q$, the same amount of charge on the plates $Q$, would generate a much higher voltage than the original $V$ and the assumption here is that when the external field collapses, the ferroelectric dipoles will act as the charge $Q$ without the dielectric layer and with an opposite sign. This trick, if it indeed works with all ferroelectric materials, offers a technological trick that amplifies $g$ by the thousands if $\varepsilon \gg 1$.

Effectively, a single-side-band sine wave between 0 and $V$ should generate an average gravitational pseudo acceleration smaller by a factor of 4 because only half of the wave should be usable and not all molecular dipoles will be oppositely aligned. So, assuming the amplification trick works, one calculates,

$$g_{effective} \approx \frac{1}{4}(1 - \varepsilon)\,\frac{V}{d}\sqrt{\pi G \varepsilon_0} \qquad (14.13)$$

Assume $\varepsilon = \mathbf{4000}$, $V = \mathbf{10000}\ \boldsymbol{Volts}$, $d = \mathbf{1}\ \boldsymbol{mm} = \mathbf{0.001}\ \boldsymbol{Meter}$.

We can consider $1 - \varepsilon = -3999 \sim -4000$.

$\sqrt{\pi G \varepsilon_0}$ ≈4.3087586002548416470445270690079e-11 ($C^2 \cdot kg^{-1} \cdot m^{-3} \cdot s^2$ m^3 kg^–1 s^–2)$^{(1/2)}$ ≈4.3087586002548416470445270690079e-11 ($C \cdot kg^{-1}$) then $(1 - \varepsilon)\frac{V}{d}\sqrt{\pi G \varepsilon_0} \approx 4000 * 10000 * 1000\sqrt{\pi G \varepsilon_0} \approx \mathbf{1.723503441019}\ \boldsymbol{Meter} * \boldsymbol{Sec^{-2}}$. The standard gravity is **g** ~

**9.80665 m/s²**, so we have about **0.17574844 g** if the hysteresis is perfect once the dielectric of 1mm thickness moves out of the plates.

All we need is a material with this hysteresis effect, A.K.A ferroelectric material, where most of the molecular mass is between the poles of each dipole, where dipoles in other directions other than opposite to the external field by the plates, before they are removed are negligible, with a high relative dielectric constant of about **4000** that has a breakdown voltage higher than **10000 volts / 1mm**. Plates of 10 $cm^2$, will allow easy measurement.

**Experiment 1:** See Fig. 3. C. The idea is to rotate disk-shaped plates of a capacitor with missing sectors of half of the area of the disks, such that the missing sectors of the top disk will match the ones of the bottom disk, while applying 10000 over 1mm or more. The disk area will be at least 0.01 $Meter^2$, the relative dielectric constant will be between 2000 to 4000. We should be able to achieve a weight reduction of a quarter of 0.17574844 g, about 0.04393711 g. The disk rotation can be also in the first direction, as in the following figure.

**Experiment 2** (much less efficient): Feed the capacitor with high DC spikes and use whole plates. This method requires ferroelectric materials with properties that are hard to achieve. Lead Zirconate Titanate – PZT is ferroelectric and may be suitable. Lead provides high mass density which is preferrable in a Bondi dipole because a higher mass, more than twice the density of titanium is expected to generate more pseudo-force due to the gravitational dipole.

**Why is it so hard to use ferroelectric materials? – an example is Lead Zirconate/Zirconium Titanate - PZT**

The polarization of PZT once the external field has collapsed to zero is in the range, 70 ms down to 0.07 ms.

Estimating the half-life for the PZT ferroelectricity

The half-life ($t_{1/2}$) of charge decay follows an exponential relationship:

$t_{1/2} = \tau \ln(2)$

For high-quality PZT, typical values are:

- $R_{leakage} \approx 10^9 - 10^{12}\ \Omega$ (depends on humidity and insulation),
- $C \approx 10 - 1000$ pF (varies with thickness and area).

This gives a **relaxation time (τ) in the range of microseconds to seconds**. In typical conditions:

- At $10^9\ \Omega$ and 100 pF → $\tau \approx 0.1$ ms, so **$t_{1/2} \approx 0.07$ milliseconds**.

- At $10^{12}$ Ω and 100 pF → $\tau \approx 100$ms, so **$t_{1/2} \approx 70$ milliseconds**.

It means that polarized DC pulses of 15 KHz can make an engine and result in **4%-5% weight loss for 10KV / 1mm**, however, a polarized pulse is hard to achieve with cutoff oscillators because the voltage graph always has a negative pulse too. With a polarized pulse **1% to 2% weight loss** is possible to achieve.

With a rotating disk with 4 missing sectors, it will be harder. 14,285.714 Hz with 0.07 milliseconds means rotation of **214,285.65 rpm**. With missing 16 disk sectors, the rotation is **53,571.42 rpm**. With these rotation speeds, **2% to 3% weight loss** is possible to achieve.

**Caution with (14)**: In reality, the charge of the induced dielectric dipoles is closer to the mass of the dipoles than the external plates. The assumptions of (14) therefore break down and the Inertial Dipole effect is much smaller. Permanent dipoles can have a stronger field than the external field. They are far more problematic than induced dipoles. One possible technological remedy to this anti-alignment is to add an Alternating Current - AC component to the DC baseline and to disrupt the anti-alignment. Still, even with such a component, a feasible propulsion system may require millions of volts as a baseline. When using voltage above 2 * 511 kV, creation of electron-positron pairs is difficult to avoid (not the Schwinger limit but accelerated electrons through parasitic leakage), and the resulting gamma rays are a serious health hazard. A dynamic voltage and/or current component renders the mathematical description of the Inertial Dipole much more difficult. The following calculations are therefore very optimistic.

Suppose we have a 1000Pf ceramic capacitor and we charge it with 10000 Volts and the area of the plates is 1 $cm^2$. The charge on the plates is then $10^{-5}$ Coulombs and its density $10^{-1}$ Coulombs per square meters. Now we want to calculate the approximate acceleration that the upper positive plate experiences due to the anti-gravity effect from the lower plate. Only a thin portion of the upper layer is affected, where the positive charge accumulates. A calculation shows: 0.48663510306 meters / $sec^2$. Dividing 0.4866351… meters/$sec^2$ by 9.81 meters / $sec^2$ we get 0.049606024776763 which is less than 5 percent relative to the gravity of the Earth. If instead of a dielectric material, an insulator with relative dielectric constant 1 is used for the same charge density of $10^{-5}$ Coulombs per 1 $cm^2$, a weight loss of the insulating slab should be measured at about 0.0496 of its weight. With a high relative dielectric constant, the affected mass could be well below 1 milligram, and it will lose 0.0496 of its weight. This renders the measurement of such an effect very hard to achieve unless the dielectric material is saturated and can no longer shield the field of the plates such as in the H4D experiment [19]. In any other case, practically no measurable thrust is expected for an area 1$cm^2$ with 10,000 Volts and scale resolution worse than $10^{-4}$ grams. In the case of saturation, at first the inertial dipole is expected to grow with the saturation of the dielectric material and with the amount of charge on the plates. [19] will be discussed later. The H4D lab [19] 69 mm radius and 2mm PMMA thickness capacitor with 20,000 volts, weight loss is at least **0.0015509 grams**, however the thickness of the metal plates is 1mm. It is sufficient to have a low frequency AC ripple from the DC power supply to churn the electrons on the plates such that not only a thin layer of the plates will be charged, also with an AC ripple, of typically 150 VAC for 20000 Volts DC, the induced gravitational field can no longer be considered static. Under such conditions (14) is no longer valid.

### 3. Thrust from 1000 Pf capacitor with two metallic plates and 10000 volts

**Caution:** The following calculations can be used if one ignores permanent dipoles and assumes that most of the molecular mass is not within permanent or induced dipoles, obviously this is not the case in real world capacitors. As mentioned before, static symmetrical capacitors are not expected to accelerate. The same is true for statically charged electrets.

**Assumptions**: Most of the dielectric mass is not completely shielded from the plate fields, permanent dipoles should not have a strong field such as in PTFE, and the attenuation of the influence of the external dipole on the mass within the induced dipole is by a factor $\varepsilon^{-1}$, where $\varepsilon$ is the relative dielectric constant. If this assumption does not hold true then (14) is invalid. Such a problem may occur at least theoretically even if in total the dielectric constant is low only because of low mass density. A second assumption is that dielectric dipoles are evenly distributed within the dielectric layer. A third assumption is a low alternating current – AC component in the power supply and that the influence of the Inertial Dipole on the metal plates is negligible due to the charge concentrating on the metallic surfaces which are in contact with the dielectric material. A high AC component might disrupt electrons alignment on the plates and if the plate's thickness is not negligible then (14) is no longer valid. Also, if the dielectric material reaches saturation and the metallic plates are thick in relation to the dielectric layer, the charge distribution on the plates can no longer be limited to the contact surfaces with the dielectric layers which also results in (14) being no longer valid.

Suppose we have a high voltage ceramic capacitor of 1000Pf of **Ta2O5** [20] with each plate area $1cm^2$ which is charged by 10,000 volts. The permittivity of vacuum is about 8.8541878128 * $10^{-12}$ Farads*$meter^{-1}$.  So we can calculate the distance d between the plates, 8.8541878128 * $10^{-12}$ Farads * $metere^{-1}$ * $10^{-4}$ $meters^2$ * $d^{-1}$ * 25 = $10^{-9}$ Farads. That means d ~ 0.22135469532 * $10^{-1}$ mm or d ~ 0.22135469532 * $10^{-2}$ cm. Now we take into account the weight density of the Ta2O5 which is 8.2 grams perm $1cm^3$ volume. So we have 8.2 * 1cm * 1cm * 0.22135469532 * $10^{-2}$ cm = 0.01815108501624 grams.  At 10000 volts the weight loss is of a portion of 0.0496060247767631571141158821638 8 of the weight of the dielectric material and the inertial dipole is attenuated by the relative dielectric constant 25 just as the electric field is.  So we have 0.01815108501624 grams * 0.04960602477676315711411588216388 * $25^{-1}$ ~ **~3.60161*$10^{-5}$ grams weight loss**. This estimate can be much lower in a multilayered capacitor where fields cancel out or when the dielectric constant is higher and the dipoles density is not uniform.

## 4. Martin Tajmar experimental null results analysis

**Caution:** The following calculations can be used if one ignores permanent dipoles and assumes that most of the molecular mass is not within permanent or induced dipoles, obviously this is not the case in real world capacitors. As mentioned before, static symmetrical capacitors are not expected to accelerate. The same is true for statically charged electrets.

### The no-go argument against static high voltage capacitors

The argument is that since charged molecules are averagely stationary in the non-covariant classical limit, and since the pseudo gravitational acceleration due to charge is parallel to the non-covariant classical electric field $E$ and is $acceleration \approx -E\sqrt{\pi G \varepsilon_0}$ where $G$ is Newton's gravity constant and $\varepsilon_0$ is the permittivity of vacuum, the molecular mass m when multiplied with the sum of $E\sqrt{\pi G \varepsilon_0}$ must be zero, i.e. $\sum mE\sqrt{\pi G \varepsilon_0} = 0$, otherwise charge would move in the dielectric layer. The exception is the conducting plates. The negative plate ideally has -2q charge near the dielectric later. +q is in the dielectric next to the plate and +q in the plate next to the -2q charge. So although the plate is charged and although the electric dipole of the capacitor is not zero, the mass * pseudo gravitational acceleration by the charge, must sum to zero. This argument means that there are only dynamic ways to generate a gravitational dipole with electric fields. Even without the idealized -2q charge, the affected layer of atoms in the plate is unfortunately negligible.

Martin Tajmar [21] used a capacitor of a relative dielectric constant 4500 and a Teflon [24] capacitor with radius 50 mm and Teflon thickness d=1.5 mm and 10,000 Volts. The highly dielectric capacitor weight loss is way below the experiment **scale resolution 3 * $10^{-4}$ grams** due to division by 4500 of the charge which is $10^{-5}$ per 1000Pf capacitance. With a radius of 0.5cm, such a capacitor with say 6.02 grams * $cm^{-3}$ density will lose about **2.077389 * $10^{-5}$ grams**. Next focus is on one of the Teflon capacitors. The gravitational acceleration on the face of the Earth, about g=9.80665 meter * $sec^{-2}$. By (14), the result is **7.5917876115 * $10^{-6}$ grams**. This result is smaller than the resolution of 3 * $10^{-4}$ grams. The results assume $\beth = 1$ in (4), (7), (13). It is important to say that unlike Martin Tajmar (sounds as Taymar), the Brazilian H4D experiment [19] used much greater capacitor areas. A significant AC ripple cannot be ruled out.

**Positive results:** Other results are the experiments and patents by Andrew Neil Aurigema and Charles Raymond Buhler [22] and by James W. Purvis [23].

Explanation to these results are as follows:

1) In the first device an asymmetry in a dielectric capacitor does not allow opposite alignment of dielectric dipoles to totally negate the external gravitational dipole. Still the resulting acceleration seems to be several orders of magnitude more than predicted by (13), (14).

2) In the second device, addition of a dynamic field to the expected gravitational dipole due to (13) is indeed expected to change the predictions of (14).

There is however a need to solve (4) analytically for both [22] and [23] and from (10), to use the classical limit $\|a^\mu\| = \sqrt{4\pi K\varepsilon_0 \beth}\|E\|$, $\left\|\frac{a^\mu}{c^2}\right\| = \left\|\frac{u^\mu}{2}\right\|$.

**Important:** General research directions for finding astronomical evidence for charge-based gravity and anti-gravity:

1) If a small galaxy collided with a large dust cloud or with another galaxy, it should be positively charged. Due to near electrostatic repulsion, star formation must be very low, however, a Dark Matter effect i.e. unexpected gravity must be higher in comparison to other galaxies, hopefully by more than 10%. On the other hand, collisions between galaxies can emit electrons to neighboring galaxies in the same cluster, which will then manifest a very weak Dark Matter effect due to the excess in negative charge. With a strong magnetic field, the collision center cannot lose electrons, also when the gas is ionized and therefore in such cases only heavy atoms, rocks and dust can escape while positively charged. The magnetic field in collision of clusters therefore causes the Dark Matter effect to be separated from the volume where there is a strong magnetic field.

2) A large, isolated galaxy with billions of light years of minimal distance to other galaxies should have sufficient time for the electrons it lost to fall back as the galaxy gets older and cooler. If there are such galaxies, then they must have a weak Dark Matter effect or no Dark Matter effect at all, despite the fact that they are big, e.g. the size of the Milky Way.

3) Electrons have a light weight and are easily accelerated to relativistic speeds which helps them to escape the galactic pull. In intergalactic space, they should cause a gravitational repulsion and the expansion of the cosmos. Therefore, cosmic expansion should be faster when there are more free electrons.

## 5. Particle mass ratios – a reverse engineering Ansatz approach

**Motivation:** solving (4) analytically is extremely hard, let alone, the more general Lagrangians that will be presented in (64.b), (64.01) and (65) for complex Reeb class vectors. One possible way to tackle this challenge is to rely on a theorem by Georges Reeb, according to which the restriction of the field to the three-dimensional foliation perpendicular to $\frac{P_\mu}{\sqrt{Z}}$, must have a zero rotor, see theorem 3. If the reader is patient, the reader will be rewarded with the muon/electron mass ratio in (24), the remark after the inverse Fine Structure Constant assessment (40) and the calculation in (41.1). Together (24), (40), (41.1) have a probability lower than $10^{-24}$ to be a fluke of chance because the errors of these 3 calculations are independent. This is said although unlike (1) – (14.12) the following (15)-(42) are not the result of rigid mathematics, but of general theoretical considerations. Understanding some of the theoretical considerations and especially why 95 * 96 is used in (40), (41), (42) will become clearer only in (46.1)-(46.10) in "Quantization of the acceleration matrix for the field strength coefficients - Compelling evidence of the field strength coefficients".

**Synopsis: How does this paper plan to reach particle mass ratios?**

From (13.03) and the crucial remark after (13.03), what happens is that when there is a gravitational field around a charged particle, then part of the acceleration (of a unit vector of the normalized Geroch [2] gradient at rest in relation to the charge) turns into gravitational energy.

Now, the energy density and the non-geodesic acceleration $a_\mu$ is $-\frac{a_\mu a^\mu}{8\pi K}$ in (+,-,-,-) convention. A quarter of this energy is held by charge-based gravity, not by the other component of the gravitational field which is due to the energy of the charge. This gravitational energy is a usable energy because charge can be mutually annihilated as in the electron-positron interaction. This realization will be interpreted as a factor $\frac{1}{4} = \left(\frac{1}{2}\right)^2$ in the following particle mass ratio calculations.

Restricting our discussion to the real numbered version of this model, the area loss around a gravity source or gain around an anti-gravity source in the direction of time $\frac{P_\mu}{\sqrt{Z}} = \frac{P_\mu}{Norm(P_\lambda)}$

$$\delta Area(disks\ with\ radius\ r) = \frac{\pi}{24}\left(R_{\mu\nu} - \frac{1}{2}g_{\mu\nu}R\right)\frac{P^\mu P^\nu}{Z}r^4 \tag{15}$$

In relation to an Euclidean disk with a small radius $r \ll 1$,

$$\frac{\delta Area(disks\ with\ radius\ r)}{\pi r^2} = \frac{1}{24}\left(R_{\mu\nu} - \frac{1}{2}g_{\mu\nu}R\right)\frac{P^\mu P^\nu}{Z}r^2 \tag{15.1}$$

The factor $\frac{1}{4}$ turns this ratio into:

$$\frac{1}{4}\frac{\delta Area(disks\ with\ radius\ r)}{\pi r^2} = \frac{1}{96}\left(R_{\mu\nu} - \frac{1}{2}g_{\mu\nu}R\right)\frac{P^\mu P^\nu}{Z}r^2 \tag{15.2}$$

According to theories which will be mentioned, gravitational energy is in relation to the energy of the mass which generates the gravitational field. The ratio between such delta area and Euclidean area then must yield an estimate of the gravitational energy. The quantum leap here is that the paper assumes that in all forms of gravity and not only charge based gravity, the usable gravitational energy of a decaying particle, is a quarter, i.e. $\frac{1}{4}$ of the gravitational energy. This assumption leads to an intriguing possibility that the energy of the electron and neutrinos that result from the Muon's decay must be proportional to the usable gravitational energy of the Muons.

Another interpretation is a non-trivial relation between an atom of length and an atom of area, an idea by a colleague Arye Aldema who passed in 30/Dec/2022. A simple relation is $\frac{1}{2}\sqrt{Area} = Length$. This specific idea also leads to a factor $\frac{1}{4}$. The surprise is that these simple assumptions led to quite an accurate mass ratio between the Muon and the electron although the spin of these particles is not accounted for in the calculations. It is as if spin does not arouse in the Planck scale. What was missing in the theory was the relation between the Reeb class vector - not the usual Reeb vector - and a radius $r \ll 1$ for the electron, Muon and Tau lepton. These were conjectured to be $\frac{95}{96}, \frac{4}{\pi}$, and 1.556198537190348396563877031439915299... from the realization that because charge is a source/drain singularity of the Reeb class field and because it is conserved, it must arise from flow singularities in 3D foliations with zero charge due to the fact that $\frac{u_\mu}{2}$, or in the complex case $\frac{u_\mu^* + u_\mu}{4}$, is a spacelike vector. These can be one drain, and one source as expected from stable charge, center singularity which consists of circles around a central line, and a saddle point which can work in the theory only for entering planar lines and exiting two directions perpendicular to the plane. In these cases, the total charge is assumed to be zero far from the singularities.

A simple analysis of these singularities yields field strengths 95/96, 4/Pi and 1.556198537190348396563877031439915299... from which the mass ratio between the Muon and the electron is calculated surprisingly accurate, and also the mass ratio between the Tau lepton and the Muon is calculated and eventually a very good assessment of the inverse Fine Structure Constant is calculated too. To make progress with this theory of particle mass ratios beyond that ansatz stage it requires deep mathematical understanding of the dynamics of flow singularities, and it cannot be done by one person. Attacking this part of the theory as pure ansatz is in my humble opinion incorrect, especially when the claims that are made here are well understood. The theory will consider other explanations to the factor $\frac{1}{4}$ and to the field strength coefficients just for the sake of being honest as a researcher, however, the explanation in (13.03) of the factor $\frac{1}{4}$ and of the field strength coefficients coming from flow singularities with a total electric charge zero are by far the simplest and the most consistent.

**Theorem 3 (Reeb):** The rotor of $\eta$, the acceleration field or as better known as Reeb class, when restricted to the perpendicular foliation to $\alpha$ such that $d\alpha = \pm\eta \wedge \alpha,\ (D\eta)\wedge\alpha$ is zero.

**Proof:** Using exterior derivative $D\frac{P_\mu}{\sqrt{Z}}dx^\mu = D\alpha = \pm\eta\wedge\alpha = \left(\frac{U_\mu}{2}\frac{P_\nu}{\sqrt{Z}} - \frac{U_\nu}{2}\frac{P_\mu}{\sqrt{Z}}\right)dx^\mu \wedge dx^\nu$

We now take the exterior derivative of $D\alpha = \eta\wedge\alpha$ and get $DD\alpha = (D\eta)\wedge\alpha - \eta\wedge(D\alpha) = 0$ because $D\alpha$ is an exact form. $DD\alpha = (D\eta)\wedge\alpha - \eta\wedge(D\alpha) = (D\eta)\wedge\alpha - \eta\wedge\eta\wedge\alpha = 0$ but $\eta\wedge\eta = 0$ so $\eta\wedge\eta\wedge\alpha = 0$ and therefore $DD\alpha = (D\eta)\wedge\alpha = 0$ Q.E.D. Let the lower indices denote covariant vector components, not derivatives and comma will denote derivatives, then $D\eta = (\eta_{\mu,\nu} - \eta_{\nu,\mu})dx^{\mu}\wedge dx^{\nu}$ and $(D\eta)\wedge\alpha = (\eta_{\mu,\nu} - \eta_{\nu,\mu})\alpha_{\lambda}dx^{\mu}\wedge dx^{\nu}\wedge dx^{\lambda}$ which means that the restriction of the rotor of $\eta_{\mu,\nu} - \eta_{\nu,\mu}$ to the foliation perpendicular to $\alpha_{\lambda}$ is zero and therefore the projection of $\frac{U_{\mu}}{2}$ on the foliation perpendicular to $\frac{P_{\mu}}{\sqrt{Z}}$ is of a conserving field.

**Corollary 4 to theorem 3:** $(D\eta)\wedge\alpha = 0 \Longrightarrow D((D\eta)\wedge\alpha) = 0 \Longrightarrow (DD\eta)\wedge\alpha + (D\eta)\wedge D\alpha = 0 \Longrightarrow (D\eta)\wedge D\alpha = (\mathrm{D}\eta)\wedge\eta\wedge\alpha = 0$, which means that $(D\eta)$ must not span the Hodge star of $\eta\wedge\alpha$ or in other words $\frac{U_{\mu,\nu}-U_{\mu,\nu}}{2}$ is a bivector that must depend on $\frac{P_{\mu}}{\sqrt{Z}}$ if it is not zero, because if it doesn't then by theorem 3, $\frac{U_{\mu,\nu}-U_{\mu,\nu}}{2}$ would be 0, Q.E.D.

**Time reversal violation** – can be skipped after Wigner's theorem up to "From theorem 3, since this model …".

Corollary 4 implies a time reversal violation at foliations points where $\frac{U_{\mu;}{}^{\mu}}{2} \neq 0$ or $\frac{1}{2}\left(\frac{U_{\mu;}{}^{\mu}}{2} + \frac{U^{*}{}_{\mu;}{}^{\mu}}{2}\right) \neq 0$ in the complex case. $\frac{U^{\mu}}{2}$ $or$ $\frac{1}{2}\left(\frac{U^{\mu}}{2} + \frac{U^{*\mu}}{2}\right) = \frac{a^{\mu}}{c^{2}} = \frac{d^{2}x^{\mu}}{(d\tau)^{2}} = \frac{d^{2}x^{\mu}}{(-d\tau)^{2}}$ and therefore, time reversal does not change the vector $\frac{a^{\mu}}{c^{2}}$ where $c$ is the speed of light. It is obvious that to change the sign of $U^{\mu}$ it is not sufficient to change only $d\tau$ to $-d\tau$. What does change sign, is velocity, $\frac{dx^{\mu}}{(-d\tau)} = -\frac{dx^{\mu}}{d\tau}$ but the field $a^{\mu}$ does not change sign, which means integration along the same line, one with the time and one back in time should be zero. Consider the coordinates map $x^{\mu} \longrightarrow -x^{\mu}$ which combines time reversal and spatial reflection, a.k.a Parity. In that case spin direction is maintained and charge is maintained. Under time reversal alone, charge is maintained. This is not surprising because this theory is not a hermitian theory but a geometry based theroy and the discussion here is of a single source of Reeb class divergence, not of interactions between particles. The Hodge star extension of $\eta\wedge\alpha$ is the extension of the field to angular acceleration and can have two different signs in the real case, left handed and right handed. Parity transormations $x^{\mu} \longrightarrow -x^{\mu}, \mu = 1,2,3$ do affect this sign.

**Fig. 4.** – Spatial reflection (Parity) and time reversal

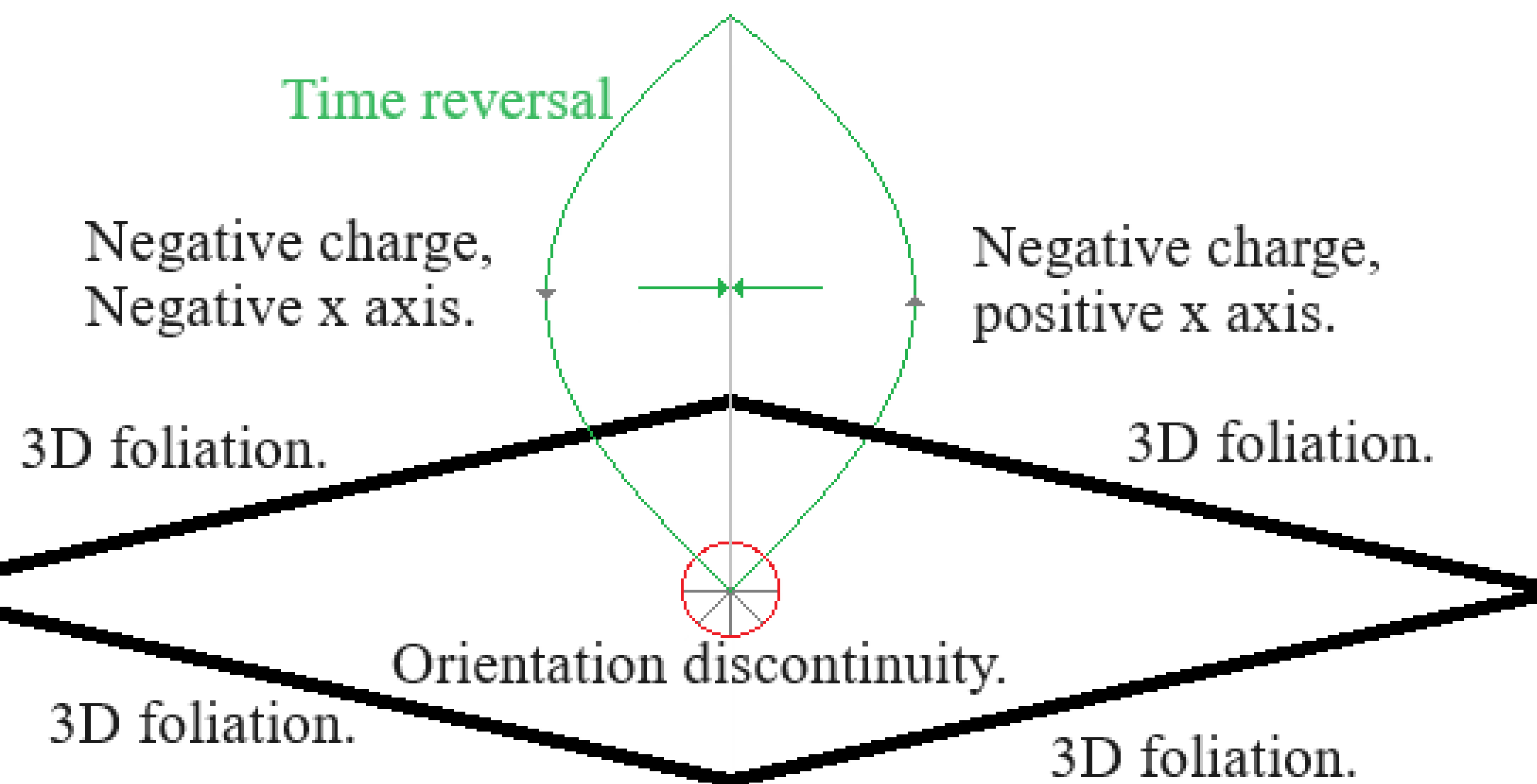

**Time asymmetry** – the difference between this model and the Hermitian representation of Quantum Mechanics

Roberts makes a two-step argument to show time reversal in Quantum Mechanics. The first is based on Uhlhorn theorem and on Wigner's theorem.

Roberts shows that time reversal can only be anti-unitary [25].

**Uhlhorn theorem:** Let T denote a linear bijection on the projection space of separable Hilbert space H with dimension greater than 2. Then if $\varphi \perp \phi \Leftrightarrow \langle \varphi, \phi \rangle = \langle T\varphi, T\phi \rangle$, there exists a unique operator $\tilde{T}$ up to a constant, which implements $T$, $\tilde{T}: H \longrightarrow H$ such that, $\tilde{\varphi} \in \varphi \Leftrightarrow \tilde{T}\tilde{\varphi} \in T\varphi$ and which satisfies $\left|\langle T\tilde{\varphi}, T\tilde{\phi} \rangle\right| = \left|\langle \tilde{\varphi}, \tilde{\phi} \rangle\right|$ for all $\tilde{\varphi}, \tilde{\phi} \in H$ [26].

Uhlhorn's idea is that independent states or mutually exclusive ones, must not depend on the direction of time.

**Wigner's Theorem:** For any operator $T$ satisfying the Uhlhorn Theorem, there is a Hilber space operator $\tilde{T}$ that implements $T$ which is either unitary or anti-unitary [27].

Then Roberts shows in sufficiently simple Hamiltonians that time reversal is equivalent to the action of an anti-unitary operator, which does not include the electroweak Hamiltonian.

Regarding this paper, the closest consideration to Roberts [26] and Wigner's theorem is to replace $f(P)$ in Appendix H by $f(P) = e^{-iP}$ or $f(P) = e^{iP}$ and see that the Lagrangian of the Reeb Class vector does not change, however, the Hodge star of $\frac{P_\mu}{\sqrt{Z}} \wedge \frac{U_\nu}{2}$ does change sign with time reversal see $B^{\mu\nu} = \frac{1}{2} E^{\mu\nu\alpha\beta} A_{\alpha\beta}$ after (3). Multiply $E^{\mu\nu\alpha\beta}$ by -1 for time reversal.

From theorem 3, since this model means the acceleration field $\frac{U_\mu}{2}$ is a representation of the electric field, in order to derive the dynamics of charge as we know it from classical mechanics, we need to contract equation (4) twice by $\frac{P_\mu}{\sqrt{Z}}$. In other words, the field must have drains and sources, by which the divergence of the field is not zero. The result of this theorem is that as the far observer $r \to 0$ in source or drain of the field, particles formation is inevitable and linearization of (4) as an approximation should be considered. This section will try to find a relationship between an acceleration as $\sqrt{|a^\mu a_\mu|} = \xi' \frac{c^2}{r}$ for some $\xi'$ and the norm of the Reeb class field $\sqrt{\frac{1}{8} |U^\mu U^*{}_\mu + U^{*\mu} U_\mu|} = \xi' \frac{1}{r}$ for some $\xi'$. In fact, this section considers (4) as $r \to 0$ as an attempt to avoid the extremely hard analytic solutions. The relationship between $\frac{1}{r}$ and $\sqrt{\frac{1}{8} |U^\mu U^*{}_\mu + U^{*\mu} U_\mu|}$ is not based on rigorous mathematics although there are rational explanations to these relationships. A better assessment is $\left\| Re(\frac{1}{2} U_\mu) \right\| = \xi' \frac{1}{r}$ for some $\xi'$.

The following section will try to reach the Reeb class field strengths of the electron, Muon and Tau Lepton. It will also try to reach the Reeb class field strength for the W and Z bosons. As we shall see, for the first 3 values, the assessment is $\frac{95}{96}, \frac{4}{\pi}$ and ~1.556198537190348396563877031439…

As we shall see $\frac{95}{96} = 1 - \frac{1}{64} + \frac{1}{192} = \frac{193}{192} + \frac{63}{64} - 1$, which can be interpreted as the summation of two fundamental states of the field. However, to keep an open mind, other possible reasons, although less plausible, are also brought into the discussion. The only value that does not come directly from this theory is $\frac{4}{\pi}$. It has a compelling Quantum Mechanics source; however, other less plausible explanations are also considered. The last value, ~1.55619853719, is derived from maximal imbalance between gravity and anti-gravity. For the W boson, two possible field strength coefficients are discussed $\frac{4}{\pi}$ and $\frac{4}{3}$. The latter yields a higher mass for the W boson although the author tends to accept $\frac{4}{\pi}$ and not $\frac{4}{3}$.

In this section, equation (4) is explored in a small infinitesimal sphere, where we assume a linear relation between a far observer radius $r$ and acceleration $\frac{a^\mu}{c^2} = \frac{U^\mu}{2} = \frac{Z^\mu}{2Z} - \frac{Z^k P_k P^\mu}{2Z^2}$, see (1), (2). Our goal is to reduce (4) from a four-dimensional Minkowsky geometry to a three-dimensional Riemannian geometry and then to a two-dimensional Riemannian geometry of surfaces.

**Critical:** We make the following assumption under which we want to see how equation (4) looks like when $r$ becomes small:

$$\frac{\|a^{\mu}\|}{c^2} = \frac{\xi}{rx} \tag{15.3}$$

Where, c is the speed of light, $\xi$ is a coefficient that depends on the field as $r \to 0$ and the variable $x$ changes with the density of the field as it passes through a two-dimensional sphere. $x$ is required because space-time curvature can cause such a sphere to be less than or more than $4\pi r^2$.

**Critical:** $r$ can also be interpreted as a small delta from a sphere that has a radius much larger than $r$.

**Note:** A natural question due to (10) is, when does the acceleration $\frac{\xi c^2}{r} = \frac{e}{4\pi\varepsilon_0 r^2}\sqrt{4\pi\varepsilon_0 k}$ , such that $e$ is the charge of the electron, $k$ is Newton's gravity constant, $\varepsilon_0$ is the permittivity of vacuum and $c$ is the speed of light? The answer is $r = \frac{e}{\xi c^2}\sqrt{\frac{k}{4\pi\varepsilon_0}}$ and for $\xi = 1$, $r = \frac{e}{c^2}\sqrt{\frac{k}{4\pi\varepsilon_0}}$ ,which by the order of the inverse of the square root of the Fine Structure Constant is smaller than the Planck length, $\frac{e}{c^2}\sqrt{\frac{k}{4\pi\varepsilon_0}}\alpha^{-\frac{1}{2}} = \frac{e}{c^2}\sqrt{\frac{k}{4\pi\varepsilon_0}}\sqrt{\frac{4\pi\varepsilon_0\hbar c}{e^2}} = \sqrt{\frac{\hbar k}{c^3}}$ , where $\hbar$ is the reduced Planck constant. This calculation of course, assumes that in such a strong field, the permittivity is that of vacuum and is not affected by virtual electric fields that attenuate the electric field. It is also limited to the far observer coordinates system.

We also make other assumptions as follows:

1) Assumption 1: In small radii, the energy of the gravitational field depends on the area around the source of gravity. This assumption is consistent with the paper of Ted Jacobson [28].

2) Assumption 2: The area ratio that has a physical meaning is between a disk to which the unit vector $\frac{P^{\mu}}{\sqrt{Z}}$ points to and the 1 weighted Euclidean sphere $ƛ * \pi r^2$ so $ƛ = 4$. The area loss of a disk is $\frac{\pi}{24}Rr^4$, where R is obtained by contracting Einstein's tensor twice with a time-like vector $\frac{P^{\mu}}{\sqrt{Z}}$ and $r$ is an infinitesimal radius. However, we consider $\frac{1}{4}\frac{\pi}{24}Rr^4 = \frac{\pi}{96}Rr^4$. As we divide this area by Euclidean disk area, we get $\frac{\pi}{96}Rr^4 * (\pi r^2)^{-1} = \frac{1}{96}Rr^2$. Following are explanations to the factor $\frac{1}{4}$.

   **Critical:** The reader may feel that bringing different explanations for the factor $\frac{1}{4}$ and for the field strength coefficients $\xi$ is tossing around all sorts of theories. For this very reason, the simplest explanation that complies with this theory is termed as "primary explanation", therefore although the approach here is an Ansatz approach, these primary explanations cannot be dismissed as "all sorts of theories".
**Critical:** A general principle for reaching these field strengths $\xi$ is that they are either a result of the sum of two fields or the result of equalities between two fields. Equalities can be due to geometrical or algebraic considerations.

**The primary explanation and its contender: Ratio between a length atom and the square root of an area atom:** It is easy to see that for (13.03) to hold, a factor $\frac{1}{4}$ must be considered, when describing electric charge as a sphere and by Occam's razor. That would be a closed case argument, however, there is an explanation which is also interesting. In the complex version of equation (4), the Geroch function [2] $PP^*$ can be replaced with a probability density function of an event $\psi\psi^*$ and except for a set of measure zero, $\psi_\mu = \frac{d\psi}{dx^\mu}$ is not zero. In that case $\int \psi\psi^* \sqrt{-g} dx^0 dx^1 dx^2 dx^3 = 1$ and the constraint should be added to the Lagrangian, $Action = Min \int_\Omega \left(R - \frac{1}{4\beth} U^k U_k + \lambda\psi\psi^*\right) \sqrt{-g}\, d\Omega$ where $\lambda$ is a constant of units $\frac{1}{Length^2}$, which implies the existence of an atom of area. The relation between a length atom and an area atom should be $\frac{1}{2}\sqrt{Area} = Length \Longrightarrow \frac{1}{4} Area = Length^2$. There is more than one reason for this relation. The simplest is that an area is measured around an event in spacetime in a sub-plane of spacetime. The time is then an exception because it has a direction and only one direction around an event can be considered. The relation between a 4-volume to such length unit is then $Area^2 = (2Length)^3 Length = 8Length^4$, providing that a time differential $cdt = dLength$ is treated as delta length.
**Caveat:** do not confuse the use of the term Area in a 4-volume relation to length, with an Area in area to length ratio, it is not the same term as in $\frac{1}{4} Area = Length^2$. The reason for the term $Area^2$ can be seen in "Hodge star spin-like field extension" and is used in (3). (3) describes two planes of acceleration, one of a boosting acceleration and one of a rotation acceleration. The relations, $\frac{1}{2}\sqrt{Area} = Length \Longrightarrow \frac{1}{4} Area = Length^2$ and $Area^2 = (2Length)^3 Length = 8Length^4$ are a result of discussions with a colleague Aryeh Aldema who unfortunately passed on 30/December/2022. This idea had led to an assessment of the inverse Fine Structure constant. Aryeh was not sure about the exact ratio between atoms of lengths, areas and 4-volumes but insisted that such ratios must be used in this paper. More specifically, by (15.3), equation (4) can be reduced to the following and will be discussed in more detail than here,

$$\frac{\pi}{24}\left(-\frac{1}{2}\frac{\xi^2}{x^2}\frac{1}{r^2}\mp\frac{\xi}{x}\frac{1}{r^2}\right)\frac{Area}{4}\frac{Area}{4}\left(\frac{Disk\ Area}{4}\right)^{-1}=\delta Area\left(\frac{Disk\ Area}{4}\right)^{-1}=x-1 \quad (15.4)$$

$$\frac{\pi}{24}\left(-\frac{1}{2}\frac{\xi^2}{x^2}\frac{1}{r^2}\mp\frac{\xi}{x}\frac{1}{r^2}\right)\frac{r^2}{4}\frac{r^2}{4}\left(\frac{\pi r^2}{4}\right)^{-1}=\frac{1}{96}\left(-\frac{1}{2}\frac{\xi^2}{x^2}\mp\frac{\xi}{x}\right)=x-1 \quad (15.5)$$

- **The simplest explanation:** The simplest explanation is that the portion of the gravitational energy that can be used when particles decay, is only from the time-like field $\frac{P(0)_\mu}{Z(0)}$ in (3.12) from which the factor $\frac{1}{4}$ comes. This explanation is however problematic even if $\frac{P(i)_\mu}{Z(i)}$ are complex functions as in (3.13) because 8 scalar functions may not be sufficient to describe gravity.

- **Blackhole thermodynamics - Bekenstein and Hawking entropy and area:** see the relation $S_{BH}=\frac{1}{4}K_B\frac{A}{{\ell_p}^2}$ [29] where $A$ is area, $\ell_p$ is the Planck length, and $K_B$ is Boltzmann's constant. We assume entropy is related to particles decay.

- **Mathematically and physically compelling explanation:** We return to the principles of the chronon field by Sam Vaknin [17] in which the time arrow is defined via spin and thus via orientation: There are two orientations to be considered. The first is the orientation of the foliation that is perpendicular to $\frac{P^\mu}{\sqrt{Z}}$. The second is the plane within that foliation which is perpendicular to $\frac{P^\mu}{\sqrt{Z}}$ and to $\frac{U^\mu}{2}$. In each case only one side of a 3D foliation and one side of a plane can be related to energy and $\frac{1}{2}*\frac{1}{2}=\frac{1}{4}$.

- **Causal triangulation explanation:** A polygonal graph is a graph in which vertices on a circle relate to edges and each vertex is also connected to the center. So, for the m vertices of the polygon and one vertex of the center, the graph has m+1 vertices. We also assume m=2n for some natural number n. The graph has 2m = 4n edges, m connecting the polygon vertices, each vertex to 2 neighbors and m connecting the polygon vertices to the center. Using graph theory techniques, it is easy to see that a random walk for a large n on such polygonal graph reaches a probability $\frac{1}{4}$ at the center and $\frac{3}{4m}$ at each polygon vertex. The probability of moving from a vertex on the polygon to one if its two neighbors is $\frac{1}{3}$ for each neighbor and to the center $\frac{1}{3}$. The probability of reaching one node of the polygon from the center is $\frac{1}{m}$. Seeing a particle as a loop with or without a center is beyond the scope of this paper, however, such a model under random walk reaches the unique probability $\frac{1}{4}$ at the center and is worth mentioning as another approach to area related to energy as $\frac{\pi}{96}Rr^4$ instead of $\frac{\pi}{24}Rr^4$, where R obtained by contracting Einstein's tensor twice with a timeline vector $\frac{P^\mu}{\sqrt{Z}}$ and $r$ is an infinitesimal radius. The python code for the random walk calculations is brought here:

```
import numpy as NP
import numpy.linalg as LA

print('Random walk on 24-Polygonal  graph with a center.')

matrix = NP.zeros((25, 25), dtype=NP.float64)
a = 1/24
b = 1/3
for i in range(1, 25):
    matrix[0, i] = b
    matrix[i, 0] = a
    k = i + 1 if i < 24 else 1
    matrix[i, k] = b
    k = i - 1 if i > 1 else 24
    matrix[i, k] = b
    w, v = LA.eig(matrix)
    scale = v[:, 0].sum()
    v[:, 0] /= scale

print('Eigenvector of probability:')

for i in range(25):
    print(f'v[{i}]={v[i, 0]}')
    print(f'Eigenvalue {w[0]}')

The output is:
Random walk on 24-Polygonal graph with a center.
Eigenvector of probability:

v[0]=0.24999999999999997
v[1]=0.03124999999999989
...
```

The Causal Set interpretation (87)-(90) and its relation to the number 96 and the Fine Structure Constant cannot be ignored!

- **Non rigid explanation:** This idea is derived from a physical principle according to which a spin of a particle always either points to an observer or in the opposite direction. In this manner, the observer can only refer to the disc which is perpendicular to the spin axis and not to an entire sphere. An area ratio $\frac{\pi r^2}{4\pi r^2} = \frac{1}{4}$ means 0 gravity.

This assumption means that the delta area of a curved sphere divided by $4\pi r^2$ is $\frac{\delta\pi r^2}{\text{ℷ}*\pi r^2}$ and not $\frac{\delta 4\pi r^2}{4\pi r^2}$. There could be other explanations to this assumption including a choice of $32\pi K$ in (7) instead of $8\pi K$ and $\frac{1}{16}$ instead of $\frac{1}{4}$ in (4), however to the author's opinion, (43) does not support such other explanations.

We revisit equation (4) and contract it twice with the unit vector $\frac{P^\mu}{\sqrt{Z}}$ which means a chosen time direction $\frac{1}{4\text{ב}}\left(U_\mu U_\nu - \frac{1}{2}g_{\mu\nu}U_\lambda U^\lambda - 2U^k{}_{;k}\frac{P_\mu P_\nu}{Z}\right)\frac{P^\mu P^\nu}{Z} = (R_{\mu\nu} - \frac{1}{2}Rg_{\mu\nu})\frac{P^\mu P^\nu}{Z}$

Since $U_\mu P^\mu = 0$, and assuming $\text{ב} = 1$, we have around an electric charge by (15.3) and an infinitesimal linearization of (15.3) such that $U_\mu$ is space-like and $\frac{U_\lambda U^\lambda}{4} \cong \xi^2\frac{1}{r^2}$, where $\xi$ is a field strength coefficient, $r \to 0$ radius and we can write $a = \xi\frac{c^2}{r}$ were $c$ is the speed of light and $a$ represents an acceleration.

$$\frac{1}{\text{ב}}\left(-\frac{1}{2}\frac{U_\lambda U^\lambda}{4} - \frac{1}{2}U^k{}_{;k}\right) = \frac{1}{\text{ב}}\left(-\frac{1}{2}\frac{\xi^2}{r^2x^2} \mp \frac{\xi}{r^2x}\right) = (R_{\mu\nu} - \frac{1}{2}Rg_{\mu\nu})\frac{P^\mu P^\nu}{Z} \qquad (16)$$

We calculated the divergence of a field of a non-geodesic acceleration from intensity $\frac{\xi}{rx}$ to 0 along the distance $r$. The divergence $U^k{}_{;k}$ can be either positive or negative and depends on the sign of the electric charge. We now refer to Seth Lloyd lecture [30],

**Fig. 5.** – Area gain or loss in the direction of a unit vector:

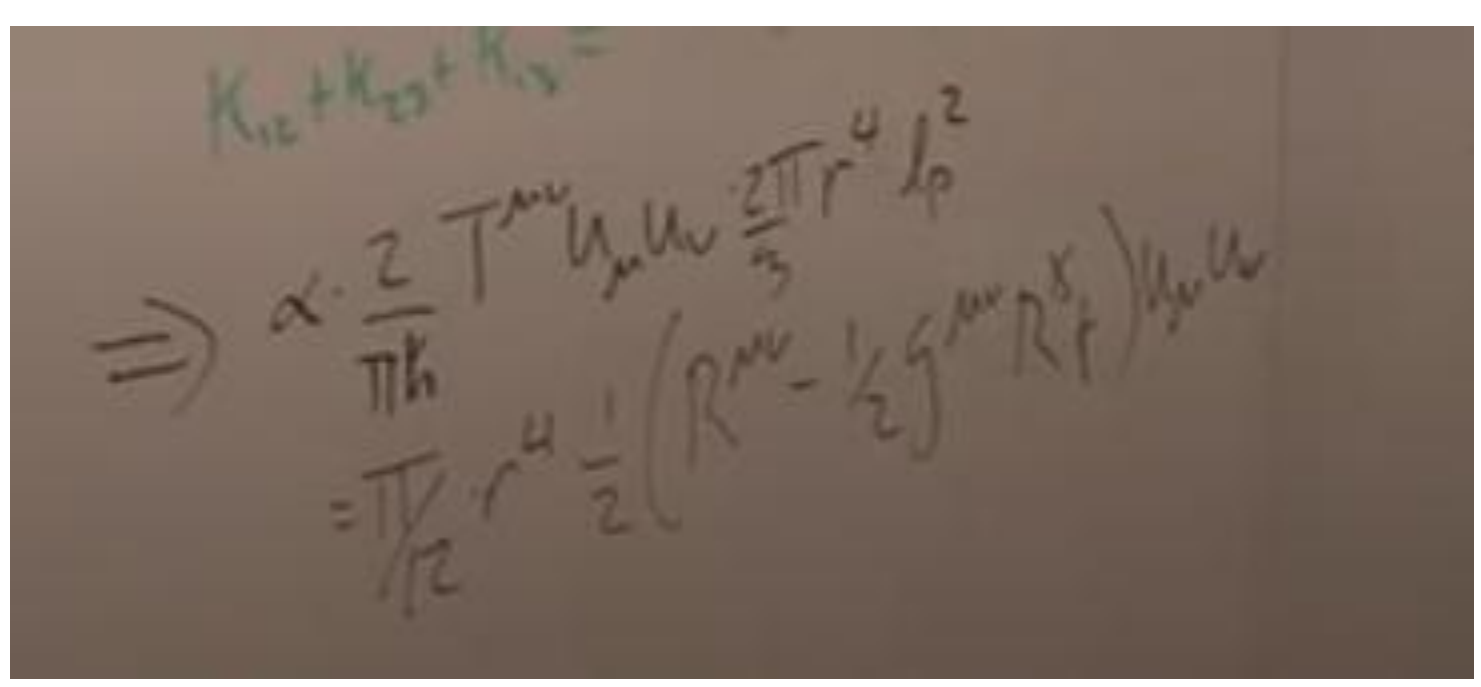

As we see, to get the area loss on a disk which is perpendicular to the unit vector $\frac{P^\mu}{\sqrt{Z}}$ due to curvature, we need to multiply (16) by $\frac{\pi}{12}\frac{1}{2}r^4 = \frac{\pi}{24}r^4$.

$$\frac{1}{\text{ב}}\left(-\frac{1}{2}\frac{\xi^2}{r^2x^2} \mp \frac{\xi}{r^2x}\right)\frac{\pi}{24}r^4 = \frac{1}{\text{ב}}\left(-\frac{1}{2}\frac{\xi^2}{x^2} \mp \frac{\xi}{x}\right)\frac{\pi}{24}r^2 = AreaLossOfADisk \qquad (17)$$

By our second assumption, the following has a physical meaning, where $\text{ℷ} = 4,\ \text{ב} * \text{ℷ} = 1 * 4 = 4$

$$\left(-\frac{1}{2}\frac{\xi^2}{x^2} \mp \frac{\xi}{x}\right)\frac{1}{96} = \frac{1}{\beth * \gimel}\frac{1}{\pi r^2}\left(-\frac{1}{2}\frac{\xi^2}{x^2} \mp \frac{\xi}{x}\right)\frac{\pi}{24}r^2 = \frac{AreaLossOfADisk}{\gimel * \pi r^2} \tag{18}$$

But x should be a ratio between an area around a charge and Euclidean area, according to assumption 2. If x is greater than 1, then by (17), the non-geodesic acceleration field density is decreased by a factor of $\frac{1}{x}$ . If the area ratio is smaller 1 then the non geodesic field density is increased by $\frac{1}{x}$. So, we must have the following equation:

$$x = 1 + \frac{AreaLossOfADisk}{4\pi r^2} \Leftrightarrow x - 1 = \frac{AreaLossOfADisk}{4\pi r^2}$$

And by (10) and (12), (18) becomes:

$$\left(-\frac{1}{2}\frac{\xi^2}{x^2} \mp \frac{\xi}{x}\right)\frac{1}{96} = \mathrm{x} - 1 \Leftrightarrow 1 + \left(-\frac{1}{2}\frac{\xi^2}{x^2} \mp \frac{\xi}{x}\right)\frac{1}{96} = x \Leftrightarrow \frac{192x^2 \mp 2\xi x - \xi^2}{192} = x^3 \tag{19}$$

The righthand side is expected to be positive around a negative charge and negative around a positive charge if we consider the H4D experimental qualitative result [19] with imprecise balance.

**Important:** Only convergent roots $\left(\frac{192{x_n}^2 \mp 2\xi x_n - \xi^2}{192}\right)^{\frac{1}{3}} = x_{n+1}$ with iteration parameter $n$ are expected to have a physical meaning. These are the roots which are closest to 1.

The values of $\xi$ that will be explored are $\frac{95}{96}$, $\frac{4}{\pi}$, ~1.556198537190348396563877031439915299.

**Zero charge in vector field singularities, not necessarily conserving, and field strength coefficients in the 3D foliations**

This section is crucial to the understanding of this paper. The field strength coefficients $\xi$ that will be discussed will be covered from different angles and other possible explanations will be covered too. The idea is simple, a flow $u_\mu$ if it appears as a perturbation in space-time must have zero charge and must appear as singularities, from which charge as we know it should emerge. Equation (4) is a result of an action without charge and therefore the sum of charge must be zero. The real part of $u_\mu$ is space-like and therefore it must be reducible to $\mathbb{R}^3$. Three of the simplest types of flow singularities are Sinks and Sources, Centers and Saddle singularities. There are three simple ways to account for the relationship between $\frac{1}{c^4}\frac{c^4}{r^2}$ and the field $-\frac{1}{4}u^\lambda u_\lambda$ or $-\frac{1}{8}(u_\lambda u^{*\lambda} + u_\lambda^* u^\lambda)$ in (+,-,-,-) metric convention. The simplest relation should be $\frac{1}{c^4}\frac{\xi^2}{x^2}\frac{c^4}{r^2}$ where x accounts for change in field strength $\xi$ due to area contraction and expansion around an electric charge as expected from (13). In this section, we will discuss singularity points of $u_\mu$ around which theorem 3 need not hold but which allow zero charge, where the summation of incoming flow and outgoing flow must be zero. $\xi$ will result from two fields, negative and positive except for the case where field lines do not enter and exit the singularity, namely in a Center singularity.

1) **Primary explanation 1, source and sink singularities $\xi = \frac{95}{96}$:**
In this case there is no way to reach zero charge except for using a separate source and a separate sink. In this case there are two $\xi$ values that can be described as follows, regarding the greatest x roots, that allow a unique equality $x = \xi$

$$\frac{1}{96}\left(-\frac{1}{2}\frac{\xi^2}{x^2}+\frac{\xi}{x}\right) = x-1 \Longrightarrow \xi = \frac{193}{192} = 1+\frac{1}{192} \tag{19.01}$$

$$\frac{1}{96}\left(-\frac{1}{2}\frac{\xi^2}{x^2}-\frac{\xi}{x}\right) = x-1 \Longrightarrow \xi = \frac{63}{64} = 1-\frac{1}{64} \tag{19.02}$$

Combining $+\frac{1}{192}$ and $-\frac{1}{64}$ to 1, we have

$$\xi = 1-\frac{1}{64}+\frac{1}{192} = \left(1+\frac{1}{192}\right)+\left(1-\frac{1}{64}\right)-1 = \frac{95}{96} \tag{19.03}$$

Since the sink and the source cannot be viewed as one singularity, the summation leads to $\xi = \frac{95}{96}$. The reader need not remember this discussion in detail right now because these considerations will be repeated.

2) **Primary explanation 2, center singularity $\xi = \frac{4}{\pi}$:**
A center singularity describes one zero eigenvalue of the flow Jacobian and two complex vectors. The Jacobian has only 2 non-zero eigenvalues and they are imaginary values.
In this case there are no lines entering and leaving the singularity, but circles around it. Along half the circles surrounding the singularity, the sign of the spacelike vector $u_\mu$ changes.
In this case we equate $\xi\frac{2c}{\frac{4r}{c}} = \frac{2c}{\frac{\pi r}{c}}$. The reason, although a unit vector acceleration in general relativity and in special relativity is not a classical acceleration, it is used here to describe a field. If a test particle moves along the diameter and returns to the same point of the circle around the singularity and changes velocity from $c$ to $-c$ then the time that is measured by an observer is $\frac{4r}{c}$ and the acceleration as an indivisible measurement is $(c-(-c))(\frac{4r}{c})^{-1}$.
Along half a circle the points where the velocity is $c$ and $-c$ are not the same, however, their distance from the mentioned observer on the diameter line is the same and the acceleration is $(c-(-c))(\frac{\pi r}{c})^{-1}$. For this reason, considering an acceleration along the diameter as a map of the circular acceleration,

$$\xi\frac{2c^2}{4r} = \frac{2c^2}{\pi r} \Longrightarrow \xi = \frac{4}{\pi} \tag{19.04}$$

**3) Primary explanation 3, saddle singularity $\xi \cong$ $\mathbf{1.556198537190348396563877031439915299}$ ...**

The saddle singularity Jacobian of the flow of the spacelike vector $u_\mu$ equates an inflow through area to two exit directions of the flow. x in the equations denotes expansion or contraction of an area but in this case, we need to equate between an area expansion/contraction and twice line expansion and contraction.

This leads to the following equations

$$\frac{1}{96}\left(-\frac{1}{2}\frac{\xi^2}{a^2}+\frac{\xi}{a}\right)=a-1 \tag{19.05}$$

$$\frac{1}{96}\left(-\frac{1}{2}\frac{\xi^2}{b^2}-\frac{\xi}{b}\right)=b-1 \tag{19.06}$$

$$\frac{1}{2}\sqrt{a-1}=1-b \Longrightarrow \xi \cong 1.556198537190348396563877031439915299 \ldots \tag{19.07}$$

The reader is not required to remember the details of this argument. The discussion that is brought here is repeated and with more possible explanations.

**Note:** A saddle point of the of the Jacobian of spacelike vector $u_\mu$ with only two eigenvectors with eigenvalues of opposite signs implies $a=b$, which cannot be satisfied by (19.05), (19.06) by a field strength other than $\xi=0$.

What are the possible values for $\xi$ if we wish to describe the electron, the Muon and the Tau lepton? It is expected that the lower value for $\xi$ and the upper value will be dictated by minimal and maximal possible values for such a field. The middle $\xi$ value which is the field strength of the Muon should not be dictated by such constraints. For example, a semi classical approach to such a field can come from the understanding that a spinning Reeb class field means that it is stronger in the spin plane and zero in the poles. This approach dictates only one possible value of $\xi=\frac{4}{\pi}$.

Consider the averaging of an acceleration field towards the center that depends on $\frac{U^{*\lambda}U_\lambda+U^\lambda U^*{}_\lambda}{8}=\xi\frac{c^2}{r}$ where c is the speed of light,

Then the intensity of the field with respect to the angle $\theta$ with the rotation plane should be $\xi\frac{c^2}{r}\cos(\theta)$. At the poles this angle is $\frac{\pi}{2}$ and $-\frac{\pi}{2}$. In the rotation plane this angle is 0. So the average field yields:

$$\frac{2\int_0^{\frac{\pi}{2}}\xi\frac{c^2}{r}\cos(\theta)2\pi r\cos(\theta)d\theta}{2\int_0^{\frac{\pi}{2}}2\pi r\cos(\theta)d\theta}=\frac{4\pi r^2\xi\frac{c^2}{r}\frac{\pi}{4}}{4\pi r^2}=\frac{c^2}{r} \tag{19.1}$$

Now we see that in order that the average field will be $\frac{c^2}{r}$ then $\xi=\frac{4}{\pi}$ which means that the field strength in the rotation plane must be stronger than the factor 1 by $\xi=\frac{4}{\pi}$. Without other constraints this should be field strength the defines the Muon, not as the minimal electron and the maximal Tauon in terms of mass. Of course, there are other possible explanations which will all be mentioned but this explanation is by far the simplest. There is another simple explanation.

**Repeating the explanation for $\xi = \frac{4}{\pi}$.** Another explanation is that changing velocity of from -c to c along half a circle and changing velocity from -c to c along the diameter of the same circle when moving forth and back requires $\xi = \frac{4}{\pi}$ to equate the accelerations.

$$\frac{c-(-c)}{\pi r} = \xi\frac{c-(-c)}{4r} \Longrightarrow \frac{2c}{\pi r} = \xi\frac{2c}{4r} \Longrightarrow \xi = \frac{4}{\pi} \tag{19.2}$$

which explains why the field strength $\xi = \frac{4}{\pi}$ should exist. See the following figure.

**Fig. 6.** – Field strength $\frac{4}{\pi}$ matches linear to circular acceleration

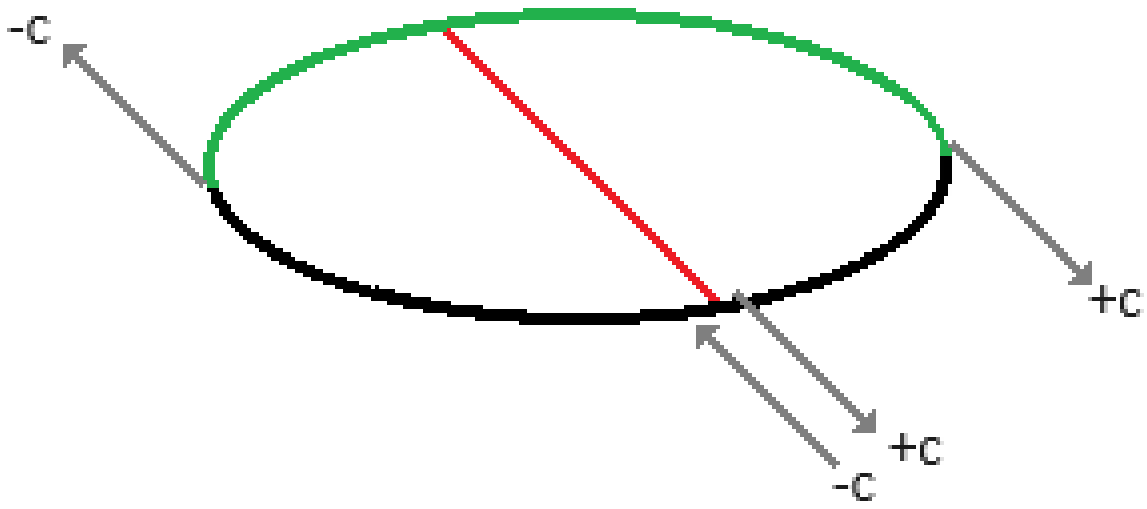

**Caveat:** When describing such acceleration as an upper limit on unit vector accelerations, the speed of light does not describe a real physical object.

**Caveat:** There is a critical remark after (15.3). $r$ can also be interpreted as a small delta from a sphere that has a radius much larger than $r$. In this case, the explanation in (19.2) can still hold but needs to be modified.

We will first start with an assumption $\xi = \frac{4}{\pi}$. This assumption is also based on Ettore Majorana's notebook [31] and on the compelling assessment of the critical strength of the Coulomb and the Yukawa potentials [32]. It is also the well-known ratio between a star graph and a Steiner star in Euclidean spaces – *star Steiner ratio* in $\mathbb{R}^d$ [33]. The addition of a middle point in a ball can reduce the length of a star graph in relation to a star where the star graph is defined as straight lines between n-1 points and a single point on the sphere. And a Steiner star connects the points to the center. A physical meaning of such a ratio is that where there is a middle point, divergence of an acceleration field can be defined, where there is no such point, no such divergence can be defined. For such a case, a different value of $\xi$ should be defined. This fact is brought here as thought, not as any proof to why $\xi = \frac{4}{\pi}$. The calculation in (19.1) is much more compelling.

Then (19) yields two solutions as follows,

$$\frac{192{x_1}^2+2\xi x_1-\xi^2}{192} = {x_1}^3 \Longrightarrow \frac{1}{x_1-1} \cong \mathbf{206.75133988502202} \tag{20}$$

This value is surprisingly very close to the mass ratio between the Muon and the electron!

$$\frac{105.6583745\text{MeV}}{0.5109989461\text{MeV}} \cong 206.7682826 \tag{21}$$

The following is an area ratio around a positive charge. The discussion about its meaning is postponed for now.

$$\frac{192{x_2}^2-2\xi x_2-\xi^2}{192} = {x_2}^3 \Longrightarrow \frac{1}{1-x_2} \cong \mathbf{44.63955017596401} \tag{22}$$

Before we continue, we need to prove another theorem which has important implications to Quantum Gravity. The factor $\frac{96}{95}$ is, however, not final in what will be described as Steiner Trees.

**Theorem 5**: In Riemannian geometry, a computational model for the connection of a finite connected set of points on a sphere S2 and the center with radius r can converge in polynomial time only to a minimal graph of S2 not within radius $r$ but within radius $r\frac{96}{95}$.

**Proof:** The proof of this theorem is a direct result of the complexity limit of the Minimum Steiner Tree. Finding the minimal length of such a graph is in polynomial time only above $\frac{96}{95}$ of the minimal graph length due to [34]. As a result, to connect all the points in the sphere and its center is possible in polynomial time only for $r\frac{96}{95}$ and we are done. The meaning of this theorem is very deep for most Quantum Gravity theories. For this specific theory, if acceleration depends on $r^{-1}$ then physically the dependence must be on $\frac{95}{96}r^{-1}$. As a caveat, $\frac{96}{95}$ is not believed by the author to be an absolute limit to the hardness of the Steiner Tree problem. Before continuing, a much more compelling explanation for the choice of $\xi = \frac{95}{96}$ for the electron's field strength will be brought. Right now, different options are described.

**Repeating the explanation for a field strength $\xi = \frac{95}{96}$**. By the principle of parsimony, the electron field strength should be explainable by ground state roots of (19). Consider $\xi = x_1$ and $\xi = x_2$ in the following area ratio equations.

$$\xi_1 = x_1 \wedge \left(\frac{1}{2}\frac{{\xi_1}^2}{{x_1}^2} + \frac{\xi_1}{x_1}\right)\frac{1}{96} = x_1 - 1 \Longrightarrow x_1 = \frac{193}{192} \Leftrightarrow \delta x_1 = x_1 - 1 = \frac{1}{192} \tag{22.1}$$

$$\xi_2 = x_2 \wedge \left(\frac{1}{2}\frac{{\xi_2}^2}{{x_2}^2} - \frac{\xi_2}{x_2}\right)\frac{1}{96} = x_2 - 1 \Longrightarrow x_2 = \frac{63}{64} \Leftrightarrow \delta x_2 = x_2 - 1 = \frac{-1}{64} \tag{22.2}$$

Adding these two delta area ratios yields

$$\delta x_1 + \delta x_2 = \frac{1}{192} + \frac{-1}{64} = -\frac{1}{96} \tag{22.3}$$

$$\xi = 1 + \delta x_1 + \delta x_2 = x_1 + x_2 - 1 = \frac{193}{192} + \frac{63}{64} - 1 = \frac{95}{96} \tag{22.4}$$

Definition: $\xi_1 = \frac{193}{192}$ and $\xi_2 = \frac{63}{64}$ will be called Stability Field Strengths and $\xi = \frac{95}{96}$ is called Joint Stability Field Strength. $\xi = \frac{95}{96}$ is the first candidate for the electron field strength that will be used in the Muon/electron mass ratio assessment. It is not difficult to see that for the choice of $\xi = \frac{95}{96}$, also see motivation in Appendix E, (74), (75), (79), and the surprising relation between the Fine Structure Constant and exponential perturbations of $\xi = \frac{95}{96}$ in (81)-(86), the following polynomials yield,

$$\left(-\frac{1}{2}\frac{\left(\frac{95}{96}\right)^2}{a^2} + \frac{\frac{95}{96}}{a}\right)\frac{1}{96} = a - 1 \Longrightarrow \frac{192a^2 + 2\frac{95}{96}a - \left(\frac{95}{96}\right)^2}{192} = a^3 \; and \; \left(-\frac{1}{2}\frac{\left(\frac{95}{96}\right)^2}{b^2} - \frac{\frac{95}{96}}{b}\right)\frac{1}{96} = b - 1 \Longrightarrow$$

$$\frac{192b^2 - 2\frac{95}{96}b - \left(\frac{95}{96}\right)^2}{192} = b^3 \; and \; \frac{1}{(a-1)(1-b)} \cong \mathbf{12202.8887406646790623199} \tag{23}$$

**Exponential stability of the field strength near $\xi = \frac{95}{96}$:** From (22.4), $x_1 + x_2 - 1 = a + b - 1$ for two different $\xi$ values, $\xi = \frac{193}{192}$, $\xi = \frac{63}{64}$, consider replacing $\frac{95}{96}$ in (23) by $(a + b - 1)^{\xi} = \xi$. The result is very surprising, $\xi \cong 1 - 95.9560893107845913618803 02^{-1}$ instead of $\xi = \frac{95}{96} = 1 - 96^{-1}$ and in that case $\frac{1}{(a-1)(1-b)} \cong 12202.9706958707520243478938 94$ which is a small error.

**Note:** Serendipity is not preferable to rigid math but can be an indication. Now consider the following expression:

$$\frac{x}{96}\left(2^{\frac{x}{95*96}} - 1\right)^{-1} \tag{23.1}$$

It is easy to see that this expression approximates the inverse Fine Structure Constant for values of x between 1 and 4. Here 2 is considered as an upper limit on the field strength $\xi < 2$. The actual maximal value will be explored as $\xi < \frac{\pi}{2}$. When $\xi = 2$ we have $\left(-\frac{1}{2}\frac{\xi^2}{a^2} + \frac{\xi}{a}\right)\frac{1}{96} = a - 1 \Rightarrow a = 1$ and then $(a-1)^{-1}$ is undefined.

From (20) and (23.1) we choose $x = log_2\sqrt{\frac{1}{x_1-1}} = -\frac{1}{2}log_2(x_1-1)$. Inserting $x$ into (23.1) we get for $(x_1-1)^{-1} \cong 206.751339885022019871030352078378200531005859375$, a surprising result:

$$x = -\frac{1}{2}log_2(x_1-1) \Longrightarrow x*(96*(2^{\frac{x}{95*96}}-1))^{-1} = \frac{\frac{x}{96}}{(2^{\frac{x}{95*96}}-1)} \cong \quad (23.2)$$

$$137.03599925379157298.$$

With $(x_1-1)^{-1}$ high sensitivity of two digits after the decimal point, which is surprising but the result in (23.2) depends on negative charge only and therefore cannot be an accurate result.

This value is very close to the accepted inverse Fine Structure Constant, but what can we make of this guess which is not any rigid proof? The expression $(2^{\frac{x}{95*96}}-1)$ is a small delta between 1 and a small power of 2. It is an exponential perturbation around 1. The expression $2^x = 2^{-\frac{1}{2}log_2(x_1-1)} = \sqrt{\frac{1}{x_1-1}}$ is the square root of the inverse delta area around a negative charge. It means approximately inverse length. But why in the power expression in the denominator of (23.2), we use $\frac{x}{95*96}$ while in the numerator the expression is $\frac{x}{96}$? This is not a well understood expression. (23.2) can, however, indicate that the Fine Structure Constat is related to 9120 = 95*96 and to powers of 2 or other expressions. A more detailed discussion of these ideas will take place later in this paper.

$(a-1)(1-b)$ from (23) answers the question of what happens when a particle is neutral. To better understand the above expression, it is best to contract the acceleration matrix $A_{\alpha\beta}$ (3), [11] with the Levi-Civita tensor (not symbol), $E^{\mu\nu\alpha\beta}$ but with a possible orientation change from $B_{\mu\nu} = \frac{1}{2}E^{\mu\nu\alpha\beta}A_{\alpha\beta}$. This description is of a second complex plane in which the divergence of a Reeb class-like acceleration vector can be of an opposite sign, $U_\mu;^\mu + {U^*}_\mu;^\mu = 0$ where $V_\upsilon$ is a unit vector perpendicular to both $\frac{U^\mu}{2}$ and to $\frac{P^\mu}{\sqrt{Z}}$. One field is then of a positive charge and one of a negative charge which is the explanation for the term $(a-1)(1-b)$ where $a$ denotes the area addition ratio around a negative charge and b is the area loss ratio around a positive charge. One would expect to see $\sqrt{(a-1)(1-b)}$ however, roots will be discussed regarding spin 1 mass ratios while (a-1)(1-b) is apparently suitable to describe mass portions of spin 1/2 particles.

Roots of such a value also have a meaning, see appendix C, (64.b), (64.01). Combining (20) and (23), the following holds:

$$\frac{(x_1-1)\mathbf{105.65837455}MeV}{1+(a-1)(1-b)} \cong \mathbf{0.5109989461}MeV$$

$$1+\frac{1}{96}\left(-\frac{1}{2}(1-\frac{1}{96})^2a^{-2}+\left(1-\frac{1}{96}\right)a^{-1}\right)=a$$
$$1+\frac{1}{96}\left(-\frac{1}{2}(1-\frac{1}{96})^2b^{-2}-\left(1-\frac{1}{96}\right)b^{-1}\right)=b$$
$$1+\frac{1}{96}\left(-\frac{1}{2}\left(\frac{4}{\pi}\right)^2c^{-2}+\frac{4}{\pi}c^{-1}\right)=c$$
$$MuonMass*(c-1)=ElectronMass+ElectronMass*(a-1)(1-b) \qquad (24)$$

By (23) the ratio is **~206.768282704414616546273464336991310119628906 25**

By (13.11) the term $(a-1)(1-b)$ is best interpreted as a coupling between a negative charge and a positive charge and is therefore a neutral term.

Where $ElectronMass*(a-1)(1-b)=\sim\mathbf{41.8752442118608}\ \boldsymbol{eV/c^2}$ looks like a new particle or resonance. Corroboration requires to detect excess in cosmic $\sim20.9\ \boldsymbol{eV}/c^2$ photons. Verification of this theory by Muon decays can be done by observing rare excess of **20.9376221059304 eV photons**. With electron energy **0.5109989500 MeV** the Muon energy is **~105.658375355 MeV**.

We only needed a small correction to the 2014 Muon energy from 105.6583745 MeV to 105.65837455 MeV with electron energy 0.5109989461055 MeV to arrive at the energy ratio and therefore mass ratio of the Muon and the electron. Is that a mere coincidence? The extremely small ratio error and the choices of $\xi=\frac{4}{\pi}$ and $\xi=\frac{95}{96}$ highly disfavor a mere coincidence. It is important to remember that $1+\frac{1}{96}\left(-\frac{1}{2}(\frac{193}{192})^2a^{-2}+\left(\frac{193}{192}\right)a^{-1}\right)=a$ has a biggest root $a=\frac{193}{192}=1+\frac{1}{192}$ and $1+\frac{1}{96}\left(-\frac{1}{2}(\frac{63}{64})^2b^{-2}-\left(\frac{63}{64}\right)b^{-1}\right)=b$ has a biggest root $b=\frac{63}{64}=1-\frac{1}{64}$. The delta $-\frac{1}{64}+\frac{1}{192}=-\frac{1}{96}$ is a delta of energy ratios between the two stable states with field strength coefficients $\xi=\frac{193}{192}$ and $\xi=\frac{63}{64}$ and roots $a=\frac{193}{192}$ and $b=\frac{63}{64}$. 1 plus this delta yields $\frac{95}{96}$, which shows that our choice of $\xi=\frac{95}{96}$ was not at random but is the result of the summation of negative and positive area ratios for which the field strengths are equal to the biggest roots.

Let us define the electron field strength as $\xi=\frac{95}{96}$ and consider a perturbation of this value and its close link to the Fine Structure Constant. Recall (23),

$$1+\frac{1}{96}\left(-\frac{1}{2}(1-\frac{1}{96})^2a^{-2}+\left(1-\frac{1}{96}\right)a^{-1}\right)=a \qquad (24.1)$$
$$1+\frac{1}{96}\left(-\frac{1}{2}(1-\frac{1}{96})^2b^{-2}-\left(1-\frac{1}{96}\right)b^{-1}\right)=b$$

Consider a little more accurate result than the one used before in (23):

$$\frac{1}{(a-1)}\frac{1}{(1-b)}\cong 12202.8887406646790623199 \qquad (24.2)$$

Now consider the following perturbation on $\xi = \frac{95}{96} = 1 - \frac{1}{96}$, and raise this $\frac{95}{96}$ to the power $1 + \alpha$ where $\alpha$ is the Fine Structure Constant - FSC. We will take the assessment from (40) of the inverse FSC, about 137.035999036827007557 8… and not the larger assessment 137.0359992990990 from the remark after (41). Our new field strength will be $\xi' = (\frac{95}{96})^{1+\alpha}$, which is an exponential perturbation with the help of the Fine Structure Constant. We want to calculate the new value of the neutral area ratio of addition and subtraction in two acceleration planes, one positive and one negative, as expected from the electron Neutrino because it is electrically neutral and see what we get. Before that, please view the following illustration, in reality, it is a 4-dimensional model with two perpendicular planes. The neutral charge is of one positive two-dimensional plane and one negative two-dimensional plane, both defined by two acceleration matrices and two generalizations of two Reeb class vectors to 4 dimensions.

**Fig. 7.** – This is an over-simplification of two Reeb class fields, not the usual Reeb vectors, in two Symplectic Lagrangian planes

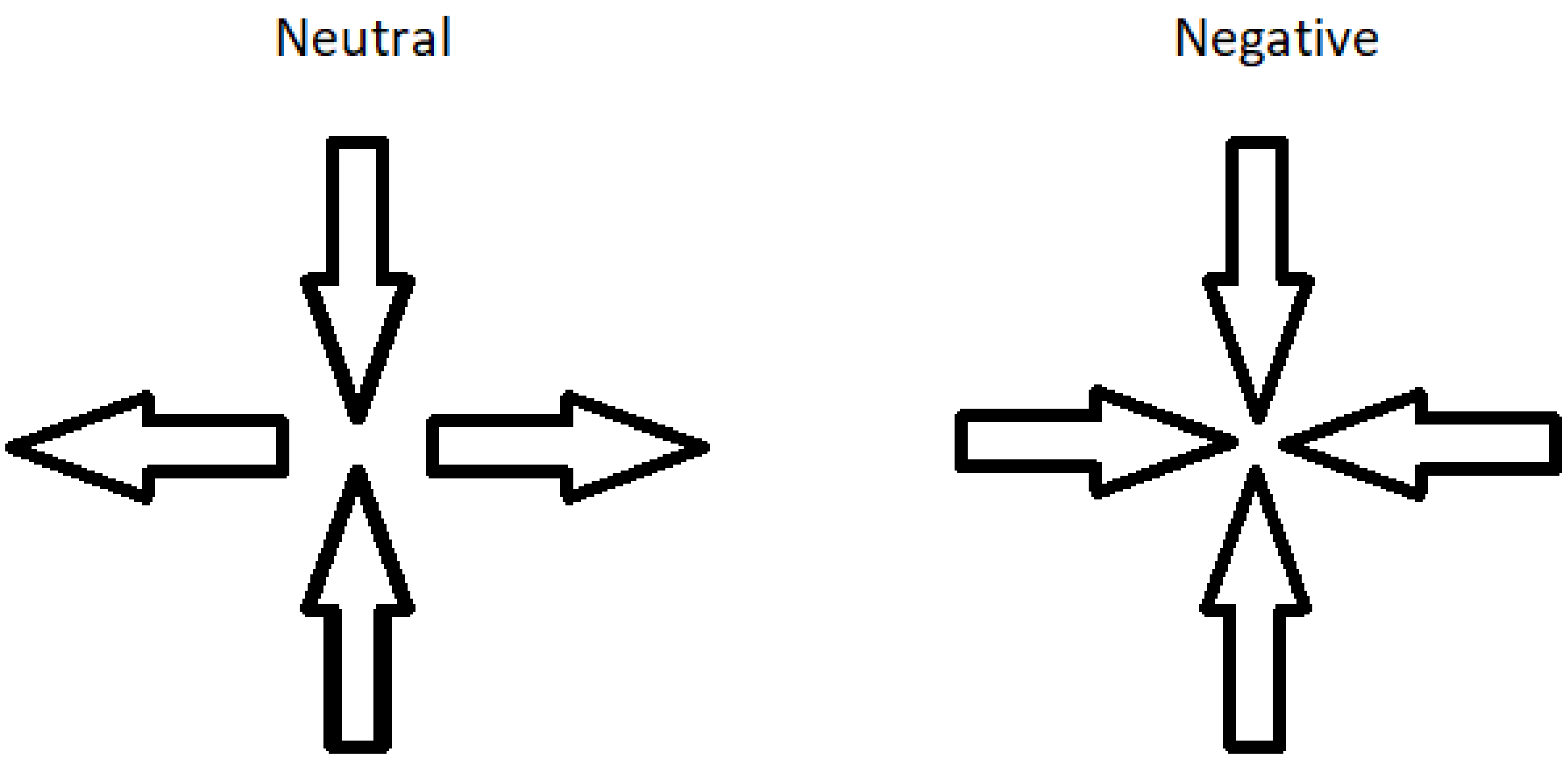

And now we have for $\xi' = (\frac{95}{96})^{1+\alpha}$,

$$\frac{1}{(a'-1)}\frac{1}{(1-b')} \cong 12204.188931677483196836 \tag{24.2}$$

And the following shows up:

$$\left(1 - \frac{(a'-1)(1-b')}{(a-1)(1-b)}\right)^{-\frac{1}{2}} \cong 96.883736818613229502 2617 \tag{24.3}$$

which is remarkably close to $\alpha^{-1}2^{-\frac{1}{2}} \cong 96.89908418561357450471405 2$, from (40).

The fact that perturbations of the field strength $\frac{95}{96}$ yield the Fine Structure Constant is easy to see in other cases other than (81)-(86) and (24.3). Using a simple datasheet without the accuracy of the Python math libraries, consider $\xi' = \left(\frac{95}{96}\right)^{1+\beta}$ where $\beta \approx 1.00370694$ for which $(\frac{95}{96} - \xi')(\frac{95}{96})^{-1} \cong 25762.75334^{-1}$ then (23) yields $\frac{1}{(a'-1)}\frac{1}{(1-b')} \cong 12203.54919$. Now consider the following function $\xi'' = 1 + \ln(\xi')$ for which (23) yields $\frac{1}{(a''-1)}\frac{1}{(1-b'')} \cong 12204.49567$,

$$\ln\left(\frac{\ln\left((a''-1)(1-b'')\right)}{\ln\left((a'-1)(1-b')\right)} - 1\right)^2 \approx 137.0359991 \tag{24.4}$$

In other cases, the inverse Fine Structure Constant can emerge from trigonometric perturbations of a higher field strength. In both cases, fractional powers of roots are involved. This is not surprising if we consider that the Fine Structure Constant must be related to electromagnetic waves, and these should be a result of perturbations of the field strength of elementary particles such as the electron or even of the Tau lepton as an upper limit of an allowed leptonic field strength. Although (24) is not a rigid mathematical proof of the mass ratios between the Muon and the electron, and although only $\xi = \frac{4}{\pi}$ is a well understood field strength, not directly from this paper, one can argue that the result in (24) is too accurate to be ignored, especially if (24.3), (24.4) and (81)-(86) in "Appendix E" are taken into account.

$\xi = 2$ as a field strength is a critical value much higher than the highest field strength for leptons which is offered in this paper, simply because for a negative charge, the gravitational field vanishes.

$\frac{192x^2+2*2*x+2^2}{192} = x^3$ with a stable root x=1, $x(n+1) = \left(\frac{192x(n)^2+2*2*x(n)-2^2}{192}\right)^{\frac{1}{3}}$. But then $\frac{1}{x-1}$ as an added area portion around the negative charge is undefined but with a left limit 0. So, asking whether a logarithmic scale that starts at 2 has a physical meaning is legitimate. We choose our scale to be:

$\{2^{\frac{95*96}{95*96}}, 2^{\frac{95*96-1}{95*96}}, \ldots, 2^{\frac{1}{95*96}}\}$, now consider $y = \left(\left(2^{\frac{1}{95*96}} - 1\right) * 96\right)^{-1} \cong 137.050820617$ and

$\frac{95}{\ln(2)} \cong 137.05602888445$ and it is easy to show that as n grows, $\left(\left(2^{\frac{1}{(n-1)*n}} - 1\right) * n\right)^{-1} \approx \frac{n-1}{\ln(2)}$. It is easy to see a nice result, $\frac{y-137.0359990368270075578}{137.0359990368270075578} \cong 96.1546032^{-2}$ so the relative error to one of the assessment of the inverse Fine Structure Constant, see (40), is nearly expressible as a power of 96. This is one good reason to search for a relation that involves 2 and powers of 96 or of 95*96 as the mathematical term that will yield the Fine Structure Constant, however, such a term should appear out of a perturbation of a field strength because the Fine Structure Constant defines the Quantum electric strength, but which field strength?

**Important:** A leading idea is that the Fine Structure Constant should be related to perturbations of a maximal allowed field strength for leptons, i.e., the Tau lepton field strength. Any perturbation exceeding this limit must be dissipated as waves.

### An upper limit $\xi < \frac{\pi}{2}$ from ordinary differential equations as an approximation of non-geodesic acceleration that depends on $\frac{c^2}{r}$, a qualitative result

The relationship between length and acceleration of a unit vector requires scaling of the ordinary acceleration to depend on time only. This goal can be achieved by dividing $r$ by a starting length $r_0$ at $t = 0$. We suppose there is a local field of non-geodesic acceleration towards point 0 which is confined in small volume around an event such that (15.3) holds true.

$$\ddot{r} \approx -\frac{\xi}{x}\frac{c^2}{r} \Leftrightarrow \ddot{r}r \approx -\frac{\xi}{x}c^2 \Leftrightarrow \frac{\ddot{r}r}{{r_0}^2} \approx -\frac{\xi}{x}\frac{c^2}{{r_0}^2} \tag{24.5}$$

Where $\ddot{r} \equiv \frac{d^2r}{dt^2}$, $c$ is the speed of light and $\xi$ is the field strength coefficient.

The caution here is that $-\frac{\xi}{x}\frac{c^2}{r}$ looks like classical acceleration while the theory in this paper is about unit vector accelerations which should have the dimensions of length$^{-1}$ and not of length * time$^{-2}$.

We can write the last equation as a time dependent regular differential equation and as an approximation of an event where such acceleration appears in spacetime and is a function of a local time t. Bearing in mind the relation $\frac{\xi c^2}{r} = \frac{e}{4\pi\varepsilon_0 r^2}\sqrt{4\pi\varepsilon_0 k}$ which appears at a certain r only,

$$\ddot{y}(t)y(t) = \frac{\ddot{r}r}{{r_0}^2} \approx -\frac{\xi}{x}\frac{c^2}{{r_0}^2} = -k \tag{24.6}$$

Where $k \equiv \frac{\xi}{x}\frac{c^2}{{r_0}^2}$. At t=0, we should have $y(0) = \frac{r}{r_0} = 1$.

Using Wolfram Math, we get two solutions with constants $c1$ and $c2$:

$$y(t) = \exp\left(\frac{c1 - 2k * erf^{-1}\left(\mp\sqrt{k\frac{2}{\pi}\exp\left(-\frac{c1}{k}\right)(c2+t)^2}\right)^2}{2k}\right) \tag{24.7}$$

$$c2 = \sqrt{\frac{\pi}{2}\frac{1}{k}\exp(\frac{c1}{k})}\ \mathrm{erf}(\sqrt{\frac{c1}{2k}}) \tag{24.8}$$

Where $erf$ is the error function and $erf^{-1}$ is the inverse error function which explodes at $erf^{-1}(-1)$ with $-\infty$ and at $erf^{-1}(1)$ with $+\infty$, and which means:

$$k\frac{2}{\pi}\exp\left(-\frac{c1}{k}\right)(c2+t)^2 < 1 \tag{24.9}$$

Now omitting the term $\exp\left(-\frac{c1}{k}\right)$ from $k\frac{2}{\pi}\exp\left(-\frac{c1}{k}\right)(c2+t)^2$ and setting $(c2+t)^2 = \frac{{r_0}^2}{c^2}$,

Where $t = \frac{r_0}{c}$ is the minimal time to travel along distance $r_0$, We have,

$$k\frac{2}{\pi}(c2+t)^2 = \frac{2}{\pi}\frac{\xi}{x}\frac{c^2}{{r_0}^2}(c2+t)^2 \tag{24.10}$$

$$(c2+t)^2 = \frac{{r_0}^2}{c^2} \Longrightarrow k\frac{2}{\pi}(c2+t)^2 = \frac{2}{\pi}\frac{\xi}{x}$$

So $exp\left(-\frac{c1}{k}\right) \longrightarrow 1$ sets a limit on the maximal possible field strength $\xi$ where $x$ is near 1,

$$\xi < \frac{\pi}{2} \tag{24.11}$$

It is easy to see that $c1 = 0$ and from $y(0) = 1$ and from an important insight. $\ddot{r} \approx -\frac{\xi}{x}\frac{c^2}{r}$ is a description of an acceleration that appears near a series of events where r is small. That is why we can start with $\ddot{y}(0) = 0$ if and only if the acceleration field is restricted to have a compact support only where $y < 1$ and in this case $t^2 = \frac{{r_0}^2}{c^2}$ sets a strong limit $\frac{\xi}{x} < \frac{\pi}{2}$,

$$t^2 = \frac{{r_0}^2}{c^2}, y(t) = \exp\left(-erf^{-1}\left(\mp\sqrt{k\frac{2}{\pi}t^2}\right)^2\right) \Longrightarrow \frac{\xi}{x} < \frac{\pi}{2} \tag{24.12}$$

Expecting the $\xi$ field strength coefficient of the Tau lepton to be maximal, it should be a value lower than $\frac{\pi}{2}$ and near $\frac{\pi}{2}$. Perturbations above such a maximal value must be accompanied by emission of electromagnetic radiation. This is a good reason for why the Fines Structure Constant must emerge out of such perturbations.

**Fig. 8.** – PDE solution with courtesy of Wolfram Alfa:

$$y(t) = e^{\frac{c_1}{2k} - \operatorname{erf}^{-1}\left(\sqrt{\frac{2}{\pi}}\sqrt{k\,e^{-c_1/k}\left(t - \frac{\sqrt{\frac{\pi}{2}}\sqrt{e^{c_1/k}}\,\operatorname{erf}\left(\frac{\sqrt{\log\left(e^{c_1/k}\right)}}{\sqrt{2}}\right)}{\sqrt{k}}\right)}\right)^2}$$

$$y(t) = e^{\frac{c_1}{2k} - \operatorname{erf}^{-1}\left(-\sqrt{\frac{2}{\pi}}\sqrt{k\,e^{-c_1/k}\left(t - \frac{\sqrt{\frac{\pi}{2}}\sqrt{e^{c_1/k}}\,\operatorname{erf}\left(\frac{\sqrt{\log\left(e^{c_1/k}\right)}}{\sqrt{2}}\right)}{\sqrt{k}}\right)}\right)^2}$$

The exploration which is performed here is not out of analytic solutions to (4) or a complex version of (4) or to the further-on mentioned (64.b), (64.01), which may take many years to yield fully analytic solutions. It is a “reverse engineering” of Nature by assessment of (4) and field strengths in an infinitesimal limit. It will require more discussion to reach more comprehensible terms for the inverse Fine Structure Constant.

Another clue to where the Fine Structure constant comes from is the following:

Consider a search for the number 96 and to keep the idea simple and related to the roots of the third order Gravity and Anti-gravity area ratio polynomials.

Consider the following known equation: $\frac{\pi^4}{96} = \sum_{k=0}^{\infty} \frac{1}{(2k+1)^4}$ which results from Parseval’s identity when developing the Fourier series of the function $f(x) = |x|$ in $(-\pi, \pi)$. Notice that the fourth root of $\frac{\pi^4}{96}$ is $\frac{\pi}{96^{\frac{1}{4}}} \cong 1.00364948118 \ldots$ . We can see what the error of this value in relation to 1 is. $\frac{\pi}{96^{\frac{1}{4}}} - 1 \cong \frac{1}{274.01155134419542\ldots} = \frac{1}{2*137.00577567209771179617192026613}$ . We may therefore search for an expression in which twice the inverse Fine Structure Constant appears. If we choose the value from (40), and not the higher value 137.0359992990990 after (41) we get the following error assessment

$$\left(1 - \frac{137.005775672097711796171920266613}{137.035999036827007557803881354 6}\right) \cong \frac{2}{(95.2271807264060288800404363625 12)^2}$$

The residual error is related to a number between 95 and 96 and the factor 2 appears again. Although it is not any mathematical proof, it is still difficult to ignore such a lead in the search for where the inverse Fine Structure comes from.

It is worthy of mentioning that getting the exact Fine Structure Constant in (81)-(86) requires a very small addition $\xi = \frac{95}{96} + \varepsilon$ for some small $\varepsilon$. Then (24) would require the mass of the Muon to be slightly higher than 105.65837455 MeV, about 105.658375 MeV.

The following Python code was used to reach the result in (24),

```
import numpy as np

x1 = 1
third = 1 / 3
f = 4 / np.pi # Ettore Majorana's ring of a disk, potential factor.
f2 = f * f

# Iterate to most stable root.
for i in range(2000):
    x1 = np.power((192 * x1 * x1 + 2 * x1 * f - f2) / 192, third)

a = 1/(x1 - 1) # Negative charge.

print('Xi = 4/Pi, a = %.48f' % a)

x3 = 1
x4 = 1
f = 95 / 96
f2 = f * f

# Iterate to most stable roots.
for i in range(2000):
    x3 = np.power((192 * x3 * x3 + 2 * x3 * f - f2) / 192, third)
    x4 = np.power((192 * x4 * x4 - 2 * x4 * f - f2) / 192, third)

c = 1/(x3 - 1) # Negative charge.
d = 1/(1 - x4) # Positive charge.
print('Xi = 95/96, c = %.48f, d = %.48f' % (c, d))
print('Xi = 95/96, c * d = %.48f' % (c * d))

print('Approximated mass ratio between the Muon and the electron %.48f'
      % (a * (1 + (x3-1)*(1-x4))))
```

Few words about $\xi = 1 - \frac{1}{96}$. What is so special about $\xi = 1 - \frac{1}{96}$? It is twice the average of an ideal area loss ratio and an ideal area addition ratio + 1. $\frac{1}{192} - \frac{1}{64} = -\frac{1}{96}$ where $x = 1 + \frac{1}{192}$ is the biggest root of $1+\frac{1}{96}\left(-\frac{1}{2}(1+\frac{1}{192})^2x^{-2}-\left(1+\frac{1}{192}\right)x^{-1}\right)=x$ and $x=1-\frac{1}{64}$ is the biggest root of $1+\frac{1}{96}\left(-\frac{1}{2}(1-\frac{1}{64})^2x^{-2}-\left(1-\frac{1}{64}\right)x^{-1}\right)$.

**Null Reeb vectors instead of gauge fields:** The mass mechanism in (3.19) offers a way to give mass to null fields while keeping the null fields locally massless and having mass only in the total summation of the Lagrangian of (3.16). How about null Reeb class vectors $\frac{U_\mu}{2}\frac{U^\mu}{2}=0$? It is not difficult to see that in this case, the unit vector $\frac{P_\mu}{\sqrt{Z}}$ should be space-like at least in the near vicinity of a charged particle as $r\to 0$ and $U^\mu$ may not be all 0 at the center of a sphere but can be a null vector. With $\xi=\frac{4}{\pi}$ and $\xi=\frac{95}{96}$ we have in this case:

$$1+\frac{1}{96}\left(\pm\frac{4}{\pi}c^{-1}\right)=1\pm\frac{c^{-1}}{24\pi}=c \tag{25}$$

and

$$1+\frac{1}{96}\left(\pm\frac{95}{96}b^{-1}\right)=1\pm\frac{95b^{-1}}{96^2}=b \tag{26}$$

From (25)

$$c_1=\frac{1+\left(1+\frac{1}{6\pi}\right)^{\frac{1}{2}}}{2}\cong 1.0130915\ldots,\ c_2=\frac{1+\left(1-\frac{1}{6\pi}\right)^{\frac{1}{2}}}{2}\cong 0.986556\ldots \text{ and} \tag{27}$$

From (26)

$$b_1=\frac{1+\left(1+\frac{95}{96*24}\right)^{\frac{1}{2}}}{2}\cong 1.010204037\ldots,\ b_2=\frac{1+\left(1-\frac{95}{96*24}\right)^{\frac{1}{2}}}{2}=\frac{95}{96}\ \text{ and}$$

$$\frac{1}{c_1-1}\cong 76.38530,\ \frac{1}{1-c_2}\cong 74.3845968,\ \frac{1}{\sqrt{(c_1-1)(1-c_2)}}\cong 75.3783115, \tag{28}$$

With $\xi=\frac{4}{3}$, (28) is a bit different:

$$\frac{1}{c_1-1}\cong 72.98648402,\ \frac{1}{1-c_2}\cong 70.98571137,\ \frac{1}{\sqrt{(c_1-1)(1-c_2)}}\cong 71.9791462, \tag{28.1}$$

**Important:** Where does this $\xi=\frac{4}{3}$ come from? The reader is advised to check that the average distance between two points on the Euclidean ring is $\xi=\frac{4}{\pi}$. The average distance between two points on the Euclidean sphere is $\xi=\frac{4}{3}$ and is left as an exercise to the reader. We may say that $\xi=\frac{4}{\pi}$ means a geometric ring field strength and $\xi=\frac{4}{3}$ is a geometric sphere field strength. If we take into account particle decay through Bosons with two different field strengths, $\xi=\frac{4}{\pi}$ if the Muon is involved and $\xi=\frac{4}{3}$ in other cases, then there is a new interaction that is not covered by the W Boson alone!

$$\frac{1}{b_1-1} \cong 98.00042535,\ \frac{1}{1-b_2} = 96,\ \ \frac{1}{\sqrt{(b_1-1)(1-b_2)}} \cong 96.99505572 \tag{29}$$

We now look at:

$$\frac{\sqrt{(b_1-1)(1-b_2)}}{\sqrt{(c_1-1)(1-c_2)}} \cong 1.134361808^{-2} \tag{30}$$

Roots are attributed in this case to spin 1 or 2. It is easy to see that also:

$$(1+(c_1-1)(1-c_2))\left(\frac{(c_1-1)(1-c_2)}{(b_1-1)(1-b_2)}\right)^{1/4} \cong 1.134561453 \tag{31}$$

$$\approx \frac{91.1876\ GeV}{80.3725\ GeV}$$

Which is remarkably close to the ratio between the energy of the Z boson and the energy of the W boson and for W Boson of $80.3725\ GeV$ the relative error of this ratio is about 1/1528961.689. For where the idea of 4th roots came from, please refer to Appendix C, (65).

Another research direction is to use the inverted value of $\xi = \frac{4}{\pi}$, i.e., $\xi = \frac{\pi}{4}$ in the negative and positive charge area ratio equations as in (24). That yields two new maximal roots ${a_1}^2 + \frac{1}{96}\left(-\frac{1}{2}\left(\frac{\pi}{4}\right)^2 + \frac{\pi}{4}a_1\right) = {a_1}^3$ and ${a_2}^2 + \frac{1}{96}\left(-\frac{1}{2}\left(\frac{\pi}{4}\right)^2 - \frac{\pi}{4}a_2\right) = {a_2}^3$ along with the older ones ${b_1}^2 + \frac{1}{96}\left(-\frac{1}{2}\left(\frac{4}{\pi}\right)^2 + \frac{4}{\pi}b_1\right) = {b_1}^3$ and ${b_2}^2 + \frac{1}{96}\left(-\frac{1}{2}\left(\frac{4}{\pi}\right)^2 - \frac{4}{\pi}b_2\right) = {b_2}^3$. Quite like the ratio in (30), we have, $\frac{\sqrt{(b_1-1)(1-b_2)}}{\sqrt{(a_1-1)(1-a_2)}} \cong \sqrt{\frac{201.6240447 * 86.46523917}{206.7513399 * 44.63955018}}$ ≅1.374383282 which is close to the following mass ratio between a Higgs Boson of 125.3267 GeV and a Z Boson of 91.1876 GeV which yields, 1.37438314, close to 1.374383282. It is interesting though not sufficiently accurate to draw any conclusion at this stage. The idea behind using charge equations without null Reeb class vectors is because the Higgs boson is supposedly responsible for non-zero mass. From (31) and using s instead of c, $\sqrt{(s_1-1)(1-s_2)} \cong 75.3783115 \ldots^{-1}$ and 91.1876 GeV * $\frac{\sqrt{(b_1-1)(1-b_2)}}{\sqrt{(a_1-1)(1-a_2)}} * (1+(s_1-1)(1-s_2))$ ≅ 125.3487702 GeV. A similar $(1+(s_1-1)(1-s_2))$ value was used in (29) as $(1+(c_1-1)(1-c_2))$. If the reasoning here is correct, the Higgs boson interacts as an electric dipole.

Returning to (22) $\frac{1}{1-x_2} \cong \mathbf{44.63955017596401}$ and written as $\frac{1}{1-c}$,

$$\frac{80372.88\ \text{MeV}\ (1-c)}{1+\sqrt{(c_1-1)(1-c_2)}} \approx \mathbf{1776.91}\ \boldsymbol{MeV} \tag{32}$$

With $\xi = \frac{4}{3}$ as in (28.1), (32) gets the same result for a higher value of the W Boson mass,

$$\frac{80422.57\ \text{MeV}\ (1-c)}{1+\sqrt{(c_1-1)(1-c_2)}} \approx \mathbf{1776.91}\ \boldsymbol{MeV} \tag{32.1}$$

The value **23.57325 MeV** is 1776.91 $MeV * \sqrt{(c_1-1)(1-c_2)}$ . It is too close to the fusion energy between two deuterium atoms into a Helium atom, 23.6 MeV, to ever be detected.
The root, $\sqrt{(c_1-1)(1-c_2)}$ can be better understood as a result of taking the root of a determinant of a Gram matrix of two Reeb class vectors in Appendix C or is related to spin 1. The value 1776.91 MeV will be discussed in (36) with a reference. A very surprising relation between Quarks and Leptons with the same $\frac{1}{1-c} \cong 44.63955017596401$ as in (22) is the relation between the **pole energy of the Bottom/Beauty Quark** [35], [36] and the anti-**Muon**, this time we take the Muon value that yields in (24) along with the denominator of (23), the exact mass ratio between the Muon and the electron $105.65837455\ MeV$ instead of the 2014 value $105.6583745\ MeV$,

$$\frac{105.65837455\ MeV}{(1-c)(1+(a-1)(1-b))}(1+\sqrt{(c_1-1)(1-c_2)})$$
$$= 44.63955017596401 * 105.65837455\ MeV$$
$$* (1+75.37831150257286827786078900969 3^{-1}) *$$

$$(1+12202.888740664679^{-1})^{-1} \cong 4{,}778.7223164425585113299 \text{ MeV} \approx \mathbf{4.78\ GeV}$$

Which is equivalent to:

$$\frac{PoleEnergyOfBottomQuark*(1-c)}{(1+\sqrt{(c_1-1)(1-c_2)})} = \frac{MuonEnergy}{(1+(a-1)(1-b))} \quad (33)$$

The portion $\sqrt{(c_1-1)(1-c_2)} * 105.6583755\ MeV \cong \mathbf{1.40170791\ MeV} \cong \mathbf{1.40\ MeV}$ could be a resonance of interest especially in neutrino energies which can originate in supernova explosions. These can produce bottom quark decay processes. In which the root in the left denominator is attributed to spin 1. New physics? Looks like it! A.M. Badalian's prediction 4,778 MeV [36] is too close to 4,778.72 MeV to be ignored. The outcome of the Muon being the electro-gravitational energy of the pole energy of the Bottom Quark is as follows:
a) Lepton universality should be broken in decays of anti-Bottom Quark that involve Muons.
b) High energy p-p collisions can no longer be considered for the calculation of the W Boson mass.
Before we proceed, it is worthy of mentioning the following Simon Plouffe identity [37]: consider the functions, $S_n(r) = \sum_{k=1}^{\infty} \frac{1}{k^n e^{\pi r k}-1}$ then there is a well-known relation between $\pi$ and 96, $\pi = 72S_1(1) - 96S_1(2) + 24S_1(4)$. Notice that the sum of the positive coefficients 72 + 24 = 96 and the negative coefficient is -96. While this identity is not a direct relation between $\xi = \frac{4}{\pi}$ and $\xi = \frac{95}{96} = 1 - \frac{1}{96}$, it does show an example of how $\pi$ and 96 can be related to each other through Zeta functions in a simple and straight forward manner. A deeper and a very surprising relation will be seen in a note after (40). In an Ansatz approach, any such clue can help although it can turn out later to be wrong.

## 6. The exact inverse Fine Structure constant – critical imbalance between gravity and anti-gravity

The following endeavor originated in the search for a field strength coefficient near $\frac{\pi}{2}$ as dictated by (24.11) and (24.12) for quite a simple reason. If a motion in a small circle is with the constant velocity c, then after half a circle the velocity will be -c. The difference c-(-c) is 2c and the time between the two velocity measurements is $\frac{\pi r}{c}$ so $2c(\frac{\pi r}{c})^{-1} = \frac{2}{\pi}\frac{c^2}{r}$ while the acceleration of the motion is $\frac{c^2}{r}$. The inferred acceleration $\frac{2}{\pi}\frac{c^2}{r}$ can be interpreted only when the velocity can take one of two values c or -c, or in other words when velocity itself is quantized. The correction in this situation is by a factor $\frac{\pi}{2}$ and $\frac{\pi}{2}\frac{2}{\pi}\frac{c^2}{r} = \frac{c^2}{r}$. Given a radius r and an upper speed limit c, the correction coefficient $\frac{\pi}{2}$ should be considered as a possible upper field strength coefficient. The way a coefficient near $\frac{\pi}{2}$ was found will be discussed along with its relation to the inverse Fine Structure Constant. The fine structure constant is surprisingly reached through the mass ratio between the Tau lepton and the Muon and an interesting perturbation of the field strength of the Tau lepton that will be found in this section. Recommended reading for this section is Appendix E, (70) - (79).

**Note:** The more advanced parts of this section require basic knowledge of electrical engineering and especially a good understanding of the trivial subject of Dissipation Factor and Loss Tangent and especially of Power Factor [38].

**Note:** Why dissipation factor? The reason is that any perturbation of the Reeb class field, which behaves as acceleration, above a maximal allowed limit, must be emitted and in mainstream physics, the electromagnetic field is dissipated as photons.

The denominator $1 + \sqrt{(c_1 - 1)(1 - c_2)}$ in (32), (33) and $(1 + (a - 1)(1 - b))$ in (24) can be used together to yield a nice result that seems to be more than just a mathematical coincidence. Consider the following imbalance equation as in (23) of negative and positive charge:

$$1 + \frac{1}{96}\left(-\frac{1}{2}\xi^2 {g_1}^{-2} + \xi {g_1}^{-1}\right) = g_1$$

$$1 + \frac{1}{96}\left(-\frac{1}{2}\xi^2 {g_2}^{-2} - \xi {g_2}^{-1}\right) = g_2$$

$$\text{Such that } (g_1 - 1)^{-\frac{1}{2}} = \frac{1}{2}(1 - g_2)^{-1} \tag{34}$$

With biggest roots $g_1 \cong 1.003629541$ and $g_2 \cong 0.969877163$. $g_1$ means an area portion $\sim\mathbf{275.51693^{-1}}$ is added around a negative charge and $\mathbf{\sim 33.19740^{-1}}$ of the area is subtracted around a positive charge, which reflects a possibly maximal allowed gravitational imbalance between negative and positive charge.

A calculation that uses an electronic datasheet, yields,

$$\xi \cong \mathbf{1.5561985371903484},\ (g_1 - 1)^{-\frac{1}{2}} \cong \mathbf{16.59870203} \tag{35}$$

which is close to the known mass ratio between the Tauon and the Muon, ≅16.817 where $\xi$ denotes a maximal allowed coefficient.

**Other explanations for the existence of a field strength $\xi \cong 1.5561985371903484$.**
The following regards "Primary explanation: Ratio between a length atom and the square root of an area atom" after (15.3) and before (16), which is a result of a discussion with a colleague, Aryeh Aldema. The Aldema interpretation as an explanation to (34): (34) can be written in a more illuminating way which reflects an idea of a colleague of mine, Aryeh Aldema, who is unfortunately no longer with us. Aryeh discussed with me the possibility that there exist atoms of length and of area that are not related to each other with simple [root] relationships. For example, one can consider $\sqrt{\delta Area} \neq \delta Length$. Considering that a line segment must have a center means that in Quantum Mechanics, half the wavelength has a physical meaning in length scales and therefore it follows that,

$$\sqrt{\frac{\delta Area_1}{L^2}} = 2\frac{\delta Length}{L} = 2\frac{\delta Area_2}{L^2} \tag{35.1}$$

which is consistent with $\frac{\delta Area_1}{L^2} < \frac{\delta Area_2}{L^2}$, with $\frac{\delta Area_1}{L^2}$ as a lower limit, with a condition,

$$\frac{\delta Length}{L} = \frac{\delta Area_2}{L^2} \tag{35.2}$$

(35.2) with Aryeh Aldema's interpretation from (35.1) in (34), yields,

$$(g_1 - 1)^{\frac{1}{2}} = 2(1 - g_2) \Longrightarrow (\delta g_1)^{\frac{1}{2}} = 2(-\delta g_2) \Longrightarrow$$

$$\left(\frac{\delta Area_1}{L^2}\right)^{\frac{1}{2}} = 2\left(\frac{\delta Length}{L}\right) = 2\frac{\delta Area_2}{L^2} \tag{35.3}$$

for some minimal length unit $L$. Notice that (35.3) implies an upper limit on the possible field strength and well explains that (34) should describe the greatest possible field strength!

Multiplying this value $\xi \cong \mathbf{1.5561985371903484}$ by $1 + \sqrt{(c_1 - 1)(1 - c_2)}$ from (32), (33) and dividing by $\big(1 + (a - 1)(1 - b)\big)$ from (24) yields,

$$\frac{Muon\ 105.6583745\ MeV}{\big(1+(a-1)(1-b)\big)} \cong \frac{\sqrt{g_1 - 1}\ Tauon\ 1776.9127923826\ MeV}{\big(1+\sqrt{(c_1-1)(1-c_2)}\big)} \tag{36}$$

Which is $\frac{\big(1+\sqrt{(c_1-1)(1-c_2)}\big)}{\big(1+(a-1)(1-b)\big)\sqrt{g_1-1}}$ ≅16.81752914. So, this calculation predicts a Tauon energy of about **1776.9127923826 MeV** which agrees with [39]. Please note the remark after (28.1) for a possible additional W Boson. We now need to check the consistency of (36) with (32) as a test to this theory. We take $\frac{1}{1-x_2}$ from (22) and $\frac{1}{\sqrt{(c_1-1)(1-c_2)}}$ from (28) and check the following:

$$\frac{1776.91279322344\ldots \text{MeV}*\big(1+\sqrt{(c_1-1)(1-c_2)}\big)}{1-x_2} \cong 80372.8876666694 MeV \tag{36.1}$$

Which is consistent with (32) but less with (32.1) of a higher W Boson energy as the approximation of the W Boson's energy with $\xi = \frac{4}{\pi}$ and a null Reeb class vector. For $\xi = \frac{4}{3}$ the W Boson energy is a bit higher. (35) is strikingly related to (20) and (22). $\frac{192{y_1}^2+2\left(\frac{4}{\pi}\right)y_1-\left(\frac{4}{\pi}\right)^2}{192}={y_1}^3$ and $\frac{192{y_2}^2-2\left(\frac{4}{\pi}\right)y_2-\left(\frac{4}{\pi}\right)^2}{192}={y_2}^3$ in the following way:
Assessing the following yields,

$$-\frac{1}{\log(y_1)}\frac{1}{\log(y_2)}\cong 9147.5718747432856616796925 66 \tag{36.2}$$

and on the other hand from (35),

$$\frac{1}{(g_1-1)}\frac{1}{(1-g_2)}\cong 9146.4461480441150342812761 66 \tag{36.3}$$

The relative error in these two values in relation to $\frac{1}{(g_1-1)}\frac{1}{(1-g_2)}$ is Relative error $\cong$ $8124.9260187105719523970037 70^{-1}$. Please note that for a small d the following holds. $\frac{1}{d}\approx\frac{1}{\log(d+1)}$ and also $\frac{1}{d}\approx-\frac{1}{\log(1-d)}$. This relation alludes to a possible exponential relation between the roots of (20), (22) and the roots of (35) but before we actually check an exponential perturbation on the field strength $\xi$ ≅1.5561985371903483965638770314399 from (35) we notice the following for the same field strength coefficient of (35):

$$\frac{2}{\cos(\xi)}\cong\frac{2}{\cos(1.5561985371903484)}\cong 137.011909869, \tag{37}$$

$$\tan^{-1}(95^2 96^2(1-g_2)^{+4})\cong 1.5561948778250207190765973767615 \tag{38}$$

remarkably approximate $\xi$ ≅1.5561985371903484 from (34), (35).

$$Error=\frac{\xi-(95^2 96^2(1-g_2)^4)}{\xi}\cong 425{,}263.6013281679051795882415 7133^{-1} \tag{39}$$

**Paul Levy Isoperimetric Theorem and Levy – Gromov Isoperimetric Theorem** – Can be skipped up to “Dissipation factor interpretation”.

Define $X, M, \mu, d$ a probability space $X$ with Borel $\sigma - Algebra\ M$, measure $\mu$ and a metric distance function $d(x_1, x_2)$ where $x_1, x_2 \in X$. Define $M_\alpha=\{\alpha\ |A\in M,\ \mu(A)=\alpha, \alpha\in(0,1)\}$ now consider $A_\epsilon=\{x\in X, d(x,A)\le\epsilon\}$. We now consider the Borel $\sigma - Algebra\ M$ of the sphere $S(N)\subset\mathbb{R}^{N+1}$. Consider $A\in M_{\frac{1}{2}}$ such that $\mu(A_\epsilon)$ is minimal when $\epsilon$ is sufficiently small, then by Paul Levy’s Isoperimetric Theorem for S(N) [40] $A$ is minimal when $A$ is half a sphere. There is a stronger result for domains with smooth boundary on Riemannian manifolds with positive Ricci curvature, see theorem 2.4 in [40] and especially (2.6) in [40]. Contraction of Einstein’s tensor twice with the accelerated time-like vector $\frac{p_\lambda}{\sqrt{Z}}$ reduces equation (4) in this paper to an equation in an ordinary Riemannian and not in a Lorentzian geometry, which is the reason why [40] of great interest for this paper. A spherical cap of the sphere $S(N)$ has a maximal measure 1 because $\mu(S(N)) = 1$. The sphere $S(N) \subset \mathbb{R}^{N+1}$ is embedded in the simplest case in Euclidean spaces.

$$\mu\big(Cap(\theta)\big) = \frac{S(N-1,r=1)}{S(N,r=1)} \int_0^\theta \sin(x)^{N-1}\, dx, \tag{39.1}$$

$$\mu\big(Cap(\pi)\big) = \mu(S(N)) = 1$$

In two dimensions $N-1=1$, $\frac{S(N-1,r=1)}{S(N,r=1)} = \frac{2\pi}{4\pi} = \frac{1}{2}$,

$$\mu(Cap(\theta)) = \frac{1}{2}(1-\cos(\theta)) \tag{39.2}$$

And for half a sphere,

$$\mu\left(Cap\left(\frac{\pi}{2}\right)\right) = \frac{1}{2}(1-0) = \frac{1}{2} \tag{39.3}$$

It is immediately apparent that $\frac{\cos(\theta)}{2}$ is the difference between the measure of half a sphere and the cap of angle $\theta$ from its geodesic center which is on the sphere.

Why consider half sphere caps? The reason is that the principal circles that pass through the middle of such caps can describe a maximal acceleration. Half these circles have length $\pi r$ and motion at the speed of light along these half circles enter the plane that cuts the cap at speed of light $c$ and leave the plane at speed $-c$ or vice-versa. If only two speed measurements are done, the acceleration that is measured is $\frac{2c^2}{\pi r}$. In terms of Special Relativity, this acceleration is not of any unit vector because $\sqrt{1-\frac{v^2}{c^2}}$ becomes zero, however, $\frac{2c^2}{\pi r}$ can still be a value of a non-geodesic acceleration field without $c$ describing any classical or even General Relativistic motion. $\mu\left(Cap\left(\frac{\pi}{2}\right)\right) = \frac{1}{2}$ is of interest because it is the probability measure of a cap which is half a sphere. The term $\frac{2c^2}{\pi r}$ requires a coefficient of $\xi = \frac{\pi}{2}$ and then $\xi\frac{2c^2}{\pi r} = \frac{c^2}{r}$. Values near $\xi = \frac{\pi}{2}$, $\xi < \frac{\pi}{2}$ are therefore of interest as an upper limit for a non-geodesic acceleration field.
$\mu(Cap(\theta)) = \frac{1}{2}(1-\cos(\theta))$ can be written as

$$\mu(Cap(\theta)) = \frac{1}{2}\left(1-\cos\left(\frac{\pi}{2}-\epsilon\right)\right) \tag{39.4}$$

for some small $\epsilon$. The area difference from half a sphere, in far observer coordinates, can then be normalized in relation to the entire sphere area,

$$\frac{\cos(\theta)}{2} = \frac{\cos\left(\frac{\pi}{2}-\epsilon\right)}{2} = \frac{\delta Area}{4\pi r^2} \tag{39.5}$$

Area loss or addition in a gravitational field is equivalent to energy. In a relation between an area to the entire sphere it is therefore equivalent to an energy quotient which is smaller or bigger than one. If $\theta$ ≈1.5562011034975267 then $\frac{\cos(\theta)}{2}$ becomes about ~137.0359992349584$^{-1}$, which is close to the inverse Fine Structure Constant. And $\frac{\pi}{2} \cong 1.5707963267948966$. For this reason, if the Fine Structure Constant can be viewed as a coefficient of energy ratio, e.g. energy dissipation due to area fluctuation near a principal circle one the sphere, when an electron is accelerated, then the equation from which the Fine Structure Constant comes, must have a term $\frac{\cos(\theta)}{2}$ on one of the sides of the equation. For the inverse Fine Structure Constant, the term $\frac{2}{\cos(\theta)}$ must appear in the equation.

**Dissipation factor interpretation:** In terms of electrical engineering Dissipation Factor and Loss Tangent, we can write, $DF = \frac{95^2 96^2}{\frac{1}{2}(1-g_2)^{-4}} \approx \tan(\xi)$ where the numerator is known as the Resistive Power Loss and the denominator as the Reactive Power Oscillation. It is expected that an oscillating charge will generate oscillation in area due gravity changes, however, it is not expected that the area portion that is lost due to gravity will appear as the power of 4. This is a very rare property that connects trigonometry and the electro-gravity polynomials (34). We can get from this relation two insights, the first is that if (37) is not a mathematical coincidence, then the inverse Fine Structure constant should come out of a trigonometric function and a numbers relation. The second is that $95^2 96^2 (1-g_2)^4$ should be part of this equation. We may think that perhaps scaling of the value of $\xi$ in a rational way, will yield the exact inverse Fine Structure Constant. So we want to find some d such that $\frac{2}{\cos\left(1.5561985371903484*(1+\frac{1}{d})\right)}$ will yield the constant we are looking for. We will soon find such d, $d \cong 606400.8$ that complies with [41] and we get, $\frac{2}{\cos\left(1.5561985371903484*(1+\frac{1}{606400.8})\right)} \cong 137.0359990462475253$. The motivation for this endeavor is taken from electrical engineering [37] where the cosine term means a ratio between delivered power and measured power in motors and other electric devices. In our case, we are interested in the ratio between radiation's energy and the energy it delivers upon interaction which should define the strength of the electric interactions. Until now, d is not very interesting because we could not find d out of any new theory. Well, not very accurate. First,

$$d = \frac{1}{2}(1-g_2)^{-4} \cong 607276.5368006824282929 \tag{39.6}$$

and

$$\frac{2}{\cos(1.5561985371903484 * (1 + 2(1-g_2)^4))} \cong 137.0359643018112763$$

(39.6) is a result of the discussion in “Primary explanation: Ratio between a length atom and the square root of an area atom” after (15.3) and before (16) and especially of the Aldema interpretation of the relations between 4-volume and length atoms.

**Caveat:** do not confuse the use of the term $Area$ in a 4-volume relation to length with a surface area to length in the following term, in which $Area^2$ means 4-volume, it is not the same term as in $\frac{1}{4}Area = Length^2$.

$$Area^2 = (2Length)^3 Length = 8Length^4 \tag{39.7}$$

Which means by (34) and (35.3),

$$\left(\frac{1}{2}\sqrt{\frac{\delta Area_1}{L^2}}\right)^3 \sqrt{\frac{\delta Area_1}{L^2}} = 2\left(\frac{\delta Length}{L}\right)^4 \tag{39.8}$$

and can be summarized as the Aldema 4-volume ratio,

$$d^{-1} = 2(1-g_2)^4 = \left(\frac{1}{2}(g_1-1)^{\frac{1}{2}}\right)^3 (g_1-1)^{\frac{1}{2}} = \frac{1}{8}(g_1-1)^2 \cong 607276.5368006824282929^{-1} \tag{39.9}$$

Consider

$$2(1-g_2)^4\, 95^2 96^2 \tag{39.10}$$

as a portion of $n = 95^2 96^2$ energy emission events which occur in a 4-volume unit. In a more illuminating way,

$$n = \left(\frac{95}{96}\right)^2 96^4 \tag{39.11}$$

which from (23) is a multiplication of $96^4$ with the smallest field strength coefficient two times, with the electron $\xi = \frac{95}{96}$. In this case, each plane is multiplied by $\xi$, which has a physical meaning of the field strength depending on the number of events in a 4-volume with a fundamental reference of $96^4$ events. (39.9) and (39.11) will both be used in a very surprising assessment of the inverse Fine Structure constant. A resolution of 96 events per each dimension is implied in the factors $\frac{\xi}{96}$, see the left-hand side of (19). Also see (23.1) and (23.2). A portion of energy emission events can be considered as a dissipation factor, which is the motivation in this discussion.

**An expected relation:** If in (39.11) $96^4$ is a maximal, not minimal number of events, and the Tau energy is the maximal energy for leptons, then the event of two annihilating Tau leptons Must yield $96^4$ multiplied by this basic energy. Then by (36),

$$2 * 1776.9127923826\ \mathrm{MeV/c^2} = 96^4 * Basic\ neutral\ particle's\ energy/c^2 \tag{39.12}$$

It turns out that such an energy is **~41.8418788293579 eV/c²**, but this value is very close to the result of (24) $ElectronMass * (a-1)(1-b) =$ **~41.8752442118608** $\boldsymbol{eV/c^2}$ with a relative error ~1255.050626$^{-1}$. The author’s opinion is that, as is, a relative error, slightly smaller than $10^{-3}$ is not sufficient to be considered as a finding, however, when considered along with

(24) and (36), and the fact that (39.12) was expected to yield a fundamental unit of mass, (39.12) cannot be ignored. 41.875244211860eV-41.8418788293579eV seems to set a strict neutrino mass bound of ~0.0333653825021eV/$c^2$ about, **0.033eV/$c^2$**.

If we test the following values for $d \cong 606400.8$, which complies with [41], we get: $\frac{95^4}{d} \cong 134.3181357940161..$, $\frac{96^4}{d} \cong 140.0635619214..$ and the geometric average of these two values is $\left(\frac{95^4}{d}\frac{96^4}{d}\right)^{\frac{1}{2}} = \frac{95^2 96^2}{d} \cong 137.1607689$. It is not difficult to see the following:

As a result of the conclusions of (38), (39), the exact inverse Fine Structure Constant was found by the following, although some aspects of the following calculation are not resolved yet. We put together (20), (22), (34), (35), (37), $\frac{1}{2}(1-g_2)^{-4} \cong 607276.5368006824282929$ from (39.9), and from (35) $\xi \cong 1.5561985371903483965638770314399 15299$,

$$1+\frac{1}{96}\left(-\frac{1}{2}\left(\frac{4}{\pi}\right)^2 a^{-2}+\frac{4}{\pi}a^{-1}\right)=a \Longrightarrow \frac{1}{a-1} \cong 206.75133988502202$$

$$1+\frac{1}{96}\left(-\frac{1}{2}\left(\frac{4}{\pi}\right)^2 b^{-2}-\frac{4}{\pi}b^{-1}\right)=b \Longrightarrow \frac{1}{1-b} \cong 44.63955017596401$$

$$d=\left(\frac{1}{2}(1-g_2)^{-4}\right)^{\frac{1}{1+(a-1)(1-b)}} \cong 606401.0372 \approx 606400.8$$

$$\frac{2}{\cos\left(1.5561985371903484*\left(1+\frac{1}{d}\right)\right)} \cong 137.0359990368270076 \approx 137.035999037 \qquad (40)$$

$1.5561985371903483965638770314399 * \left(1+\frac{1}{d}\right)$ exceeds the maximal allowed value of the field strength $\xi = 1.5561985371903483965638770314399$ and therefore must account for emission of what we know in mainstream physics as photons.

**Note:** $p = ((a-1)(1-b))^{-\frac{1}{2}} \cong 96.0691772148863$ is a very special number in the following property that bridges between area ratios and powers as follows, denote $s = \frac{1}{2}(1-g_2)^{-4}$ then $s^{\left(\frac{1}{1+(a-1)(1-b)}\right)} \approx s\left(2-\frac{1}{96^2(a-1)(1-b)}\right)$ or written as numbers 606401.0372 ~ 606401.0194 with a relative error of about 34,109,836.56$^{-1}$. An exact equality, $s^{\frac{1}{1+p^{-2}}} = s\left(2-\frac{p^2}{96^2}\right) \cong 606401.0371$, follows from replacing $p = \sim 96.0691772148863$ with p = ~96.06917582 with a relative error in $96.0691772148863 = ((a-1)(1-b))^{-1/2}$ of ~1.45953 * $10^{-8}$. If the reader still thinks (40) is a fluke of chance, then this note does not agree with such a hypothesis. Also note that p comes from (20), (22) which resulted in (24). See Python code and it’s more exact output in Appendix F. Slightly different values are obtained for

$\xi\left(1-\frac{1}{s^{\left(\frac{1}{1+(a-1)(1-b)}\right)}}\right)^{-1}$, the reason for a term $\xi\left(1-\frac{1}{d}\right)^{-1}$ instead of $\xi\left(1+\frac{1}{d}\right)$ is important and is discussed in (42.1).

Another result is by finding the variable s where a and b are given in (40) and consider using 95*96 from (23.1), (23.2), and in its second power in (39.11) as follows:

$$\left(\frac{95^2*96^2}{s}\right)^{1+(a-1)(1-b)} = \frac{2}{\cos(\xi\left(1+\frac{1}{s^{\left(\frac{1}{1+(a-1)(1-b)}\right)}}\right))} \Longrightarrow \quad (41)$$

$$\left(\frac{95^2 * 96^2}{s}\right)^{1+(a-1)(1-b)} \cong 137.035999036428876252$$

**Very important**: Another more accurate algorithm reached 137.03599903642884783039335161447525 and s ~= 607276.54683397442568093538284301757812 and comparing this value to $\frac{1}{2}(1-g_2)^{-4} \cong$ 607276.536800682428292930126190185546875 from (39.9), it is statistically impossible that (41) is a coincidence. The inverse relative error in relation to $\frac{1}{2}(1-g_2)^{-4}$ is,

$$Err^{-1} = \left(\frac{s}{\frac{1}{2}(1-g_2)^{-4}} - 1\right)^{-1} \cong 60526150.002596460282802581787109 \approx 6*10^8 \quad (41.1)$$

This result means that any claim that the results of this paper are by chance is ridiculous and possibly irresponsible.

Where the term $\frac{1}{1+(a-1)(1-b)}$ is taken from (24) but with $\xi = \frac{4}{\pi}$ as in (40) and not $\xi = \frac{95}{96}$.

A $\left(\frac{95^2*96^2}{s}\right)^{1+\frac{1}{95*96}}$ ≅137.0359992990990, $s$ ≅607280.4243559269234538 and $s^{1/(1+(96*95)^{-1})} \cong$ 606394.43614689458627253770 and $(a-1)(1-b)$ replaced with $(96*95)^{-1}$.

With $a = g1, b = g2$ from (34), $1+(a-1)(1-b)$ yields in (41), $s \cong$ 607279.477540519786998629570007 and $\left(\frac{95^2*96^2}{s}\right)^{1+(a-1)(1-b)} \cong$ **137.035999234958467241085600**

Slightly different values are obtained for $\xi\left(1-\frac{1}{s^{\left(\frac{1}{1+(a-1)(1-b)}\right)}}\right)^{-1}$, the reason for a term $\xi\left(1-\frac{1}{d}\right)^{-1}$ instead of $\xi\left(1+\frac{1}{d}\right)$ is important and is discussed in (42.1).

The latter choice can be explained as a result of some functional relations. Following is an example. Consider the Airy function Bi, Airy functions Ai and Bi are very popular in Quantum Mechanics [42],

$$Bi(x) = \frac{1}{\pi}\int_0^{\infty}\left[\exp(-\frac{t^3}{3}+xt)+\sin(\frac{t^3}{3}+xt)\right]dt$$

The reason Bi(x) is tested here is because it maps the highest $\xi$ value from (34) near 2 and along with Ai(x) it is used to describe the wave function of a particle in a triangular potential well.

**Table 1.**

| $\xi$ | Airy function $Bi(\xi)$ | $\frac{1}{2-Bi(\xi)}$ | $\frac{1}{\sqrt{((a-1)(1-b))}}$, see (19), a, b are the largest roots, a>1, b<1. |
|---|---|---|---|
| Electron, $\frac{95}{96}$ | 1.197775825963505749636 | 1.246534363291960163318 | 110.466686112441522027438622899353504180908203125 |
| Muon, $\frac{4}{\pi}$ | 1.51534445151380815816 | 2.06332105992280961894 | 96.069177214886309457142488099634647369384765625 |
| Tauon, 1.5561985371903483965638770314399… | 1.989537211925930116713 | 95.5768188097318024431364884 | **95.63705426268688825075514614582061767578125** |
| $\arccos\left(\frac{2}{137.0359990368270078..}\right)$ | 1.989542478676644522794 | **95.624954430323135605547** | |
| $\arccos\left(\frac{2}{137.0359992349584672..}\right)$ | 1.989542478719955590708 | **95.624954826365255405784** | |

We can see a surprising possible relation between $\frac{1}{2-Bi(\xi)}$, $\frac{1}{((a-1)(1-b))}$ and the Fine Structure Constant. This relation is one of the motivations to try the value $1+(a-1)(1-b)$ in (41), where $a = g_1$ and $b = g_2$ in (34).

Another idea is to solve the following equation where s is given by (34) and p is a variable and where $\frac{95^2*96^2}{s} = 95^2 * 96^2 * 2(1-g_2)^4$ is from (39.10):

$$s = (\frac{1}{2}(1-g_2)^{-4}) \cong 607276.536800682428292930$$

$$\left(\frac{95^2*96^2}{s}\right)^{1+1/(p*p)} = \frac{2}{\cos(\xi\left(1+\frac{1}{s\left(\frac{1}{1+1/(p*p)}\right)}\right))} \Longrightarrow \tag{42}$$

$$\left(\frac{95^2*96^2}{s}\right)^{1+1/(p*p)} \cong \mathbf{137.035999035747181551}, p \cong 96.070666670305840285$$

$1+\frac{1}{s^{\left(\frac{1}{1+1/(p*p)}\right)}}$ approximates $(1-\frac{1}{s^{\left(\frac{1}{1+\frac{1}{p*p}}\right)}})^{-1}$ and

$$\left(\frac{95^2*96^2}{s}\right)^{1+\frac{1}{p*p}}=\frac{2}{\cos\left(\xi\left(1-\frac{1}{s^{\left(\frac{1}{1+\frac{1}{p*p}}\right)}}\right)^{-1}\right)}\Longrightarrow$$

$$\left(\frac{95^2*96^2}{s}\right)^{1+\frac{1}{p*p}}\cong 137.035999075497272770007839426398 28,$$

$$p\cong 96.0706405302390749056939966976 64,$$

$$d=s^{\left(\frac{1}{1+\frac{1}{p*p}}\right)}\cong 606401.0638207901502028107643127441 4062$$

The meaning of the term $\xi\left(1-\frac{1}{s^{\left(\frac{1}{1+\frac{1}{p*p}}\right)}}\right)^{-1}$ is that a 4-volume unit is contracted by a factor of $1-\frac{1}{s^{\left(\frac{1}{1+\frac{1}{p*p}}\right)}}$ and that the field strength $\xi$ is increased as a result.

For comparison, if we set p=96 in the right-hand side of (42) we get the value 137.0359990869357602605305 15. Combining (41) and (42) we find a numerical attractor at (42) with $s\cong\mathbf{607276.5368006824282929301262}\cong(\frac{1}{2}(1-g_2)^{-4})$, $s^{\left(\frac{1}{1+1/(p*p)}\right)}\cong$ $606401.064296812633983791$, $\xi\cong\mathbf{1.5561985371903484}$ from (35). Before we close this discussion, it is nice to mention another relation $(1-\ln\left(\left(1+\frac{1}{137.035999035747181551}\right)^{137.035999035747181551}\right))^{-1}\cong 275.4045237287\ \approx 275.51693\cong$ $(g_1-1)^{-1}$ in (43.10). That is not a total surprise because $(1-\ln\left(\left(1+\frac{1}{z}\right)^{z}\right))^{-1}\approx 2z$ for big z.

**Reverse engineering Nature** – Looking for simple but not random relations

In this section a much less significant result than (24), (40), remark after (40), (41), (42), will be considered as an interesting course of research. This time, an approximation of the inverse Fine

Structure Constant will not be as nearly as accurate and will not be a result of exponential perturbations of a Reeb class field strength.

The search for meaningful field strength coefficients for the electron, Muon and Tau lepton reached the following $\xi \in \{\frac{95}{96}, \frac{4}{\pi}, \sim 1.556198537190348396563877031439 9\}$

But these field strength coefficients did not appear out of solutions to equation (4). In fact, there has been no collaboration with mainstream physics to reach such solutions and especially to the complex form of (4). The analytic solutions of such an equation make take decades and without collaboration on solving the Lagrangians in (4), (64.b), (64.01), (65), other approaches are required in order to convince the reader that the choices of field strength coefficients are not a mere mathematical pareidolia. The assessment of the mass ratio between the Muon and the electron in (24) is already with a sufficiently small error to trigger interest, especially when considering the simplicity of (24) and that the choice of $\frac{4}{\pi}$ came out of an existing theory [32]. (40), the remark after (40), (41) and (42) are also strong indicators that this research is on the right path. It will be wrong not to mention other findings which are straight forward from the method which had been presented in (16), (17), (18), (19) and the first interesting result (20). In this method, the Reeb class vector term was collapsed with the non-geodesic or accelerated time direction $\frac{P^\mu}{\sqrt{|Z|}}$ and we saw the contraction $\frac{1}{4}\left(U_\mu U_\nu - \frac{1}{2} g_{\mu\nu} U_\lambda U^\lambda - 2U^k{}_{;k} \frac{P_\mu P_\nu}{Z}\right) \frac{P^\mu P^\nu}{Z}$ that resulted in (20), (22).

With acceleration field $\frac{\xi c^2}{r}$, where $\xi = \frac{4}{\pi}$ denotes the field strength and $x$ is the adjustment factor of the acceleration field because of area loss, we used the term $\left(-\frac{1}{2}\frac{\xi^2}{r^2 x^2} \mp \frac{\xi}{r^2 x}\right)\frac{\pi}{24} r^4 = \left(-\frac{1}{2}\frac{\xi^2}{x^2} \mp \frac{\xi}{x}\right)\frac{\pi}{24} r^2$ to express area loss due to a gravitational field at small $r$ in the far observer coordinates. Now it is time to look at area loss in a direction perpendicular to the direction of time, namely the momentum direction in spacetime, or as expressed through a bivector derived from a unit vector, consider $\frac{U^\mu U^\nu}{U^\lambda U_\lambda}$ and for the sake simplicity, the contraction is not with a complex bivector $\frac{2U^{*\mu} U^{*\nu}}{U^{*\lambda} U_\lambda + U^\lambda U^*{}_\lambda}$. From $U^\mu P_\mu = 0$, it is easy to see the following,

$$\frac{1}{4}\left(U_\mu U_\nu - \frac{1}{2} g_{\mu\nu} U_\lambda U^\lambda - 2U^k{}_{;k} \frac{P_\mu P_\nu}{Z}\right) \frac{U^\mu U^\nu}{U^\lambda U_\lambda} = \frac{1}{8} U_\lambda U^\lambda = \frac{1}{2}\frac{\xi^2}{r^2 x^2} \tag{42.2}$$

The latter is to achieve a reduction of the curvature calculation from Lorentzian to Riemannian geometry.

**Caveat**: Notice that using $\frac{U^\mu U^\nu}{U^\lambda U_\lambda}$ and not $\frac{U^\mu U^\nu}{|U^\lambda U_\lambda|}$ is done here in order to achieve $g_{\mu\nu}\frac{U^\mu U^\nu}{U^\lambda U_\lambda} = +1$ as expected from a unit vector in (+,-,-,-) metric convention. The reader may criticize this choice of a bivector because Reeb class vectors in this paper are space-like and not time-like because they

represent non-geodesic acceleration as a result of misaligned events in an observer spacetime object.

Multiplying by $\frac{\pi}{24}r^4$ due to [30], see lecture of Seth Lloyd, and dividing by 4 times the area of an Euclidean disk, due to assumption 2 after the note after (15.3), yields,

$$-\frac{1}{192}\frac{\xi^2}{{x_3}^2} = \frac{1}{2}\frac{\xi^2}{r^2{x_3}^2}\frac{\pi}{24}r^4\frac{1}{4\pi r^2} = x_3 - 1 \tag{42.3}$$

From which

$$\frac{192{x_3}^2-\xi^2}{192} = {x_3}^3 \tag{42.4}$$

Which is an iterative equation that converges to the most stable root, a technique that had been used in all previous third order polynomial equations. Solving for $\xi = \frac{4}{\pi}$ as in (20), (22), yields,

$$(x_3 - 1)^{-1} \cong 120.410611116112391982824192382395267486572265625 \tag{42.5}$$

$$(x_1 - 1)^{-1} \cong 206.7513398850220198710303520783782005310058 59375$$

$$(1 - x_2)^{-1} \cong 44.63955017596401120272275875322520732879638671875$$

And the following calculation yields an interesting result,

$$\ln((x_1 - 1)^{-1}(1 - x_2)^{-1}(x_3 - 1)^{-1})^2 2^{-\frac{1}{2}}$$

$$\cong \ln(1111304.0650477090384811162948608398437)^2\, 2^{-\frac{1}{2}} \cong$$

$$137.034102324667713901362731121480464 9353 \tag{42.6}$$

Which is surprisingly simple and unexpected approximation of the inverse Fine Structure Constant. The relative error of (42.6) in relation to the result in (40) is about $72249.23316^{-1}$ which is not even closely significant as (40), (41), (42) or the remark after (40) and yet, if this result joins other approximations of the inverse Fine Structure Constant in this paper, it is not wise to ignore (42.6). In (24.3), (40), (41), (42), (81)-(86), the inverse Fine Structure Constant comes out of exponential field perturbations as in (24.3), (40), (41), (42) or as exponential functions of 2 or $\frac{4}{\pi}$ with coefficients $(95*96)^{-1}$ or 95 and 96 as seen in (81)-(86). Notice that both (24.3) and the last result, involve the square root of 2.

**Hypergeometric tests - Dr. Sam Vaknin's suggestion from 2013**

A suggestion from Dr. Sam Vaknin regarding the possible solutions of the equations of the Geometric Chronon Field Theory was that they are related to Hypergeometric functions [43]. His idea was lately checked regarding the stable roots of third order polynomials of gravity and anti-gravity, area ratio loss and gain, see (22.1) and (22.2). The stable field strength coefficients were defined as $\xi = \frac{193}{192} = 1 + \frac{1}{192}$ for negative charge and $\xi = \frac{63}{64} = 1 - \frac{1}{64} = 1 - \frac{3}{192}$ for positive charge. The summation of the two deltas $+\frac{1}{192} - \frac{1}{64}$ to 1 yields the field strength coefficient $\xi = 1 + \frac{1}{192} - \frac{1}{64} = 1 + \frac{1}{192} - \frac{3}{192} = 1 - \frac{1}{96} = \frac{95}{96}$. The question is what do these values $\frac{1}{192}$ and $\frac{3}{192}$ teach us about any possible grand theory of particle physics? 1,3 and 192 with 192 in the denominator should hint us about such a theory. As we saw in (40), (41), (42) a key number in the calculation of the positive perturbation over $\xi$ was

$s = 0.5/(1 - g_2)^4 \cong 607276.536800682428292930126190185546875$, see (42).

Can this number be a result of combinatorial mixing by the Gauss hypergeometric function 2F1?

The question is if ${}_2\mathrm{f}_1(\mathrm{a}, \mathrm{b}, \mathrm{c}, \mathrm{z}) = \sum_{k=0}^{\infty} \frac{(a)_n (\mathrm{b})_n}{(c)_n} \frac{z^n}{n!}$, such that $(q)_n = \begin{Bmatrix} 1 \\ q(q+1)(q+2) \ldots (q+n-1) \end{Bmatrix} \begin{vmatrix} n = 0 \\ n \geq 1 \end{vmatrix}$ can yield such a number in a meaningful way.

If Dr. Sam Vaknin was right, we may be able to find a meaningful z that solves

$$ {}_2\mathrm{f}_1(-3, 1, 192, \mathrm{z}) = 1 - 2(1 - g_2)^4 \tag{42.6} $$

This is exactly what was done numerically. The result was very surprising, and it is very unlikely that it is a fluke of chance:

$$ z \cong \frac{2}{(137.03627140261694705714035080 74)^2} \tag{42.7} $$

The relative error of $137.0362714026169470571403508074$ from the assessment $137.035990368270075578\ldots$ in (40) is about $\frac{137.035990368270075578 - 137.035990368270075578\ldots}{137.035990368270075578\ldots} \cong 503132.199783077405299913692^{-1}$

Also, quite near the higher value 137.0359992990990… after (41).

It is quite compelling to say that Dr. Sam Vaknin was right already back then in 2013. There is even stronger evidence in his favor.

Consider a second order perturbation on the hypergeometric coefficients -3, 1:

$$ {}_2\mathrm{f}_1\left(-3 * \frac{63}{64}, 1 * \frac{193}{192}, 192, \frac{2}{(137.035990368270075578\ldots)^2}\right) \cong $$

$$ 1 - \frac{1}{607299.792079592822119593620300292968750} \tag{42.8} $$

and

$$ {}_2f_1\left(-3*\frac{193}{192},1*\frac{63}{64},192,\frac{2}{(137.035999036827007557 8\ldots)^2}\right) $$

$$ \cong 1-\frac{1}{607299.8060423402348533272743225097 65625} \quad (42.9) $$

Comparing the right hand side denominator to $\frac{1}{2}\frac{1}{(1-g_2)^4} \cong$ $607276.536800682428292930126190185546875$ from the remark before (40) and from (42), the results in (42.8) and (42.9) are very interesting although not within the ranges of (40)-(42) with a highest value of 137.0359992990990… .

Instead of $\frac{63}{64}$ if we consider all the powers of $-\frac{1}{64}$ we have $\frac{64}{65}=\sum_{k=0}^{\infty}\left(-\frac{1}{64}\right)^k$ and with $+\frac{1}{192}$ we have $\frac{192}{191}=\sum_{k=0}^{\infty}\left(\frac{1}{192}\right)^k$

Consider a second order perturbation on the hypergeometric coefficients -3, 1:

$$ {}_2f_1\left(-3*\frac{64}{65},1*\frac{192}{191},192,\frac{2}{(137.0359990368270075578\ldots)^2}\right) \cong $$

$$ 1-\frac{1}{607135.0557247012620791792869567871 09375} \quad (42.8.1) $$

and

$$ {}_2f_1\left(-3*\frac{192}{191},1*\frac{64}{65},192,\frac{2}{(137.0359990368270075578\ldots)^2}\right) $$

$$ \cong 1-\frac{1}{607135.0695571019314229488372802734 37500} \quad (42.9.1) $$

Here is the code in Python for (42.6) and (42.7):

```
import numpy as NP
from scipy.special import hyp2f1 as SCIPY_SPECIAL_hyp2f1

a = 137.035999036827007557803881354629993438720703
q = 607276.536800682428292930126190185546875

#s = NP.power(q * 2, 0.25)
s = NP.sqrt(NP.sqrt(q * 2))
s = NP.sqrt(s * s * s * 0.25)

print(f's={s:.42f}')

# Was a numerical analysis output:
```

```
w = 137.0362714026169470571403508074
u = 1/(w/a - 1)

print(f'u={u:.42f}')

r = SCIPY_SPECIAL_hyp2f1(-3, 1, 192, 2/(w ** 2))
r = 1/(1-r)
r /= 607276.536800682428292930126190185546875
r = 1/(1-r)
r /= s
print(f'r={r:.42f}')

r = SCIPY_SPECIAL_hyp2f1(-3, 1, 192,
2/137.035999036827007557803881354629993438720703 ** 2)
r = 1/(1-r)
r /= 607276.536800682428292930126190185546875
r = 1/(1-r)
print(f'r={r:.42f}')
```

Here is the code in Python for (42.8), (42.9)

```
import numpy as NP
from scipy.special import hyp2f1 as SCIPY_SPECIAL_hyp2f1

Xi = 1.556198537190348396563877031439915299415588378906

r1 = \
    SCIPY_SPECIAL_hyp2f1(-3 * 63/64, 1 * 193/192, 192,
                         2/137.035999036827007557803881354629993438720703 **
                         2)
r1 = 1/(1-r1)
print(f'1/(1-r1)={r1:.33f} compared to'
      f' 607276.536800682428292930126190185546875')

inverse_alpha1 = 2/NP.cos(Xi*(1+1/NP.power(r1, 1/(1+1/(95*96)))))
print(f'Inverse alpha(r1) {inverse_alpha1:.33f}')

r2 = \
    SCIPY_SPECIAL_hyp2f1(-3 * 193/192, 1 * 63/64, 192,
                         2/137.035999036827007557803881354629993438720703 **
                         2)
r2 = 1/(1-r2)
print(f'1/(1-r2)={r2:.33f} compared to'
      f' 607276.536800682428292930126190185546875')

inverse_alpha2 = 2/NP.cos(Xi*(1+1/NP.power(r2, 1/(1+1/(95*96)))))
print(f'Inverse alpha(r2) {inverse_alpha2:.33f}')

r = r1 / r2
r = 1/(1-r)
print(f'1/(1-r1/r2)={r:.33f}')

r3 = \
    SCIPY_SPECIAL_hyp2f1(-3 * 64/65, 1 * 192/191, 192,
                         2/137.035999036827007557803881354629993438720703 **
```

```
                         2)
r3 = 1/(1-r3)
print(f'1/(1-r3={r3:.33f} compared to'
      f' 607276.536800682428292930126190185546875')

inverse_alpha3 = 2/NP.cos(Xi*(1+1/NP.power(r3, 1/(1+1/(95*96)))))
print(f'Inverse alpha(r3) {inverse_alpha3:.33f}')

r4 = \
    SCIPY_SPECIAL_hyp2f1(-3 * 192/191, 1 * 64/65, 192,
                         2/137.035999036827007557803881354629993438720703 **
                         2)
r4 = 1/(1-r4)
print(f'1/(1-r4)={r4:.33f} compared to'
      f' 607276.536800682428292930126190185546875')

inverse_alpha4 = 2/NP.cos(Xi*(1+1/NP.power(r4, 1/(1+1/(95*96)))))
print(f'Inverse alpha(r4)
{2/NP.cos(Xi*(1+1/NP.power(r4,1/(1+1/(96*96))))):.33f}')

r = r3 / r4
r = 1/(1-r)
print(f'1/(1-r3/r4)={r:.33f}')

print(f'(1/Alpha1+1/Alpha3)/2: '
      f'{(inverse_alpha1+inverse_alpha3)/2:.33f}')
print(f'(1/Alpha2+1/Alpha4)/2: '
      f'{(inverse_alpha2+inverse_alpha4)/2:.33f}')
```

## 7. The mass hierarchy – a possible exponential relationship

By (13) and considering the Planck mass $\sqrt{\frac{\hbar c}{K}}$ and the Fine structure constant Alpha:

$$\sqrt{\frac{\hbar c}{K} * \frac{e^2}{4\pi\varepsilon_0 \hbar c}} = \frac{2e}{2\sqrt{4\pi K\varepsilon_0}} = \frac{2e}{\sqrt{16\pi K\varepsilon_0}} = PlanckMass * \sqrt{Alpha} \tag{43}$$

So, multiplication of the Plank mass by the square root of the Fine Structure Constant yields twice the electro-gravitational mass of a charge e! If we take $\xi \cong 1.5561985371903484$ from (35) to be the maximal allowed field coefficient of an electric charge, then the field around a single charge as a normalized quantity is obtained as

$$\frac{1}{\xi}\frac{1}{2} PlanckMass * \sqrt{Alpha} = \frac{1}{\xi}\frac{e}{\sqrt{16\pi K\varepsilon_0}} \tag{44}$$

Now we recall from (24) the following root 'a' around a negative charge:

$$1 + \frac{1}{96}\left(-\frac{1}{2}\left(\frac{95}{96}\right)^2 a^{-2} + \left(\frac{95}{96}\right) a^{-1}\right) = a \cong 1 + 192.0463944^{-1} \tag{45}$$

We take from (24), (40), $(a-1)(1-b) \cong \frac{1}{206.75133988502202 * 44.63955017596401}$ and calculate

$$\left(\frac{\frac{11}{\xi_2}PlanckMass*\sqrt{Alpha}}{M_e}\right)^{(a-1)(1-b)} \cong 1+192.04864774452^{-1} \tag{46}$$

Where $M_e \cong 0.5109989461\ MeV$, e is the electron's charge $1.602176634\times10^{-19}$ Coulombs, K is Newton's constant of gravity $6.674\times10^{-11}$ $m^3\cdot kg^{-1}\cdot s^{-2}$, Planck mass $1.22091\times10^{22}$ MeV, from (40) $Alpha \cong 137.0359990368270076^{-1}$. The relative error between (46) and (45) is $\frac{192.04864774452-192.0463944}{192.0463944} \cong 85{,}227.266539382^{-1}$. We are also led to the following conclusion that $\xi = \frac{95}{96}$ is the Reeb class field strength coefficient of the electron field, $\xi = \frac{4}{\pi}$ is the Muon field strength and from the solution to (35) $\xi = 1.5561985371903484$ ... is the field strength coefficient of the Tau lepton. Of course, a lot of work has to be done to achieve exact analytic solutions to (4) and as we shall see also to (64.b), (64.01), because only $\xi = \frac{4}{\pi}$ has a compelling source [32]. Serendipity is part of physics and mathematical rigor must follow.

**Quantization of the acceleration matrix for the field strength coefficients - Compelling evidence of the field strength coefficients**

The following is statistical proof that the field strength coefficients are indeed of interest.

The idea is that when the acceleration field acts on itself even times, i.e. even power, it scales a time vector. If the power is 4m+2 then the time-like vector is reversed. If the power is 4m then the time-like vector is scaled. An acceleration which is a unit in the complex plane does not change the gradient of time but changes its phase.

An odd power of the acceleration matrix is again an acceleration matrix.

The scaling 96 that appears in (24), (34), (35) and in other terms is considered also as a maximal power of an acceleration matrix but the maximal acceleration matrix is the power 95 of a characteristic acceleration of the acceleration field of the Muon and of Tau Lepton.

$(A^{95})^{96} = A^{9120}$ should be able to recover the field of the Muon, Tau Lepton.

We will check $\xi = 2$ as an upper limit – see remark after (23.1), $\xi \cong 1.5561985371903483965638770314399$ from (34), (35), $\xi = \frac{4}{\pi}$ from (24) and $\xi = \frac{95}{96}$ from (24).

The acceleration matrix in one boost plane spanned by $\frac{P_\mu}{Z}$ and by $\frac{U_\mu}{2}$ can be written as

$$A = \begin{pmatrix} 0 & -a \\ a & 0 \end{pmatrix} = \begin{pmatrix} 1 & 1 \\ i & -i \end{pmatrix}\begin{pmatrix} -ai & 0 \\ 0 & ai \end{pmatrix}\begin{pmatrix} \frac{1}{2} & -\frac{1}{2}i \\ \frac{1}{2} & \frac{1}{2}i \end{pmatrix} \tag{46.1}$$

Where $i = \sqrt{-1}$.

$$A^n = \begin{pmatrix} 0 & -a \\ a & 0 \end{pmatrix} = \begin{pmatrix} 1 & 1 \\ i & -i \end{pmatrix} \begin{pmatrix} (-ai)^n & 0 \\ 0 & (ai)^n \end{pmatrix} \begin{pmatrix} \frac{1}{2} & -\frac{1}{2}i \\ \frac{1}{2} & \frac{1}{2}i \end{pmatrix} = \begin{pmatrix} \frac{(-ia)^n+(ia)^n}{2} & \frac{-i(-ia)^n+i(ia)^n}{2} \\ \frac{i(-ia)^n-i(ia)^n}{2} & \frac{(-ia)^n+(ia)^n}{2} \end{pmatrix} \tag{46.2}$$

$A^n$ is an acceleration matrix where $n = 2m + 1$ and a scaling matrix where $n = 2m$.

Consider that the resolution of acceleration matrix is dictated by $A^n$ for some $a$ that depends on the field strength $\xi$. Assuming time is quantized with $a\delta\tau = 1 + \alpha$, the following is a dimensionless matrix,

$$A = \begin{pmatrix} 0 & -a\delta\tau \\ a\delta\tau & 0 \end{pmatrix}, \text{ also } \; Exp(A) = \begin{pmatrix} \cos(a\delta\tau) & -\sin(a\delta\tau) \\ \sin(a\delta\tau) & \cos(a\delta\tau) \end{pmatrix} \in SO(2) \tag{46.3}$$

And the powers of the acceleration matrix become

$$A^n = \begin{pmatrix} \frac{(-ia\delta\tau)^n+(ia\delta\tau)^n}{2} & \frac{-i(-ia\delta\tau)^n+i(ia\delta\tau)^n}{2} \\ \frac{i(-ia\delta\tau)^n-i(ia\delta\tau)^n}{2} & \frac{(-ia\delta\tau)^n+(ia\delta\tau)^n}{2} \end{pmatrix} \tag{46.4}$$

Now consider a maximal $n = 96$ then a maximal acceleration in absolute value must result from an odd power of $A$ and in this case the maximal odd $n$ is 95.

$$A_{max} = A^{95} = \begin{pmatrix} 0 & (a\delta\tau)^{95} \\ -(a\delta\tau)^{95} & 0 \end{pmatrix} \tag{46.5}$$

Notice the negative sign of the acceleration components when $n = 4m + 2 + 1$.

$$A^{4m+2+1} = \begin{pmatrix} 0 & (a\delta\tau)^{4m+2+1} \\ -(a\delta\tau)^{4m+2+1} & 0 \end{pmatrix} \tag{46.6}$$

$$A^{4m+1} = \begin{pmatrix} 0 & -(a\delta\tau)^{4m+1} \\ (a\delta\tau)^{4m+1} & 0 \end{pmatrix}$$

Also notice $A^{96} = \begin{pmatrix} (a\delta\tau)^{96} & 0 \\ 0 & (a\delta\tau)^{96} \end{pmatrix}$, $A^{4m+2} = \begin{pmatrix} -(a\delta\tau)^{4m+2} & 0 \\ 0 & -(a\delta\tau)^{4m+2} \end{pmatrix}$

The maximal scaling with $A_{max}(Acceleration) = A^{95}$ is then

$$A_{max}(Scaling) = (A^{95})^{96} = A^{9120}$$

9120 = 95*96. Setting an upper limit of the acceleration matrix to 2 results in

$$(a\delta\tau)^{95} < 2$$

It is remarkable that setting $a\delta\tau = Alpha$ where $Alpha$ is the Fine Structure Constant yields for an approximate choice $a\delta\tau \sim 1 + 137.0359990368270075578^{-1}$

$(1 + 137.0359990368270075578^{-1})^{95} \cong$ 1.99517405768451506744077268406650665 . Indeed very close to 2.

Is $1 + 137.0359990368270075578^{-1}$ the minimal positive acceleration or there are other choices, depending on the field strength? What could be the values of $A^{9120}$ or $a\delta\tau^{9120}$?

**Analysis of $\xi \cong 1.5561985371903483965638770314399$**

Consider the following:

$$(\xi^{\frac{1}{9120}} - 1)96\xi\alpha^{-1} \tag{46.7}$$

Such that $\alpha^{-1} \cong 137.0359990368270075578$

With $\xi \cong 1.5561985371903483965638770314399$

$$S = (\xi^{\frac{1}{9120}} - 1)96\xi\alpha^{-1} \cong 0.99277429348500763011970583955727 \;\approx\; 1$$

Checking the residual,

$$J = (S^{-1} - 1)^{-1} \cong 138.03338778624051385484569471361$$

$$R = \left(\frac{J}{\alpha^{-1}} - 1\right)^{-1} \cong 137.39477121374005441058563601124$$

Continuing the residual calculation,

$$W = \left(\frac{R}{\alpha^{-1}} - 1\right)^{-1} \cong 381.95826726563437220242441367347$$

It is important to note that 384 = 96 * 4 and that 95.5 * 4 = 382.

$$T = \left(\frac{382}{W} - 1\right)^{-1} \cong 9152.4860058109557729045419653489$$

Surprisingly $\sqrt{T} \cong 95.668626026566074923030595362229$ which is very close to $((g_1 - 1)(1 - g2))^{-\frac{1}{2}} \cong 95.637054262686888250755146145820617675 78125$ in (34), (35) in the paper.

**Analysis of $\xi = 2$**

Consider $\xi = 2$ as an upper limit.

$$S = (\xi^{\frac{1}{9120}} - 1)96\xi\alpha^{-1} \cong 1.99978370680015988811348859554 91 \approx 2 \quad (46.8)$$

$$R = \left(\frac{2}{S} - 1\right)^{-1} \cong 9245.70771655529925480997965809 67$$

Now consider

$$J = \left(\frac{R}{96^2} - 1\right)^{-1} \cong (\frac{9245.7077165552992548099796580967}{9216} - 1)^{-1}$$

$J \cong 310.2224293423875521373176068660 7$

Now this residual number is also interesting,

$$2 * (log_2(J))^2 \cong 137.0227284464476667414899823607 3 \quad (46.8.1)$$

Which is very close to the inverse Fine Structure Constant. Unlike the other residual numbers which did require any effort from the author to reach, $2 * (log_2(J))^2$ is not trivial and therefore the significance of $2 * (log_2(J))^2$ is not clear.

Going back to (46.7), with $\xi \cong 1.5561985371903483965638770314399$ consider

$$\mathrm{R} = \ln(\zeta)/\ln(1 + \alpha * 96^{-1} * \zeta^{-1}) \cong \quad (46.8.2)$$

$$9054.103154327295460220465932271 6 \approx 9025 = 95^2$$

$$J = (\frac{\mathrm{R}}{95^2} - 1)^{-1} \cong 310.1038429891282571207972593295 3$$

$$2 * (log_2(J))^2 \cong 137.0044665486939751482933527490 4$$

The same term $2 * (log_2(J))^2$ as in (46.8.1) leads to the approximation of the inverse Fine Structure Constant. Residual calculations lead to a deep theory between the field strength coefficients and the Fine Structure Constant.

## Analysis of $\xi = \frac{4}{\pi}$

Unlike the previous two cases, where $(\xi^{\frac{1}{9120}} - 1)96\xi\alpha^{-1}$ yielded interesting residuals, for the upper limit $\xi = 2$ and for $\xi \cong 1.5561985371903483965638770314399$ the term that works for $\xi = \frac{4}{\pi}$ is

$$S = (\xi^{\frac{1}{9120}} - 1)(2 * \alpha^{-2}) \cong 0.9948172709426247156245398797525 \cong 1 \quad (46.9)$$

$(S^{-1} - 1)^{-1} \cong$ 191.94853906687436314121512871427 ≈ 192 = 2 * 96.

**Analysis of $\xi = \frac{95}{96}$**

$$\text{In this case there is no } (\xi^{\frac{1}{9120}} - 1) < 1. \text{ In fact, } (1 - \xi^{\frac{1}{9120}}) * 9120^{\frac{3}{2}} \approx 1 \quad (46.10)$$

The author is aware of one's own cognitive limits and as a work principle, tries to avoid any metaphysical assertion, however, to be brutally honest, as required by every scientist, the residual results, whether in relation to $1, 2, 96^2, 95^2, 2 * 96, 4 * 95.5$, and the inverse Fines Structure Constant, appear as Intelligent Design!

**8. Interesting acceleration to radius coefficients relation – the field strength coefficients**

Consider the coefficients $\frac{95}{96}$, from (23) $\frac{4}{\pi}$, from (24) and $\xi =$ 1.5561985371903483965638770314399152994155 88 from (34), (35). Note the following table

**Table 2.**

| $\xi$ of Electron, Muon, Tau | $\xi\left(\frac{4}{\pi}\right)^{-1}$ | $\xi \cdot 9 \cdot \left(\frac{4}{\pi}\right)^{-1}$ |
|---|---|---|
| $\frac{95}{96}$ | 0.7772169325287248897$\sim\frac{7}{9}$ | 6.994952392758524 |
| $\frac{4}{\pi}$ | $1 = \frac{9}{9}$ | 9 |
| 1.556198537190348396563877031439 9 | 1.22223547299109529$\sim\frac{11}{9}$ | 11.0001192569 |

The numbers in the left column suggest that the field strength coefficients are related to natural numbers. Models of natural numbers occur for example in Heisenberg's XXZ model of spin chains. The normalization factor for L=4 and $\Delta = 1$ in such a spin chain is remarkably close to $\frac{4}{\pi}$, and remarkably close to 1.556198537190348… with L=11 and $\Delta = 1$. The normalization factor is close to $\frac{95}{96}$ when L=13 and $\Delta = 0$. Here the model is not of any spin chain and spin chains are only brought as an example of how natural numbers can be related to numbers such as this model's field strength coefficients.

The reader can check that 7 * 9 * 11 = 693 is the integer floor of $\left(\frac{1}{96^2(a-1)(1-b)} - 1\right)^{-1} \cong$ 693.634239847 , see the note after (40) with $2 - \frac{1}{96^2(a-1)(1-b)}$. We have yet to show more compelling evidence the choice of $\xi$ is not by chance. Some readers will remain skeptical no matter what evidence is brought in this paper. This section is not meant for such readers but for readers who agree that serendipity is important for new discoveries in physics. The coefficient $\frac{4}{\pi}$ from (22), (80) is well understood [32], however, $\frac{95}{96}$ from (23), (79), (86) and

1.5561985371903483965638770314399 from the solution to (35) are not well understood. Evidence, except from the previous table and the note after (40), can be found if we look at the polynomial term that means loss or addition of area in relation to 4 times the area of a disk. The factor 4 was thoroughly discussed before (16) and led to the number 96 from $\frac{\frac{1}{4}*\frac{\pi}{24}*R(3)*r^4}{\pi r^2} = \frac{R(3)*r^2}{96}$ where R(3) is obtained by double contraction of the Einstein tensor with a direction of time. We return to (18), $\left(-\frac{1}{2}\frac{\xi^2}{x^2} \mp \frac{\xi}{x}\right)\frac{1}{96} = \frac{\delta Area}{4\pi r^2}$ and consider the following polynomials $\left(-\frac{1}{2}\xi^2 \mp \xi\right)\frac{1}{96}$ of the field strength coefficient $\xi$. Like before in (23), we consider the terms $\frac{1}{a-1}$ and $\frac{1}{1-b}$ from the biggest and stable roots a, b. Not too surprisingly, these terms are approximated by $\alpha = p1(\xi) = \left(\left(-\frac{1}{2}\xi^2 + \xi\right)\frac{1}{96}\right)^{-1}$ and $\beta = p2(\xi) = \left(\left(-\frac{1}{2}\xi^2 - \xi\right)\frac{1}{96}\right)^{-1}$ which only depend on the field strength coefficients. Consider the following relative error terms $RatioA = (\frac{a-1}{\alpha} - 1)^{-1}$ and $RatioB = (\frac{1-b}{\beta} - 1)^{-1}$or as an output of a python code:

```
Field strength coefficient analysis:

Xi=0.9895833333333334, p1=192.02083559413998, p2=64.89902805794975

Xi=0.9895833333333334, 1/(a-1)=192.04639436012951, 1/(1-b)=63.54135822920768

Xi=0.9895833333333334, RatioA=-7513.91496909199486, RatioB=46.80177527998884

--------------------------------------------------

Xi=1.2732395447351628, p1=207.49126659259227, p2=46.06948110927548

Xi=1.2732395447351628, 1/(a-1)=206.75133988502202, 1/(1-b)=44.63955017596401

Xi=1.2732395447351628, RatioA=279.42137750905704, RatioB=31.21797643232097

--------------------------------------------------

Xi=1.5561985371903484, p1=278.00172875202145, p2=34.69366870085835

Xi=1.5561985371903484, 1/(a-1)=275.51690891864394, 1/(1-b)=33.19740405023536

Xi=1.5561985371903484, RatioA=110.88003452715024, RatioB=22.18685313217340

--------------------------------------------------
```

This output shows proximities between functions of the field strength coefficient, $\xi$ or Xi in the Python output. The proximities are p1 of the next $\xi$ to RatioA and p2 of the next $\xi$ to RatioB. The first value ~-7513.91496909199486 is in red as an exception because it is not matched to the value of p1 for the next field strength coefficient $\xi = \frac{4}{\pi}$. Following is the code in Python that was used for the last calculations,

```
import numpy as NP

def function_p(p_x):

    return (-0.5 * p_x * p_x + p_x)/96, -(-0.5 * p_x * p_x - p_x)/96

def function_cubic_viete(a, b, c, d): # If all roots are real.

     # Viete's formula when all roots are real.

     b2 = NP.longdouble(b * b)

     b3 = NP.longdouble(b2 * b)

     a2 = NP.longdouble(a * a)

     a3 = a2 * a

     p = (3 * a * c – b2) / (3 * a2)

     q = (2 * b3 – 9 * a * b * c + 27 * a2 * d) / (27 * a3)

     offset = b / (3 * a)

     t1 = 2 * NP.sqrt(-p / 3) * NP.cos(NP.arccos(NP.sqrt(-3 / p) \

                                                 * (3 * q) / (2 * p)) / 3)

     t2 = 2 * NP.sqrt(-p / 3) * NP.cos(NP.arccos(NP.sqrt(-3 / p) * \

                                                 (3 * q) / (2 * p)) / 3 - NP.pi / 3)

     t3 = 2 * NP.sqrt(-p / 3) * NP.cos(NP.arccos(NP.sqrt(-3 / p) * \

                                                 (3 * q) / (2 * p)) / 3 - 2 * NP.pi /
3)

     x1 = t1 - offset

     x2 = t2 - offset

     x3 = t3 - offset

     return (x1, x2, x3)

ma_list = [95/96, 4/NP.pi, 1.5561985371903484]

print('Field strength coefficient analysis:')
```

```
for ma_x in ma_list:
    ma_tuple = function_p(ma_x)

    ma_a,_,_ = function_cubic_viete(1, -1, -ma_x / 96,
                                    (ma_x * ma_x) / 192)

    ma_b,_,_ = function_cubic_viete(1, -1, ma_x / 96,
                                    (ma_x * ma_x) / 192)

    print('Xi={}, p1={:.14f}, p2={:.14f}'.format(ma_x, 1/ma_tuple[0], 1/ma_tuple[1]))
    print('Xi={}, 1/(a-1)={:.14f}, 1/(1-b)={:.14f}'.format(ma_x, 1/(ma_a-1), 1/(1-
ma_b)))

    ma_a = (ma_a - 1) / ma_tuple[0]
    ma_b = (1 - ma_b) / ma_tuple[1]
    ma_a = 1 / (ma_a - 1)
    ma_b = 1 / (ma_b - 1)

    print('Xi={}, RatioA={:.14f}, RatioB={:.14f}'.format(ma_x, ma_a, ma_b))
    print('-----------------------------------------------------')
```

We now return to the field which is smaller than 1, namely to $\xi = \frac{95}{96}$. It is easy to see that if we pick $\xi_1 = \frac{193}{192} = 1 + \frac{1}{192}$ and $\xi_2 = \frac{63}{64} = 1 - \frac{1}{64}$ we get rational roots for the following anti-gravity equation $x_1^2 + \frac{1}{96}\xi_1 x_1 - \frac{1}{192}\xi_1^2 = x_1^3$ and gravity equation $x_2^2 - \frac{1}{96}\xi_2 x_2 - \frac{1}{192}\xi_2^2 = x_2^3$ for which $x_1 = \xi_1$ and $x_2 = \xi_2$, interestingly $1 + \frac{\xi_2 - \xi_1}{2} = \frac{95}{96}$ and $\frac{\xi_1 - \xi_2}{2} = \frac{1}{96}$.

Some nice relation between the roots of gravity and anti-gravity of area ratio polynomials with field strength coefficients $\xi = \frac{95}{96}$ and $\xi \cong 1.5561985371903483965638770314399$ as in (35) is considered. We saw that for $\xi \cong 1.5561985371903483965638770314399$ the following holds: $\frac{2(1-x_2)}{(x_1-1)^{\frac{1}{2}}} = 1$. There is another relation that is not less illuminating, $\frac{4(1-x_2)^{\frac{1}{2}}}{x_1-1}$. With low accuracy of a simple datasheet we can see that for $\xi = \frac{95}{96}$ we get $\frac{4(1-x_2)^{\frac{1}{2}}}{x_1-1} \cong 96.36912199$ and for $\xi \cong 1.5561985371903483965638770314399$ as in (35), we get $\frac{4(1-x_2)^{\frac{1}{2}}}{x_1-1} \cong 191.2741085$ which is almost 192=2*96. Multiplying these two values together we have $96.36912199 \ldots * 191.2741085 \ldots \cong 18432.9179 \approx 18432 = 2 * 96^2$, and we can see $(\frac{18432.9179}{2})^{\frac{1}{2}} \cong 96.00239033$.

## Conclusion

The presented model predicts gravity not only by the energy of electric charge but also by the electric charge itself. It offers a technological breakthrough by generating gravitational dipoles and it offers mass ratios between particles that are not accessible through the Standard Model. (33) and (65) can only be interpreted as the existence of a fifth force of Nature with symmetry SU(4), or by (3.12) is related to gravity, while (24) results in a new neutrally charged particle of energy ~41.8752442118608 eV and (39.12) seems to set a strict electron neutrino mass bound of ~0.033eV/c2. The muon field strength coefficient is different from the electron's and Tauon field strength, which implies different physics. (33) indicates a deep relation between leptons and hadrons and especially between the Muon and the Bottom Quark. An energy of 1.40 MeV should be searched for in neutrinos and antineutrinos in processes in which bottom quarks are generated and decay. A resonance of 23.573 MeV should exist in processes in which a Tau lepton is created.

As for the theoretical approach that this paper took, not any Gauge fields are a blessing. There was a big expectation from Albert Einstein that the Palatini action, which is identical to Einstein-Hilbert action, would be a great insight into Quantum Gravity, especially since spinor equations require tetrads because they are limited to an orthogonal reference frame. However, this paper took a very different approach, to leave the metric tensor as is and instead of using tetrads or Ashtekar variables, to consider the metric as of a reference manifold, like coordinates but as an entire geometric reference object, not as a physically accessible object. Then in this framework, the idea was that time must be the engine of the model and that acceleration of that time in the sense of a generalized Reeb class field - not limited to contact manifolds - will describe the possibility of non-geodesic curves and will predict the electric force. In (64.b), (64.01) it becomes an electro-weak-strong action, using 5 fields, but unlike tetrads, time is a meaningful Geroch function, while the other fields are Gauge fields. There is a redundancy in the system because this time can be accounted for by 3 vectors just as Ashtekar variables. This redundancy is cancelled out in action (64.b), (64.01), instead of using an ADM formalism or Ashtekar variables, and orthogonality is no longer needed, which renders spin connections redundant. 4 out of the 5 scalar fields describe additional geometric information to the metric as foliations. The same theory can be written with tetrads and generalized Reeb class vectors of these tetrad fields, but the Einstein-Hilbert action will be the same. On the other hand, action (64.b), (64.01), in this case, does add geometric information as non-geodesic alignment of curves and thus of forces. It is a far simpler approach than that of Abhay Ashtekar and it achieves new results. Adding a summation constraint to the action of (64.b), (64.01), e.g. that each chronon probability sums to 1, keeps the same action but then PP* is replaced by an event function and the integration of PP* becomes 1. That requires the only constant in the theory except for the speed of light to be with the units of $Length^{-2}$ and then the action is defined *almost everywhere* in terms of measure theory.

**Appendix A: Euler Lagrange minimum action equations**

We assume $\sigma = 8\pi$ (from the previously discussed term, $-a_\mu a^\mu / 8\pi K$ as an energy density).

$$Z = N^2 = P_\mu P^\mu \text{ and } U_\lambda = \frac{Z_\lambda}{Z} - \frac{Z_k P^k P_\lambda}{Z^2} \text{ and } L = \frac{1}{4} U^k U_k$$

$$R = Ricci \text{ curvature.}$$

$$Min\ Action = Min \int_\Omega \left( R - \frac{8\pi}{\sigma} L \right) \sqrt{-g} d\Omega =$$

$$Min \int_\Omega \left( R - \frac{1}{4} U^k U_k \right) . \sqrt{-g} d\Omega \text{ s.t. } \sigma = 8\pi \tag{47}$$

The variation of the Ricci scalar is well known. It uses the Platini identity and Stokes theorem to calculate the variation of the Ricci curvature and reaches the Einstein tensor [44], as follows,

$\delta R = R_{\mu\nu} \delta g^{\mu\nu}$ and $\delta \sqrt{-g} = -\frac{1}{2} g_{\mu\nu} \delta g^{\mu\nu} \sqrt{-g}$ by which we infer $\delta(R\sqrt{-g}) = (R_{\mu\nu} - \frac{1}{2} R g_{\mu\nu}) \delta g^{\mu\nu}$ which will be later added to the variation of $\left( (R - \frac{1}{4} U^j U_j) \right) \sqrt{-g}$ by $\delta g^{\mu\nu}$. The following Euler Lagrange equations must hold,

$$\frac{\partial}{\partial g^{\mu\nu}} - \frac{d}{dx^m} \frac{\partial}{\partial g^{\mu\nu}{}_{,m}} + \frac{d^2}{dx^m dx^s} \frac{\partial}{\partial g^{\mu\nu}{}_{,m,s}} \left( (R - \frac{1}{4} U^j U_j) \sqrt{-g} \right) = 0,$$

$$\frac{\partial}{\partial P} - \frac{d}{dx^m} \frac{\partial}{\partial P_{,m}} + \frac{d^2}{dx^m dx^s} \frac{\partial}{\partial P_{,m,s}} \left( (R - \frac{1}{4} U^j U_j) \sqrt{-g} \right) = 0$$

$U^k U_k = \frac{Z_\mu Z^\mu}{Z^2} - \frac{(Z_s P^s)^2}{Z^3}$ which we obtain from the minimum Euler Lagrange equation because $U_\lambda P^\lambda = \frac{Z_\lambda P^\lambda}{Z} - \frac{Z_k P^k P_\lambda P^\lambda}{Z^2} = 0$. In order to calculate the minimum action Euler-Lagrange equations, we will separately treat the Lagrangians, $L = \frac{Z_\mu Z^\mu}{Z^2}$ and $L = \frac{(Z_s P^s)^2}{Z^3}$ to derive the Euler Lagrange equations of the Lagrangian $L = \frac{Z_\mu Z^\mu}{Z^2} - \frac{(Z_s P^s)^2}{Z^3} = U_\mu U^\mu$. The Euler Lagrange operator of the Ricci scalar $\left( \frac{\partial}{\partial g^{\mu\nu}} - \frac{d}{dx^m} \frac{\partial}{\partial (g^{\mu\nu}{}_{,m})} + \frac{d^2}{dx^m dx^s} \frac{\partial}{\partial (g^{\mu\nu}{}_{,m,s})} \right.$.

The reader may skip the following equations up to equation (53). Equations (53), (54) and (55) are, however, crucial. Note: the relation $\frac{d}{dx^{\nu}}\sqrt{|g|} = \Gamma^{\lambda}_{\lambda\nu}\sqrt{|g|}$ is used in the next equations.

$$L = \frac{\left(P_{\lambda}Z^{\lambda}\right)^{2}}{Z^{3}} \; s.t. \; Z = P_{\mu}P^{\mu} \; and \; Z_{s} \equiv Z_{,s} = \frac{dZ}{dx^{s}}$$

$$\begin{aligned}
&\frac{\partial(L\sqrt{-g}}{\partial g^{\mu\nu}} - \frac{d}{dx^{m}}\frac{\partial(L\sqrt{-g}\,)}{\partial g^{\mu\nu}{}_{,m}} \\
&= \Bigg(-2\left(\frac{Z_{,s}\,P^{s}}{Z^{3}}P_{\mu}P_{\nu}P^{m}\right)_{;m} + 2\left(\frac{Z_{,s}\,P^{s}}{Z^{3}}\right)\left(\Gamma^{\mathrm{i}}_{\mathrm{\mu m}}P_{i}P_{\nu}P^{m} + \Gamma^{\mathrm{i}}_{\mathrm{\nu m}}P_{\mu}P_{i}P^{m}\right) \\
&+ 2\left(\frac{Z_{,s}\,P^{s}}{Z^{3}}\right)\left(P_{\mu}P_{\nu}\right)_{;m}P^{m} - 2\left(\frac{Z_{,s}\,P^{s}}{Z^{3}}\right)\left(\Gamma^{\mathrm{i}}_{\mathrm{\mu m}}P_{i}P_{\nu}P^{m} + \Gamma^{\mathrm{i}}_{\mathrm{\nu m}}P_{\mu}P_{i}P^{m}\right) \\
&+ 2\left(\frac{Z_{,s}\,P^{s}}{Z^{3}}\right)Z_{\mu}P_{\nu} - 3\frac{(Z_{,s}\,P^{s})^{2}}{Z^{4}}P_{\mu}P_{\nu} - \frac{1}{2}\frac{(Z_{,s}\,P^{s})^{2}}{Z^{3}}g_{\mu\nu}\Bigg)\sqrt{-g} =
\end{aligned}$$

$$\left(-2\left(\frac{Z_{,s}P^{s}}{Z^{3}}P^{k}\right)_{;k}P_{\mu}P_{\nu} - 2\frac{(Z_{,s}P^{s})^{2}}{Z^{3}}\frac{P_{\mu}P_{\nu}}{Z} - \frac{(Z_{,s}P^{s})^{2}}{Z^{3}}\frac{P_{\mu}P_{\nu}}{Z} + 2\left(\frac{Z_{,s}P^{s}}{Z^{3}}\right)Z_{\mu}P_{\nu} - \frac{1}{2}\frac{(Z_{,s}P^{s})^{2}}{Z^{3}}g_{\mu\nu}\right)\sqrt{-g} \quad (48)$$

$$L = \frac{Z^{\lambda}Z_{\lambda}}{Z^{2}} \; s.t. \; Z = P_{\mu}P^{\mu}, s.t. \; Z = P_{\mu}P^{\mu} \; and \; Z_{s} \equiv Z_{,s} = \frac{dZ}{dx^{s}}$$

$$\begin{aligned}
&\frac{\partial(L\sqrt{-g}}{\partial g^{\mu\nu}} - \frac{d}{dx^{m}}\frac{\partial(L\sqrt{-g}\,)}{\partial g^{\mu\nu}{}_{,m}} = \left(-2\left(\frac{Z^{m}P_{\mu}P_{\nu}}{Z^{2}}\right)_{;m} + 2\frac{(\Gamma^{\mathrm{i}}_{\mathrm{\mu m}}\mathrm{P_{i}P_{\nu}Z^{m}} + \Gamma^{\mathrm{i}}_{\mathrm{\nu m}}P_{i}P_{\mu}Z^{m})}{Z^{2}} + 2\frac{(P_{\mu}P_{\nu})_{;m}Z^{m}}{Z^{2}} - \right. \\
&2\frac{(\Gamma^{\mathrm{i}}_{\mathrm{\mu m}}\mathrm{P_{i}P_{\nu}Z^{m}} + \Gamma^{\mathrm{i}}_{\mathrm{\nu m}}P_{i}P_{\mu}Z^{m})}{Z^{2}} + \frac{Z_{\mu}Z_{\nu}}{Z^{2}} - 2\frac{Z_{s}Z^{s}}{Z^{3}}P_{\mu}P_{\nu} - \left.\frac{1}{2}\frac{Z_{m}Z^{m}}{Z^{2}}g_{\mu\nu}\right)\sqrt{-g} = \left(-2\left(\frac{Z^{m}}{Z^{2}}\right)_{;m}P_{\mu}P_{\nu} - \right. \\
&\left. 2\frac{Z_{s}Z^{s}}{Z^{3}}P_{\mu}P_{\nu} - \frac{1}{2}\frac{Z_{m}Z^{m}}{Z^{2}}g_{\mu\nu} + \frac{Z_{\mu}Z_{\nu}}{Z^{2}}\right)\sqrt{-g} \qquad (49)
\end{aligned}$$

We subtract (48) from (49)

$$Z = P_{\mu}P^{\mu}, s.t. \; Z = P_{\mu}P^{\mu} \; and \; Z_{s} \equiv Z_{,s} = \frac{dZ}{dx^{s}}, U_{\lambda} = \frac{Z_{\lambda}}{Z} - \frac{Z_{k}P^{k}P_{\lambda}}{Z^{2}}, L = U^{k}U_{k} = \frac{Z_{\lambda}Z^{\lambda}}{Z^{2}} - \frac{(Z_{k}P^{k})^{2}}{Z^{3}}$$

$$\begin{aligned}
&\frac{\partial(L\sqrt{-g}}{\partial g^{\mu\nu}} - \frac{d}{dx^{m}}\frac{\partial(L\sqrt{-g}\,)}{\partial g^{\mu\nu}{}_{,m}} = \left(+2\left(\frac{Z_{m}P^{m}}{Z^{3}}P^{k}\right)_{;k}P_{\mu}P_{\nu} + 2\frac{(Z_{m}P^{m})^{2}}{Z^{3}}\frac{P_{\mu}P_{\nu}}{Z} - 2\frac{Z_{m}P^{m}}{Z^{3}}Z_{\mu}P_{\nu} + \frac{1}{2}\frac{(Z_{m}P^{m})^{2}}{Z^{3}}g_{\mu\nu} + \frac{(Z_{m}P^{m})^{2}}{Z^{3}}\frac{P_{\mu}P_{\nu}}{Z} + \right. \\
&\left.(-2\left(\frac{Z^{m}}{Z^{2}}\right)_{;m}P_{\mu}P_{\nu} - 2\frac{Z_{\lambda}Z^{\lambda}}{Z^{2}}\frac{P_{\mu}P_{\nu}}{Z} - \frac{1}{2}\frac{Z_{\lambda}Z^{\lambda}}{Z^{2}}g_{\mu\nu} + \frac{Z_{\mu}Z_{\nu}}{Z^{2}})\right)\sqrt{-g} = \left(\left(+2\left(\frac{Z_{m}P^{m}}{Z^{3}}P^{k}\right)_{;k} - 2\left(\frac{Z^{m}}{Z^{2}}\right)_{;m}\right)P_{\mu}P_{\nu} + \right. \\
&\left. 2\frac{\left(P^{\lambda}Z_{\lambda}\right)^{2}}{Z^{3}}\frac{P_{\mu}P_{\nu}}{z} - 2\frac{Z^{\lambda}Z_{\lambda}}{Z^{2}}\frac{P_{\mu}P_{\nu}}{Z} + \frac{1}{2}\frac{\left(P^{\lambda}Z_{\lambda}\right)^{2}}{Z^{3}}g_{\mu\nu} - \frac{1}{2}\frac{Z_{k}Z^{k}}{Z^{2}}g_{\mu\nu} + \frac{Z_{\mu}Z_{\nu}}{Z^{2}} - 2\left(\frac{Z_{s}P^{s}}{Z^{3}}\right)Z_{\mu}P_{\nu} + \frac{\left(P^{\lambda}Z_{\lambda}\right)^{2}}{Z^{3}}\frac{P_{\mu}P_{\nu}}{Z}\right)\sqrt{-g} = \\
&\left(\left(+2\left(\frac{Z_{m}P^{m}}{Z^{3}}P^{k}\right)_{;k} - 2\left(\frac{Z^{m}}{Z^{2}}\right)_{;m}\right)P_{\mu}P_{\nu} + 2\frac{\left(P^{\lambda}Z_{\lambda}\right)^{2}}{Z^{3}}\frac{P_{\mu}P_{\nu}}{z} - 2\frac{Z^{\lambda}Z_{\lambda}}{Z^{2}}\frac{P_{\mu}P_{\nu}}{z} + U_{\mu}U_{\nu} - \frac{1}{2}U^{\lambda}U_{\lambda}g_{\mu\nu}\right)\sqrt{-g} =
\end{aligned}$$

$$\left(U_{\mu}U_{\nu} - \frac{1}{2}U^{\lambda}U_{\lambda}g_{\mu\nu} - 2U^{k}{}_{;k}\frac{P_{\mu}P_{\nu}}{z}\right)\sqrt{-\mathrm{g}} \qquad (50)$$

$$
\begin{array}{l}
L = \dfrac{(Z^{s} P_{s})^{2}}{Z^{3}} \quad s.t.\ Z = P^{\lambda} P_{\lambda}\ \ and\ \ Z_{m} = (P^{\lambda} P_{\lambda})_{,m} \\
\dfrac{\partial (L\sqrt{-g})}{\partial P_{\mu}} - \dfrac{d}{dx^{\nu}} \dfrac{\partial (L\sqrt{-g})}{\partial P_{\mu},_{\nu}} = \\
\left( \begin{array}{l}
-4\left(\dfrac{(Z_{s} P^{s})}{Z^{3}} P^{\mu} P^{\nu}\right);_{\nu} + 4\dfrac{(Z_{s} P^{s})}{Z^{3}} \Gamma_{i}{}^{\mu}{}_{\nu} P^{i} P^{\nu} + \\
+4\dfrac{(Z_{s} P^{s})}{Z^{3}} P^{\mu};_{\nu} P^{\nu} - 4\dfrac{(Z_{s} P^{s})}{Z^{3}} \Gamma_{i}{}^{\mu}{}_{k} P^{i} P^{k} + \\
+2\dfrac{Z_{m} P^{m} Z^{\mu}}{Z^{3}} - 6\dfrac{(Z_{m} P^{m})^{2}}{Z^{4}} P^{\mu}
\end{array} \right) \sqrt{-g} = \\
\left( -4\left(\dfrac{(Z_{s} P^{s}) P^{\nu}}{Z^{3}}\right);_{\nu} P^{\mu} + 2\dfrac{Z_{m} P^{m} Z^{\mu}}{Z^{3}} - 6\dfrac{(Z_{m} P^{m})^{2}}{Z^{4}} P^{\mu} \right) \sqrt{-g}
\end{array}
\tag{51}
$$

$$
\begin{array}{l}
L = \dfrac{Z^{s} Z_{s}}{Z^{2}} \quad s.t.\ Z = P^{\lambda} P_{\lambda}\ \ and\ \ Z_{m} = (P^{\lambda} P_{\lambda})_{,m} \\
\dfrac{\partial (L\sqrt{-g})}{\partial P_{\mu}} - \dfrac{d}{dx^{\nu}} \dfrac{\partial (L\sqrt{-g})}{\partial P_{\mu},_{\nu}} = \\
\left( \begin{array}{l}
-4\left(\dfrac{P^{\mu} Z^{\nu}}{Z^{2}}\right);_{\nu} + \dfrac{4}{Z^{2}} \Gamma_{i}{}^{\mu}{}_{k} P^{i} Z^{k} + \\
+\dfrac{4}{Z^{2}} P^{\mu};_{\nu} Z^{\nu} - \dfrac{4}{Z^{2}} \Gamma_{i}{}^{\mu}{}_{k} P^{i} Z^{k} + \\
-4\dfrac{Z_{m} Z^{m}}{Z^{3}} P^{\mu} \sqrt{-g}
\end{array} \right) \sqrt{-g} = \\
\left( -4\left(\dfrac{Z^{\nu}}{Z^{2}}\right);_{\nu} - 4\dfrac{Z_{m} Z^{m}}{Z^{3}} \right) P^{\mu} \sqrt{-g}
\end{array}
\tag{52}
$$

We subtracted the Euler Lagrange operators of $\frac{(Z^{s}P_{s})^{2}}{Z^{3}}\sqrt{-g}$ in (48) from the Euler Lagrange operators of $\frac{Z^{\lambda}Z_{\lambda}}{Z^{2}}\sqrt{-g}$ in (49) and got (50) and we will subtract (51) from (52) to get two tensor equations of gravity, these will be (53), and (55). Assuming $\sigma = 8\pi$, where the metric variation equations (47), (48), (49) and (50) yield

$$
\begin{array}{l}
\mathrm{Z} = N^{2} = \mathrm{P}_{\mu}\mathrm{P}^{\mu}\ ,\ U_{\lambda} = \dfrac{Z_{\lambda}}{Z} - \dfrac{Z_{k} P^{k} P_{\lambda}}{Z^{2}}\ ,\ L = \dfrac{1}{4} U_{i} U^{i}\ \ \text{and}\ \mathrm{Z} = \mathrm{P}^{\mathrm{k}}\mathrm{P}_{\mathrm{k}} \\
\dfrac{8\pi}{\sigma} \dfrac{1}{4} \left( \begin{array}{l}
+2\left(\left(\dfrac{(\mathrm{P}^{\lambda}\mathrm{P}_{\lambda}),_{\mathrm{m}} P^{m}}{Z^{3}} P^{k}\right);_{k} - 2\left(\dfrac{Z^{m}}{Z^{2}}\right);_{m}\right) P_{\mu} P_{\nu} + \\
+2\dfrac{(P^{\lambda} Z_{\lambda})^{2}}{Z^{3}} \dfrac{P_{\mu} P_{\nu}}{Z} - 2\dfrac{Z^{\lambda} Z_{\lambda}}{Z^{2}} \dfrac{P_{\mu} P_{\nu}}{Z} + \\
+U_{\mu} U_{\nu} - \dfrac{1}{2} U_{\mathrm{k}} U^{k} g_{\mu\nu}
\end{array} \right) = \\
\dfrac{8\pi}{\sigma} \dfrac{1}{4} \left( U_{\mu} U_{\nu} - \dfrac{1}{2} U_{\mathrm{k}} U^{k} g_{\mu\nu} - 2U^{k};_{k} \dfrac{P_{\mu} P_{\nu}}{Z} \right) = R_{\mu\nu} - \dfrac{1}{2} R g_{\mu\nu} \\
\text{s.t.}\ \ R = R_{\mu\upsilon} g^{\mu\upsilon} \\
\text{s.t.}\ R_{kj} = (\Gamma_{jk}{}^{P}),_{p} - (\Gamma_{pk}{}^{P}),_{j} + \Gamma_{p\mu}{}^{P} \Gamma_{jk}{}^{\mu} - \Gamma_{pj}{}^{\mu} \Gamma_{k\mu}{}^{p}
\end{array}
\tag{53}
$$

$R_{\mu\nu}$ is the Ricci tensor and $R_{\mu\nu} - \frac{1}{2} R g_{\mu\nu}$ is the Einstein tensor [44]. In general, by (4) and $\sigma = 8\pi$, (53) can be written in $(-1, +1, +1, +1)$ metric convention, so $Z = |P_\mu P^\mu|$ as,

$$\frac{1}{4}(U_\mu U_\nu - \frac{1}{2} U_{\mathrm{k}} U^k g_{\mu\nu} - 2U^k;_k \frac{P_\mu P_\nu}{Z}) = R_{\mu\nu} - \frac{1}{2} R g_{\mu\nu} \tag{54}$$

**Charge-less field:** The term $-2U^k;_k \dfrac{P_\mu P_\nu}{Z}$ in (54) can be generalized to:

$-2((U^k;_k + U^{*k};_k)/2) \dfrac{(P_\mu P^*_\nu + P^*_\mu P_\nu)/2}{Z}$ and can be zero under the following condition,

$$4(A_{\mu\nu};^\mu \frac{P^{*\nu}}{\sqrt{Z}} + A^*_{\mu\nu};^\mu \frac{P^\nu}{\sqrt{Z}}) = U_\mu U^{*\mu} + U^*_\mu U^\mu \Rightarrow U^k;_k + U^{*k};_k = 0$$

**Note:** The complimentary matrix $B_{\mu\nu} = \frac{1}{\sqrt{2}} E^{\mu\nu\alpha\beta} A_{\alpha\beta}$, see few lines before (3), can be transformed to a real matrix due to the SU(2) x U(1) degrees of freedom and also be imaginary.

From (51), (52) we have, $\dfrac{d}{dx^\mu}(\dfrac{\partial}{\partial P_\mu} - \dfrac{d}{dx^\nu}\dfrac{\partial}{\partial P_{\mu,\nu}})(U_k U^k \sqrt{-g}) = W^\mu;_\mu \sqrt{-g} = 0$

We recall, $W^\mu = (\dfrac{\partial}{\partial P_\mu} - \dfrac{d}{dx^\nu}\dfrac{\partial}{\partial P_{\mu,\nu}})(U_k U^k \sqrt{-g})$

$$\begin{aligned}
&W^\mu = \\
&(-4(\frac{Z^\nu}{Z^2});_\nu - 4\frac{Z_{\mathrm{m}} Z^m}{Z^3})\mathrm{P}^\mu + 4(\frac{(Z_s P^s)P^\nu}{Z^3});_\nu P^\mu - 2\frac{Z_{\mathrm{m}} P^m Z^\mu}{Z^3} + 6\frac{(Z_m P^m)^2}{Z^4} P^\mu = \\
&-4(\frac{Z^\nu}{Z^2});_\nu \mathrm{P}^\mu - 4\frac{Z_{\mathrm{m}} Z^m}{Z^3}\mathrm{P}^\mu + \\
&+4(\frac{(Z_s P^s)P^\nu}{Z^3});_\nu P^\mu + 4\frac{(Z_m P^m)^2}{Z^4} P^\mu \\
&-2\frac{Z_{\mathrm{m}} P^m}{Z^2}(\frac{Z^\mu}{Z} - \frac{Z_m P^m P^\mu}{Z^2}) = \\
&-4((\frac{U^k}{Z});_k + \frac{U^k U_k}{Z})P^\mu - 2\frac{Z_{\mathrm{m}} P^m}{Z^2} U^\mu = 0
\end{aligned}$$

$$W^\mu;_\mu = \left(-4U^\nu;_\nu \frac{P^\mu}{Z} - 2\frac{(Z_m P^m)}{Z^2} U^\mu\right);_\mu = 0 \tag{55}$$

**Appendix B: Proof of conservation**

**Theorem**: Conservation law of the real Reeb class vector.

From the vanishing of the divergence of Einstein tensor and (54), we have to prove the following in $(-1,+1,+1,+1)$ metric convention:

$$\frac{1}{4}\left(U_\mu U_\nu - \frac{1}{2}U_k U^k g_{\mu\nu} - 2U^k;_k \frac{P_\mu P_\nu}{Z}\right);^\mu = G_{\mu\nu};^\mu = (R_{\mu\nu} - \frac{1}{2}Rg_{\mu\nu});^\mu = 0 \tag{56}$$

**Proof:**

From the zero variation by the scalar time field (55)

$$W^\mu;_\mu = \left(-4U^\nu;_\nu \frac{P^\mu}{Z} - 2\frac{(Z_m P^m)}{Z^2}U^\mu\right);_\mu = 0 \tag{57}$$

$$-\left(2U^\nu;_\nu \frac{P^\mu}{Z}\right);_\mu = \left(\frac{(Z_m P^m)}{Z^2}U^\mu\right);_\mu \tag{58}$$

$$\left(-2U^k;_k \frac{P^\mu P^\nu}{Z}\right);_\mu = \left(\frac{(Z_m P^m)}{Z^2}U^\mu\right);_\mu P^\nu - \left(2U^k;_k \frac{P^\mu}{Z}\right)P^\nu;_\mu = \left(\frac{(Z_m P^m)}{Z^2}U^\mu\right);_\mu P^\nu - U^k;_k \frac{Z^\nu}{Z} \tag{59}$$

Now let $t \equiv Z_m P^m$

And we have $\left(\frac{t}{Z^2}U^\mu\right);_\mu P^\nu - U^k;_k \frac{Z^\nu}{Z} = (\frac{t}{Z^2});_\mu U^\mu P^\nu + \frac{t}{Z^2}U^\mu;_\mu P^\nu - U^k;_k \frac{Z^\nu}{Z} =$

$$-U^\mu;_\mu U^\nu + (\frac{t}{Z^2});_\mu U^\mu P^\nu$$

This is because $-U^\nu = -\frac{Z^\nu}{Z} + \frac{t}{Z^2}P^\nu \Rightarrow -U^\mu;_\mu \frac{Z^\nu}{Z} + \frac{t}{Z^2}U^\mu;_\mu P^\nu = -U^\mu;_\mu U^\nu$. Note that $-U^\nu$ is minus twice the real numbered Reeb class vector. So,

$$(-2U^k;_k \frac{P^\mu P^\nu}{Z});_\mu = -U^\mu;_\mu U^\nu + (\frac{t}{Z^2});_\mu U^\mu P^\nu \tag{60}$$

Returning to the theorem we have to prove and using equation (60), we have to show,

$$\left(U^{\mu}U^{\nu}-\frac{1}{2}U_{k}U^{k}g^{\mu\nu}-2U^{k};_{k}\frac{P^{\mu}P^{\nu}}{Z}\right);^{\mu}=$$

$$U^{\mu};_{\mu}U^{\nu}+U^{\mu}U^{\nu};_{\mu}-\frac{1}{2}(U_{k};_{\mu}U_{s}+U_{k}U_{s};_{\mu})g^{ks}g^{\mu\nu}-$$

$$U^{\mu};_{\mu}U^{\nu}+(\frac{t}{Z^{2}});_{\mu}U^{\mu}P^{\nu}=$$

$$U^{\mu}U^{\nu};_{\mu}-\frac{1}{2}(U^{s}U_{s});^{\nu}+(\frac{t}{Z^{2}});_{\mu}U^{\mu}P^{\nu}=0 \tag{61}$$

Notice that

$$U^{\mu}U^{\nu};_{\mu}-\frac{1}{2}U^{s}U_{s};^{\nu}=$$

$$U^{\mu}\left((\frac{Z_{k}}{Z});_{\mu}-(\frac{t}{Z^{2}});_{\mu}P_{k}-(\frac{t}{Z^{2}})P_{k};_{\mu}\right)g^{k\nu}-$$

$$U^{s}\left((\frac{Z_{s}}{Z});_{k}-(\frac{t}{Z^{2}});_{k}P_{s}-(\frac{t}{Z^{2}})P_{s};_{k}\right)g^{k\nu}=$$

$$-U^{\mu}(\frac{t}{Z^{2}});_{\mu}P^{\nu} \tag{62}$$

Since $-(\frac{t}{Z^{2}});_{k}P_{s}U^{s}=0$ because the Reeb class vector is perpendicular to the foliation kernel $\frac{P_{\lambda}}{\sqrt{Z}}$, $\frac{P^{k}}{\sqrt{Z}}\frac{U_{k}}{2}=0$.

Equation (62) is also a result of $\ln(Z),_{k};_{\mu}U^{\mu}g^{k\nu}=\ln(Z),_{s};_{k}U^{s}g^{k\nu}$ and of $P_{k};_{\mu}U^{\mu}g^{k\nu}=P_{s};_{k}U^{s}g^{k\nu}$.

$$U^{\mu}U^{\nu};_{\mu}-\frac{1}{2}(U^{s}U_{s});^{\nu}+(\frac{t}{Z^{2}});_{\mu}U^{\mu}P^{\nu}=-U^{\mu}(\frac{t}{Z^{2}});_{\mu}P^{\nu}+(\frac{t}{Z^{2}});_{\mu}U^{\mu}P^{\nu}=0 \tag{63}$$

and we are done.

### Appendix C: Generalization to more than one generalized Reeb class vector

The idea of generalization of the Lagrangian to more than one acceleration is based on the idea that $\frac{P_{\lambda}}{\sqrt{Z}}=\frac{P_{\lambda}(0)}{\sqrt{Z(0)}}$ represents local time. Therefore, accelerations of unit vector fields should be perpendicular to $\frac{P_{\lambda}}{\sqrt{Z}}$, i.e. spacelike. This requirement does not align well with a Lagrangian which is based on embedding curvatures (64.2) - (64.12). You may read (64.2)-(64.12) first. In this appendix, a Lagrangian which is not based on embedding curvature will be developed and is compatible with the explanation of Fig. 2. The reference to all acceleration field will be $\frac{P_{\lambda}}{\sqrt{Z}}$ and by using Gram matrices, components of the acceleration fields that are parallel to $\frac{P_{\lambda}}{\sqrt{Z}}$ are nullified. This approach is contrary to the very definition of embedding curvature.

**Crucial:** This appendix is the continuation of "Adhering to the methodology of physics" that was discussed before the introduction. In fact The probability wave function $f = P(2)P(3)P(4)P(5)P(6)$ such that $ff^*$ is a probability density is indeed multiplicative if and only if $P(2),\ P(3),\ P(4),\ P(5),\ P(6)$ are independent functions. $ff^*$ then measures the probability of a reachable event in an embedding spacetime manifold.

**Crucial:** To reach a parsimonious theory, we may assume that $U(2)_\mu = U(3)_\mu = 0$ and that $\frac{\mathrm{P_k}}{\sqrt{\mathrm{Z}}}$ and $\frac{\mathrm{P(1)_k}}{\sqrt{\mathrm{Z}(1)}}$ are dependent as functions, and orthogonal as vectors. In his case we end up with 4 independent functions $P, P(1), P(4), P(5), P(6)$, which are suitable to include also gravity.

Considering (3.2.1), (3.2.2), $G_{\mu\nu};_\lambda \frac{P^\lambda}{\sqrt{Z}} = 0$ (13.11) which means other Reeb class vectors can rotate or be constant around $P^\lambda$ and given the previous fields $\frac{\mathrm{P(0)_k}}{\sqrt{\mathrm{Z}}} \equiv \frac{\mathrm{P_k}}{\sqrt{\mathrm{Z}}}$ and $\frac{U(0)_\mu}{2} \equiv \frac{U_\mu}{2}$ and additional Reeb class vector fields, not the usual Reeb vectors, $\frac{U(1)_\mu}{2}$ derived form $\frac{U(1)_\mu}{2} = \frac{\mathrm{Z}(1)_\mu}{2Z} - \frac{\mathrm{Z}(1)_\lambda P(1)^{*\lambda} P_\mu}{2z^2}$, i.e. $\frac{U(2)_\mu}{2}$ derived from $P(2)$, $\frac{U(3)_\mu}{2}$ derived from $P(3)$, $\frac{S_\mu}{2}$ derived from $P(4)$, $\frac{W_\mu}{2}$ derived from $P(5)$, $\frac{T_\mu}{2}$ derived from $P(6)$, such that these fields are perpendicular to $\frac{\mathrm{P_k}}{\sqrt{\mathrm{Z}}}$, the following Lagrangian can be defined with the determinant of the metric g:

$G(3) \equiv$

$$\equiv \begin{vmatrix} 1 & \frac{P(0)_\mu P(1)^{*\mu} + P(0)^*_\mu P(1)^\mu}{2\sqrt{Z(0)Z(1)}} & 0 \\ \frac{P(0)_\mu P(1)^{*\mu} + P(0)^*_\mu P(1)^\mu}{2\sqrt{Z(0)Z(1)}} & 1 & \frac{U_\mu(0)P(1)^{*\mu} + U^*_\lambda(0)P^\lambda(1)}{2\sqrt{\frac{1}{2}\left|U^{*\beta}(0)U_\beta(0) + U^\alpha(0)U^*_\alpha(0)\right| Z(1)}} \\ 0 & \frac{U_\mu(0)P(1)^{*\mu} + U^*_\lambda(0)P^\lambda(1)}{2\sqrt{\frac{1}{2}\left|U^{*\beta}(0)U_\beta(0) + U^\alpha(0)U^*_\alpha(0)\right| Z(1)}} & 1 \end{vmatrix}$$

$D(4)$

$$\equiv \begin{vmatrix} 1 & \frac{P(0)_\mu P(4)^{*\mu} + P(0)^*_\mu P(4)^\mu}{\sqrt{Z(0)Z(4)}} & \frac{P(0)_\mu P(5)^{*\mu} + P(0)^*_\mu P(5)^\mu}{\sqrt{Z(0)Z(5)}} & \frac{P(0)_\mu P(6)^{*\mu} + P(0)^*_\mu P(6)^\mu}{\sqrt{Z(0)Z(6)}} \\ \frac{P(4)_\mu P(0)^{*\mu} + P(4)^*_\mu P(0)^\mu}{\sqrt{Z(4)Z(0)}} & 1 & \frac{P(4)_\mu P(5)^{*\mu} + P(4)^*_\mu P(5)^\mu}{\sqrt{Z(4)Z(5)}} & \frac{P(4)_\mu P(6)^{*\mu} + P(4)^*_\mu P(6)^\mu}{\sqrt{Z(4)Z(6)}} \\ \frac{P(5)_\mu P(0)^{*\mu} + P(5)^*_\mu P(0)^\mu}{\sqrt{Z(5)Z(0)}} & \frac{P(5)_\mu P(4)^{*\mu} + P(5)^*_\mu P(4)^\mu}{\sqrt{Z(5)Z(4)}} & 1 & \frac{P(5)_\mu P(6)^{*\mu} + P(5)^*_\mu P(6)^\mu}{\sqrt{Z(5)Z(6)}} \\ \frac{P(6)_\mu P(0)^{*\mu} + P(6)^*_\mu P(0)^\mu}{\sqrt{Z(6)Z(0)}} & \frac{P(6)_\mu P(4)^{*\mu} + P(6)^*_\mu P(4)^\mu}{\sqrt{Z(6)Z(4)}} & \frac{P(6)_\mu P(5)^{*\mu} + P(6)^*_\mu P(5)^\mu}{\sqrt{Z(6)Z(5)}} & 1 \end{vmatrix}$$

$$L=\begin{vmatrix}1 & 0\\ 0 & \frac{U(0)^{k}U(0)_{k}^{*}+U(0)^{*k}U(0)_{k}}{8}\end{vmatrix}\sqrt{-\mathrm{g}}+$$

$$+\begin{vmatrix}1 & \frac{\mathrm{P}_{\mathrm{k}}P(1)^{*k}+\mathrm{P}^{*}{}_{\mathrm{k}}P(1)^{k}}{2\sqrt{\mathrm{ZZ}(1)}} & 0 & \frac{\mathrm{P}_{\mathrm{k}}U(1)^{*k}+\mathrm{P}^{*}{}_{\mathrm{k}}U(1)^{k}}{2\sqrt{2\mathrm{Z}}}\\ \frac{\mathrm{P}_{\mathrm{k}}P(1)^{*k}+\mathrm{P}^{*}{}_{\mathrm{k}}P(1)^{k}}{2\sqrt{\mathrm{ZZ}(1)}} & 1 & \frac{\mathrm{P}(1)_{\mathrm{k}}U^{*k}+\mathrm{P}(1)^{*}{}_{\mathrm{k}}U^{k}}{2\sqrt{2\mathrm{Z}(1)}} & 0\\ 0 & \frac{\mathrm{P}(1)_{\mathrm{k}}U^{*k}+\mathrm{P}(1)^{*}{}_{\mathrm{k}}U^{k}}{2\sqrt{2\mathrm{Z}}} & \frac{U^{k}U_{k}^{*}+U^{*k}U_{k}}{8} & \frac{U(1)^{k}U_{k}^{*}+U(1)^{*k}U_{k}}{8}\\ \frac{\mathrm{P}_{\mathrm{k}}U(1)^{*k}+\mathrm{P}^{*}{}_{\mathrm{k}}U(1)^{k}}{2\sqrt{2\mathrm{Z}}} & 0 & \frac{U(1)^{k}U_{k}^{*}+U(1)^{*k}U_{k}}{8} & \frac{U(1)^{k}U(1)_{k}^{*}+U(1)^{*k}U(1)_{k}}{8}\end{vmatrix}^{\frac{1}{2}}\mathrm{G}(3)^{-\frac{1}{2}}\sqrt{-g}$$

$$+\begin{vmatrix}1 & \frac{\mathrm{P}(0)_{\mathrm{k}}U(2)^{*k}+\mathrm{P}(0)^{*}{}_{\mathrm{k}}U(2)^{k}}{2\sqrt{2\mathrm{Z}}} & \frac{\mathrm{P}(0)_{\mathrm{k}}U(3)^{*k}+\mathrm{P}(0)^{*}{}_{\mathrm{k}}U(3)^{k}}{2\sqrt{2\mathrm{Z}}}\\ \frac{\mathrm{P}(0)_{\mathrm{k}}U(2)^{*k}+\mathrm{P}(0)^{*}{}_{\mathrm{k}}U(2)^{k}}{2\sqrt{2\mathrm{Z}}} & \frac{U(2)^{k}U(2)_{k}^{*}+U(2)^{*k}U(2)_{k}}{8} & \frac{U(2)^{k}U(3)_{k}^{*}+U(2)^{*k}U(3)_{k}}{8}\\ \frac{\mathrm{P}(0)_{\mathrm{k}}U(3)^{*k}+\mathrm{P}(0)^{*}{}_{\mathrm{k}}U(3)^{k}}{2\sqrt{2\mathrm{Z}}} & \frac{U(3)^{k}U(2)_{k}^{*}+U(3)^{*k}U(2)_{k}}{8} & \frac{U(3)^{k}U(3)_{k}^{*}+U(3)^{*k}U(3)_{k}}{8}\end{vmatrix}^{\frac{1}{2}}\mathrm{D}(3)^{-\frac{1}{2}}\sqrt{-g}$$

$$+\begin{vmatrix}1 & \frac{p_{\mu}S^{*\mu}+p_{\mu}^{*}S^{\mu}}{2\sqrt{2\mathrm{Z}}} & \frac{p_{\mu}W^{*\mu}+p_{\mu}^{*}W^{\mu}}{2\sqrt{2\mathrm{Z}}} & \frac{p_{\mu}T^{*\mu}+p_{\mu}^{*}T^{\mu}}{2\sqrt{2\mathrm{Z}}}\\ \frac{p_{\mu}S^{*\mu}+p_{\mu}^{*}S^{\mu}}{2\sqrt{2\mathrm{Z}}} & \frac{S_{\mu}S^{*\mu}+S_{\mu}^{*}S^{\mu}}{8} & \frac{S_{\mu}W^{*\mu}+S_{\mu}^{*}W^{\mu}}{8} & \frac{S_{\mu}T^{*\mu}+S_{\mu}^{*}T^{\mu}}{8}\\ \frac{p_{\mu}W^{*\mu}+p_{\mu}^{*}W^{\mu}}{2\sqrt{2\mathrm{Z}}} & \frac{W_{\mu}S^{*\mu}+W_{\mu}^{*}S^{\mu}}{8} & \frac{W_{\mu}W^{*\mu}+W_{\mu}^{*}W^{\mu}}{8} & \frac{W_{\mu}T^{*\mu}+W_{\mu}^{*}T^{\mu}}{8}\\ \frac{p_{\mu}T^{*\mu}+p_{\mu}^{*}T^{\mu}}{2\sqrt{2\mathrm{Z}}} & \frac{T_{\mu}S^{*\mu}+T_{\mu}^{*}S^{\mu}}{8} & \frac{T_{\mu}W^{*\mu}+T_{\mu}^{*}W^{\mu}}{8} & \frac{T_{\mu}T^{*\mu}+T_{\mu}^{*}T^{\mu}}{8}\end{vmatrix}^{\frac{1}{3}}\mathrm{D}(4)^{-\frac{1}{3}}\sqrt{-\mathrm{g}} \qquad (64.a)$$

**Important:** The role of the Gram determinants G(3), D(3), D(4) is to prevent trivial extremums of (64.a). The power the Gram determinants are raised to is kept compatible with the power of Gram determinants of the acceleration vectors / Reeb class vectors, not the usual Reeb vectors. G(3) could be easily turned into another Gram determinant G(4) that uses an addition normalized $U_{\mu}(1)$ vector just as it uses a normalized $U_{\mu}(0)$ vector, however, the simplest denominator G(3) only increases the Lagrangian when $P_{\mu}(1)$ is not perpendicular to the $U_{\lambda}(0)$ and $P_{\nu}(0)$ plane. In the same manner, D(4) increases the Lagrangian when $P_{\nu}(4), P_{\alpha}(5), P_{\beta}(6)$ are not perpendicular to $P_{\mu}(0)$. So D(4) and G(3) are the simplest denominators with a geometric meaning.

**Crucial:** An exception is the less successful D(3) which does not lead to a well-defined theory unless $U_{\nu}(2)=0$ and $U_{\nu}(3)=0$. Where D(3) is the Gram determinant $D(3)\equiv \left|Gram(\frac{P_{\mu}(0)}{\sqrt{Z(0)}},\frac{P_{\nu}(2)}{\sqrt{Z(2)}},\frac{P_{\lambda}(3)}{\sqrt{Z(3)}})\right|$. The definition can work with $\tilde{G}(4)\equiv$

$$\left|Gram\left(\frac{P_\mu(0)}{\sqrt{Z(0)}},\frac{P_\nu(2)}{\sqrt{Z(2)}},\frac{P_\lambda(3)}{\sqrt{Z(3)}},\frac{U(2)}{\sqrt{\frac{1}{2}\left|U_\alpha(2)U^{*\alpha}(2)+U^*_\beta(2)U^\beta(2)\right|}}\right)\right|$$ instead of D(3) but will force $U_\alpha(2)$ not to be zero.

**Important:** If by using two acceleration fields, one is supposed to retrieve the electroweak force, there is a problem that the electric field cannot be zero although its total electric charge can be zero. That leaves an inconvenient option, which is that the Higgs boson should behave as an electric dipole. An electric dipole interaction in the UV range falls very rapidly and is very short lived. There is currently no experimental evidence that the Higgs boson behaves as an electric dipole in the UV distance range.

There is another way to try to get a Lagrangian that allows a weak charge independent of the electric charge as a second Lagrangian plane action.

To allow the properties of the weak force, one needs to allow charge but also electroweak charge without mutual charge exclusion as seen in the chromatic charge of Quarks.

This can be done by removing the component of the acceleration field $\frac{U_\mu(1)}{2}$ in the plane which is spanned by the unit vectors $\frac{P_\mu(0)}{\sqrt{Z(0)}}$ and $\frac{U_\mu(0)}{\sqrt{\frac{1}{2}\left|U^{*\beta}(0)U_\beta(0)+U^\alpha(0)U^*_\alpha(0)\right|}}$ and since these two components are perpendicular to each other, the independent acceleration component $\frac{U_\mu(1)}{2}$ has the following square acceleration perpendicular component which can exist independently of whether $\frac{U_\mu(0)}{2}$ is zero or not with the caution that $\frac{U_\mu(0)}{2}=0 \Longrightarrow \sqrt{\frac{1}{2}\left|U^{*\beta}(0)U_\beta(0)+U^\alpha(0)U^*_\alpha(0)\right|}=0.$

We define the Gram determinant of the 3 unit vectors $\frac{P(0)_\mu}{\sqrt{Z(0)}},\frac{P(1)_\mu}{\sqrt{Z(1)}},\frac{U_\mu(0)}{\sqrt{\frac{1}{2}\left|U^{*\beta}(0)U_\beta(0)+U^\alpha(0)U^*_\alpha(0)\right|}}$

$$\mathrm{L}_{Weak} \equiv \left(\frac{1}{8}\left(U^{*\mu}(1)U_\mu(1)+U^\lambda(1)U^*_\lambda(1)\right)-\frac{1}{16}\left(\frac{P_\mu(0)}{\sqrt{Z(0)}}U^{*\mu}(1)+\frac{P^*_\nu(0)}{\sqrt{Z(0)}}U^\nu(1)\right)^2 \right.$$
$$\left. -\frac{1}{16}\left(\frac{U_\mu(0)U^{*\mu}(1)+U^*_\lambda(0)U^\lambda(1)}{\sqrt{\frac{1}{2}\left|U^{*\beta}(0)U_\beta(0)+U^\alpha(0)U^*_\alpha(0)\right|}}\right)^2\right)\mathrm{G}(3)^{-\frac{1}{2}}\sqrt{-g}$$

Notice that the power $-\frac{1}{2}$ of G(3) was chosen because $L_{Weak}$ involves two acceleration vectors. The logic here is kept compatible with (64.a).

And we have for this choice,

$$L = \begin{vmatrix} 1 & 0 \\ 0 & \frac{U(0)^k U(0)^*_k + U(0)^{*k} U(0)_k}{8} \end{vmatrix} \sqrt{-g}$$

$$+ \left( \left( \frac{1}{8} \left( U^{*\mu}(1) U_\mu(1) + U^\lambda(1) U^*_\lambda(1) \right) - \frac{1}{16} \left( \frac{P_\mu(0)}{\sqrt{Z(0)}} U^{*\mu}(1) + \frac{P^*_\nu(0)}{\sqrt{Z(0)}} U^\nu(1) \right)^2 \right.\right.$$

$$\left.\left. - \frac{1}{16} \left( \frac{U_\mu(0) U^{*\mu}(1) + U^*_\lambda(0) U^\lambda(1)}{\sqrt{\frac{1}{2} \left| U^{*\beta}(0) U_\beta(0) + U^\alpha(0) U^*_\alpha(0) \right|}} \right)^2 \right) G(3)^{-\frac{1}{2}} \sqrt{-g}$$

$$+ \begin{vmatrix} 1 & \frac{p_\mu S^{*\mu} + p^*_\mu S^\mu}{2\sqrt{2Z}} & \frac{p_\mu W^{*\mu} + p^*_\mu W^\mu}{2\sqrt{2Z}} & \frac{p_\mu T^{*\mu} + p^*_\mu T^\mu}{2\sqrt{2Z}} \\ \frac{p_\mu S^{*\mu} + p^*_\mu S^\mu}{2\sqrt{2Z}} & \frac{S_\mu S^{*\mu} + S^*_\mu S^\mu}{8} & \frac{S_\mu W^{*\mu} + S^*_\mu W^\mu}{8} & \frac{S_\mu T^{*\mu} + S^*_\mu T^\mu}{8} \\ \frac{p_\mu W^{*\mu} + p^*_\mu W^\mu}{2\sqrt{2Z}} & \frac{W_\mu S^{*\mu} + W^*_\mu S^\mu}{8} & \frac{W_\mu W^{*\mu} + W^*_\mu W^\mu}{8} & \frac{W_\mu T^{*\mu} + W^*_\mu T^\mu}{8} \\ \frac{p_\mu T^{*\mu} + p^*_\mu T^\mu}{2\sqrt{2Z}} & \frac{T_\mu S^{*\mu} + T^*_\mu S^\mu}{8} & \frac{T_\mu W^{*\mu} + T^*_\mu W^\mu}{8} & \frac{T_\mu T^{*\mu} + T^*_\mu T^\mu}{8} \end{vmatrix}^{\frac{1}{3}} D(4)^{-\frac{1}{3}} \sqrt{-g} \qquad (64.b)$$

Notice that G(3) and D(4) are the determinants of Gram matrices of 3 unit vectors and of 4 unit vectors from which the Reeb class vectors, which are accelerations of unit vectors, are derived. The division by the root of G(3) and the root of D(4) is typically used in the theory of foliations when the embedding curvature of the foliations are calculated. In (64.b) the terms do not involve the square root because the Gram determinants where the Reeb class vectors are components have the second power of accelerations, which are square norms of Reeb class vectors. The use of G(3), D(3), and of D(4) is therefore not a very creative step.

There is a redundancy between the following terms:

$\begin{vmatrix} 1 & 0 \\ 0 & \frac{U(0)^k U(0)^*_k + U(0)^{*k} U(0)_k}{8} \end{vmatrix} \sqrt{-g}$ and two perpendicular planes Lagrangian,

$$\begin{vmatrix} 1 & \frac{\mathrm{P_k}P(1)^{*k} + \mathrm{P^*}_{\mathrm{k}}P(1)^k}{2\sqrt{\mathrm{ZZ(1)}}} & 0 & \frac{\mathrm{P_k}U(1)^{*k} + \mathrm{P^*}_{\mathrm{k}}U(1)^k}{2\sqrt{2\mathrm{Z}}} \\ \frac{\mathrm{P_k}P(1)^{*k} + \mathrm{P^*}_{\mathrm{k}}P(1)^k}{2\sqrt{\mathrm{ZZ(1)}}} & 1 & \frac{\mathrm{P(1)_k}U^{*k} + \mathrm{P(1)^*}_{\mathrm{k}}U^k}{2\sqrt{2\mathrm{Z(1)}}} & 0 \\ 0 & \frac{\mathrm{P(1)_k}U^{*k} + \mathrm{P(1)^*}_{\mathrm{k}}U^k}{2\sqrt{2\mathrm{Z}}} & \frac{U^k U_k^* + U^{*k}U_k}{8} & \frac{U(1)^k U_k^* + U(1)^{*k}U_k}{8} \\ \frac{\mathrm{P_k}U(1)^{*k} + \mathrm{P^*}_{\mathrm{k}}U(1)^k}{2\sqrt{2\mathrm{Z}}} & 0 & \frac{U(1)^k U_k^* + U(1)^{*k}U_k}{8} & \frac{U(1)^k U(1)_k^* + U(1)^{*k}U(1)_k}{8} \end{vmatrix}^{\frac{1}{2}} \mathrm{G(3)}^{-\frac{1}{2}}\sqrt{-g}$$

Because with a correct choice of $\mathrm{P(1)_k}$ they are both identical, see (3). Therefore (64.b) can be written with the omission of $\begin{vmatrix} 1 & 0 \\ 0 & \frac{U(0)^k U(0)_k^* + U(0)^{*k}U(0)_k}{8} \end{vmatrix}\sqrt{-\mathrm{g}}$ .

The denominators $2\sqrt{2\mathrm{Z}}$ in the first 0 index row and the first 0 index column in (64.b) are required because if any term in the determinant calculation is taken from positions other than 0 of the first row, then one such term must be taken from the zero column. For example, $\frac{p_\mu S^{*\mu}+p_\mu^* S^\mu}{2\sqrt{2\mathrm{Z}}}\frac{p_\mu W^{*\mu}+p_\mu^* W^\mu}{2\sqrt{2\mathrm{Z}}}$ is a multiplication of two projections on the non-geodesic time direction $p_\mu$. This multiplication has a denominator $8\sqrt{\mathrm{Z}}$ as required. A projection on the time direction in general, reduces the determinant absolute value, which means that acceleration fields are preferred to be perpendicular to the time direction by the Lagrangian in (64.b). Each of the determinants of (64.b) agrees with the multiplicative rule of (13.12) but unlike (13.12), higher values are achieved when the acceleration vectors are perpendicular. The roots equate units of length to units of area and to units of volume and the components of (64.b) are higher when the accelerations, which are Reeb class vectors, are also perpendicular to the time-like vector $\frac{p_\mu}{\sqrt{Z}}$ as expected from an acceleration vector of a unit field, to be perpendicular to a time-like vector.

The last term of (64.b) has SU(3) * reflections symmetry, however, when considering the space-like foliation which is perpendicular to $\frac{p_\mu}{\sqrt{Z}}$, extremal solutions, not saddle variations, have a physical meaning of rotating fields. See Fig. 9. There could be better action operators than (64.b), after all, (64.b) is no more than a research offer although it has its own logic which is not fully explained in this paper.

**Theorem 6 - chromodynamic mutual charge exclusion:** Due to (13.01), given that where $S_\mu;^\mu \neq 0,\ S_\mu = 0$, where $W_\mu;^\mu \neq 0,\ W_\mu = 0, T_\mu;^\mu \neq 0,\ T_\mu = 0$ such that $S_\mu, W_\mu, T_\mu$ are perpendicular when independent, charge terms of (64.b) exclusively allow only one charge term to be non-zero. Without loss of generality, a non-zero charge term of $S_\lambda$ is defined as

$\frac{W_\zeta W^{*\zeta}+W_\zeta^* W^\zeta}{8}\cdot\frac{T_\nu T^{*\nu}+T_\nu^* T^\nu}{8}\frac{1}{4}({S^*}_\lambda+S_\lambda);^\lambda \neq 0$ and because $S_\mu, W_\mu, T_\mu$ are perpendicular, (64.b) is reduced to $\sqrt[3]{\frac{S_\mu S^{*\mu}+S_\mu^* S^\mu}{8}\frac{W_\zeta W^{*\zeta}+W_\zeta^* W^\zeta}{8}\frac{T_\nu T^{*\nu}+T_\nu^* T^\nu}{8}}\sqrt{-g}.$

**Proof:** The proof is trivial because the last term requires both $W_\zeta \neq 0$ and $T_\nu \neq 0$ which by the assumption that $S_\mu;^\mu \neq 0 \Longrightarrow S_\mu = 0$, $W_\mu;^\mu \neq 0 \Longrightarrow W_\mu = 0$, $T_\mu;^\mu \neq 0 \Longrightarrow T_\mu = 0$, the following must be true $W_\zeta;^\zeta = 0$, $T_\nu;^\nu = 0$, Q.E.D .

**Volumetric chromodynamic action of the electric field – an alternative formalism to (4)**

Here is the calculation that subtracts the action of the electric field due to its projection on the normal to the foliation spanned by $p(4)_\mu, p(5)_\mu, p(6)_\mu$ this span can be written as $\langle p(4)_\mu, p(5)_\mu, p(6)_\mu \rangle$. Notice that unlike in spin connections, the orthogonality of $p(0)_\mu$ and $\langle p(4)_\mu, p(5)_\mu, p(6)_\mu \rangle$ is not imposed, however, such orthogonality reduces to the ordinary action that was used in (4), namely $\mathrm{L} = \frac{N(1)}{D(4)} = \frac{U_\mu U^{*\mu} + U_\mu^* U^\mu}{8}\sqrt{-g}$. In the general case,

$$N(1) \equiv \Bigg(\frac{U_\mu U^{*\mu} + U_\mu^* U^\mu}{8} -$$

$$\begin{vmatrix} \frac{U_\mu U^{*\mu}+U_\mu^* U^\mu}{8} & \frac{p_\mu(4)U^{*\mu}+p_\mu^*(4)U^\mu}{2\sqrt{2\mathrm{Z}(4)}} & \frac{p_\mu(5)U^{*\mu}+p_\mu^*(5)U^\mu}{2\sqrt{2\mathrm{Z}(5)}} & \frac{p_\mu(6)U^{*\mu}+p_\mu^*(6)U^\mu}{2\sqrt{2\mathrm{Z}(6)}} \\ \frac{p(4)_\mu U^{*\mu}+p(4)_\mu^* U^\mu}{2\sqrt{2\mathrm{Z}(4)}} & 1 & \frac{P(4)_\mu P(5)^{*\mu}+P(4)_\mu^* P(5)^\mu}{\sqrt{Z(4)Z(5)}} & \frac{P(4)_\mu P(6)^{*\mu}+P(4)_\mu^* P(6)^\mu}{\sqrt{Z(4)Z(6)}} \\ \frac{p(5)_\mu U^{*\mu}+p(5)_\mu^* U^\mu}{2\sqrt{2\mathrm{Z}(5)}} & \frac{P(5)_\mu P(4)^{*\mu}+P(5)_\mu^* P(4)^\mu}{\sqrt{Z(5)Z(4)}} & 1 & \frac{P(5)_\mu P(6)^{*\mu}+P(5)_\mu^* P(6)^\mu}{\sqrt{Z(5)Z(6)}} \\ \frac{p(6)_\mu U^{*\mu}+p(6)_\mu^* U^\mu}{2\sqrt{2\mathrm{Z}(6)}} & \frac{P(6)_\mu P(4)^{*\mu}+P(6)_\mu^* P(4)^\mu}{\sqrt{Z(6)Z(4)}} & \frac{P(6)_\mu P(5)^{*\mu}+P(6)_\mu^* P(5)^\mu}{\sqrt{Z(6)Z(5)}} & 1 \end{vmatrix} \sqrt{-g}\Bigg)$$

$$D(4) \equiv \begin{vmatrix} 1 & \frac{P(0)_\mu P(4)^{*\mu} + P(0)_\mu^* P(4)^\mu}{\sqrt{Z(0)Z(4)}} & \frac{P(0)_\mu P(5)^{*\mu} + P(0)_\mu^* P(5)^\mu}{\sqrt{Z(0)Z(5)}} & \frac{P(0)_\mu P(6)^{*\mu} + P(0)_\mu^* P(6)^\mu}{\sqrt{Z(0)Z(6)}} \\ \frac{P(4)_\mu P(0)^{*\mu} + P(4)_\mu^* P(0)^\mu}{\sqrt{Z(4)Z(0)}} & 1 & \frac{P(4)_\mu P(5)^{*\mu} + P(4)_\mu^* P(5)^\mu}{\sqrt{Z(4)Z(5)}} & \frac{P(4)_\mu P(6)^{*\mu} + P(4)_\mu^* P(6)^\mu}{\sqrt{Z(4)Z(6)}} \\ \frac{P(5)_\mu P(0)^{*\mu} + P(5)_\mu^* P(0)^\mu}{\sqrt{Z(5)Z(0)}} & \frac{P(5)_\mu P(4)^{*\mu} + P(5)_\mu^* P(4)^\mu}{\sqrt{Z(5)Z(4)}} & 1 & \frac{P(5)_\mu P(6)^{*\mu} + P(5)_\mu^* P(6)^\mu}{\sqrt{Z(5)Z(6)}} \\ \frac{P(6)_\mu P(0)^{*\mu} + P(6)_\mu^* P(0)^\mu}{\sqrt{Z(6)Z(0)}} & \frac{P(6)_\mu P(4)^{*\mu} + P(6)_\mu^* P(4)^\mu}{\sqrt{Z(6)Z(4)}} & \frac{P(6)_\mu P(5)^{*\mu} + P(6)_\mu^* P(5)^\mu}{\sqrt{Z(6)Z(5)}} & 1 \end{vmatrix}$$

$$\mathrm{L} = \frac{N(1)}{D(4)^{\frac{1}{2}}} \tag{64.0.1}$$

The division by $D(4)$ as in (64.b) is obvious because it eliminates an increase or a decrease in the action due to deviations from orthogonality. Such a technique is similar to the calculation of the embedding curvature of codim-1 foliations. It was hoped by the author that a volumetric formalism of (4) can get rid of the Einstein – Hilber action, however, even without such elimination, the formalism is a generalization of (4). The complexity of (64.0.1) is an obvious disadvantage.

**Why spacetime with an arrow of time must have 4 dimensions?**

If you look carefully at theorem 0 (Suchard - Vaknin), you immediately see that for an arrow of time to exist, such that time is considered as an orthogonality referenced as in (64.b), conservation of the action due to Reeb class vectors (not the usual Reeb vector) dictates exactly 4 dimensions. In 2 dimensions, the orientation of the span $P_\mu$^$U_\nu$ can be maintained also when $p_\mu$ flips direction because then $u_\nu$ is flipped simultaneously so 2 dimensions are not possible. In 3 dimensions, the conservation of the orientation of $P_\mu(0)$^$U_\nu(0)$ still allows the flipping of $P_\mu(0)$ and of $U_\nu(0)$ simultaneously, however, one can demand $U_\mu(1)$^$U_\nu(2)$ to also keep orientation as they define a codim-1 action. Flipping both $U_\mu(1)$and $U_\nu(2)$ does not change the orientation of $U_\mu(1)$^$U_\nu(2)$ where $U_\mu(1)$^$U_\nu(2)$ define the field of the foliation perpendicular to $P_\lambda(0)$. Now observe that $U_\nu(0)$ is contained in the foliation spanned by $U_\mu(1)$^$U_\nu(2)$. If as a result, $U_\mu(0)$ is flipped, then $P_\mu(0)$ is also flipped and $P_\mu(0)$ does not maintain the arrow of time.

**Theorem 7**: We now repeat Theorem 0 (Suchard - Vaknin) to prove a new theorem. Both Theorem 0 and theorem 7 are significant and are not trivial due to "Appendix H – Causality conservation theorem".

In 4 dimensions there are 5 fields, that define two types of orientations, as mentioned in (2.22) and in (2.23). We can start with $U_\mu(0), U_\mu(1), \ldots, U_\mu(4)$. $U_\mu(1)$ is defined in the Lagrangian plane perpendicular to span $P_\mu(0)$^$U_\nu(0)$. The field of the 3D foliation perpendicular to $P_\mu(0)$ is $U_\mu(2)$^$U_\nu(3)$^$U_\lambda(4)$ and also as an alternative we can use $P_\mu(2)$^$P_\nu(3)$^$P_\lambda(4)$ as in (2.23) and this basis must keep orientation to be a regular field for which (64.b) is non-degenerate, i.e. not zero. $P_\mu(1)$^$U_\nu(1)$ defines the perpendicular Lagrangian plane to $P_\mu(0)$^$U_\nu(0)$. The important realization is that $P_\mu(1)$, $U_\nu(1)$, $U_\lambda(0)$ are contained in the foliation spanned by $P_\mu(2)$^$P_\nu(3)$^$P_\lambda(4)$ and that since the Levi-Civita tensor is smooth, the relative orientation between $P_\mu(0)$^$U_\nu(0)$ and $P_\mu(1)$^$U_\nu(1)$ cannot change because $P_\mu(0)$^$U_\nu(0)$ and $P_\mu(1)$^$U_\nu(1)$ describe a field as two perpendicular acceleration planes and there are two such possible orientations, $P_\mu(1)$^$U_\nu(1)$ or $-P_\mu(1)$^$U_\nu(1)$, which can also be $U_\nu(1)$^$P_\mu(1)$.

We can summarize 4 possibilities of flipping two vector signs which maintain both the foliation orientation and the relative orientation between the two Lagrangian planes, the third is trivial:

$$-U_\mu(1), -P_\mu(1) \Rightarrow +U_\mu(0) \Rightarrow +P_\mu(0)$$

$$+U_\mu(1), -P_\mu(1) \Rightarrow -U_\mu(0) \Rightarrow +P_\mu(0)$$

$$+U_\mu(1), +P_\mu(1) \Rightarrow +U_\mu(0) \Rightarrow +P_\mu(0)$$

$$-U_\mu(1), P_\mu(1) \Rightarrow -U_\mu(0) \Rightarrow +P_\mu(0) \qquad (64.02)$$

Of course, we needed all 5 acceleration fields / Reeb class vectors, not the usual Reeb vectors, or at the very least well defined $P_\mu(2)^\wedge P_\nu(3)^\wedge P_\lambda(4)$ to be alive and kicking for this theorem to work. This demand is tantamount to a particle / worldline that has $P_\mu(0)$, $U_\nu(0)$, $P_\mu(1)$, $U_\nu(1)$, $P_\mu(2)$, $P_\nu(3)$, $P_\lambda(4)$ non-degenerate. The Geroch splitting theorem [2] dictates that in a causal spacetime there must be a 3D Cauchy surface. Demanding that $P_\mu(0)$ will be non-degenerate on this Cauchy surface except for a subset of Lebesgue measure zero means that the orientation of $P_\mu(0)$ must be the same on this Cauchy surface, however, replacing the Cauchy surface with any other surface ensures that if $P_\mu(0)$ is non-degenerate on this surface, the time orientation of the worldlines that run through this surface is maintained as long as locally the orientation between $P_\mu(0)^\wedge U_\nu(0)$ and $P_\mu(1)^\wedge U_\nu(1)$ is distinguishable and . Theorem 0 dictates that once there is such a particle possibly a proton, time can only have one orientation, so the minimal dimension in which Lagrangians as in (64.b) can be valid along with the arrow of time is 4. Q.E.D.

We relied on two types of fields, a chromodynamic acceleration field which is a codim-1 field and on an acceleration field in independent Lagrangian planes a.k.a symplectic Lagrangian planes. It means the observable Nature exploits the maximal number of fields it can while using an existing dimensionality.

**Uniform gravity and forces formalism, challenges, and an open problem:** This section uses a similar idea as in (13.14) - (13.18). Apparently the first and the last additives of (64), 2x2 and 4x4 matrices imply that the Einstein-Hilbert action can be written in a tetradic formulation with 4 scalar functions where:

$$e^J{}_\mu = \frac{p^J{}_\mu}{\sqrt{Z(J)}} = \frac{p_{(J),\mu}}{\sqrt{Z(J)}} \text{ and } \eta^{KJ} e_{K\mu} e_{J\nu} = e^J{}_\mu e_{J\nu} = g_{\mu\nu} \text{ or } \frac{1}{2}\left(e^{*J}{}_\mu e_{J\nu} + e^J{}_\mu e^*{}_{J\nu}\right) = g_{\mu\nu} \text{ and} \tag{64.1}$$

$$K \neq J \Rightarrow \eta^{KJ} = 0,\ \eta^{00} = -1,\ \eta^{11} = \eta^{22} = \eta^{33} = 1$$

**Caveat:** $\frac{p^J{}_\mu}{\sqrt{Z(J)}}$ may not be related to any force fields but only to gravity. Please refer to the remark after (3.13).

**Clarification:** Notice that P(0)P*(0) is a Geroch function. Adopting Sam Vaknin's methods [17] in a scalar function formalism, P(0)P*(0), can be a probability density of a single event in 4-volume, instead of time. The good news is that $P(0), P(1), P(2), P(3)$ yield 4 gradients $P(0)_\mu, P(1)_\mu, P(2)_\mu, P(3)_\mu$ and 4 non-geodesic accelerations as generalized Reeb class vectors - not limited to contact manifolds, unlike the usual Reeb vectors- $\frac{U_\mu}{2}, \frac{S_\mu}{2}, \frac{W_\mu}{2}, \frac{T_\mu}{2}$ as explained in (64.b). The much less good news is that 4 complex functions $P(0), P(1), P(2), P(3)$ are 8 real functions which mean that the 2 degrees of freedom in the 10 independent General Relativity equations are gone (without considering the 4 vanishing divergence equations). This problem can be partially mitigated by multiplication of the Tetradic metric tensor by a scalar function $\phi$, however, such a solution to a unified formalism of forces and gravity is awkward, $\frac{1}{2}\phi\phi^*(e^{*J}{}_{\mu}e_{J\nu} + e^{J}{}_{\mu}e^{*}{}_{J\nu}) = g_{\mu\nu}$ is still incomplete. The problem is the following condition:

$\frac{P*_{\mu}(0)P_{\nu}(0)+P_{\mu}(0)P*_{\nu}(0)}{2\sqrt{Z(0)}} - \sum_{i=1}^{3}\frac{P*_{\mu}(i)P_{\nu}(i)+P_{\mu}(i)P*_{\nu}(i)}{2\sqrt{Z(i)}} = g_{\mu\nu}$ and from the vanishing of the ordinary covariant derivative: $\frac{1}{2}(e^{*J}{}_{\mu}e_{J\nu} + e^{J}{}_{\mu}e^{*}{}_{J\nu});_{\lambda} = g_{\mu\nu};_{\lambda} = 0$. This is not a good place to begin with. For the Riemann tensor in the real case, we have:

$R^{J}{}_{KLM} = e^{J}((\nabla_L\nabla_M - \nabla_M\nabla_L - [e_L, e_M])e_K)$ and $R = R^{JM}{}_{JM}$ and the volume element is $\sqrt{|Deteminant(e^{J}{}_{\mu}e_{J\nu})|}$ with Greek letters denoting the ordinary Riemann indices.

There are, however, some caveats. $p^{J=0}{}_{\mu=1} = p^{0}{}_{2} = p^{0}{}_{3} = 0$ means that P(0)P*(0) must be a Geroch time function [2]. The rest of the tetrads need not be perpendicular to each other, $e^{J}{}_{\mu}e^{*K\mu} + e^{*J}{}_{\mu}e^{K\mu} \neq 0$ but using spin connections they may be formulated in such a way that $e^{J}{}_{\mu}e^{*K\mu} + e^{*J}{}_{\mu}e^{K\mu} = 0$.

**Fig. 9.** – Spin property of extremal solutions to (64.b), shearing vs rotation.

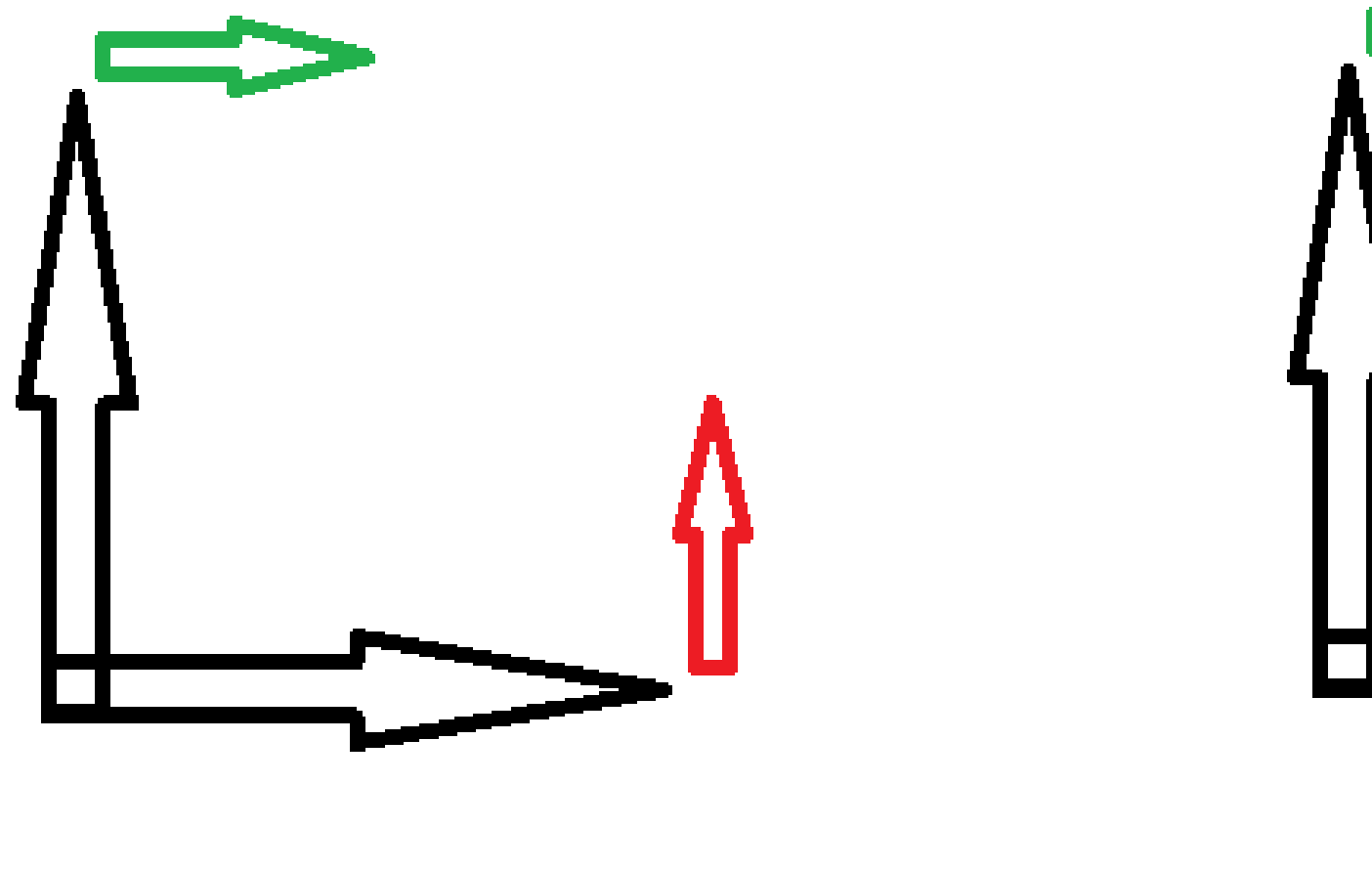

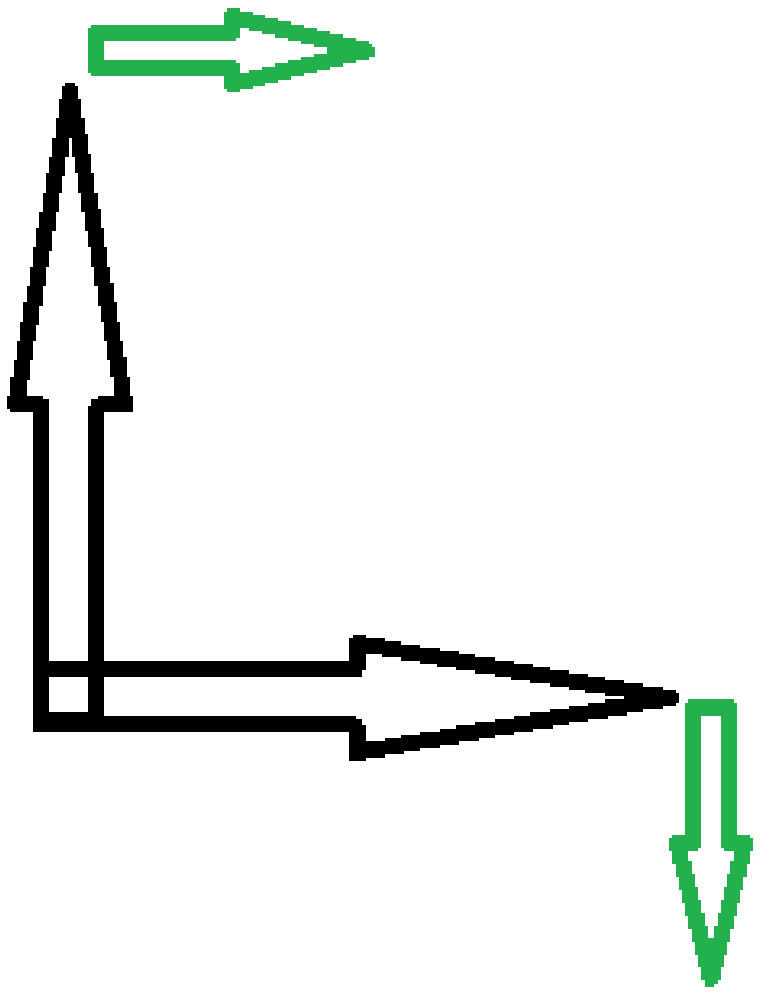

**The Aryeh Aldema's offer of a Relative curvature / Embedding curvature action and it's meaning**

Let $\frac{p_{\mu}(0)}{\sqrt{|Z(0)|}} = \frac{p_{\mu}}{\sqrt{|Z|}}$ and 3 other scalar fields are defined $\frac{p_{\mu}(1)}{\sqrt{|Z(1)|}}, \frac{p_{\mu}(2)}{\sqrt{|Z(2)|}}, \frac{p_{\mu}(3)}{\sqrt{|Z(3)|}}$ with real numbers. Taking the Gaussian curvature but without demanding that the embedding manifold will be flat, consider

$$K\sqrt{-g} \equiv \frac{\left(\frac{p_i(0)}{\sqrt{|Z(0)|}}\right)_{;\mu}\frac{p^{\mu}(1)}{\sqrt{|Z(1)|}}\left(\frac{p_j(0)}{\sqrt{|Z(0)|}}\right)_{;\nu}\frac{p^{\nu}(2)}{\sqrt{|Z(2)|}}\left(\frac{p_k(0)}{\sqrt{|Z(0)|}}\right)_{;\zeta}\frac{p^{\zeta}(3)}{\sqrt{|Z(3)|}}\frac{p_s(0)}{\sqrt{|Z(0)|}}\epsilon^{ijks}}{\frac{p_a(0)}{\sqrt{|Z(0)|}}\frac{p_b(1)}{\sqrt{|Z(1)|}}\frac{p_c(2)}{\sqrt{|Z(2)|}}\frac{p_d(3)}{\sqrt{|Z(3)|}}\epsilon^{abcd}}\sqrt{-g} \tag{64.2}$$

where $\epsilon^{ijks}$ is the Levi-Civita alternating symbols. It is not difficult to prove that this definition does not depend on the choice of $\frac{p_\mu(1)}{\sqrt{|Z(1)|}}, \frac{p_\mu(2)}{\sqrt{|Z(2)|}}, \frac{p_\mu(3)}{\sqrt{|Z(3)|}}$ as long as they are mutually independent and independent of $\frac{p_s(0)}{\sqrt{|Z(0)|}}$. Obviously, $\left(\frac{p_s(0)}{\sqrt{|Z(0)|}}\right)_{;i} p^i(0) = \frac{p_{s;i}(0)}{\sqrt{|Z(0)|}}p^i(0) - \frac{1}{2}\frac{p_s(0)}{|Z(0)|^{\frac{3}{2}}}Z_i(0)p^i(0) = 0$ so there is no need to worry about $p_\mu(1), p_\mu(2), p_\mu(3)$ being within the foliation. A full proof replaces $\epsilon^{ijks}$ with the Levi-Civita tensor at first $E^{ijks} = sgn(g)\frac{\epsilon^{ijks}}{\sqrt{-g}} = -\frac{\epsilon^{ijks}}{\sqrt{-g}}$ so we immediately see that

$$K \equiv \frac{\left(\frac{p_i(0)}{\sqrt{|Z(0)|}}\right)_{;\mu}\frac{p^{\mu}(1)}{\sqrt{|Z(1)|}}\left(\frac{p_j(0)}{\sqrt{|Z(0)|}}\right)_{;\nu}\frac{p^{\nu}(2)}{\sqrt{|Z(2)|}}\left(\frac{p_k(0)}{\sqrt{|Z(0)|}}\right)_{;\zeta}\frac{p^{\zeta}(3)}{\sqrt{|Z(3)|}}\frac{p_s(0)}{\sqrt{|Z(0)|}}E^{ijks}}{\frac{p_a(0)}{\sqrt{|Z(0)|}}\frac{p_b(1)}{\sqrt{|Z(1)|}}\frac{p_c(2)}{\sqrt{|Z(2)|}}\frac{p_d(3)}{\sqrt{|Z(3)|}}E^{abcd}} \tag{64.3}$$

Which means that K is the quotient of two scalar functions and is therefore a scalar function.

Without loss of generality,

$$\begin{aligned}
&\frac{\left(\frac{p_i(0)}{\sqrt{|Z(0)|}}\right)_{;\mu}(\frac{p^{\mu}(1)+rp^{\mu}(2)}{\sqrt{\|p^{\mu}(1)+rp^{\mu}(2)\|}}\left(\frac{p_j(0)}{\sqrt{|Z(0)|}}\right)_{;\nu}\frac{p^{\nu}(2)}{\sqrt{|Z(2)|}}\left(\frac{p_k(0)}{\sqrt{|Z(0)|}}\right)_{;\zeta}\frac{p^{\zeta}(3)}{\sqrt{|Z(3)|}}\frac{p_s(0)}{\sqrt{|Z(0)|}}E^{ijks}}{\frac{p_a(0)}{\sqrt{|Z(0)|}}\frac{p_b(1)+rp_b(2)}{\sqrt{\|p_b(1)+rp_b(2)\|}}\frac{p_c(2)}{\sqrt{|Z(2)|}}\frac{p_d(3)}{\sqrt{|Z(3)|}}E^{abcd}} \\
&= \frac{\left(\frac{p_i(0)}{\sqrt{|Z(0)|}}\right)_{;\mu}(\frac{p^{\mu}(1)}{\sqrt{\|p^{\mu}(1)+rp^{\mu}(2)\|}}\left(\frac{p_j(0)}{\sqrt{|Z(0)|}}\right)_{;\nu}\frac{p^{\nu}(2)}{\sqrt{|Z(2)|}}\left(\frac{p_k(0)}{\sqrt{|Z(0)|}}\right)_{;\zeta}\frac{p^{\zeta}(3)}{\sqrt{|Z(3)|}}\frac{p_s(0)}{\sqrt{|Z(0)|}}E^{ijks}}{\frac{p_a(0)}{\sqrt{|Z(0)|}}\frac{p_b(1)}{\sqrt{\|p_b(1)+rp_b(2)\|}}\frac{p_c(2)}{\sqrt{|Z(2)|}}\frac{p_d(3)}{\sqrt{|Z(3)|}}E^{abcd}} \\
&= \frac{\left(\frac{p_i(0)}{\sqrt{|Z(0)|}}\right)_{;\mu}(\frac{p^{\mu}(1)}{\sqrt{\|p^{\mu}(1)\|}}\left(\frac{p_j(0)}{\sqrt{|Z(0)|}}\right)_{;\nu}\frac{p^{\nu}(2)}{\sqrt{|Z(2)|}}\left(\frac{p_k(0)}{\sqrt{|Z(0)|}}\right)_{;\zeta}\frac{p^{\zeta}(3)}{\sqrt{|Z(3)|}}\frac{p_s(0)}{\sqrt{|Z(0)|}}E^{ijks}}{\frac{p_a(0)}{\sqrt{|Z(0)|}}\frac{p_b(1)}{\sqrt{\|p_b(1)\|}}\frac{p_c(2)}{\sqrt{|Z(2)|}}\frac{p_d(3)}{\sqrt{|Z(3)|}}E^{abcd}}
\end{aligned} \tag{64.4}$$

This is because the Levi-Civita symbols are alternating so if the same vector occurs twice with two different indices, this component vanishes.

For example, in the numerator $\left(\frac{p_i(0)}{\sqrt{|Z(0)|}}\right)_{;\mu}\frac{rp^{\mu}(2)}{\sqrt{\|p^{\mu}(1)+rp^{\mu}(2)\|}}$ is annihilated with $\left(\frac{p_j(0)}{\sqrt{|Z(0)|}}\right)_{;\nu}\frac{p^{\nu}(2)}{\sqrt{|Z(2)|}}$ when these are multiplied by the Levi-Civita tensor. So, we find that K does not depend on the choice of $p_\mu(1), p_\mu(2), p_\mu(3)$ and is therefore intrinsic to the foliation perpendicular to $p_\mu(0)$.

Our argument was about the importance of the "relative curvature". My opinion was clear, that the need for $p_\mu(1), p_\mu(2), p_\mu(3)$ is only if they can reveal a value which is not anticipated by the geometry of the 3D foliation because the Reeb class vector $\frac{U_\mu}{2}$ already contains geometric information about the foliation which is perpendicular to the vector $p_\mu = p_\mu(0)$. This information is of how the fields $p_\mu(1), p_\mu(2), p_\mu(3)$ are misaligned and do not make geodesic curves. This was the reason behind (64.b). Aryeh's input clarified the importance of (64.b). Unfortunately, Aryeh Aldema, who was my colleague, passed away in Dec/30/2022.

**Aldema's action - orthogonal projectors in the Geometric Chronon Field Theory**

Here, a full mathematical interpretation of Aryeh Aldema's "relative curvature of foliations" is presented by using the basic notion of embedding curvature as a relative curvature of foliations of spacetime.

In the following V1, V2, … Vn can be also denoted as V(1), V(2),…,V(n) and are chosen to be unit vectors on a Lorentzian manifold.

Let $\epsilon^{\mu\alpha\beta\gamma}$ denote the Levi-Civita alternating symbol, not tensor.

Then the orthogonal projector to $\frac{P_\mu}{\sqrt{|Z|}}$ where $Z \equiv \frac{P^{*\lambda}P_\lambda + P^\lambda P^*_\lambda}{2}$ can be expressed in terms of the Levi-Civita tensor $Proj_\nu = \frac{1}{3!}\frac{P_\mu}{\sqrt{|Z|}}\epsilon^{\mu\alpha\beta\gamma}\epsilon_{\nu\alpha\beta\gamma}$ and

$$Proj^*_\lambda V^\lambda + Proj_\lambda V^{*\lambda} = V_\nu\left(V^{*\nu} - \left(\frac{P^{*\nu}}{\sqrt{|Z|}}\right)\right) + V^*_\nu\left(V^\nu - \left(\frac{P^\nu}{\sqrt{|Z|}}\right)\right) \tag{64.5}$$

The orthogonal projector can derive time from space, given

$$V1 = \frac{P_\mu(1)}{\sqrt{|Z(1)|}}, V2 = \frac{P_\mu(2)}{\sqrt{|Z(2)|}}, V3 = \frac{P_\mu(3)}{\sqrt{|Z(3)|}} \tag{64.6}$$

The Gram determinant is

$$G(V1, V2, V3) \equiv \begin{vmatrix} 1 & \frac{V1_\mu V2^{*\mu} + V2_\mu V1^{*\mu}}{2} & \frac{V1_\mu V3^{*\mu} + V3_\mu V1^{*\mu}}{2} \\ \frac{V2_\mu V1^{*\mu} + V1_\mu V2^{*\mu}}{2} & 1 & \frac{V2_\mu V3^{*\mu} + V3_\mu V2^{*\mu}}{2} \\ \frac{V3_\mu V1^{*\mu} + V1_\mu V3^{*\mu}}{2} & \frac{V3_\mu V2^{*\mu} + V2_\mu V3^{*\mu}}{2} & 1 \end{vmatrix} \tag{64.7}$$

We should take the absolute value of the determinant in (64.7).

Then we have, when $g$ is the determinant of the metric tensor:

$$V^{\mu}(0) = \frac{\epsilon^{\mu\alpha\beta\gamma} V1_{\alpha} V2_{\beta} V3_{\gamma}}{\sqrt{-g}\sqrt{|G|}} \qquad (64.8)$$

The definitions in (64.7) and (64.8) are easily generalized to any dimension.

**Aldema's action - the relative Gaussian curvature of m-foliations in n-manifolds**

Here an alternative to (64.2) is presented which hopefully correctly represents my deceased colleague Aryeh Aldema's idea of "relative curvature of a foliation". For simplicity we will focus on real numbers. Similar to (64.2) we first need to calculate the multi-normal vectors to the tangent space which is spanned by $V_{\alpha_1}(1), \dots, V_{\alpha_m}(m)$ in a manifold which is n dimensional. It is assumed that every square root in the following expressions always regards the absolute value. The multi-normal tensor is calculated in the following expression:

$$N^{\alpha_{n-m},\dots,\alpha_{n-1},\alpha_n} \equiv \frac{V_{\alpha_1}(1)V_{\alpha_2}(2)\dots V_{\alpha_m}(m)}{\sqrt{G(V_{\alpha_1}(1),V_{\alpha_2}(2),\dots,V_{\alpha_m}(m))}} \frac{\epsilon^{\alpha_1\alpha_2\dots\alpha_n}}{\sqrt{-g}} \qquad (64.9)$$

Notice that $N^{\alpha_{n-m},\dots,\alpha_{n-1},\alpha_n}$ is indeed a tensor because $\frac{\epsilon^{\alpha_1\alpha_2\dots\alpha_n}}{\sqrt{-g}}$ and $\sqrt{-g}\epsilon_{\alpha_1\alpha_2\dots\alpha_n}$ are the Levi-Civita alternating tensors, while $\epsilon^{\alpha_1\alpha_2\dots\alpha_n}$ and $\epsilon_{\alpha_1\alpha_2\dots\alpha_n}$ are the Levi-Civita alternating symbols.

**Caveat:** Normalizations of the multi vectors $V_{\alpha_1}(1)V_{\alpha_2}(2)\dots V_{\alpha_m}(m)$ and $V^{\alpha_1}(1),\dots,V^{\alpha_m}(m)$ are both with the same denominator $\sqrt{G(V_{\alpha_1}(1),V_{\alpha_2}(2),\dots,V_{\alpha_m}(m))}$ in normalized unit m-volumes. This case is different from the way $\sqrt{-g}$ is used in the Levi-Civita tensors to turn tensor densities to tensors and is a source of errors in basic differential geometry.

Then we need to find the following covariant derivatives of the multi-normal tensor along the vectors $V_{\alpha_1}(1), V_{\alpha_2}(2), \dots, V_{\alpha_m}(m)$,

$$W^{\alpha_{n-m},\dots,\alpha_n}(i) = V^{\mu}(i) N^{\alpha_{n-m},\dots,\alpha_{n-1},\alpha_n}{}_{;\mu} \qquad (64.10)$$

Notice that $W^{\alpha_{n-m},\dots,\alpha_n}(k)$ generalizes the covariant derivative of a single codimension 1 hyper-surface normal vector to the covariant derivative of a multi-normal tensor for codimension (n-m). The next step is to project $W^{\alpha_{n-m},\dots,\alpha_n}(i)$ back on the space spanned by $V_{\alpha_1}(1), V_{\alpha_2}(2), \dots, V_{\alpha_m}(m)$. Like in (64.2) we need to multiply all the terms and then we need to contract the result with a unit m-volume $\frac{V^{\alpha_1}(1)\dots V^{\alpha_m}(m)}{\sqrt{G(V_{\alpha_1}(1),V_{\alpha_2}(2),\dots,V_{\alpha_m}(m))}}$.

$$K \equiv \pi_{i=1}^{m} \frac{1}{(n-m)!}\left( V^{\mu}(i) N^{\alpha_{n-m},\dots,\alpha_{n-1},\alpha_n}{}_{;\mu} \sqrt{-g}\epsilon_{\alpha_1\alpha_2\dots\alpha_n} \frac{V^{\alpha_1}(1)\dots V^{\alpha_m}(m)}{\sqrt{G(V_{\alpha_1}(1),V_{\alpha_2}(2),\dots,V_{\alpha_m}(m))}} \right) \qquad (64.11)$$

The term $(n-m)!$ is because we contract the alternating indices $\alpha_{n-m}, \dots, \alpha_{n-1}, \alpha_n$ between two alternating tensors.

To get this curvature $K$ in the Aldema format, we need to use the power $\frac{2}{m}$, where the power 2 is because accelerations in Lagrangians in this paper always appear in the second power.

The Aldema Lagrangian is simply the following action:

$$Ald = K^{\frac{2}{m}}\sqrt{-g} \tag{64.12}$$

**The vorticity action of 4 Reeb class vectors**

Possibly a fifth force of Nature or by (3.12) and mapping curves to a flat spacetime, massive gravity is described by the following SU(4) symmetry Lagrangian of 4 Reeb class vectors: $\frac{\aleph_\mu}{2}, \frac{\beth_\mu}{2}, \frac{\gimel_\mu}{2}, \frac{\daleth_\mu}{2}$, with Hebrew letters Alef, Beit, Gimmel, Dalet, for D(4) see (64),

$$\begin{vmatrix} \frac{\aleph_\mu \aleph^{*\mu}+\aleph^*_\mu \aleph^\mu}{8} & \frac{\aleph_\mu \beth^{*\mu}+\aleph^*_\mu \beth^\mu}{8} & \frac{\aleph_\mu \gimel^{*\mu}+\aleph^*_\mu \gimel^\mu}{8} & \frac{\aleph_\mu \daleth^{*\mu}+\aleph^*_\mu \daleth^\mu}{8} \\ \frac{\aleph_\mu \beth^{*\mu}+\aleph^*_\mu \beth^\mu}{8} & \frac{\beth_\mu \beth^{*\mu}+\beth^*_\mu \beth^\mu}{8} & \frac{\beth_\mu \gimel^{*\mu}+\beth^*_\mu \gimel^\mu}{8} & \frac{\beth_\mu \daleth^{*\mu}+\beth^*_\mu \daleth^\mu}{8} \\ \frac{\aleph_\mu \gimel^{*\mu}+\aleph^*_\mu \gimel^\mu}{8} & \frac{\gimel_\mu \beth^{*\mu}+\gimel^*_\mu \beth^\mu}{8} & \frac{\gimel_\mu \gimel^{*\mu}+\gimel^*_\mu \gimel^\mu}{8} & \frac{\gimel_\mu \daleth^{*\mu}+\gimel^*_\mu \daleth^\mu}{8} \\ \frac{\aleph_\mu \daleth^{*\mu}+\aleph^*_\mu \daleth^\mu}{8} & \frac{\daleth_\mu \beth^{*\mu}+\daleth^*_\mu \beth^\mu}{8} & \frac{\daleth_\mu \gimel^{*\mu}+\daleth^*_\mu \gimel^\mu}{8} & \frac{\daleth_\mu \daleth^{*\mu}+\daleth^*_\mu \daleth^\mu}{8} \end{vmatrix}^{\frac{1}{4}} D(4)^{-\frac{1}{4}}\sqrt{-g} \tag{65}$$

The determinant of two Reeb class vectors can help to understand the roots in (30), (31), (32), and (33). It describes accelerations in two perpendicular planes. Three Reeb class vectors describe accelerations in the foliation perpendicular to $P_\mu$. It is not clear whether (65) is related to (3.12) in which case it does not represent a new field but a massive gravitational field.

**Appendix D: Another way to derive the Reeb class vector**

We may now write the Lie derivative [45] of $\frac{P_i}{\sqrt{Z}}$ with respect to the vector field $\frac{P^{*m}}{\sqrt{Z}}$,

$$Lie\left(\frac{P^{*m}}{\sqrt{Z}}, \frac{P_i}{\sqrt{Z}}\right) = \frac{P^{*m}}{\sqrt{Z}}\left(\frac{P_i}{\sqrt{Z}}\right)_{,m} + \left(\frac{P^{*m}}{\sqrt{Z}}\right)_{,i}\frac{P_m}{\sqrt{Z}} \tag{66}$$

In which the second term is positive because the differentiated $\frac{P_i}{\sqrt{Z}}$ vector has a low index.

The first term becomes,

$$\frac{P^{*m}}{\sqrt{Z}}\left(\frac{P_i}{\sqrt{Z}}\right)_{,m} = \frac{P^{*m}P_{i,m}}{Z} - \frac{P^{*m}}{\sqrt{Z}}\frac{P_i Z_m}{2Z^{3/2}} = \frac{P^{*m}P_{i,m}}{Z} - \frac{P^{*m}Z_m P_i}{2Z^2} \tag{67}$$

The second term is,

$$\left(\frac{P^{*m}}{\sqrt{Z}}\right)_{,i}\frac{P_m}{\sqrt{Z}} = \frac{P^{*m}{}_{,i}P_m}{Z} - \frac{P^{*m}P_m Z_i}{2Z^2} = \frac{P^{*m}{}_{,i}P_m}{Z} - \frac{Z_i}{2Z} \tag{68}$$

We add (67) and (68) to get (66) and notice that $\frac{P^{*m}P_{i,m}}{Z} + \frac{P^{*m}{}_{,i}P_m}{Z} = \frac{P^{*m}P_{m,i}}{Z} + \frac{P^{*m}{}_{,i}P_m}{Z} = \frac{Z_i}{Z}$ from which (66) becomes

$$Lie\left(\frac{P^{*m}}{\sqrt{Z}}, \frac{P_i}{\sqrt{Z}}\right) = \frac{Z_i}{Z} - \frac{Z_i}{2Z} - \frac{P^{*m}Z_m P_i}{2Z^2} = \frac{Z_i}{2Z} - \frac{P^{*m}Z_m P_i}{2Z^2} = \frac{U_i}{2} \tag{69}$$

**Appendix E: 95/96, the precursor of the inverse Fine Structure Constant and of the muon/electron mass ratio**

Results (24), (36), (40), (41), (42), were not reached immediately. There was one finding that was pure serendipity that later led to these results. The observation was the following, given a scaling factor 1+d of area addition with d=1 as a maximal value, 1+d = 2.

$$(1+\alpha)^{95} < 2 \ \wedge (1+\alpha)^{96} > 2 \tag{70}$$

Where α is the Fine Structure Constant. More precisely

$$\aleph = (2^{\frac{1}{96}} - 1)^{-1} \cong 137.999325615 \tag{71}$$

And

$$\beth = (2^{\frac{1}{95}} - 1)^{-1} \cong 136.5566369 \tag{72}$$

And the geometric average is:

$$\sqrt{\aleph\beth} \cong 137.27608605 \tag{73}$$

Which is close to the result from (40), 137.0359990368270076.

An immediate observation is

$$\aleph = \left(\frac{2 - 2^{\frac{95}{96}}}{2^{\frac{95}{96}}}\right)^{-1} \tag{74}$$

And

$$\beth = \left(\frac{2^{\frac{96}{95}} - 2}{2}\right)^{-1} \tag{75}$$

Where we expressed a power which is close to 1, namely $\xi = \frac{95}{96}$ and $\xi^{-1} = \frac{96}{95}$, as such, $\xi$ was nominated as polynomial coefficient because it was in the range between 0 and 2, unlike $\xi = \frac{4}{\pi}$ which has a geometric interpretation thanks to Ettore Majorana, $\xi = \frac{95}{96}$ seems to have an algebraic meaning.

We continue with a rather surprising relation

$$(2^{\frac{1}{95*96}} - 1)^{-1} \cong 13{,}156.87877924 \tag{76}$$

And it is quite easy to notice the following:

$$\frac{1}{96(1+96^{-2})}(2^{\frac{1}{95*96}} - 1)^{-1} \cong 137.03595126474 \tag{77}$$

which is very close to the inverse Fine Structure Constant. Actually, if we replace the factor $\frac{1}{96(1+96^{-2})}$ by $\frac{1}{n(1+n^{-2})}$ for some integer n, the closest result to the inverse Fine Structure Constant is when n=96

In fact

$$\frac{(2^{\frac{1}{95*96}}-1)^{-1}}{137.0359990368270076} \cong 96.010383196499723 \cong 96(1 + 96.1546032^{-2}) \tag{78}$$

See (40). The factor $\frac{1}{95*96}$ can be seen as

$$\frac{1}{95*96} = \frac{95}{96} + \frac{96}{95} - 2 \tag{79}$$

The factor $95 * 96$ found expression in (41), (42) and is the final missing piece in the puzzle. It is the bridge between trigonometry and electro-gravitational polynomials (35) which resulted in: $\xi \cong 1.5561985371903483965638770314399152994155883789 06$ and $\frac{1}{2}(1 - g_2)^{-4} \cong 607276.53680068242829293 01262$, provided here with more accuracy if required for further research.

In (78) plugging in $\frac{4}{\pi}$ from (24) instead of 2 and dividing by $2 * 137.0359990368270076^2$ instead of by $137.0359990368270076$ we get another indication of a deep theoretical relation,

$$\frac{((\frac{4}{\pi})^{\frac{1}{95*96}}-1)^{-1}}{2*137.0359990368270076^2} \cong 1 + (2 * 95.974269533437)^{-1} \tag{80}$$

We now explore another approach, exponential perturbation of the field strength coefficient $\frac{95}{96}$.

This approach was not further investigated due to numerical stability issues, but the author finds it quite interesting. The field strength coefficient $\frac{95}{96}$ that appears in (23) is the lowest among 3 coefficients $\frac{95}{96}, \frac{4}{\pi}, 1.5561985371903484 \ldots$ . At first this fact was an incentive to search for a relation between the fine structure constant and perturbations around the value $\frac{95}{96}$.

We return to (23):

$$\frac{192a^2+2\frac{95}{96}a-(\frac{95}{96})^2}{192} = a^3 \; and \; \frac{192b^2-2\frac{95}{96}b-(\frac{95}{96})^2}{192} = b^3 \tag{81}$$

And to the multiplication in (23) $\frac{1}{(a-1)(1-b)} \cong 12202.888740664679.$

We look at the following exponential $\frac{n-1}{n}$ perturbation of the coefficient $\frac{95}{96}$,

$$\frac{192c^2+2\left(\frac{95}{96}\right)^{\frac{n-1}{n}}c-(\frac{95}{96})^{2\frac{n-1}{n}}}{192} = c^3 \; and \; \frac{192d^2-2\left(\frac{95}{96}\right)^{\frac{n-1}{n}}d-(\frac{95}{96})^{2\frac{n-1}{n}}}{192} = d^3 \tag{82}$$

And we check how relatively close is $(c-1)(1-d)$ to $(a-1)(1-b)$.

The calculation is:

$$Relative\ error = \frac{(c-1)(1-d)}{(a-1)(1-b)} - 1 \tag{83}$$

The strange fact is that

$\alpha^{-1} = \frac{2}{n}\left(\frac{(c-1)(1-d)}{(a-1)(1-b)} - 1\right)$ approximates the inverse fine structure constant. Not as good as (40), (41), (42) but good enough to trigger interest. The last term can be written as in (40) $\alpha^{-1} = \frac{2}{\cos(\eta)}$ for $\eta \equiv cos^{-1}(2\alpha)$. It turns out that $\alpha^{-1}$ is maximal or locally maximal at $n = 96^4 - 805$ or if $n$ is allowed to take real values,

$$n \cong 96^4 - 805.9334 \tag{84}$$

$$\alpha^{-1} \cong 137.0158482935 \tag{85}$$

Putting the terms together:

$$\frac{192a^2+2\frac{95}{96}a-(\frac{95}{96})^2}{192} = a^3 \; and \; \frac{192b^2-2\frac{95}{96}b-(\frac{95}{96})^2}{192} = b^3 \tag{86}$$

$$\frac{192c^2+2\left(\frac{95}{96}\right)^{\frac{n-1}{n}}c-(\frac{95}{96})^{2\frac{n-1}{n}}}{192} = c^3 \; and \; \frac{192d^2-2\left(\frac{95}{96}\right)^{\frac{n-1}{n}}d-(\frac{95}{96})^{2\frac{n-1}{n}}}{192} = d^3$$

$$\max_n \frac{2}{n}\left(\frac{(c-1)(1-d)}{(a-1)(1-b)} - 1\right) \cong 137.015848292861875279413652606308460235595703,$$

$$n \cong 96^4 - 805.933$$

See appendix G for the code in Python for (81)-(86). Consider the same type of perturbation of the field strength $\xi = \frac{4}{\pi}$,

$$\frac{192a^2 + 2\frac{4}{\pi}a - (\frac{4}{\pi})^2}{192} = a^3 \ and \ \frac{192b^2 - 2\frac{4}{\pi}b - (\frac{4}{\pi})^2}{192} = b^3 \tag{86.1}$$

$$\frac{192c^2 + 2\left(\frac{4}{\pi}\right)^{\frac{n-1}{n}} c - (\frac{4}{\pi})^{2\frac{n-1}{n}}}{192} = c^3 \ and \ \frac{192d^2 - 2\left(\frac{4}{\pi}\right)^{\frac{n-1}{n}} d - (\frac{4}{\pi})^{2\frac{n-1}{n}}}{192} = d^3$$

$$\max_n \frac{2}{n}\left(\frac{(c-1)(1-d)}{(a-1)(1-b)} - 1\right) \cong 136.4^{\frac{1}{2}}$$

Which is close to the square root of the inverse Fine Structure Constant with $n \cong 96^4 - 140631.4697265625$. In both cases, numerical stability issues in (86) and (86.1) made it very difficult to check how close such exponential perturbations of the field strength coefficient can be to the inverse Fine Structure Constant through the error in the polynomial roots. Numerical stability does exist up to $n = 96^3$. Before we proceed, consider the following, $\xi = (\frac{4}{\pi})^{1+\frac{1}{151.06357822765725984}}$ which is approximately $\frac{4}{\pi}(1 + \frac{1}{624.85524})$, then it is easy to check that

$$\frac{192a^2 + 2\xi a - \xi^2}{192} = a^3 \ and \ \frac{192b^2 - 2\xi b - \xi^2}{192} = b^3$$

$$\frac{192c^2 + 2\frac{2}{\xi}c - \left(\frac{2}{\xi}\right)^2}{192} = c^3 \ and \ \frac{192d^2 - 2\frac{2}{\xi}d - \left(\frac{2}{\xi}\right)^2 \frac{4}{\pi})^{2\frac{n-1}{n}}}{192} = d^3 \Rightarrow$$

$$\frac{(c-1)(1-d)}{(a-1)(1-b)} \cong 1 \tag{86.2}$$

This result is expected from $\xi = 2^{\frac{1}{2}} = \frac{2}{\xi}$ but not from a field strength so close to $\frac{4}{\pi}$. It is easy to see that from $\xi = 1.25$ to $\xi = 1.5$, (86.2) is very close to 1 within %1 but not as close as when $\xi = (\frac{4}{\pi})^{1+\frac{1}{151.06357822765725984}}$ or when trivially $\xi = 2^{\frac{1}{2}} = \frac{2}{\xi}$.

The Fine Structure Constant because of Poisson Distribution of events within radius r:

We proceed with the methods we have discussed until now. Consider the following expression,

$$f(x) = xe^{-x} \tag{87}$$

which is the Poison distribution for one event and with $\lambda = x$.

Consider the following perturbation equations in two variables in x around 1.

$$\eta = f\left(1 - \frac{1}{a}\right) = f\left(1 + \frac{1}{b}\right) \tag{88}$$

With the following condition for a wide range of $\eta > 10000$,

$$\alpha^{-2} = (-\ln(\eta) - 1)^{-1} \text{ and } 2\left(\frac{1}{b} + \frac{1}{a}\right)^{-1} \cong \alpha^{-1}2^{-\frac{1}{2}} \tag{89}$$

Then the system of equations (88), (89) approximates the Fine Structure Constant with the following approximated solution:

$$a \cong 97.2332790992 \tag{90}$$

$$b \cong 96.56660927693$$

$$\alpha^{-2} = (-\ln(\eta) - 1)^{-1} \cong 18778.86503$$

$$2\left(\frac{1}{b} + \frac{1}{a}\right)^{-1} = \alpha^{-1}2^{-\frac{1}{2}} \cong 96.89879752$$

$$\text{With } \alpha^{-1} \approx 137.03559363$$

These estimates can be greatly improved with better numerical precision than that of an Excel datasheet, however, this paper does not deal with the Causal Set interpretation of the presented theory and chooses to focus on other subjects. Also, (90) depends on the choice of $\eta$.

The Causal Set interpretations can be written as Probability(n=k) = $\frac{\xi^k e^{-\xi}}{k!}$ where k is the number of events within a sphere of some small radius r and n is the number of events if this number has the Poisson distribution.

### Appendix F: The Python code for (40) and for the remark after (40) and its output

```
import numpy as NP

def function_cubic_viete(a, b, c, d): # If all roots are real.

    # Viete's formula when all roots are real.
```

```
    b2 = NP.longdouble(b * b)

    b3 = NP.longdouble(b2 * b)

    a2 = NP.longdouble(a * a)

    a3 = a2 * a

    p = (3 * a * c - b2) / (3 * a2)

    q = (2 * b3 - 9 * a * b * c + 27 * a2 * d) / (27 * a3)

    offset = b / (3 * a)

    t1 = 2 * NP.sqrt(-p / 3) * NP.cos(NP.arccos(NP.sqrt(-3 / p) \
                                                  * (3 * q) / (2 * p)) / 3)

    t2 = 2 * NP.sqrt(-p / 3) * NP.cos(NP.arccos(NP.sqrt(-3 / p) * \
                                                  (3 * q) / (2 * p)) / 3 -
                                        NP.pi / 3)

    t3 = 2 * NP.sqrt(-p / 3) * NP.cos(NP.arccos(NP.sqrt(-3 / p) * \
                                                  (3 * q) / (2 * p)) / 3 -
                                        2 * NP.pi / 3)

    x1 = t1 - offset

    x2 = t2 - offset

    x3 = t3 - offset

    return (x1, x2, x3)

def function_fsc_polynomials(): # If all roots are real.

    fp_f, fp_a, fp_b = 1, 1, 1

    fp_start,  fp_end = 1.556, NP.pi / 2
```

```
for i in range(2000):
    # Get the biggest roots. These are the closest to 1.
    # One is above 1 and one is below 1.

    fp_f = (fp_start + fp_end) * 0.5

    fp_a, _, _ = function_cubic_viete(1, -1, -fp_f / 96,
                                      (fp_f * fp_f) / 192)

    fp_b, _, _ = function_cubic_viete(1, -1, fp_f / 96,
                                      (fp_f * fp_f) / 192)

    fp_result_middle = 1/NP.sqrt(fp_a-1) - 0.5/(1-fp_b)

    if fp_result_middle >= 0:
        fp_end = fp_f
    else:
        fp_start = fp_f

fp_s = 1/(1 - fp_b)
fp_s *= fp_s
fp_s *= fp_s * 0.5
fp_xi = fp_f

print('1/(x1-1): %.42lf\n1/(1-x2): %.42lf' %(1/(fp_a-1), 1/(1-fp_b)))
print('Xi: %.42lf\ns=0.5/(1-x2)^4: %.42lf' %(fp_f, fp_s))

fp_f = 4 / NP.pi
```

```
# Get the biggest roots. These are the closest to 1.
# One is above 1 and one is below 1.
fp_a, _, _ = function_cubic_viete(1, -1, -fp_f / 96, (fp_f * fp_f) / 192)
fp_b, _, _ = function_cubic_viete(1, -1, fp_f / 96, (fp_f * fp_f) / 192)
fp_mul = (fp_a - 1) * (1 - fp_b)

fp_inv_fsc = 2 / NP.cos( fp_xi * (1 + 1/NP.power(fp_s,1/(1+fp_mul))))

print('Inv FSC: %.42lf' %(fp_inv_fsc))

fp_p2 = fp_mul
fp_start,  fp_end = fp_mul, fp_mul + 0.00001

for i in range(2000):
    # Get the biggest roots. These are the closest to 1.
    # One is above 1 and one is below 1.

    fp_f = (fp_start + fp_end) * 0.5

    fp_result_middle = \
        fp_s * (2 - 1/(96*96*fp_f)) - NP.power(fp_s, 1/(1+fp_f))
    if fp_result_middle >= 0:
        fp_end = fp_f
    else:
        fp_start = fp_f

fp_p = 1/NP.sqrt(fp_mul)
fp_miracle_p = 1/NP.sqrt(fp_f)
```

```
    fp_relative_p_error = fp_p / (fp_p - fp_miracle_p)

    print('P: %.42lf\nMiracle P: %.48lf\nRelative error in P: %.48lf^-1'
          % (fp_p, fp_miracle_p, fp_relative_p_error))

function_fsc_polynomials()

'''
Output when run from PyCharm and Python 3.6:
1/(x1-1): 275.516908918643935066938865929841995239257812
1/(1-x2): 33.197404050235356010034593055024743080139160
Xi: 1.556198537190348396563877031439915299415588
s=0.5/(1-x2)^4: 607276.536800682428292930126190185546875000000000
Inv FSC: 137.035999036827007557803881354629993438720703
P: 96.069177214886295246287772897630929946899414
Miracle P: 96.069175812725177365791751071810722351074218750000
Relative error in P:
68515077.183215767145156860351562500000000000000000000000^-1
'''
```

## Appendix G: The Python code for (81)-(86)

```
import numpy as NP

def function_cubic_viete(a, b, c, d): # If all roots are real.

    # Viete's formula when all roots are real.
```

```
    b2 = NP.longdouble(b * b)
    b3 = NP.longdouble(b2 * b)
    a2 = NP.longdouble(a * a)
    a3 = a2 * a

    p = (3 * a * c - b2) / (3 * a2)

    q = (2 * b3 - 9 * a * b * c + 27 * a2 * d) / (27 * a3)

    offset = b / (3 * a)
    t1 = 2 * NP.sqrt(-p / 3) * NP.cos(NP.arccos(NP.sqrt(-3 / p) \
                                              * (3 * q) / (2 * p)) / 3)
    t2 = 2 * NP.sqrt(-p / 3) * NP.cos(NP.arccos(NP.sqrt(-3 / p) * \
                                              (3 * q) / (2 * p)) / 3 -
                                      NP.pi / 3)
    t3 = 2 * NP.sqrt(-p / 3) * NP.cos(NP.arccos(NP.sqrt(-3 / p) * \
                                              (3 * q) / (2 * p)) / 3 -
                                      2 * NP.pi / 3)
    x1 = t1 - offset
    x2 = t2 - offset
    x3 = t3 - offset

    return (x1, x2, x3)

def function_f_polynomials(fp_n=96*96*96*96): # If all roots are real.

    fp_f = 95/96
```

```
    fp_a, _, _ = function_cubic_viete(1, -1, -fp_f / 96,
                                      (fp_f * fp_f) / 192)

    fp_b, _, _ = function_cubic_viete(1, -1, fp_f / 96,
                                      (fp_f * fp_f) / 192)

    fp_mul1 = (fp_a - 1)*(1 - fp_b)

    fp_f = NP.power(fp_f, (fp_n-1)/fp_n)

    fp_a, _, _ = function_cubic_viete(1, -1, -fp_f / 96,
                                      (fp_f * fp_f) / 192)

    fp_b, _, _ = function_cubic_viete(1, -1, fp_f / 96,
                                          (fp_f * fp_f) / 192)

    fp_mul2 = (fp_a - 1)*(1 - fp_b)

    fp_combine = 2/(fp_n *(fp_mul2/fp_mul1-1))

    #print('%.42lf' %fp_combine)

    return fp_combine

def main():
    ma_best_val = 0
    ma_best_m = 0
```

```
#function_f_polynomials(96 * 96 * 96 * 96 - 1)

#function_f_polynomials(96 * 96 * 96 * 96)

#function_f_polynomials(96 * 96 * 96 * 96 + 1)

print('Coarse search:')

for i in range(-1000, 1000):

    ma_r =  function_f_polynomials(96 * 96 * 96 * 96 - i)

    if ma_best_val < ma_r:

        ma_best_val = ma_r

        ma_best_m = i

print('Best value %.42lf' %ma_best_val)

print('Best m = 96^4-%d' % ma_best_m)

print('Fine search:')

ma_best_val = 0.0

ma_best_m = 0.0

for i in range(8050000-10000, 8050000+10000):

    ma_d = i/10000

    ma_r = function_f_polynomials(96 * 96 * 96 * 96 - ma_d)

    if ma_best_val < ma_r:

     Fappen   ma_best_val = ma_r

        ma_best_m = ma_d

print('Best value %.42lf' %ma_best_val)

print('Best m = 96^4-%.42lf' % ma_best_m)
```

```
    '''
    Coarse search:
    Best value 137.015846787740116496934206224977970123291016
    Best m = 96^4-805
    Fine search:
    Best value 137.015848292861875279413652606308460235595703
    Best m = 96^4-805.93299999999992724042385816574096679687500
    '''
if __name__ == '__main__':
    main()
```

## Appendix H – Causality conservation theorem

**Theorem:** If p is real, any monotone function $f(p)$, called causality function will yield the same Reeb class vector. The reader is advised to check the case when p is an imaginary function. Then the Reeb class vector is defined as $\frac{u_\nu}{2} = \frac{z_\nu}{2z} - \frac{z_k}{2z^2} p *^k p_\nu$.

**Proof:**

We will use capital letters for $P = f(p)$ and as in previous pages, $z = p_\lambda p^\lambda$ and here $Z = P_\lambda P^\lambda$.

$$P = f(p)$$

$$P_\mu = f'(p) p_\mu$$

$$Z = f'(p) p_\mu f'(p) p^\mu = f'(p)^2 z$$

$$\frac{Z_\nu}{Z} = \frac{2f'(p)f''(p)p_\nu z}{f'(p)^2 z} + \frac{f'(p)^2 z_\nu}{f'(p)^2 z} = \frac{2f''(p)p_\nu}{f'(p)} + \frac{z_\nu}{z}$$

$$U_\nu = \frac{2f''(p)p_\nu}{f'(p)} + \frac{z_\nu}{z} - \left(\frac{2f''(p)p_k}{f'(p)} + \frac{z_k}{z}\right)\frac{f'(p)p^k f'(p)p_\nu}{f'(p)^2 z}$$

$$U_\nu = \frac{2f''(p)p_\nu}{f'(p)} - \frac{2f''(p)p_\nu}{f'(p)} + \frac{z_\nu}{z} - \frac{z_k}{z^2}p^k p_\nu = \frac{z_\nu}{z} - \frac{z_k}{z^2}p^k p_\nu = u_\nu$$

$$\frac{U_\nu}{2} = \frac{u_\nu}{2} \tag{91}$$